\documentclass[aps,showpacs,preprintnumbers]{revtex4}
\usepackage{feynmp,epsf,psfig}
\usepackage{amsmath}
\usepackage{supercite}
\newcommand{\nc}{\newcommand}
\nc{\Gz}{\fullg}
\nc{\Gvc}{\boldsymbol{G}^c}
\nc{\Gam}{\boldsymbol{\Gamma}}
\nc{\Sig}{\boldsymbol{\Sigma}}
\nc{\TS}{\tilde{\Sig}}
\nc{\TG}{\tilde{\mbox{\boldmath $G$}}}
\nc{\gam}{\boldsymbol{\gamma}}
\nc{\alp}{\boldsymbol{\alpha}}
\nc{\hs}{\hspace*{1mm}}
\nc{\fullg}{\boldsymbol{G}}
\nc{\scs}{\scriptstyle}
\nc{\beq}{\begin{eqnarray}}
\nc{\eeq}{\end{eqnarray}}
\nc{\la}{\label}
\nc{\no}{\nonumber}
\nc{\ci}{\cite}
\nc{\trace}{{\rm Tr\,}}
\nc{\setval}{\fmfset{wiggly_len}{1.5mm}\fmfset{arrow_len}{1.5mm}\fmfset{arrow_ang}{13}\fmfset{dash_len}{1.5mm}\fmfpen{0.125mm}\fmfset{dot_size}{1thick}}
\nc{\dphi}[3]{\frac{\delta #1}{\delta
\parbox{12mm}{\centerline{
\begin{fmfgraph*}(7,3)
\fmfpen{0.125mm}
\fmfleft{v1}
\fmfright{v2}
\fmf{plain}{v2,v1}
\fmfv{decor.size=0,label={\footnotesize #2},l.dist=0.5mm}{v1}
\fmfv{decor.size=0,label={\footnotesize #3},l.dist=0.5mm}{v2}
\end{fmfgraph*}
}}}}
\nc{\ddphi}[1]{\frac{\delta^2 #1}{\delta
\parbox{12mm}{\centerline{
\begin{fmfgraph*}(7,3)
\fmfset{arrow_len}{1.5mm}
\fmfpen{0.125mm}
\fmfleft{v1}
\fmfright{v2}
\fmf{plain}{v2,v1}
\fmfv{decor.size=0,label={\footnotesize 1},l.dist=0.5mm}{v1}
\fmfv{decor.size=0,label={\footnotesize 2},l.dist=0.5mm}{v2}
\end{fmfgraph*}
}}\,\delta
\parbox{12mm}{\centerline{
\begin{fmfgraph*}(7,3)
\fmfpen{0.125mm}
\fmfleft{v1}
\fmfright{v2}
\fmf{plain}{v2,v1}
\fmfv{decor.size=0,label={\footnotesize 3},l.dist=0.5mm}{v1}
\fmfv{decor.size=0,label={\footnotesize 4},l.dist=0.5mm}{v2}
\end{fmfgraph*}
}}}}
\begin{document}
\setlength{\unitlength}{1mm}
\begin{fmffile}{graph1}
\title{Recursive Graphical Solution of Closed Schwinger-Dyson Equations in $\phi^4$-Theory --
Part1: Generation of Connected and One-Particle Irreducible Feynman Diagrams}
\author{Axel Pelster and Konstantin Glaum}
\affiliation{Institut f\"ur Theoretische Physik, Freie Universit\"at Berlin, 
Arnimallee 14, 14195 Berlin, Germany\\
{\tt pelster@physik.fu-berlin.de,glaum@physik.fu-berlin.de} }
\date{\today}
\begin{abstract}
Using functional derivatives with respect to the free correlation function we derive a closed set of
Schwinger-Dyson equations in $\phi^4$-theory. Its conversion to graphical recursion relations allows us to
systematically generate all connected and one-particle irreducible Feynman diagrams for the two- and four-point
function together with their weights. 
\end{abstract}
\maketitle
\section{Introduction}
Quantum and statistical field theory investigate the influence of field fluctuations on the $n$-point functions.
Interactions lead to an infinite hierarchy of Schwinger-Dyson equations for the $n$-point functions
\ci{Drell,Amit,Bellac,Zuber,Zinn,Peskin}. These integral equations can only be closed approximately,
for instance, by the well-established the self-consistent method of Kadanoff and Baym \ci{Baym}.\\

Recently, it has been shown that the Schwinger-Dyson equations of QED can be closed in a certain functional-analytic
sense \ci{SDQED1}. Using functional derivatives with respect to the free propagators and the interaction 
\ci{SDQED1,QED,SYM,ASYM1,ASYM2,GINZ,Devreese}  
two closed sets of equations were derived.
The first one involves the connected electron and two-point function
as well as the connected three-point function, whereas the second one determines 
the electron and photon self-energy as well as the one-particle irreducible
three-point function. Their conversion to graphical recursion relations leads to a systematic graphical generation of all 
connected and one-particle irreducible Feynman diagrams in QED, respectively.\\

The purpose of the present paper is to apply this functional-analytic approach to the $\phi^4$-theory of second-order phase
transitions in the disordered, symmetric phase. A short ouline of this program was already published in  Ref. \cite{FEST}.
To this end we derive in Section \ref{PHI} a closed set of equations for the
connected two- and four-point function. Analogously, we determine in Section \ref{IRRED} a closed set of Schwinger-Dyson equations
for the self-energy and the one-particle irreducible four-point function. In both cases, 
the closed set of Schwinger-Dyson
equations can be converted into graphical recursion relations for the 
corresponding connected and one-particle irreducible Feynman diagrams in $\phi^4$-theory. 
From these the respective connected vacuum diagrams follow by short-circuiting external legs. 
Thus our present approach is complementary to Ref. \ci{SYM} which was  
based on the observation that the complete knowledge of the vacuum energy implies the
knowledge of the entire theory (``the vacuum is the world'') \cite{Streater,Schwinger}.
In that paper the vacuum diagrams were constructed in a first step, together
with their weights, as solutions of a graphical recursion relation derived from a nonlinear functional differential equation.
In a second step, all diagrams with external lines are obtained from functional derivatives of the connected vacuum diagrams
with respect to the free correlation function.
\section{Scalar $\phi^4$-Theory} \la{PHI}
Euclidean $\phi^4$-theories in $d$ dimensions are useful models for
a large family of universality classes of continuous phase transitions \ci{Verena}. In particular,
the O$(N)$-symmetric $\phi^4$-theory serves to describe
the critical phenomena in dilute polymer solutions ($N=0$), Ising- and
Heisenberg-like magnets ($N=1,3$), and superfluids ($N=2$). In all these
systems, the thermal fluctuations of a self-interacting scalar order parameter field $\phi$ with $N$ components
are controlled by the Ginzburg-Landau energy functional 
\beq
\la{GL}
E[\phi] =  \int d^d x \left\{ \frac{1}{2} \,\sum_{\alpha=1}^N
\phi_{\alpha} (x)\, (-\partial^2_x + m^2) \, \phi_{\alpha} (x) + \frac{g}{24} 
\left[ \sum_{\alpha=1}^N \phi^2_{\alpha} (x)\right]^2  \right\}  \, ,
\eeq
where the mass $m^2$ is proportional to the temperature deviation
from the critical point, and $g$ denotes the coupling constant. In the following it turns out to be advantageous to rewrite the
Ginzburg-Landau energy functional (\ref{GL}) as
\beq
\la{EF}
E [ \phi ] = \frac{1}{2} \int_{12} G^{-1}_{12} \phi_1 \phi_2 
+ \frac{1}{24} \int_{1234} V_{1234}  \phi_1 \phi_2 \phi_3 \phi_4 \,.
\eeq
In this short-hand notation, the spatial and tensorial arguments of the order parameter
field $\phi$, the bilocal kernel $G^{-1}$, and the quartic interaction $V$ are indicated by simple number indices, i.e.,
\beq
1 \equiv \{ x_1 , \alpha_1 \} \, , \hspace*{0.5cm}
\int_1 \equiv \sum_{\alpha_1=1}^N \int d^d x_1 \, , \hspace*{0.5cm}
\phi_1 \equiv \phi_{\alpha_1} ( x_1 ) \, .
\eeq
The kernel $G^{-1}$ represents the functional matrix
\beq
\la{PH1} 
G^{-1}_{12} \equiv G_{\alpha_1 , \alpha_2}^{-1} ( x_1 , x_2 )  = 
\delta_{\alpha_1 , \alpha_2} \, \left( - \partial_{x_1}^2 + m^2 
\right) \delta ( x_1 - x_2 ) \, , 
\eeq
while the interaction $V$ is given by the functional tensor 
\beq
\la{PH2}
V_{1234} \equiv V_{\alpha_1,\alpha_2,\alpha_3,\alpha_4} ( x_1 , x_2 , x_3 , x_4 ) = \frac{g}{3} \left(
\delta_{\alpha_1 , \alpha_2} \delta_{\alpha_3 , \alpha_4} +
\delta_{\alpha_1 , \alpha_3} \delta_{\alpha_2 , \alpha_4} +
\delta_{\alpha_1 , \alpha_4} \delta_{\alpha_2 , \alpha_3} \right) \,
\delta ( x_1 - x_2 ) \delta ( x_1 - x_3 ) \delta ( x_1 - x_4 ) \, ,
\eeq
both being symmetric in their indices. 
For the purpose of this paper we shall leave the kernel $G^{-1}$ in the energy functional
(\ref{EF}) completely general, 
except for the symmetry with respect to its indices. By doing so, we regard the energy functional (\ref{EF})
as a functional of the kernel $G^{-1}$:
\beq
\la{EFF}
E[\phi] = E[\phi,G^{-1}] \, .
\eeq
As a consequence, all global and local statistical quantities derived from (\ref{EFF})
are also functionals of the bilocal kernel $G^{-1}$. In particular, we are interested in studying the functional dependence of  
the partition function, defined by a functional integral over a Boltzmann weight in natural units
\beq
\la{PF}
Z [G^{-1}]= \int {\cal D} {\bf \phi} \, e^{- E [ {\bf \phi},G^{-1} ]} \, ,
\eeq
and the (negative) vacuum energy
\beq
\la{VAS}
W [G^{-1}]= \ln Z [G^{-1}] \, .
\eeq
For the sake of simplicity we restrict ourselves in the present paper to study the disordered, symmetric phase 
of the $\phi^4$-theory where the $n$-point functions 
\beq
\la{NPO}
\fullg_{1 \cdots n} [G^{-1}] = \frac{1}{Z[G^{-1}]} \int {\cal D} {\bf \phi} \, 
\phi_1 \cdots \phi_n \, e^{- E [ {\bf \phi} ,G^{-1}]} \, ,
\eeq
with odd $n$ vanish. Thus the first nonvanishing $n$-point functions are the two-point function
\beq
\la{2P}
\fullg_{12}  [G^{-1}]= \frac{1}{Z[G^{-1}]}\,
\int {\cal D} \phi \, \phi_1 \phi_2 \, e^{- E [ \phi , G^{-1}]} 
\eeq
and the four-point function
\beq
\la{4P}
\fullg_{1234}  [G^{-1}]= \frac{1}{Z[G^{-1}]}\,
\int {\cal D} \phi \, \phi_1 \phi_2 \phi_3 \phi_4 \, e^{- E [ \phi , G^{-1}]} \, .
\eeq
Further important statistical quantities are the correlation functions, i.e. the connected $n$-point functions.
In the disordered, symmetric phase, the connected two-point function coincides with the two-point function
\beq
\fullg_{12}^{\rm c}  [G^{-1}] = \fullg_{12}  [G^{-1}] \, ,
\eeq
whereas the connected four-point function is defined by
\beq
\la{C4P}
\fullg_{1234}^{\rm c}[G^{-1}]
\equiv \fullg_{1234} [G^{-1}]- \fullg_{12} [G^{-1}]\fullg_{34}[G^{-1}]
- \fullg_{13} [G^{-1}]\fullg_{24}[G^{-1}]- \fullg_{14}[G^{-1}] \fullg_{23}[G^{-1}] \, .
\eeq
By expanding the functional integrals (\ref{PF}) and (\ref{NPO}) in powers of the coupling constant $g$, the expansion
coefficients of the partition function and the $n$-point functions consist of free-field expectation values. These
are evaluated with the help of Wick's rule as a sum of Feynman integrals, which are pictured as diagrams constructed from
lines and vertices. Thereby the free correlation function $G_{12}$, which is the functional inverse of the kernel $G^{-1}$ 
in the energy functional (\ref{EF})
\beq
\la{FP}
\int_{2} G_{12} \, G^{-1}_{23} = \delta_{13} 
\eeq
with $\delta_{13} = \delta_{\alpha_1,\alpha_3} \delta ( x_1 - x_3)$, is represented by a  line in a Feynman diagram  
\beq
\la{PRO}
G_{12} \hspace*{2mm}  \equiv \hspace*{2mm} 
\parbox{20mm}{\centerline{
\begin{fmfgraph*}(8,3)
\setval
\fmfleft{v1}
\fmfright{v2}
\fmf{plain,width=0.2mm}{v2,v1}
\fmfv{decor.size=0, label=${\scs 1}$, l.dist=1mm, l.angle=-180}{v1}
\fmfv{decor.size=0, label=${\scs 2}$, l.dist=1mm, l.angle=0}{v2}
\end{fmfgraph*} }}  
\, ,
\eeq
and the interaction $V$ is pictured as a vertex 
\beq
\la{VE}
 - \hspace*{2mm} V_{1234} \hspace*{2mm}  \equiv  \hspace*{2mm}  
 \parbox{7mm}{\centerline{
  \begin{fmfgraph*}(4.5,4.5)
  \setval
  \fmfforce{0w,0h}{v1}
  \fmfforce{0w,1h}{v2}
  \fmfforce{1w,1h}{v3}
  \fmfforce{1w,0h}{v4}
  \fmfforce{1/2w,1/2h}{v5}
  \fmf{plain}{v1,v3}
  \fmf{plain}{v2,v4}
  \fmfv{decor.size=0, label=${\scs 1}$, l.dist=1mm, l.angle=-135}{v1}
  \fmfv{decor.size=0, label=${\scs 2}$, l.dist=1mm, l.angle=135}{v2}
  \fmfv{decor.size=0, label=${\scs 3}$, l.dist=1mm, l.angle=45}{v3}
  \fmfv{decor.size=0, label=${\scs 4}$, l.dist=1mm, l.angle=-45}{v4}
  \fmfdot{v5}
  \end{fmfgraph*} } } 
\hspace*{5mm} . \\ \no
\eeq
The graphical elements (\ref{PRO}) and (\ref{VE}) are combined by an integral which graphically corresponds to the 
gluing prescription
\beq
  - \int_4 V_{1234} \hs G_{45}
  \hspace*{3mm} \equiv  \hspace*{3mm}
  \parbox{8mm}{\centerline{
  \begin{fmfgraph*}(4.5,4.5)
  \setval
  \fmfforce{0w,0h}{v1}
  \fmfforce{0w,1h}{v2}
  \fmfforce{1w,1h}{v3}
  \fmfforce{1w,0h}{v4}
  \fmfforce{1/2w,1/2h}{v5}
  \fmf{plain}{v1,v3}
  \fmf{plain}{v2,v4}
  \fmfv{decor.size=0, label=${\scs 1}$, l.dist=1mm, l.angle=-135}{v1}
  \fmfv{decor.size=0, label=${\scs 2}$, l.dist=1mm, l.angle=135}{v2}
  \fmfv{decor.size=0, label=${\scs 3}$, l.dist=1mm, l.angle=45}{v3}
  \fmfv{decor.size=0, label=${\scs 4}$, l.dist=1mm, l.angle=-45}{v4}
  \fmfdot{v5}
  \end{fmfgraph*} } } 
  \hspace*{3mm}
  \parbox{9mm}{\centerline{
  \begin{fmfgraph*}(6,7.5)
  \setval
  \fmfforce{0w,0h}{v1}
  \fmfforce{1w,0h}{v2}
  \fmf{plain,width=0.2mm}{v2,v1}
  \fmfv{decor,size=0, label=${\scs 4}$, l.dist=1mm, l.angle=-180}{v1}
  \fmfv{decor,size=0, label=${\scs 5}$, l.dist=1mm, l.angle=0}{v2}
  \end{fmfgraph*} } } 
  \hspace*{3mm} \equiv \hspace*{3mm}  
  \parbox{8mm}{\centerline{
  \begin{fmfgraph*}(4.5,4.5)
  \setval
  \fmfforce{0w,0h}{v1}
  \fmfforce{0w,1h}{v2}
  \fmfforce{1w,1h}{v3}
  \fmfforce{8/4.5w,-1/4.5h}{v4}
  \fmfforce{1/2w,1/2h}{v5}
  \fmf{plain}{v1,v3}
  \fmf{plain}{v2,v5}
  \fmf{plain,right=0.15}{v5,v4}
  \fmfv{decor.size=0, label=${\scs 1}$, l.dist=1mm, l.angle=-135}{v1}
  \fmfv{decor.size=0, label=${\scs 2}$, l.dist=1mm, l.angle=135}{v2}
  \fmfv{decor.size=0, label=${\scs 3}$, l.dist=1mm, l.angle=45}{v3}
  \fmfv{decor.size=0, label=${\scs 5}$, l.dist=1mm, l.angle=-20}{v4}
  \fmfdot{v5}
  \end{fmfgraph*} } } \hspace*{5mm} .
\label{5}
\eeq
In this paper we analyze the resulting diagrams for the statistical quantities (\ref{PF})--(\ref{C4P}) by using their functional
dependence on the kernel $G^{-1}_{12}$. To this end we
introduce the functional derivative with respect to the kernel $G^{-1}_{12}$ whose
basic rule reflects the symmetry of its indices:
\beq
\la{DR1}
\frac{\delta G^{-1}_{12}}{\delta G^{-1}_{34}} = \frac{1}{2} \left( 
\delta_{13} \delta_{42} + \delta_{14} \delta_{32} \right) \, .
\eeq
From the identity (\ref{FP}) and the functional product rule
we find the effect of this derivative on the free correlation function
\beq
\la{ACT}
-  \frac{\delta G_{12}}{\delta G^{-1}_{34}} = \frac{1}{2} \left(
G_{13} G_{42} + G_{14} G_{32} \right) \, ,
\eeq
which has the graphical representation
\beq
-  \frac{\delta}{\delta G^{-1}_{34}}
\parbox{15mm}{\centerline{
\begin{fmfgraph*}(8,3)
\setval
\fmfleft{v1}
\fmfright{v2}
\fmf{plain,width=0.2mm}{v2,v1}
\fmfv{decor.size=0, label=${\scs 1}$, l.dist=1mm, l.angle=-180}{v1}
\fmfv{decor.size=0, label=${\scs 2}$, l.dist=1mm, l.angle=0}{v2}
\end{fmfgraph*} }} 
= \frac{1}{2} \,\, \bigg(
\parbox{15mm}{\centerline{
\begin{fmfgraph*}(8,3)
\setval
\fmfleft{v1}
\fmfright{v2}
\fmf{plain,width=0.2mm}{v2,v1}
\fmfv{decor.size=0, label=${\scs 1}$, l.dist=1mm, l.angle=-180}{v1}
\fmfv{decor.size=0, label=${\scs 3}$, l.dist=1mm, l.angle=0}{v2}
\end{fmfgraph*}}}
\parbox{15mm}{\centerline{
\begin{fmfgraph*}(8,3)
\setval
\fmfleft{v1}
\fmfright{v2}
\fmf{plain,width=0.2mm}{v2,v1}
\fmfv{decor.size=0, label=${\scs 4}$, l.dist=1mm, l.angle=-180}{v1}
\fmfv{decor.size=0, label=${\scs 2}$, l.dist=1mm, l.angle=0}{v2}
\end{fmfgraph*}}}
+
\parbox{15mm}{\centerline{
\begin{fmfgraph*}(8,3)
\setval
\fmfleft{v1}
\fmfright{v2}
\fmf{plain,width=0.2mm}{v2,v1}
\fmfv{decor.size=0, label=${\scs 1}$, l.dist=1mm, l.angle=-180}{v1}
\fmfv{decor.size=0, label=${\scs 4}$, l.dist=1mm, l.angle=0}{v2}
\end{fmfgraph*}}}
\parbox{15mm}{\centerline{
\begin{fmfgraph*}(8,3)
\setval
\fmfleft{v1}
\fmfright{v2}
\fmf{plain,width=0.2mm}{v2,v1}
\fmfv{decor.size=0, label=${\scs 3}$, l.dist=1mm, l.angle=-180}{v1}
\fmfv{decor.size=0, label=${\scs 2}$, l.dist=1mm, l.angle=0}{v2}
\end{fmfgraph*}}}
\bigg) \, .
\eeq
Thus a functional derivative with respect to the kernel $G^{-1}_{12}$ is represented by a graphical operation which
cuts a line of a Feynman diagram in all possible ways \ci{SDQED1,QED,SYM,ASYM1,ASYM2,GINZ,Devreese}. 
For practical purposes it is convenient to use also
functional derivatives with respect to
the free correlation function $G_{12}$ whose basic rule reads
\beq
\la{DR2}
\frac{\delta G_{12}}{\delta G_{34}} \equiv  
\dphi{ 
  \parbox{12mm}{\centerline{
  \begin{fmfgraph*}(7,3)
  \setval
  \fmfforce{0w,1/2h}{v1}
  \fmfforce{1w,1/2h}{v2}
  \fmf{plain,width=0.2mm}{v2,v1}
  \fmfv{decor,size=0, label=${\scs 1}$, l.dist=0.5mm, l.angle=-180}{v1}
  \fmfv{decor,size=0, label=${\scs 2}$, l.dist=0.5mm, l.angle=0}{v2}
  \end{fmfgraph*} } } 
}{3}{4}
= \frac{1}{2} \left( 
\delta_{13} \delta_{42} + \delta_{14} \delta_{32} \right) \, .
\eeq
Such functional derivatives are represented graphically by removing a line
of a Feynman diagram in all possible ways  \ci{SDQED1,QED,SYM,ASYM1,ASYM2,GINZ,Devreese}. 
The functional derivatives with respect to the kernel $G^{-1}_{12}$ and the correlation function $G_{12}$  are related via
the functional chain rule
\beq
\frac{\delta}{\delta G^{-1}_{12}} \hs = - \hs \int_{34} G_{13} G_{24}
\hs \frac{\delta}{\delta G_{34}}  \hspace*{3mm} .
\label{13}
\eeq
These functional derivatives are used in Subsection \ref{CSE}
to derive a closed set of Schwinger-Dyson equations for the connected two- and four-point functions. In
Subsection \ref{GRC} they are converted into graphical recursion relations for the corresponding connected Feynman diagrams.
Finally, the connected vacuum diagrams contributing to the vacuum energy are constructed in a graphical way in 
Subsection \ref{VAA}.
\subsection{Closed Set of Equations for Connected Two- and Four-Point Function}\la{CSE}
In this subsection we apply the functional derivatives introduced so far to a
functional identity which immediately follows from the definition of the
functional integral. By doing so, we derive a closed set of Schwinger-Dyson equations
determining the connected two- and four-point function.
\subsubsection{Connected Two-Point Function}
In order to derive an equation for the connected two-point function, we start with the trivial identity
\beq
\int {\cal D} \phi \, \frac{\delta}{\delta \phi_1} \left( \phi_2 
\, e^{- E [ \phi ]} \right) = 0 \, ,
\eeq
which follows via direct functional integration from the vanishing of 
the exponential at infinite fields. Taking
into account the explicit form of
the energy functional (\ref{EF}), we perform
the functional derivative with respect to the field
and obtain
\beq
\la{FD1}
\int {\cal D} \phi \left( \delta_{12} 
- \int_3 G^{-1}_{13} \phi_2 \phi_3
- \frac{1}{6} \int_{345} V_{1345} \, \phi_2 \phi_3 \phi_4 \phi_5 
\right) e^{- E [ \phi ]} = 0 \, .
\eeq
Applying the definitions (\ref{PF})--(\ref{4P}) and (\ref{C4P}), this
equation can be expressed in terms of the connected two- and four-point 
function as follows\\
\beq
\delta_{12} - \int_3 G^{-1}_{13} \fullg_{23} = \frac{1}{2}
\int_{345} V_{1345} \fullg_{34} \fullg_{52} + \frac{1}{6} \int_{345}
V_{1345} \fullg_{2345}^{\rm c} \, .
\la{Idy}
\eeq
Multiplying this equation with $G_{16}$ and integrating with respect to the 
index $1$ finally leads to the Schwinger-Dyson equation which
determines the connected two-point function:
\beq
\la{SD1}
\fullg_{12} = G_{12} - \frac{1}{2}
\int_{3456} G_{13} V_{3456} \fullg_{45} \fullg_{62} - \frac{1}{6} \int_{3456}
G_{13} V_{3456} \fullg_{2456}^{\rm c} \, .
\eeq
When the connected two-point function is graphically represented in Feynman diagrams by a double line 
\beq
\label{FS}
\fullg_{12} \hspace*{2mm} \equiv \hspace*{0.5mm}
\parbox{20mm}{\centerline{
\begin{fmfgraph*}(8,3)
\setval
\fmfleft{v1}
\fmfright{v2}
\fmf{double,width=0.2mm}{v2,v1}
\fmfv{decor.size=0, label=${\scs 1}$, l.dist=1mm, l.angle=-180}{v1}
\fmfv{decor.size=0, label=${\scs 2}$, l.dist=1mm, l.angle=0}{v2}
\end{fmfgraph*}}}  
\hspace*{1mm} \, ,
\eeq
and the connected four-point function is pictured by a vertex with an open dot with four legs\\
\beq
\fullg_{1234}^{\rm c} \hspace*{2mm} \equiv \hspace*{2mm}
\parbox{12mm}{\centerline{
  \begin{fmfgraph*}(9,9)
  \setval
  \fmfforce{0w,0h}{v1}
  \fmfforce{0w,1h}{v2}
  \fmfforce{1w,1h}{v3}
  \fmfforce{1w,0h}{v4}
  \fmfforce{3.5/9w,3.5/9h}{v5}
  \fmfforce{3.5/9w,5.5/9h}{v6}
  \fmfforce{5.5/9w,5.5/9h}{v7}
  \fmfforce{5.5/9w,3.5/9h}{v8}
  \fmf{plain}{v1,v5}
  \fmf{plain}{v2,v6}
  \fmf{plain}{v3,v7}
  \fmf{plain}{v4,v8}
  \fmf{plain,width=0.2mm,left=1}{v5,v7,v5}
  \fmfv{decor.size=0, label=${\scs 1}$, l.dist=1mm, l.angle=-135}{v1}
  \fmfv{decor.size=0, label=${\scs 2}$, l.dist=1mm, l.angle=135}{v2}
  \fmfv{decor.size=0, label=${\scs 3}$, l.dist=1mm, l.angle=45}{v3}
  \fmfv{decor.size=0, label=${\scs 4}$, l.dist=1mm, l.angle=-45}{v4}
  \end{fmfgraph*} } }
\hspace*{5mm} , \\ \no
\eeq
this Schwinger-Dyson equation reads graphically:
\beq
\la{SDG1}
\parbox{10mm}{\centerline{
\begin{fmfgraph*}(7,3)
\setval
\fmfleft{v1}
\fmfright{v2}
\fmf{double,width=0.2mm}{v2,v1}
\fmfv{decor.size=0, label=${\scs 1}$, l.dist=1mm, l.angle=-180}{v1}
\fmfv{decor.size=0, label=${\scs 2}$, l.dist=1mm, l.angle=0}{v2}
\end{fmfgraph*}}}
\hspace*{3mm} = \hspace*{3mm}
\parbox{10mm}{\centerline{
\begin{fmfgraph*}(7,3)
\setval
\fmfleft{v1}
\fmfright{v2}
\fmf{plain}{v2,v1}
\fmfv{decor.size=0, label=${\scs 1}$, l.dist=1mm, l.angle=-180}{v1}
\fmfv{decor.size=0, label=${\scs 2}$, l.dist=1mm, l.angle=0}{v2}
\end{fmfgraph*}}}
\hspace*{3mm} + \frac{1}{2} \hspace*{3mm} 
\parbox{13mm}{\centerline{
\begin{fmfgraph*}(10,5)
\setval
\fmfforce{0w,0h}{v1}
\fmfforce{1/2w,0h}{v2}
\fmfforce{1w,0h}{v3}
\fmfforce{1/4w,1/2h}{v4}
\fmfforce{3/4w,1/2h}{v5}
\fmf{plain}{v2,v1}
\fmf{double,width=0.2mm}{v2,v3}
\fmf{plain}{v2,v1}
\fmf{double,width=0.2mm,left=1}{v4,v5,v4}
\fmfv{decor.size=0, label=${\scs 1}$, l.dist=1mm, l.angle=-180}{v1}
\fmfv{decor.size=0, label=${\scs 2}$, l.dist=1mm, l.angle=0}{v3}
\fmfdot{v2}
\end{fmfgraph*}}}
\hspace*{0.3cm} + \frac{1}{6} \hspace*{0.3cm}
\parbox{22mm}{\centerline{
\begin{fmfgraph*}(18,6)
\setval
\fmfforce{0w,1/2h}{v1}
\fmfforce{5/18w,1/2h}{v2}
\fmfforce{10/18w,1/2h}{v3}
\fmfforce{13/18w,1/2h}{v4}
\fmfforce{1w,1/2h}{v5}
\fmfforce{11.2/18w,2.2/3h}{v6}
\fmfforce{11.2/18w,0.8/3h}{v7}
\fmf{plain,width=0.2mm,left=1}{v3,v4,v3}
\fmf{plain,left=0.8}{v2,v6}
\fmf{plain,right=0.8}{v2,v7}
\fmf{plain}{v2,v3}
\fmf{plain}{v5,v4}
\fmf{plain}{v2,v1}
\fmfv{decor.size=0, label=${\scs 1}$, l.dist=1mm, l.angle=-180}{v1}
\fmfv{decor.size=0, label=${\scs 2}$, l.dist=1mm, l.angle=0}{v5}
\fmfdot{v2}
\end{fmfgraph*} }} 
\hspace*{3mm} .
\eeq
It represents an integral equation for the connected two-point function on the left-hand side which turns out to
appear also on the right-hand side. Iteratively solving the integral equation (\ref{SDG1}) for the connected 
two-point function $\fullg_{12}$ necessitates, however, the knowledge of the
connected four-point function $\fullg_{1234}^{\rm c}$. 
\subsubsection{Connected Four-Point Function}
In principle, one could determine the connected four-point function (\ref{C4P}) from the connected two-point function (\ref{2P})
as follows. We obtain from (\ref{EF}) and (\ref{DR1})
\beq
\label{SR}
\phi_1 \phi_2 = 2 \, \frac{\delta E [ \phi ]}{\delta G^{-1}_{12}} \, ,
\eeq
so that we yield from (\ref{PF}), (\ref{4P}), and (\ref{C4P})
\beq
\label{GDE}
\fullg_{1234}^{\rm c}
= - 2 \, \frac{\delta \fullg_{12}}{\delta G_{34}^{-1}} 
- \fullg_{13} \fullg_{24}- \fullg_{14} \fullg_{23} \, .
\eeq
This result reads graphically

\beq
  \parbox{12mm}{\centerline{
  \begin{fmfgraph*}(9,9)
  \setval
  \fmfforce{0w,0h}{v1}
  \fmfforce{0w,1h}{v2}
  \fmfforce{1w,1h}{v3}
  \fmfforce{1w,0h}{v4}
  \fmfforce{3.5/9w,3.5/9h}{v5}
  \fmfforce{3.5/9w,5.5/9h}{v6}
  \fmfforce{5.5/9w,5.5/9h}{v7}
  \fmfforce{5.5/9w,3.5/9h}{v8}
  \fmf{plain}{v1,v5}
  \fmf{plain}{v2,v6}
  \fmf{plain}{v3,v7}
  \fmf{plain}{v4,v8}
  \fmf{plain,width=0.2mm,left=1}{v5,v7,v5}
  \fmfv{decor.size=0, label=${\scs 1}$, l.dist=1mm, l.angle=-135}{v1}
  \fmfv{decor.size=0, label=${\scs 2}$, l.dist=1mm, l.angle=135}{v2}
  \fmfv{decor.size=0, label=${\scs 3}$, l.dist=1mm, l.angle=45}{v3}
  \fmfv{decor.size=0, label=${\scs 4}$, l.dist=1mm, l.angle=-45}{v4}
  \end{fmfgraph*} } }
  \hspace*{2mm} = \hspace*{3mm} - \, 2 \hspace*{3mm} 
  \frac{\delta \hs
  \parbox{10mm}{\centerline{
  \begin{fmfgraph*}(7,3)
  \setval
  \fmfforce{0w,1/2h}{v1}
  \fmfforce{1w,1/2h}{v2}
  \fmf{double,width=0.2mm}{v2,v1}
  \fmfv{decor,size=0, label=${\scs 1}$, l.dist=0.5mm, l.angle=-180}{v1}
  \fmfv{decor,size=0, label=${\scs 2}$, l.dist=0.5mm, l.angle=0}{v2}
  \end{fmfgraph*} } }
  }{\delta G^{-1}_{34}} 
\hspace*{3mm} - \hspace*{3mm}
  \parbox{10mm}{\centerline{
  \begin{fmfgraph*}(7,3)
  \setval
  \fmfforce{0w,1/2h}{v1}
  \fmfforce{1w,1/2h}{v2}
  \fmf{double,width=0.2mm}{v2,v1}
  \fmfv{decor,size=0, label=${\scs 1}$, l.dist=1mm, l.angle=-180}{v1}
  \fmfv{decor,size=0, label=${\scs 3}$, l.dist=1mm, l.angle=0}{v2}
  \end{fmfgraph*} } } 
\hspace*{4mm}
  \parbox{10mm}{\centerline{
  \begin{fmfgraph*}(7,3)
  \setval
  \fmfforce{0w,1/2h}{v1}
  \fmfforce{1w,1/2h}{v2}
  \fmf{double,width=0.2mm}{v2,v1}
  \fmfv{decor,size=0, label=${\scs 2}$, l.dist=1mm, l.angle=-180}{v1}
  \fmfv{decor,size=0, label=${\scs 4}$, l.dist=1mm, l.angle=0}{v2}
  \end{fmfgraph*} } } 
\hspace*{3mm} - \hspace*{3mm}
  \parbox{10mm}{\centerline{
  \begin{fmfgraph*}(7,3)
  \setval
  \fmfforce{0w,1/2h}{v1}
  \fmfforce{1w,1/2h}{v2}
  \fmf{double,width=0.2mm}{v2,v1}
  \fmfv{decor,size=0, label=${\scs 1}$, l.dist=1mm, l.angle=-180}{v1}
  \fmfv{decor,size=0, label=${\scs 4}$, l.dist=1mm, l.angle=0}{v2}
  \end{fmfgraph*} } } 
\hspace*{4mm}
  \parbox{10mm}{\centerline{
  \begin{fmfgraph*}(7,3)
  \setval
  \fmfforce{0w,1/2h}{v1}
  \fmfforce{1w,1/2h}{v2}
  \fmf{double,width=0.2mm}{v2,v1}
  \fmfv{decor,size=0, label=${\scs 2}$, l.dist=1mm, l.angle=-180}{v1}
  \fmfv{decor,size=0, label=${\scs 3}$, l.dist=1mm, l.angle=0}{v2}
  \end{fmfgraph*} } } 
\hspace*{5mm} . \la{GDEb} \\ \no
\eeq
However, such a procedure would have the disadvantage that cutting a line in the diagrams of the
connected two-point function $\fullg_{12}$ would also lead to disconnected
diagrams which are later on removed by the second and the third term on the right-hand side
of (\ref{GDEb}). As the number of undesired disconnected diagrams occuring
at an intermediate step of the calculation increases with the loop
order, this procedure is quite inefficient to determine the 
connected four-point function $\fullg_{1234}^{\rm c}$. Therefore we aim
at deriving another equation for $\fullg_{1234}^{\rm c}$ whose iterative
solution only involves connected diagrams. To this end we insert the
Schwinger-Dyson equation (\ref{SD1}) into relation (\ref{GDE}) and obtain
\beq
\la{ZW}
  \fullg_{1234}^{\rm c} &=& \frac{1}{3} \int_{5678} G_{15} V_{5678} \, 
  \frac{\delta \fullg_{6782}^{\rm c}}{\delta G^{-1}_{34}} \hspace*{2mm} 
  - \int_{5678} G_{15} V_{5678} \Gz_{62} \Gz_{73} \Gz_{84}
  - \hs \frac{1}{2} \int_{5678} G_{15} V_{5678} \Gz_{6734}^{\rm c} \Gz_{82} 
  \hs  \no  \\*[3mm]  & &
  - \hs \frac{1}{2} \int_{5678} G_{15} V_{5678} \Gz_{8234}^{\rm c} \Gz_{67}
  + \hs \frac{1}{6} \int_{5678} G_{15} V_{5678} \Gz_{6784}^{\rm c} \Gz_{23} 
  \hs + \hs \frac{1}{6} \int_{5678} G_{15} V_{5678} \Gz_{6783}^{\rm c} \Gz_{24}
  \hspace*{2mm} .
  \label{59}
\eeq 
The last two terms in (\ref{ZW}) are still disconnected and cancel the disconnected
diagrams which are generated by the functional derivative of the 
connected four-point function with respect to the kernel
in the first term. In order to eliminate all disconnected terms from (\ref{ZW})
we use the commutator relation
\beq
\la{KO1}
\frac{\delta \fullg_{1234}^{\rm c}}{\delta G^{-1}_{56}} - 
\frac{\delta \fullg_{1256}^{\rm c}}{\delta G^{-1}_{34}} &=& 
\frac{1}{2} \left( 
\fullg_{13} \fullg_{2456}^{\rm c} + \fullg_{14} \fullg_{2356}^{\rm c} +
\fullg_{23} \fullg_{1456}^{\rm c} + \fullg_{24} \fullg_{1356}^{\rm c} 
\right. \no \\ && \left. - 
\fullg_{15} \fullg_{2346}^{\rm c} - \fullg_{16} \fullg_{2345}^{\rm c} -
\fullg_{25} \fullg_{1346}^{\rm c} - \fullg_{26} \fullg_{1345}^{\rm c} 
\right) \, ,
\eeq
which directly follows from (\ref{GDE}) because of the identity that mixed second functional derivatives with respect to the
kernel $G^{-1}$ can be interchanged:
\beq
\frac{\delta^2 \fullg_{12}}{\delta G^{-1}_{56} \delta G^{-1}_{34}} = 
\frac{\delta^2 \fullg_{12}}{\delta G^{-1}_{34} \delta G^{-1}_{56}}  \, .
\eeq
The commutator relation (\ref{KO1}) states that within the functional derivative of the 
connected four-point function with respect to the kernel
indices might be interchanged at the expense of the additional terms on the
right-hand side. Inserting (\ref{KO1}) in (\ref{ZW}) by taking into account
the Schwinger-Dyson equation (\ref{SD1}) and the functional chain rule (\ref{13}),
we yield the following functional integrodifferential equation for the connected four-point function:
\beq
  \Gz_{1234}^{\rm c} &=& -  \int_{5678} G_{15} V_{5678} \Gz_{62} \Gz_{73} 
  \Gz_{84} \hs - \frac{1}{3} \int_{5678} G_{15} V_{5678} \hs G_{69} G_{70} 
  \hs \frac{\delta \Gz_{8234}^{\rm c}}{\delta G_{90}} 
  - \hs \frac{1}{6} \int_{5678} G_{15} V_{5678} \Gz_{8234}^{\rm c} \Gz_{67} \hspace*{2mm}    \no  \\*[3mm]  & & 
  \hs - \hs \frac{1}{6} \int_{5678} G_{15} V_{5678} \Gz_{6723}^{\rm c} \Gz_{84}
  - \hs \frac{1}{6} \int_{5678} G_{15} V_{5678} \Gz_{6724}^{\rm c} \Gz_{83} 
  \hs - \hs \frac{1}{6} \int_{5678} G_{15} V_{5678} \Gz_{6734}^{\rm c} \Gz_{82}
  \hspace*{3mm}.
\la{SD2} 
\eeq 
Although Eq. (\ref{SD2}) has the disadvantage of being more complex than
Eq. (\ref{GDE}), it has the advantage that it does not lead to disconnected
diagrams at an intermediate stage of the calculation. This can be immediately
seen in its graphical representiation which reads
\beq
  \parbox{12mm}{\centerline{
  \begin{fmfgraph*}(9,9)
  \setval
  \fmfforce{0w,0h}{v1}
  \fmfforce{0w,1h}{v2}
  \fmfforce{1w,1h}{v3}
  \fmfforce{1w,0h}{v4}
  \fmfforce{3.5/9w,3.5/9h}{v5}
  \fmfforce{3.5/9w,5.5/9h}{v6}
  \fmfforce{5.5/9w,5.5/9h}{v7}
  \fmfforce{5.5/9w,3.5/9h}{v8}
  \fmf{plain}{v1,v5}
  \fmf{plain}{v2,v6}
  \fmf{plain}{v3,v7}
  \fmf{plain}{v4,v8}
  \fmf{plain,width=0.2mm,left=1}{v5,v7,v5}
  \fmfv{decor.size=0, label=${\scs 1}$, l.dist=1mm, l.angle=-135}{v1}
  \fmfv{decor.size=0, label=${\scs 2}$, l.dist=1mm, l.angle=135}{v2}
  \fmfv{decor.size=0, label=${\scs 3}$, l.dist=1mm, l.angle=45}{v3}
  \fmfv{decor.size=0, label=${\scs 4}$, l.dist=1mm, l.angle=-45}{v4}
  \end{fmfgraph*} } } 
\hspace*{2mm} = \hspace*{5mm}
  \parbox{7mm}{\centerline{
  \begin{fmfgraph*}(7,7)
  \setval
  \fmfforce{0w,0h}{v1}
  \fmfforce{0w,1h}{v2}
  \fmfforce{1w,1h}{v3}
  \fmfforce{1w,0h}{v4}
  \fmfforce{1/2w,1/2h}{v5}
  \fmf{double,width=0.2mm}{v2,v4}
  \fmf{plain}{v1,v5}
  \fmf{double,width=0.2mm}{v5,v3}
  \fmfv{decor.size=0, label=${\scs 1}$, l.dist=1mm, l.angle=-135}{v1}
  \fmfv{decor.size=0, label=${\scs 2}$, l.dist=1mm, l.angle=135}{v2}
  \fmfv{decor.size=0, label=${\scs 3}$, l.dist=1mm, l.angle=45}{v3}
  \fmfv{decor.size=0, label=${\scs 4}$, l.dist=1mm, l.angle=-45}{v4}
  \fmfdot{v5}
  \end{fmfgraph*} } }
\hspace*{3mm} + \hspace*{2mm} \frac{1}{3} \hspace*{3mm} 
  \parbox{12mm}{\centerline{
  \begin{fmfgraph*}(10,8)
  \setval
  \fmfforce{0w,4/8h}{v1}
  \fmfforce{1/2w,4/8h}{v2}
  \fmfforce{1w,4/8h}{v3}
  \fmfforce{1w,0h}{v4}
  \fmfforce{6.5/10w,7/8h}{v5}
  \fmf{plain}{v1,v2}
  \fmf{plain}{v2,v3}
  \fmf{plain,right=0.3}{v2,v4}
  \fmf{plain}{v2,v5}
  \fmfv{decor.size=0, label=${\scs 1}$, l.dist=1mm, l.angle=-180}{v1}
  \fmfv{decor.size=0, label=${\scs 6}$, l.dist=1mm, l.angle=0}{v3}
  \fmfv{decor.size=0, label=${\scs 7}$, l.dist=1mm, l.angle=0}{v4}
  \fmfv{decor.size=0, label=${\scs 5}$, l.dist=1mm, l.angle=35}{v5}
  \fmfdot{v2}
  \end{fmfgraph*} } } 
\hspace*{3mm}
\raisebox{2mm}{\begin{minipage}{1.8cm}
\beq
\dphi{
  \parbox{14mm}{\centerline{
  \begin{fmfgraph*}(9,9)
  \setval
  \fmfforce{0w,2.5/9h}{v1}
  \fmfforce{0w,11.5/9h}{v2}
  \fmfforce{1w,11.5/9h}{v3}
  \fmfforce{1w,2.5/9h}{v4}
  \fmfforce{3.5/9w,6/9h}{v5}
  \fmfforce{3.5/9w,8/9h}{v6}
  \fmfforce{5.5/9w,8/9h}{v7}
  \fmfforce{5.5/9w,6/9h}{v8}
  \fmf{plain}{v1,v5}
  \fmf{plain}{v2,v6}
  \fmf{plain}{v3,v7}
  \fmf{plain}{v4,v8}
  \fmf{plain,width=0.2mm,left=1}{v5,v7,v5}
  \fmfv{decor.size=0, label=${\scs 5}$, l.dist=0.5mm, l.angle=-135}{v1}
  \fmfv{decor.size=0, label=${\scs 2}$, l.dist=0.5mm, l.angle=135}{v2}
  \fmfv{decor.size=0, label=${\scs 3}$, l.dist=0.5mm, l.angle=45}{v3}
  \fmfv{decor.size=0, label=${\scs 4}$, l.dist=0.5mm, l.angle=-45}{v4}
  \end{fmfgraph*} } } 
}{6}{7}
\no  \eeq 
\end{minipage}}
\hspace*{2mm}+ \hspace*{2mm} \frac{1}{6} \hspace*{2mm}
  \parbox{23mm}{\centerline{
  \begin{fmfgraph*}(18,13)
  \setval
  \fmfforce{0w,1/2h}{v1}
  \fmfforce{1w,1/2h}{v2}
  \fmfforce{11.5/18w,1h}{v3}
  \fmfforce{11.5/18w,0h}{v4}
  \fmfforce{5/18w,1/2h}{v5}
  \fmfforce{11.5/18w,8/13h}{v6}
  \fmfforce{13/18w,1/2h}{v7}
  \fmfforce{11.5/18w,5/13h}{v8}
  \fmfforce{10/18w,1/2h}{v9}
  \fmfforce{5/18w,11.5/13h}{v10}
  \fmf{plain}{v1,v5}
  \fmf{double,width=0.2mm,left=1}{v5,v10,v5}
  \fmf{plain}{v5,v9}
  \fmf{plain}{v2,v7}
  \fmf{plain,width=0.2mm,left=1}{v9,v7,v9}
  \fmf{plain}{v3,v6}
  \fmf{plain}{v4,v8}
  \fmfv{decor.size=0, label=${\scs 1}$, l.dist=1mm, l.angle=180}{v1}
  \fmfv{decor.size=0, label=${\scs 2}$, l.dist=1mm, l.angle=90}{v3}
  \fmfv{decor.size=0, label=${\scs 3}$, l.dist=1mm, l.angle=0}{v2}
  \fmfv{decor.size=0, label=${\scs 4}$, l.dist=1mm, l.angle=-90}{v4}
  \fmfdot{v5}
  \end{fmfgraph*} } } 
\no
\eeq
\beq
\hspace*{2.3cm}+ \hspace*{2mm} \frac{1}{6} \hspace*{3mm}
  \parbox{17mm}{\centerline{
  \begin{fmfgraph*}(14.5,9)
  \setval
  \fmfforce{0w,1/9h}{v1}
  \fmfforce{0w,8/9h}{v2}
  \fmfforce{1w,1h}{v3}
  \fmfforce{1w,0h}{v4}
  \fmfforce{8.5/14w,3.5/9h}{v5}
  \fmfforce{8.5/14w,5.5/9h}{v6}
  \fmfforce{10.5/14w,5.5/9h}{v7}
  \fmfforce{10.5/14w,3.5/9h}{v8}
  \fmfforce{3.5/14w,1/2h}{v9}
  \fmf{plain}{v1,v9}
  \fmf{double,width=0.2mm}{v2,v9}
  \fmf{plain}{v3,v7}
  \fmf{plain}{v4,v8}
  \fmf{plain,width=0.2mm,left=1}{v5,v7,v5}
  \fmf{plain,left=0.8}{v9,v6}
  \fmf{plain,right=0.8}{v9,v5}
  \fmfv{decor.size=0, label=${\scs 1}$, l.dist=1mm, l.angle=-135}{v1}
  \fmfv{decor.size=0, label=${\scs 4}$, l.dist=1mm, l.angle=135}{v2}
  \fmfv{decor.size=0, label=${\scs 2}$, l.dist=0.5mm, l.angle=0}{v3}
  \fmfv{decor.size=0, label=${\scs 3}$, l.dist=0.5mm, l.angle=0}{v4}
  \fmfdot{v9}
  \end{fmfgraph*} } } 
+ \hspace*{2mm} \frac{1}{6} \hspace*{3mm}
  \parbox{17mm}{\centerline{
  \begin{fmfgraph*}(14.5,9)
  \setval
  \fmfforce{0w,1/9h}{v1}
  \fmfforce{0w,8/9h}{v2}
  \fmfforce{1w,1h}{v3}
  \fmfforce{1w,0h}{v4}
  \fmfforce{8.5/14w,3.5/9h}{v5}
  \fmfforce{8.5/14w,5.5/9h}{v6}
  \fmfforce{10.5/14w,5.5/9h}{v7}
  \fmfforce{10.5/14w,3.5/9h}{v8}
  \fmfforce{3.5/14w,1/2h}{v9}
  \fmf{plain}{v1,v9}
  \fmf{double,width=0.2mm}{v2,v9}
  \fmf{plain}{v3,v7}
  \fmf{plain}{v4,v8}
  \fmf{plain,width=0.2mm,left=1}{v5,v7,v5}
  \fmf{plain,left=0.8}{v9,v6}
  \fmf{plain,right=0.8}{v9,v5}
  \fmfv{decor.size=0, label=${\scs 1}$, l.dist=1mm, l.angle=-135}{v1}
  \fmfv{decor.size=0, label=${\scs 3}$, l.dist=1mm, l.angle=135}{v2}
  \fmfv{decor.size=0, label=${\scs 4}$, l.dist=0.5mm, l.angle=0}{v3}
  \fmfv{decor.size=0, label=${\scs 2}$, l.dist=0.5mm, l.angle=0}{v4}
  \fmfdot{v9}
  \end{fmfgraph*} } } 
\hspace*{3mm}+ \hspace*{2mm} \frac{1}{6} \hspace*{3mm}
  \parbox{17mm}{\centerline{
  \begin{fmfgraph*}(14.5,9)
  \setval
  \fmfforce{0w,1/9h}{v1}
  \fmfforce{0w,8/9h}{v2}
  \fmfforce{1w,1h}{v3}
  \fmfforce{1w,0h}{v4}
  \fmfforce{8.5/14w,3.5/9h}{v5}
  \fmfforce{8.5/14w,5.5/9h}{v6}
  \fmfforce{10.5/14w,5.5/9h}{v7}
  \fmfforce{10.5/14w,3.5/9h}{v8}
  \fmfforce{3.5/14w,1/2h}{v9}
  \fmf{plain}{v1,v9}
  \fmf{double,width=0.2mm}{v2,v9}
  \fmf{plain}{v3,v7}
  \fmf{plain}{v4,v8}
  \fmf{plain,width=0.2mm,left=1}{v5,v7,v5}
  \fmf{plain,left=0.8}{v9,v6}
  \fmf{plain,right=0.8}{v9,v5}
  \fmfv{decor.size=0, label=${\scs 1}$, l.dist=1mm, l.angle=-135}{v1}
  \fmfv{decor.size=0, label=${\scs 2}$, l.dist=1mm, l.angle=135}{v2}
  \fmfv{decor.size=0, label=${\scs 3}$, l.dist=0.5mm, l.angle=0}{v3}
  \fmfv{decor.size=0, label=${\scs 4}$, l.dist=0.5mm, l.angle=0}{v4}
  \fmfdot{v9}
  \end{fmfgraph*} } } 
\hspace*{5mm} .  
\label{SDG2}              
\eeq
Thus the closed set of Schwinger-Dyson equations for the connected two- and four-point function
is given by (\ref{SDG1}) and (\ref{SDG2}).
\subsection{Graphical Recursion Relations}\la{GRC}
Now we demonstrate how the diagrams of the connected two- and four-point function are recursively generated 
in a graphical way. To this end we perform for both quantities a perturbative expansion
\beq
\la{G2P}
  \parbox{10mm}{\centerline{
  \begin{fmfgraph*}(7,3)
  \setval
  \fmfforce{0w,1/2h}{v1}
  \fmfforce{1w,1/2h}{v2}
  \fmf{double,width=0.2mm}{v2,v1}
  \fmfv{decor,size=0, label=${\scs 1}$, l.dist=1mm, l.angle=-180}{v1}
  \fmfv{decor,size=0, label=${\scs 2}$, l.dist=1mm, l.angle=0}{v2}
  \end{fmfgraph*} } } 
\hspace*{3mm} &\equiv &\hspace*{2mm} \sum_{p=0}^{\infty} \hspace*{3mm} 
  \parbox{10mm}{\centerline{
  \begin{fmfgraph*}(7,3)
  \setval
  \fmfforce{0w,1/2h}{v1}
  \fmfforce{1w,1/2h}{v2}
  \fmfforce{1/2w,1/2h}{v3}
  \fmf{double,width=0.2mm}{v2,v1}
  \fmfv{decor,size=0, label=${\scs 1}$, l.dist=1mm, l.angle=-180}{v1}
  \fmfv{decor,size=0, label=${\scs 2}$, l.dist=1mm, l.angle=0}{v2}
  \fmfv{decor,size=0, label=${\scs (p)}$, l.dist=1.5mm, l.angle=90}{v3}
  \end{fmfgraph*} } } 
\hspace*{5mm}, \\[3mm]  \hspace*{8mm}
\la{G4P}
  \parbox{12mm}{\centerline{
  \begin{fmfgraph*}(9,9)
  \setval
  \fmfforce{0w,0h}{v1}
  \fmfforce{0w,1h}{v2}
  \fmfforce{1w,1h}{v3}
  \fmfforce{1w,0h}{v4}
  \fmfforce{3.5/9w,3.5/9h}{v5}
  \fmfforce{3.5/9w,5.5/9h}{v6}
  \fmfforce{5.5/9w,5.5/9h}{v7}
  \fmfforce{5.5/9w,3.5/9h}{v8}
  \fmf{plain}{v1,v5}
  \fmf{plain}{v2,v6}
  \fmf{plain}{v3,v7}
  \fmf{plain}{v4,v8}
  \fmf{plain,width=0.2mm,left=1}{v5,v7,v5}
  \fmfv{decor.size=0, label=${\scs 1}$, l.dist=1mm, l.angle=-135}{v1}
  \fmfv{decor.size=0, label=${\scs 2}$, l.dist=1mm, l.angle=135}{v2}
  \fmfv{decor.size=0, label=${\scs 3}$, l.dist=1mm, l.angle=45}{v3}
  \fmfv{decor.size=0, label=${\scs 4}$, l.dist=1mm, l.angle=-45}{v4}
  \end{fmfgraph*} } } 
\hs &\equiv &\hspace*{2mm} \sum_{p=0}^{\infty} \hspace*{3mm}
  \parbox{12mm}{\centerline{
  \begin{fmfgraph*}(10.5,10.5)
  \setval
  \fmfforce{0w,0h}{v1}
  \fmfforce{0w,1h}{v2}
  \fmfforce{1w,1h}{v3}
  \fmfforce{1w,0h}{v4}
  \fmfforce{1/3w,1/3h}{v5}
  \fmfforce{1/3w,2/3h}{v6}
  \fmfforce{2/3w,2/3h}{v7}
  \fmfforce{2/3w,1/3h}{v8}
  \fmfforce{1/2w,1/2h}{v9}
  \fmf{plain}{v1,v5}
  \fmf{plain}{v2,v6}
  \fmf{plain}{v3,v7}
  \fmf{plain}{v4,v8}
  \fmf{plain,width=0.2mm,left=1}{v5,v7,v5}
  \fmfv{decor.size=0, label=${\scs 1}$, l.dist=1mm, l.angle=-135}{v1}
  \fmfv{decor.size=0, label=${\scs 2}$, l.dist=1mm, l.angle=135}{v2}
  \fmfv{decor.size=0, label=${\scs 3}$, l.dist=1mm, l.angle=45}{v3}
  \fmfv{decor.size=0, label=${\scs 4}$, l.dist=1mm, l.angle=-45}{v4}
  \fmfv{decor.size=0, label=${\scs p}$, l.dist=0mm, l.angle=90}{v9}
  \end{fmfgraph*} } } 
  \hspace*{3mm} , \label{64} \\ && \no
\eeq
where $p$ denotes the number of interactions $V$ which contribute. With these  we obtain from (\ref{SDG1}) and (\ref{SDG2})
the following closed set of graphical recursion relations:
\beq
\la{REK1}
\hspace*{-5.5cm}  
\parbox{10mm}{\centerline{
  \begin{fmfgraph*}(7,3)
  \setval
  \fmfforce{0w,1/2h}{v1}
  \fmfforce{1w,1/2h}{v2}
  \fmfforce{1/2w,1/2h}{v3}
  \fmf{double,width=0.2mm}{v2,v1}
  \fmfv{decor.size=0, label=${\scs 1}$, l.dist=1mm, l.angle=-180}{v1}
  \fmfv{decor.size=0, label=${\scs 2}$, l.dist=1mm, l.angle=0}{v2}
  \fmfv{decor.size=0, label=${\scs {(p+1)}}$, l.dist=1.5mm, l.angle=90}{v3}
  \end{fmfgraph*}}}
\hspace*{5mm} = \hspace*{2mm} \frac{1}{2} \hs
\sum_{q=0}^p \hspace*{2mm} 
  \parbox{13mm}{\centerline{
  \begin{fmfgraph*}(10,5)
  \setval
  \fmfforce{0w,0h}{v1}
  \fmfforce{1/2w,0h}{v2}
  \fmfforce{1w,0h}{v3}
  \fmfforce{1/2w,1h}{v4}
  \fmfforce{10/10w,0h}{v5}
  \fmf{plain}{v2,v1}
  \fmf{double,width=0.2mm}{v2,v3}
  \fmf{double,width=0.2mm,left=1}{v4,v2,v4}
  \fmfv{decor.size=0, label=${\scs 1}$, l.dist=1mm, l.angle=-180}{v1}
  \fmfv{decor.size=0, label=${\scs 2}$, l.dist=1mm, l.angle=0}{v3}
  \fmfv{decor.size=0, label=${\scs {(p-q)}}$, l.dist=2mm, l.angle=90}{v4}
  \fmfv{decor.size=0, label=${\scs (q)}$, l.dist=1.5mm, l.angle=90}{v5}
  \fmfdot{v2}
  \end{fmfgraph*}}}
\hspace*{0.4cm} + \hspace*{0.2cm} \frac{1}{6} \hspace*{0.3cm}
  \parbox{23mm}{\centerline{
  \begin{fmfgraph*}(20,9)
  \setval
  \fmfforce{0w,1/2h}{v1}
  \fmfforce{5/20w,1/2h}{v2}
  \fmfforce{10/20w,1/2h}{v3}
  \fmfforce{15/20w,1/2h}{v4}
  \fmfforce{1w,1/2h}{v5}
  \fmfforce{12.2/20w,7.1/9h}{v6}
  \fmfforce{12.2/20w,1.9/9h}{v7}
  \fmfforce{12.5/20w,1/2h}{v8}
  \fmf{plain}{v1,v2}
  \fmf{plain}{v4,v5}
  \fmf{plain,left=0.65}{v2,v6}
  \fmf{plain}{v2,v3}
  \fmf{plain,right=0.65}{v2,v7}
  \fmf{plain,width=0.2mm,right=1}{v3,v4,v3}
  \fmfv{decor.size=0, label=${\scs 1}$, l.dist=0.5mm, l.angle=-180}{v1}
  \fmfv{decor.size=0, label=${\scs 2}$, l.dist=0.5mm, l.angle=0}{v5}
  \fmfv{decor.size=0, label=${\scs p}$, l.dist=0mm, l.angle=90}{v8}
  \fmfdot{v2}
  \end{fmfgraph*} } }
\hspace*{2mm} , 
\eeq
\beq
\la{REK2}
  \parbox{14mm}{\centerline{
  \begin{fmfgraph*}(11,11)
  \setval
  \fmfforce{0w,0h}{v1}
  \fmfforce{0w,1h}{v2}
  \fmfforce{1w,1h}{v3}
  \fmfforce{1w,0h}{v4}
  \fmfforce{3.5/11w,3.5/11h}{v5}
  \fmfforce{3.5/11w,7.5/11h}{v6}
  \fmfforce{7.5/11w,7.5/11h}{v7}
  \fmfforce{7.5/11w,3.5/11h}{v8}
  \fmfforce{1/2w,1/2h}{v9}
  \fmf{plain}{v1,v5}
  \fmf{plain}{v2,v6}
  \fmf{plain}{v3,v7}
  \fmf{plain}{v4,v8}
  \fmf{plain,width=0.2mm,left=1}{v5,v7,v5}
  \fmfv{decor.size=0, label=${\scs 1}$, l.dist=1mm, l.angle=-135}{v1}
  \fmfv{decor.size=0, label=${\scs 2}$, l.dist=1mm, l.angle=135}{v2}
  \fmfv{decor.size=0, label=${\scs 3}$, l.dist=1mm, l.angle=45}{v3}
  \fmfv{decor.size=0, label=${\scs 4}$, l.dist=1mm, l.angle=-45}{v4}
  \fmfv{decor.size=0, label=${\scs {p+1}}$, l.dist=0mm, l.angle=90}{v9}
  \end{fmfgraph*} } } 
\hspace*{2mm} = \hspace*{2mm} \sum_{q=0}^{p} \sum_{r=0}^q \hspace*{8mm}
  \parbox{7mm}{\centerline{
  \begin{fmfgraph*}(7,7)
  \setval
  \fmfforce{0w,0h}{v1}
  \fmfforce{0w,1h}{v2}
  \fmfforce{1w,1h}{v3}
  \fmfforce{1w,0h}{v4}
  \fmfforce{1/2w,1/2h}{v5}
  \fmfforce{1/3w,2/3h}{v6}
  \fmfforce{2/3w,2/3h}{v7}
  \fmfforce{1/2w,0h}{v8}
  \fmf{double,width=0.2mm}{v2,v4}
  \fmf{plain}{v1,v5}
  \fmf{double,width=0.2mm}{v5,v3}
  \fmfv{decor.size=0, label=${\scs 1}$, l.dist=1mm, l.angle=-135}{v1}
  \fmfv{decor.size=0, label=${\scs 2}$, l.dist=1mm, l.angle=135}{v2}
  \fmfv{decor.size=0, label=${\scs 3}$, l.dist=1mm, l.angle=45}{v3}
  \fmfv{decor.size=0, label=${\scs 4}$, l.dist=1mm, l.angle=-45}{v4}
  \fmfv{decor.size=0, label=${\scs {(p-q)}}$, l.dist=1.5mm, l.angle=180}{v6}
  \fmfv{decor.size=0, label=${\scs {(q-r)}}$, l.dist=1.5mm, l.angle=0}{v7}
  \fmfv{decor.size=0, label=${\scs (r)}$, l.dist=0mm, l.angle=90}{v8}
  \fmfdot{v5}
  \end{fmfgraph*} } }
\hspace*{6mm} + \hspace*{2mm} \frac{1}{3} \hspace*{3mm} 
  \parbox{12mm}{\centerline{
  \begin{fmfgraph*}(10,8)
  \setval
  \fmfforce{0w,4/8h}{v1}
  \fmfforce{1/2w,4/8h}{v2}
  \fmfforce{1w,4/8h}{v3}
  \fmfforce{1w,0h}{v4}
  \fmfforce{6.5/10w,7/8h}{v5}
  \fmf{plain}{v1,v2}
  \fmf{plain}{v2,v3}
  \fmf{plain,right=0.3}{v2,v4}
  \fmf{plain}{v2,v5}
  \fmfv{decor.size=0, label=${\scs 1}$, l.dist=1mm, l.angle=-180}{v1}
  \fmfv{decor.size=0, label=${\scs 6}$, l.dist=1mm, l.angle=0}{v3}
  \fmfv{decor.size=0, label=${\scs 7}$, l.dist=1mm, l.angle=0}{v4}
  \fmfv{decor.size=0, label=${\scs 5}$, l.dist=1mm, l.angle=35}{v5}
  \fmfdot{v2}
  \end{fmfgraph*} } } 
\hspace*{3mm}
\raisebox{3mm}{\begin{minipage}{1.8cm}
\beq
\dphi{
  \parbox{14mm}{\centerline{
  \begin{fmfgraph*}(10.5,10.5)
  \setval
  \fmfforce{0w,2/10.5h}{v1}
  \fmfforce{0w,12.5/10.5h}{v2}
  \fmfforce{1w,12.5/10.5h}{v3}
  \fmfforce{1w,2/10.5h}{v4}
  \fmfforce{1/3w,5.5/10.5h}{v5}
  \fmfforce{1/3w,9/10.5h}{v6}
  \fmfforce{2/3w,9/10.5h}{v7}
  \fmfforce{2/3w,5.5/10.5h}{v8}
  \fmfforce{1/2w,7.25/10.5h}{v9}
  \fmf{plain}{v1,v5}
  \fmf{plain}{v2,v6}
  \fmf{plain}{v3,v7}
  \fmf{plain}{v4,v8}
  \fmf{plain,width=0.2mm,left=1}{v5,v7,v5}
  \fmfv{decor.size=0, label=${\scs 5}$, l.dist=0.5mm, l.angle=-180}{v1}
  \fmfv{decor.size=0, label=${\scs 2}$, l.dist=0.5mm, l.angle=180}{v2}
  \fmfv{decor.size=0, label=${\scs 3}$, l.dist=0.5mm, l.angle=0}{v3}
  \fmfv{decor.size=0, label=${\scs 4}$, l.dist=0.5mm, l.angle=-0}{v4}
  \fmfv{decor.size=0, label=${\scs p}$, l.dist=0mm, l.angle=90}{v9}
  \end{fmfgraph*} } } 
}{6}{7}
\no  \eeq 
\end{minipage}}
\hspace*{3.3mm} + \hspace*{2mm} \frac{1}{6} \hs \sum_{q=1}^p \hspace*{2mm}
  \parbox{23mm}{\centerline{
  \begin{fmfgraph*}(20,15)
  \setval
  \fmfforce{0w,1/2h}{v1}
  \fmfforce{1w,1/2h}{v2}
  \fmfforce{12.5/20w,1h}{v3}
  \fmfforce{12.5/20w,0h}{v4}
  \fmfforce{5/20w,1/2h}{v5}
  \fmfforce{12.5/20w,10/15h}{v6}
  \fmfforce{15/20w,1/2h}{v7}
  \fmfforce{12.5/20w,5/15h}{v8}
  \fmfforce{10/20w,1/2h}{v9}
  \fmfforce{5/20w,12.5/15h}{v10}
  \fmfforce{12.5/20w,1/2h}{v11}
  \fmf{plain}{v1,v5}
  \fmf{double,width=0.2mm,left=1}{v5,v10,v5}
  \fmf{plain}{v5,v9}
  \fmf{plain}{v2,v7}
  \fmf{plain,width=0.2mm,left=1}{v9,v7,v9}
  \fmf{plain}{v3,v6}
  \fmf{plain}{v4,v8}
  \fmfv{decor.size=0, label=${\scs 1}$, l.dist=1mm, l.angle=180}{v1}
  \fmfv{decor.size=0, label=${\scs 2}$, l.dist=1mm, l.angle=90}{v3}
  \fmfv{decor.size=0, label=${\scs 3}$, l.dist=1mm, l.angle=0}{v2}
  \fmfv{decor.size=0, label=${\scs 4}$, l.dist=1mm, l.angle=-90}{v4}
  \fmfv{decor.size=0, label=${\scs {(p-q)}}$, l.dist=1.5mm, l.angle=90}{v10}
  \fmfv{decor.size=0, label=${\scs q}$, l.dist=0mm, l.angle=90}{v11}
  \fmfdot{v5}
  \end{fmfgraph*} } } 
\no
\eeq
\beq
 \hspace*{2cm} + \hspace*{2mm} \frac{1}{6} \hs \sum_{q=1}^p \hspace*{2mm}
  \parbox{18mm}{\centerline{
  \begin{fmfgraph*}(15.5,10.5)
  \setval
  \fmfforce{0w,1.75/10.5h}{v1}
  \fmfforce{0w,8.75/10.5h}{v2}
  \fmfforce{1w,1h}{v3}
  \fmfforce{1w,0h}{v4}
  \fmfforce{8.5/15.5w,1/3h}{v5}
  \fmfforce{8.5/15.5w,2/3h}{v6}
  \fmfforce{12/15.5w,2/3h}{v7}
  \fmfforce{12/15.5w,1/3h}{v8}
  \fmfforce{3.5/15.5w,1/2h}{v9}
  \fmfforce{10.25/15.5w,1/2h}{v10}
  \fmf{plain}{v1,v9}
  \fmf{double,width=0.2mm}{v2,v9}
  \fmf{plain}{v3,v7}
  \fmf{plain}{v4,v8}
  \fmf{plain,width=0.2mm,left=1}{v5,v7,v5}
  \fmf{plain,left=0.6}{v9,v6}
  \fmf{plain,right=0.6}{v9,v5}
  \fmfv{decor.size=0, label=${\scs 1}$, l.dist=1mm, l.angle=-135}{v1}
  \fmfv{decor.size=0, label=${\scs 4}$, l.dist=1mm, l.angle=135}{v2}
  \fmfv{decor.size=0, label=${\scs 2}$, l.dist=0.5mm, l.angle=0}{v3}
  \fmfv{decor.size=0, label=${\scs 3}$, l.dist=0.5mm, l.angle=0}{v4}
  \fmfv{decor.size=0, label=${\scs {(p-q)}}$, l.dist=4.5mm, l.angle=90}{v9}
  \fmfv{decor.size=0, label=${\scs q}$, l.dist=0mm, l.angle=90}{v10}
  \fmfdot{v9}
  \end{fmfgraph*} } } 
\hspace*{3mm} + \hspace*{2mm} \frac{1}{6} \hs \sum_{q=1}^p \hspace*{2mm}
  \parbox{18mm}{\centerline{
  \begin{fmfgraph*}(15.5,10.5)
  \setval
  \fmfforce{0w,1.75/10.5h}{v1}
  \fmfforce{0w,8.75/10.5h}{v2}
  \fmfforce{1w,1h}{v3}
  \fmfforce{1w,0h}{v4}
  \fmfforce{8.5/15.5w,1/3h}{v5}
  \fmfforce{8.5/15.5w,2/3h}{v6}
  \fmfforce{12/15.5w,2/3h}{v7}
  \fmfforce{12/15.5w,1/3h}{v8}
  \fmfforce{3.5/15.5w,1/2h}{v9}
  \fmfforce{10.25/15.5w,1/2h}{v10}
  \fmf{plain}{v1,v9}
  \fmf{double,width=0.2mm}{v2,v9}
  \fmf{plain}{v3,v7}
  \fmf{plain}{v4,v8}
  \fmf{plain,width=0.2mm,left=1}{v5,v7,v5}
  \fmf{plain,left=0.6}{v9,v6}
  \fmf{plain,right=0.6}{v9,v5}
  \fmfv{decor.size=0, label=${\scs 1}$, l.dist=1mm, l.angle=-135}{v1}
  \fmfv{decor.size=0, label=${\scs 3}$, l.dist=1mm, l.angle=135}{v2}
  \fmfv{decor.size=0, label=${\scs 4}$, l.dist=0.5mm, l.angle=0}{v3}
  \fmfv{decor.size=0, label=${\scs 2}$, l.dist=0.5mm, l.angle=0}{v4}
  \fmfv{decor.size=0, label=${\scs {(p-q)}}$, l.dist=4.5mm, l.angle=90}{v9}
  \fmfv{decor.size=0, label=${\scs q}$, l.dist=0mm, l.angle=90}{v10}
  \fmfdot{v9}
  \end{fmfgraph*} } } 
\hspace*{2mm} + \hspace*{2mm} \frac{1}{6} \hs \sum_{q=1}^p \hspace*{2mm}
  \parbox{18mm}{\centerline{
  \begin{fmfgraph*}(15.5,10.5)
  \setval
  \fmfforce{0w,1.75/10.5h}{v1}
  \fmfforce{0w,8.75/10.5h}{v2}
  \fmfforce{1w,1h}{v3}
  \fmfforce{1w,0h}{v4}
  \fmfforce{8.5/15.5w,1/3h}{v5}
  \fmfforce{8.5/15.5w,2/3h}{v6}
  \fmfforce{12/15.5w,2/3h}{v7}
  \fmfforce{12/15.5w,1/3h}{v8}
  \fmfforce{3.5/15.5w,1/2h}{v9}
  \fmfforce{10.25/15.5w,1/2h}{v10}
  \fmf{plain}{v1,v9}
  \fmf{double,width=0.2mm}{v2,v9}
  \fmf{plain}{v3,v7}
  \fmf{plain}{v4,v8}
  \fmf{plain,width=0.2mm,left=1}{v5,v7,v5}
  \fmf{plain,left=0.6}{v9,v6}
  \fmf{plain,right=0.6}{v9,v5}
  \fmfv{decor.size=0, label=${\scs 1}$, l.dist=1mm, l.angle=-135}{v1}
  \fmfv{decor.size=0, label=${\scs 2}$, l.dist=1mm, l.angle=135}{v2}
  \fmfv{decor.size=0, label=${\scs 3}$, l.dist=0.5mm, l.angle=0}{v3}
  \fmfv{decor.size=0, label=${\scs 4}$, l.dist=0.5mm, l.angle=0}{v4}
  \fmfv{decor.size=0, label=${\scs {(p-q)}}$, l.dist=4.5mm, l.angle=90}{v9}
  \fmfv{decor.size=0, label=${\scs q}$, l.dist=0mm, l.angle=90}{v10}
  \fmfdot{v9}
  \end{fmfgraph*} } } 
\hspace*{5mm} .
\\  \no
\eeq
This is solved starting from
\beq
\la{IN1}
  \parbox{10mm}{\centerline{
  \begin{fmfgraph*}(7,3)
  \setval
  \fmfforce{0w,1/2h}{v1}
  \fmfforce{1w,1/2h}{v2}
  \fmfforce{1/2w,1/2h}{v3}  
  \fmf{double,width=0.2mm}{v2,v1}
  \fmfv{decor.size=0, label=${\scs 1}$, l.dist=1mm, l.angle=-180}{v1}
  \fmfv{decor.size=0, label=${\scs 2}$, l.dist=1mm, l.angle=0}{v2}
  \fmfv{decor.size=0, label=${\scs (0)}$, l.dist=1mm, l.angle=90}{v3}
  \end{fmfgraph*}}}
\hspace*{3mm} &=& \hspace*{3mm}
  \parbox{10mm}{\centerline{
  \begin{fmfgraph*}(7,3)
  \setval
  \fmfleft{v1}
  \fmfright{v2}
  \fmf{plain}{v2,v1}
  \fmfv{decor.size=0, label=${\scs 1}$, l.dist=1mm, l.angle=-180}{v1}
  \fmfv{decor.size=0, label=${\scs 2}$, l.dist=1mm, l.angle=0}{v2}
  \end{fmfgraph*}}}
\hspace*{2mm}, \\ [4mm]
\la{IN2}
  \parbox{12mm}{\centerline{
  \begin{fmfgraph*}(10.5,10.5)
  \setval
  \fmfforce{0w,0h}{v1}
  \fmfforce{0w,1h}{v2}
  \fmfforce{1w,1h}{v3}
  \fmfforce{1w,0h}{v4}
  \fmfforce{1/3w,1/3h}{v5}
  \fmfforce{1/3w,2/3h}{v6}
  \fmfforce{2/3w,2/3h}{v7}
  \fmfforce{2/3w,1/3h}{v8}
  \fmfforce{1/2w,1/2h}{v9}
  \fmf{plain}{v1,v5}
  \fmf{plain}{v2,v6}
  \fmf{plain}{v3,v7}
  \fmf{plain}{v4,v8}
  \fmf{plain,width=0.2mm,left=1}{v5,v7,v5}
  \fmfv{decor.size=0, label=${\scs 1}$, l.dist=1mm, l.angle=-135}{v1}
  \fmfv{decor.size=0, label=${\scs 2}$, l.dist=1mm, l.angle=135}{v2}
  \fmfv{decor.size=0, label=${\scs 3}$, l.dist=1mm, l.angle=45}{v3}
  \fmfv{decor.size=0, label=${\scs 4}$, l.dist=1mm, l.angle=-45}{v4}
  \fmfv{decor.size=0, label=${\scs 0}$, l.dist=0mm, l.angle=90}{v9}
  \end{fmfgraph*} } } 
\hspace*{2mm} &= &\hspace*{3mm} 0  \hspace*{2mm} . \\ \nonumber
\eeq
Note that these graphical recursion relations (\ref{REK1})--(\ref{IN2}) allow to prove via complete induction that all
diagrams contributing to the connected two- and four-point function are, indeed,
connected \ci{Glaum}.\\

The first few perturbative contributions to $\fullg_{12}$ and $\fullg_{1234}^{\rm c}$ are determined as follows. Inserting 
(\ref{IN1}) and (\ref{IN2}) in (\ref{REK1}) and (\ref{REK2}), we obtain for $p=1$ the connected two-point function
\beq  
  \parbox{10mm}{\centerline{
  \begin{fmfgraph*}(7,3)
  \setval
  \fmfforce{0w,1/2h}{v1}
  \fmfforce{1w,1/2h}{v2}
  \fmfforce{1/2w,1/2h}{v3}
  \fmf{double,width=0.2mm}{v2,v1}
  \fmfv{decor.size=0, label=${\scs 1}$, l.dist=1mm, l.angle=-180}{v1}
  \fmfv{decor.size=0, label=${\scs 2}$, l.dist=1mm, l.angle=0}{v2}
  \fmfv{decor.size=0, label=${\scs (1)}$, l.dist=1.5mm, l.angle=90}{v3}
  \end{fmfgraph*}}}
\hspace*{5mm} = \hspace*{2mm} \frac{1}{2} \hspace*{3mm}
  \parbox{13mm}{\centerline{
  \begin{fmfgraph*}(10,5)
  \setval
  \fmfforce{0w,0h}{v1}
  \fmfforce{1/2w,0h}{v2}
  \fmfforce{1w,0h}{v3}
  \fmfforce{1/2w,1h}{v4}
  \fmf{plain}{v2,v1}
  \fmf{plain}{v2,v3}
  \fmf{plain,left=1}{v4,v2,v4}
  \fmfv{decor.size=0, label=${\scs 1}$, l.dist=1mm, l.angle=-180}{v1}
  \fmfv{decor.size=0, label=${\scs 2}$, l.dist=1mm, l.angle=0}{v3}
  \fmfdot{v2}
  \end{fmfgraph*}}}
\label{69}
\eeq
and the connected four-point function
\beq
  \parbox{14mm}{\centerline{
  \begin{fmfgraph*}(10.5,10.5)
  \setval
  \fmfforce{0w,0h}{v1}
  \fmfforce{0w,1h}{v2}
  \fmfforce{1w,1h}{v3}
  \fmfforce{1w,0h}{v4}
  \fmfforce{3.5/10.5w,3.5/10.5h}{v5}
  \fmfforce{3.5/10.5w,7/10.5h}{v6}
  \fmfforce{7/10.5w,7/10.5h}{v7}
  \fmfforce{7/10.5w,3.5/10.5h}{v8}
  \fmfforce{1/2w,1/2h}{v9}
  \fmf{plain}{v1,v5}
  \fmf{plain}{v2,v6}
  \fmf{plain}{v3,v7}
  \fmf{plain}{v4,v8}
  \fmf{plain,width=0.2mm,left=1}{v5,v7,v5}
  \fmfv{decor.size=0, label=${\scs 1}$, l.dist=1mm, l.angle=-135}{v1}
  \fmfv{decor.size=0, label=${\scs 2}$, l.dist=1mm, l.angle=135}{v2}
  \fmfv{decor.size=0, label=${\scs 3}$, l.dist=1mm, l.angle=45}{v3}
  \fmfv{decor.size=0, label=${\scs 4}$, l.dist=1mm, l.angle=-45}{v4}
  \fmfv{decor.size=0, label=${\scs 1}$, l.dist=0mm, l.angle=90}{v9}
  \end{fmfgraph*} } } 
\hspace*{3mm} = \hspace*{3mm}  
  \parbox{7mm}{\centerline{
  \begin{fmfgraph*}(7,7)
  \setval
  \fmfforce{0w,0h}{v1}
  \fmfforce{0w,1h}{v2}
  \fmfforce{1w,1h}{v3}
  \fmfforce{1w,0h}{v4}
  \fmfforce{1/2w,1/2h}{v5}
  \fmf{plain}{v2,v4}
  \fmf{plain}{v1,v3}
  \fmfv{decor.size=0, label=${\scs 1}$, l.dist=1mm, l.angle=-135}{v1}
  \fmfv{decor.size=0, label=${\scs 2}$, l.dist=1mm, l.angle=135}{v2}
  \fmfv{decor.size=0, label=${\scs 3}$, l.dist=1mm, l.angle=45}{v3}
  \fmfv{decor.size=0, label=${\scs 4}$, l.dist=1mm, l.angle=-45}{v4}
  \fmfdot{v5}
  \end{fmfgraph*} } }
\hspace*{3mm} . \label{70} \\ \no  
\eeq
With this we get from (\ref{REK1}) the second-order contribution to the connected two-point function\\
\beq
  \parbox{10mm}{\centerline{
  \begin{fmfgraph*}(7,3)
  \setval
  \fmfforce{0w,1/2h}{v1}
  \fmfforce{1w,1/2h}{v2}
  \fmfforce{1/2w,1/2h}{v3}
  \fmf{double,width=0.2mm}{v2,v1}
  \fmfv{decor.size=0, label=${\scs 1}$, l.dist=1mm, l.angle=-180}{v1}
  \fmfv{decor.size=0, label=${\scs 2}$, l.dist=1mm, l.angle=0}{v2}
  \fmfv{decor.size=0, label=${\scs (2)}$, l.dist=1.5mm, l.angle=90}{v3}
  \end{fmfgraph*}}}
\hspace*{5mm} = \hspace*{3mm} \frac{1}{4}  \hspace*{2mm} 
  \parbox{13mm}{\centerline{
  \begin{fmfgraph*}(10,10)
  \setval
  \fmfforce{0w,2/10h}{v1}
  \fmfforce{1/2w,2/10h}{v2}
  \fmfforce{1w,2/10h}{v3}
  \fmfforce{1/2w,7/10h}{v4}
  \fmfforce{1/2w,12/10h}{v5}
  \fmf{plain}{v1,v3}
  \fmf{plain,left=1}{v2,v4,v2}
  \fmf{plain,left=1}{v4,v5,v4}
  \fmfv{decor.size=0, label=${\scs 1}$, l.dist=1mm, l.angle=-180}{v1}
  \fmfv{decor.size=0, label=${\scs 2}$, l.dist=1mm, l.angle=0}{v3}
  \fmfdot{v2}
  \fmfdot{v4}
  \end{fmfgraph*}}}
\hspace*{3mm} + \hspace*{1.5mm} \frac{1}{4} \hspace*{2mm}
  \parbox{20mm}{\centerline{
  \begin{fmfgraph*}(17,5)
  \setval
  \fmfforce{0w,0h}{v1}
  \fmfforce{5/17w,0h}{v2}
  \fmfforce{12/17w,0h}{v3}
  \fmfforce{1w,0h}{v4}
  \fmfforce{5/17w,1h}{v5}
  \fmfforce{12/17w,1h}{v6}
  \fmf{plain}{v1,v4}
  \fmf{plain,left=1}{v2,v5,v2}
  \fmf{plain,left=1}{v3,v6,v3}
  \fmfv{decor.size=0, label=${\scs 1}$, l.dist=1mm, l.angle=-180}{v1}
  \fmfv{decor.size=0, label=${\scs 2}$, l.dist=1mm, l.angle=0}{v4}
  \fmfdot{v2}
  \fmfdot{v3}
  \end{fmfgraph*}}}
\hspace*{3mm} + \hspace*{1.5mm} \frac{1}{6} \hspace*{2.5mm}
 \parbox{15mm}{\centerline{
  \begin{fmfgraph*}(11,6)
  \setval
  \fmfforce{0w,1/2h}{v1}
  \fmfforce{2.5/11w,1/2h}{v2}
  \fmfforce{8.5/11w,1/2h}{v3}
  \fmfforce{1w,1/2h}{v4}
  \fmf{plain}{v1,v4}
  \fmf{plain,left=1}{v2,v3,v2}
  \fmfv{decor.size=0, label=${\scs 1}$, l.dist=1mm, l.angle=-180}{v1}
  \fmfv{decor.size=0, label=${\scs 2}$, l.dist=1mm, l.angle=0}{v4}
  \fmfdot{v2}
  \fmfdot{v3}
  \end{fmfgraph*}}}
\hspace*{3mm} .
\label{73}
\eeq
Amputating one line from (\ref{70}),
\beq
\raisebox{1mm}{\begin{minipage}[l]{1.6cm}
\beq
\dphi{
  \parbox{14mm}{\centerline{
  \begin{fmfgraph*}(10.5,10.5)
  \setval
  \fmfforce{0w,2/10.5h}{v1}
  \fmfforce{0w,12.5/10.5h}{v2}
  \fmfforce{1w,12.5/10.5h}{v3}
  \fmfforce{1w,2/10.5h}{v4}
  \fmfforce{1/3w,5.5/10.5h}{v5}
  \fmfforce{1/3w,9/10.5h}{v6}
  \fmfforce{2/3w,9/10.5h}{v7}
  \fmfforce{2/3w,5.5/10.5h}{v8}
  \fmfforce{1/2w,7.25/10.5h}{v9}
  \fmf{plain}{v1,v5}
  \fmf{plain}{v2,v6}
  \fmf{plain}{v3,v7}
  \fmf{plain}{v4,v8}
  \fmf{plain,width=0.2mm,left=1}{v5,v7,v5}
  \fmfv{decor.size=0, label=${\scs 5}$, l.dist=0.5mm, l.angle=-180}{v1}
  \fmfv{decor.size=0, label=${\scs 2}$, l.dist=0.5mm, l.angle=180}{v2}
  \fmfv{decor.size=0, label=${\scs 3}$, l.dist=0.5mm, l.angle=0}{v3}
  \fmfv{decor.size=0, label=${\scs 4}$, l.dist=0.5mm, l.angle=-0}{v4}
  \fmfv{decor.size=0, label=${\scs 1}$, l.dist=0mm, l.angle=90}{v9}
  \end{fmfgraph*} } } 
}{6}{7}
\no  \eeq 
\end{minipage}}
\hspace{3mm}&=&\hspace*{2mm} \frac{1}{2} \hspace*{2mm} \delta_{26} \hspace*{2mm} 
  \parbox{15mm}{\centerline{
  \begin{fmfgraph*}(10,10)
  \setval
  \fmfforce{0w,1/2h}{v1}
  \fmfforce{1/2w,4/5h}{v2}
  \fmfforce{1w,1/2h}{v3}
  \fmfforce{1/2w,0h}{v4}
  \fmfforce{1/2w,1/2h}{v5}
  \fmf{plain}{v2,v4}
  \fmf{plain}{v1,v3}
  \fmfv{decor.size=0, label=${\scs 5}$, l.dist=0.5mm, l.angle=180}{v1}
  \fmfv{decor.size=0, label=${\scs 7}$, l.dist=1mm, l.angle=90}{v2}
  \fmfv{decor.size=0, label=${\scs 3}$, l.dist=0.5mm, l.angle=0}{v3}
  \fmfv{decor.size=0, label=${\scs 4}$, l.dist=0.5mm, l.angle=-90}{v4}
  \fmfdot{v5}
  \end{fmfgraph*} } }
\hs + \hs \frac{1}{2} \hspace*{2mm} \delta_{36} \hspace*{2mm} 
  \parbox{15mm}{\centerline{
  \begin{fmfgraph*}(10,10)
  \setval
  \fmfforce{0w,1/2h}{v1}
  \fmfforce{1/2w,1h}{v2}
  \fmfforce{4/5w,1/2h}{v3}
  \fmfforce{1/2w,0h}{v4}
  \fmfforce{1/2w,1/2h}{v5}
  \fmf{plain}{v2,v4}
  \fmf{plain}{v1,v3}
  \fmfv{decor.size=0, label=${\scs 5}$, l.dist=0.5mm, l.angle=180}{v1}
  \fmfv{decor.size=0, label=${\scs 2}$, l.dist=0.5mm, l.angle=90}{v2}
  \fmfv{decor.size=0, label=${\scs 7}$, l.dist=1mm, l.angle=0}{v3}
  \fmfv{decor.size=0, label=${\scs 4}$, l.dist=0.5mm, l.angle=-90}{v4}
  \fmfdot{v5}
  \end{fmfgraph*} } }
\hs + \hs \frac{1}{2} \hspace*{2mm} \delta_{46} \hspace*{2mm} 
  \parbox{15mm}{\centerline{
  \begin{fmfgraph*}(10,10)
  \setval
  \fmfforce{0w,1/2h}{v1}
  \fmfforce{1/2w,1h}{v2}
  \fmfforce{1w,1/2h}{v3}
  \fmfforce{1/2w,1/5h}{v4}
  \fmfforce{1/2w,1/2h}{v5}
  \fmf{plain}{v2,v4}
  \fmf{plain}{v1,v3}
  \fmfv{decor.size=0, label=${\scs 5}$, l.dist=0.5mm, l.angle=180}{v1}
  \fmfv{decor.size=0, label=${\scs 2}$, l.dist=0.5mm, l.angle=90}{v2}
  \fmfv{decor.size=0, label=${\scs 3}$, l.dist=0.5mm, l.angle=0}{v3}
  \fmfv{decor.size=0, label=${\scs 7}$, l.dist=1mm, l.angle=-90}{v4}
  \fmfdot{v5}
  \end{fmfgraph*} } }
\hs + \hs  \frac{1}{2} \hspace*{2mm} \delta_{27} \hspace*{2mm} 
  \parbox{15mm}{\centerline{
  \begin{fmfgraph*}(10,10)
  \setval
  \fmfforce{0w,1/2h}{v1}
  \fmfforce{1/2w,4/5h}{v2}
  \fmfforce{1w,1/2h}{v3}
  \fmfforce{1/2w,0h}{v4}
  \fmfforce{1/2w,1/2h}{v5}
  \fmf{plain}{v2,v4}
  \fmf{plain}{v1,v3}
  \fmfv{decor.size=0, label=${\scs 5}$, l.dist=0.5mm, l.angle=180}{v1}
  \fmfv{decor.size=0, label=${\scs 6}$, l.dist=1mm, l.angle=90}{v2}
  \fmfv{decor.size=0, label=${\scs 3}$, l.dist=0.5mm, l.angle=0}{v3}
  \fmfv{decor.size=0, label=${\scs 4}$, l.dist=0.5mm, l.angle=-90}{v4}
  \fmfdot{v5}
  \end{fmfgraph*} } }
\no  \\*[4mm]
&& \hspace*{2mm} + \hspace*{2mm} \frac{1}{2} \hspace*{2mm} \delta_{37} \hspace*{2mm} 
  \parbox{15mm}{\centerline{
  \begin{fmfgraph*}(10,10)
  \setval
  \fmfforce{0w,1/2h}{v1}
  \fmfforce{1/2w,1h}{v2}
  \fmfforce{4/5w,1/2h}{v3}
  \fmfforce{1/2w,0h}{v4}
  \fmfforce{1/2w,1/2h}{v5}
  \fmf{plain}{v2,v4}
  \fmf{plain}{v1,v3}
  \fmfv{decor.size=0, label=${\scs 5}$, l.dist=0.5mm, l.angle=180}{v1}
  \fmfv{decor.size=0, label=${\scs 2}$, l.dist=0.5mm, l.angle=90}{v2}
  \fmfv{decor.size=0, label=${\scs 6}$, l.dist=1mm, l.angle=0}{v3}
  \fmfv{decor.size=0, label=${\scs 4}$, l.dist=0.5mm, l.angle=-90}{v4}
  \fmfdot{v5}
  \end{fmfgraph*} } }
\hs + \hs \frac{1}{2} \hspace*{2mm} \delta_{47} \hspace*{2mm} 
  \parbox{15mm}{\centerline{
  \begin{fmfgraph*}(10,10)
  \setval
  \fmfforce{0w,1/2h}{v1}
  \fmfforce{1/2w,1h}{v2}
  \fmfforce{1w,1/2h}{v3}
  \fmfforce{1/2w,1/5h}{v4}
  \fmfforce{1/2w,1/2h}{v5}
  \fmf{plain}{v2,v4}
  \fmf{plain}{v1,v3}
  \fmfv{decor.size=0, label=${\scs 5}$, l.dist=0.5mm, l.angle=180}{v1}
  \fmfv{decor.size=0, label=${\scs 2}$, l.dist=0.5mm, l.angle=90}{v2}
  \fmfv{decor.size=0, label=${\scs 3}$, l.dist=0.5mm, l.angle=0}{v3}
  \fmfv{decor.size=0, label=${\scs 6}$, l.dist=1mm, l.angle=-90}{v4}
  \fmfdot{v5}
  \end{fmfgraph*} } }
\hs + \hs  \frac{1}{2} \hspace*{2.5mm} \delta_{56} \hspace*{1.5mm} 
  \parbox{15mm}{\centerline{
  \begin{fmfgraph*}(10,10)
  \setval
  \fmfforce{1/5w,1/2h}{v1}
  \fmfforce{1/2w,1h}{v2}
  \fmfforce{1w,1/2h}{v3}
  \fmfforce{1/2w,0h}{v4}
  \fmfforce{1/2w,1/2h}{v5}
  \fmf{plain}{v2,v4}
  \fmf{plain}{v1,v3}
  \fmfv{decor.size=0, label=${\scs 7}$, l.dist=1mm, l.angle=180}{v1}
  \fmfv{decor.size=0, label=${\scs 2}$, l.dist=0.5mm, l.angle=90}{v2}
  \fmfv{decor.size=0, label=${\scs 3}$, l.dist=0.5mm, l.angle=0}{v3}
  \fmfv{decor.size=0, label=${\scs 4}$, l.dist=0.5mm, l.angle=-90}{v4}
  \fmfdot{v5}
  \end{fmfgraph*} } }
\hs + \hs \frac{1}{2} \hspace*{2.5mm} \delta_{57} \hspace*{1.5mm} 
  \parbox{15mm}{\centerline{
  \begin{fmfgraph*}(10,10)
  \setval
  \fmfforce{1/5w,1/2h}{v1}
  \fmfforce{1/2w,1h}{v2}
  \fmfforce{1w,1/2h}{v3}
  \fmfforce{1/2w,0h}{v4}
  \fmfforce{1/2w,1/2h}{v5}
  \fmf{plain}{v2,v4}
  \fmf{plain}{v1,v3}
  \fmfv{decor.size=0, label=${\scs 6}$, l.dist=1mm, l.angle=180}{v1}
  \fmfv{decor.size=0, label=${\scs 2}$, l.dist=0.5mm, l.angle=90}{v2}
  \fmfv{decor.size=0, label=${\scs 3}$, l.dist=0.5mm, l.angle=0}{v3}
  \fmfv{decor.size=0, label=${\scs 4}$, l.dist=0.5mm, l.angle=-90}{v4}
  \fmfdot{v5}
  \end{fmfgraph*} } }
\hspace*{0.3cm},  \label{71} \\ && \no
\eeq
we find the second-order contribution to the connected four-point function from (\ref{REK2}):\\
\beq
  \parbox{11mm}{\centerline{
  \begin{fmfgraph*}(10.5,10.5)
  \setval
  \fmfforce{0w,0h}{v1}
  \fmfforce{0w,1h}{v2}
  \fmfforce{1w,1h}{v3}
  \fmfforce{1w,0h}{v4}
  \fmfforce{3.5/10.5w,3.5/10.5h}{v5}
  \fmfforce{3.5/10.5w,7/10.5h}{v6}
  \fmfforce{7/10.5w,7/10.5h}{v7}
  \fmfforce{7/10.5w,3.5/10.5h}{v8}
  \fmfforce{1/2w,1/2h}{v9}
  \fmf{plain}{v1,v5}
  \fmf{plain}{v2,v6}
  \fmf{plain}{v3,v7}
  \fmf{plain}{v4,v8}
  \fmf{plain,width=0.2mm,left=1}{v5,v7,v5}
  \fmfv{decor.size=0, label=${\scs 1}$, l.dist=1mm, l.angle=-135}{v1}
  \fmfv{decor.size=0, label=${\scs 2}$, l.dist=1mm, l.angle=135}{v2}
  \fmfv{decor.size=0, label=${\scs 3}$, l.dist=1mm, l.angle=45}{v3}
  \fmfv{decor.size=0, label=${\scs 4}$, l.dist=1mm, l.angle=-45}{v4}
  \fmfv{decor.size=0, label=${\scs 2}$, l.dist=0mm, l.angle=90}{v9}
  \end{fmfgraph*} } } 
\hspace*{4mm} & = & \hspace*{2mm} \frac{1}{2} \left( \hspace*{2mm}
  \parbox{18mm}{\centerline{
  \begin{fmfgraph*}(16,10)
  \setval
  \fmfforce{0w,1/2h}{v1}
  \fmfforce{11/16w,1h}{v2}
  \fmfforce{1w,1/2h}{v3}
  \fmfforce{11/16w,0h}{v4}
  \fmfforce{5/16w,1/2h}{v5}
  \fmfforce{11/16w,1/2h}{v6}
  \fmfforce{5/16w,1h}{v7}
  \fmf{plain}{v1,v3}
  \fmf{plain,left=1}{v5,v7,v5}
  \fmf{plain}{v2,v4}
  \fmfv{decor.size=0, label=${\scs 1}$, l.dist=1mm, l.angle=180}{v1}
  \fmfv{decor.size=0, label=${\scs 2}$, l.dist=1mm, l.angle=90}{v2}
  \fmfv{decor.size=0, label=${\scs 3}$, l.dist=1mm, l.angle=0}{v3}
  \fmfv{decor.size=0, label=${\scs 4}$, l.dist=1mm, l.angle=-90}{v4}
  \fmfdot{v5}
  \fmfdot{v6}
  \end{fmfgraph*} } } 
\hspace*{2.5mm} + \hspace*{2.5mm}
  \parbox{18mm}{\centerline{
  \begin{fmfgraph*}(16,10)
  \setval
  \fmfforce{0w,1/2h}{v1}
  \fmfforce{11/16w,1h}{v2}
  \fmfforce{1w,1/2h}{v3}
  \fmfforce{11/16w,0h}{v4}
  \fmfforce{5/16w,1/2h}{v5}
  \fmfforce{11/16w,1/2h}{v6}
  \fmfforce{5/16w,1h}{v7}
  \fmf{plain}{v1,v3}
  \fmf{plain,left=1}{v5,v7,v5}
  \fmf{plain}{v2,v4}
  \fmfv{decor.size=0, label=${\scs 2}$, l.dist=1mm, l.angle=180}{v1}
  \fmfv{decor.size=0, label=${\scs 3}$, l.dist=1mm, l.angle=90}{v2}
  \fmfv{decor.size=0, label=${\scs 4}$, l.dist=1mm, l.angle=0}{v3}
  \fmfv{decor.size=0, label=${\scs 1}$, l.dist=1mm, l.angle=-90}{v4}
  \fmfdot{v5}
  \fmfdot{v6}
  \end{fmfgraph*} } } 
\hspace*{2.5mm} + \hspace*{2.5mm}
  \parbox{18mm}{\centerline{
  \begin{fmfgraph*}(16,10)
  \setval
  \fmfforce{0w,1/2h}{v1}
  \fmfforce{11/16w,1h}{v2}
  \fmfforce{1w,1/2h}{v3}
  \fmfforce{11/16w,0h}{v4}
  \fmfforce{5/16w,1/2h}{v5}
  \fmfforce{11/16w,1/2h}{v6}
  \fmfforce{5/16w,1h}{v7}
  \fmf{plain}{v1,v3}
  \fmf{plain,left=1}{v5,v7,v5}
  \fmf{plain}{v2,v4}
  \fmfv{decor.size=0, label=${\scs 3}$, l.dist=1mm, l.angle=180}{v1}
  \fmfv{decor.size=0, label=${\scs 4}$, l.dist=1mm, l.angle=90}{v2}
  \fmfv{decor.size=0, label=${\scs 1}$, l.dist=1mm, l.angle=0}{v3}
  \fmfv{decor.size=0, label=${\scs 2}$, l.dist=1mm, l.angle=-90}{v4}
  \fmfdot{v5}
  \fmfdot{v6}
  \end{fmfgraph*} } } 
\hspace*{2.5mm} + \hspace*{2.5mm}
  \parbox{18mm}{\centerline{
  \begin{fmfgraph*}(16,10)
  \setval
  \fmfforce{0w,1/2h}{v1}
  \fmfforce{11/16w,1h}{v2}
  \fmfforce{1w,1/2h}{v3}
  \fmfforce{11/16w,0h}{v4}
  \fmfforce{5/16w,1/2h}{v5}
  \fmfforce{11/16w,1/2h}{v6}
  \fmfforce{5/16w,1h}{v7}
  \fmf{plain}{v1,v3}
  \fmf{plain,left=1}{v5,v7,v5}
  \fmf{plain}{v2,v4}
  \fmfv{decor.size=0, label=${\scs 4}$, l.dist=1mm, l.angle=180}{v1}
  \fmfv{decor.size=0, label=${\scs 1}$, l.dist=1mm, l.angle=90}{v2}
  \fmfv{decor.size=0, label=${\scs 2}$, l.dist=1mm, l.angle=0}{v3}
  \fmfv{decor.size=0, label=${\scs 3}$, l.dist=1mm, l.angle=-90}{v4}
  \fmfdot{v5}
  \fmfdot{v6}
  \end{fmfgraph*} } } 
\hspace*{2mm}  \right)
\no  \\*[6mm]
& &\hspace*{2mm} + \hspace*{2mm} \frac{1}{2} \hs \Bigg(  \hs  
  \parbox{18mm}{\centerline{
  \begin{fmfgraph*}(12,7)
  \setval
  \fmfforce{0w,0h}{v1}
  \fmfforce{0w,1h}{v2}
  \fmfforce{1w,1h}{v3}
  \fmfforce{1w,0h}{v4}
  \fmfforce{3.5/12w,1/2h}{v5}
  \fmfforce{8.5/12w,1/2h}{v6}
  \fmf{plain}{v1,v5}
  \fmf{plain}{v2,v5}
  \fmf{plain}{v3,v6}
  \fmf{plain}{v4,v6}
  \fmf{plain,left=1}{v5,v6}
  \fmf{plain,right=1}{v5,v6}
  \fmfv{decor.size=0, label=${\scs 1}$, l.dist=1mm, l.angle=-135}{v1}
  \fmfv{decor.size=0, label=${\scs 2}$, l.dist=1mm, l.angle=135}{v2}
  \fmfv{decor.size=0, label=${\scs 3}$, l.dist=1mm, l.angle=45}{v3}
  \fmfv{decor.size=0, label=${\scs 4}$, l.dist=1mm, l.angle=-45}{v4}
  \fmfdot{v5}
  \fmfdot{v6}
  \end{fmfgraph*} } } 
\hs + \hs 
  \parbox{18mm}{\centerline{
  \begin{fmfgraph*}(12,7)
  \setval
  \fmfforce{0w,0h}{v1}
  \fmfforce{0w,1h}{v2}
  \fmfforce{1w,1h}{v3}
  \fmfforce{1w,0h}{v4}
  \fmfforce{3.5/12w,1/2h}{v5}
  \fmfforce{8.5/12w,1/2h}{v6}
  \fmf{plain}{v1,v5}
  \fmf{plain}{v2,v5}
  \fmf{plain}{v3,v6}
  \fmf{plain}{v4,v6}
  \fmf{plain,left=1}{v5,v6}
  \fmf{plain,right=1}{v5,v6}
  \fmfv{decor.size=0, label=${\scs 1}$, l.dist=1mm, l.angle=-135}{v1}
  \fmfv{decor.size=0, label=${\scs 3}$, l.dist=1mm, l.angle=135}{v2}
  \fmfv{decor.size=0, label=${\scs 4}$, l.dist=1mm, l.angle=45}{v3}
  \fmfv{decor.size=0, label=${\scs 2}$, l.dist=1mm, l.angle=-45}{v4}
  \fmfdot{v5}
  \fmfdot{v6}
  \end{fmfgraph*} } } 
\hs + \hs
  \parbox{18mm}{\centerline{
  \begin{fmfgraph*}(12,7)
  \setval
  \fmfforce{0w,0h}{v1}
  \fmfforce{0w,1h}{v2}
  \fmfforce{1w,1h}{v3}
  \fmfforce{1w,0h}{v4}
  \fmfforce{3.5/12w,1/2h}{v5}
  \fmfforce{8.5/12w,1/2h}{v6}
  \fmf{plain}{v1,v5}
  \fmf{plain}{v2,v5}
  \fmf{plain}{v3,v6}
  \fmf{plain}{v4,v6}
  \fmf{plain,left=1}{v5,v6}
  \fmf{plain,right=1}{v5,v6}
  \fmfv{decor.size=0, label=${\scs 1}$, l.dist=1mm, l.angle=-135}{v1}
  \fmfv{decor.size=0, label=${\scs 4}$, l.dist=1mm, l.angle=135}{v2}
  \fmfv{decor.size=0, label=${\scs 2}$, l.dist=1mm, l.angle=45}{v3}
  \fmfv{decor.size=0, label=${\scs 3}$, l.dist=1mm, l.angle=-45}{v4}
  \fmfdot{v5}
  \fmfdot{v6}
  \end{fmfgraph*} } } 
\hs  \Bigg)  \hspace*{3mm} .
\label{74} \\ && \no 
\eeq
The results (\ref{69})--(\ref{73}), and (\ref{74}) are listed in Table \ref{tab1} and \ref{tab2}
which show all diagrams of the connected two- and four-point function up to the forth perturbative order irrespective of 
their spatial indices. However, to evaluate the recursion relations in higher orders, it becomes necessary to
reassign the spatial indices to the end points of the diagrams of the connected two- and four-point function in Table \ref{tab1}
and \ref{tab2}. This involves a decomposition of the weights shown in Table \ref{tab1}
and \ref{tab2} due to symmetry considerations. To this end we characterize a diagram by its symmetry degree $N$.
Thus reassigning the spatial indices to the 
end points of the diagrams of the connected two- and four-point function leads to $24/N$ different diagrams.
As an example for determining the symmetry degree $N$ we consider diagram \#4.5 in Table \ref{tab2}.
Successively assigning the indices $1,2,3,4$ to the respective end points leads to $N=4$:\\
\beq
& & 24 \hs
  \parbox{14mm}{\centerline{
  \begin{fmfgraph}(12,13)
  \setval
  \fmfforce{0w,0h}{i1}
  \fmfforce{0w,1h}{i2}
  \fmfforce{1w,0h}{o1}
  \fmfforce{1w,1h}{o2}
  \fmfforce{3.5/12w,3.5/13h}{v1}
  \fmfforce{3.5/12w,9.5/13h}{v2}
  \fmfforce{8.5/12w,9.5/13h}{v3}
  \fmfforce{8.5/12w,3.5/13h}{v4}
  \fmf{plain}{i1,v1}
  \fmf{plain}{i2,v2}
  \fmf{plain}{o1,v4}
  \fmf{plain}{o2,v3}
  \fmf{plain,left=1}{v1,v3,v1}
  \fmf{plain,right=0.3}{v1,v2}
  \fmf{plain,right=0.3}{v3,v4}
  \fmfdot{v1,v2,v3,v4}
  \end{fmfgraph} } }
\hspace*{4mm} \mapsto \hspace*{5mm} 24 \hs
  \parbox{14mm}{\centerline{
  \begin{fmfgraph*}(12,13)
  \setval
  \fmfforce{0w,0h}{i1}
  \fmfforce{0w,1h}{i2}
  \fmfforce{1w,0h}{o1}
  \fmfforce{1w,1h}{o2}
  \fmfforce{3.5/12w,3.5/13h}{v1}
  \fmfforce{3.5/12w,9.5/13h}{v2}
  \fmfforce{8.5/12w,9.5/13h}{v3}
  \fmfforce{8.5/12w,3.5/13h}{v4}
  \fmf{plain}{i1,v1}
  \fmf{plain}{i2,v2}
  \fmf{plain}{o1,v4}
  \fmf{plain}{o2,v3}
  \fmf{plain,left=1}{v1,v3,v1}
  \fmf{plain,right=0.3}{v1,v2}
  \fmf{plain,right=0.3}{v3,v4}
  \fmfdot{v1,v2,v3,v4}
  \fmfv{decor.size=0, label=${\scs 1}$, l.dist=1mm, l.angle=-135}{i1}
  \end{fmfgraph*} } }
\hspace*{4mm} \mapsto \hspace*{5mm} 8 \hs
  \parbox{14mm}{\centerline{
  \begin{fmfgraph*}(12,13)
  \setval
  \fmfforce{0w,0h}{i1}
  \fmfforce{0w,1h}{i2}
  \fmfforce{1w,0h}{o1}
  \fmfforce{1w,1h}{o2}
  \fmfforce{3.5/12w,3.5/13h}{v1}
  \fmfforce{3.5/12w,9.5/13h}{v2}
  \fmfforce{8.5/12w,9.5/13h}{v3}
  \fmfforce{8.5/12w,3.5/13h}{v4}
  \fmf{plain}{i1,v1}
  \fmf{plain}{i2,v2}
  \fmf{plain}{o1,v4}
  \fmf{plain}{o2,v3}
  \fmf{plain,left=1}{v1,v3,v1}
  \fmf{plain,right=0.3}{v1,v2}
  \fmf{plain,right=0.3}{v3,v4}
  \fmfdot{v1,v2,v3,v4}
  \fmfv{decor.size=0, label=${\scs 1}$, l.dist=1mm, l.angle=-135}{i1}
  \fmfv{decor.size=0, label=${\scs 2}$, l.dist=1mm, l.angle=135}{i2}
  \end{fmfgraph*} } }
\hs + \hspace*{1.5mm} 8 \hs
  \parbox{14mm}{\centerline{
  \begin{fmfgraph*}(12,13)
  \setval
  \fmfforce{0w,0h}{i1}
  \fmfforce{0w,1h}{i2}
  \fmfforce{1w,0h}{o1}
  \fmfforce{1w,1h}{o2}
  \fmfforce{3.5/12w,3.5/13h}{v1}
  \fmfforce{3.5/12w,9.5/13h}{v2}
  \fmfforce{8.5/12w,9.5/13h}{v3}
  \fmfforce{8.5/12w,3.5/13h}{v4}
  \fmf{plain}{i1,v1}
  \fmf{plain}{i2,v2}
  \fmf{plain}{o1,v4}
  \fmf{plain}{o2,v3}
  \fmf{plain,left=1}{v1,v3,v1}
  \fmf{plain,right=0.3}{v1,v2}
  \fmf{plain,right=0.3}{v3,v4}
  \fmfdot{v1,v2,v3,v4}
  \fmfv{decor.size=0, label=${\scs 1}$, l.dist=1mm, l.angle=-135}{i1}
  \fmfv{decor.size=0, label=${\scs 2}$, l.dist=1mm, l.angle=45}{o2}
  \end{fmfgraph*} } }
\hs + \hspace*{1.5mm} 8 \hs
  \parbox{14mm}{\centerline{
  \begin{fmfgraph*}(12,13)
  \setval
  \fmfforce{0w,0h}{i1}
  \fmfforce{0w,1h}{i2}
  \fmfforce{1w,0h}{o1}
  \fmfforce{1w,1h}{o2}
  \fmfforce{3.5/12w,3.5/13h}{v1}
  \fmfforce{3.5/12w,9.5/13h}{v2}
  \fmfforce{8.5/12w,9.5/13h}{v3}
  \fmfforce{8.5/12w,3.5/13h}{v4}
  \fmf{plain}{i1,v1}
  \fmf{plain}{i2,v2}
  \fmf{plain}{o1,v4}
  \fmf{plain}{o2,v3}
  \fmf{plain,left=1}{v1,v3,v1}
  \fmf{plain,right=0.3}{v1,v2}
  \fmf{plain,right=0.3}{v3,v4}
  \fmfdot{v1,v2,v3,v4}
  \fmfv{decor.size=0, label=${\scs 1}$, l.dist=1mm, l.angle=-135}{i1}
  \fmfv{decor.size=0, label=${\scs 2}$, l.dist=1mm, l.angle=-45}{o1}
  \end{fmfgraph*} } }
\hspace*{4mm} \mapsto 
\no  \\*[8mm]
& & \mapsto \hspace*{4mm} 4 \hs
  \parbox{14mm}{\centerline{
  \begin{fmfgraph*}(12,13)
  \setval
  \fmfforce{0w,0h}{i1}
  \fmfforce{0w,1h}{i2}
  \fmfforce{1w,0h}{o1}
  \fmfforce{1w,1h}{o2}
  \fmfforce{3.5/12w,3.5/13h}{v1}
  \fmfforce{3.5/12w,9.5/13h}{v2}
  \fmfforce{8.5/12w,9.5/13h}{v3}
  \fmfforce{8.5/12w,3.5/13h}{v4}
  \fmf{plain}{i1,v1}
  \fmf{plain}{i2,v2}
  \fmf{plain}{o1,v4}
  \fmf{plain}{o2,v3}
  \fmf{plain,left=1}{v1,v3,v1}
  \fmf{plain,right=0.3}{v1,v2}
  \fmf{plain,right=0.3}{v3,v4}
  \fmfdot{v1,v2,v3,v4}
  \fmfv{decor.size=0, label=${\scs 1}$, l.dist=1mm, l.angle=-135}{i1}
  \fmfv{decor.size=0, label=${\scs 2}$, l.dist=1mm, l.angle=135}{i2}
  \fmfv{decor.size=0, label=${\scs 3}$, l.dist=1mm, l.angle=45}{o2}
  \fmfv{decor.size=0, label=${\scs 4}$, l.dist=1mm, l.angle=-45}{o1}
  \end{fmfgraph*} } }
\hspace*{0.5mm} + \hspace*{1.5mm} 4 \hs
  \parbox{14mm}{\centerline{
  \begin{fmfgraph*}(12,13)
  \setval
  \fmfforce{0w,0h}{i1}
  \fmfforce{0w,1h}{i2}
  \fmfforce{1w,0h}{o1}
  \fmfforce{1w,1h}{o2}
  \fmfforce{3.5/12w,3.5/13h}{v1}
  \fmfforce{3.5/12w,9.5/13h}{v2}
  \fmfforce{8.5/12w,9.5/13h}{v3}
  \fmfforce{8.5/12w,3.5/13h}{v4}
  \fmf{plain}{i1,v1}
  \fmf{plain}{i2,v2}
  \fmf{plain}{o1,v4}
  \fmf{plain}{o2,v3}
  \fmf{plain,left=1}{v1,v3,v1}
  \fmf{plain,right=0.3}{v1,v2}
  \fmf{plain,right=0.3}{v3,v4}
  \fmfdot{v1,v2,v3,v4}
  \fmfv{decor.size=0, label=${\scs 1}$, l.dist=1mm, l.angle=-135}{i1}
  \fmfv{decor.size=0, label=${\scs 2}$, l.dist=1mm, l.angle=135}{i2}
  \fmfv{decor.size=0, label=${\scs 4}$, l.dist=1mm, l.angle=45}{o2}
  \fmfv{decor.size=0, label=${\scs 3}$, l.dist=1mm, l.angle=-45}{o1}
  \end{fmfgraph*} } }
\hspace*{0.5mm} + \hspace*{1.5mm} 4 \hs
  \parbox{14mm}{\centerline{
  \begin{fmfgraph*}(12,13)
  \setval
  \fmfforce{0w,0h}{i1}
  \fmfforce{0w,1h}{i2}
  \fmfforce{1w,0h}{o1}
  \fmfforce{1w,1h}{o2}
  \fmfforce{3.5/12w,3.5/13h}{v1}
  \fmfforce{3.5/12w,9.5/13h}{v2}
  \fmfforce{8.5/12w,9.5/13h}{v3}
  \fmfforce{8.5/12w,3.5/13h}{v4}
  \fmf{plain}{i1,v1}
  \fmf{plain}{i2,v2}
  \fmf{plain}{o1,v4}
  \fmf{plain}{o2,v3}
  \fmf{plain,left=1}{v1,v3,v1}
  \fmf{plain,right=0.3}{v1,v2}
  \fmf{plain,right=0.3}{v3,v4}
  \fmfdot{v1,v2,v3,v4}
  \fmfv{decor.size=0, label=${\scs 1}$, l.dist=1mm, l.angle=-135}{i1}
  \fmfv{decor.size=0, label=${\scs 3}$, l.dist=1mm, l.angle=135}{i2}
  \fmfv{decor.size=0, label=${\scs 2}$, l.dist=1mm, l.angle=45}{o2}
  \fmfv{decor.size=0, label=${\scs 4}$, l.dist=1mm, l.angle=-45}{o1}
  \end{fmfgraph*} } }
\hspace*{0.5mm} + \hspace*{1.5mm} 4 \hs
  \parbox{14mm}{\centerline{
  \begin{fmfgraph*}(12,13)
  \setval
  \fmfforce{0w,0h}{i1}
  \fmfforce{0w,1h}{i2}
  \fmfforce{1w,0h}{o1}
  \fmfforce{1w,1h}{o2}
  \fmfforce{3.5/12w,3.5/13h}{v1}
  \fmfforce{3.5/12w,9.5/13h}{v2}
  \fmfforce{8.5/12w,9.5/13h}{v3}
  \fmfforce{8.5/12w,3.5/13h}{v4}
  \fmf{plain}{i1,v1}
  \fmf{plain}{i2,v2}
  \fmf{plain}{o1,v4}
  \fmf{plain}{o2,v3}
  \fmf{plain,left=1}{v1,v3,v1}
  \fmf{plain,right=0.3}{v1,v2}
  \fmf{plain,right=0.3}{v3,v4}
  \fmfdot{v1,v2,v3,v4}
  \fmfv{decor.size=0, label=${\scs 1}$, l.dist=1mm, l.angle=-135}{i1}
  \fmfv{decor.size=0, label=${\scs 4}$, l.dist=1mm, l.angle=135}{i2}
  \fmfv{decor.size=0, label=${\scs 2}$, l.dist=1mm, l.angle=45}{o2}
  \fmfv{decor.size=0, label=${\scs 3}$, l.dist=1mm, l.angle=-45}{o1}
  \end{fmfgraph*} } }
\hspace*{0.5mm} + \hspace*{1.5mm} 4 \hs
  \parbox{14mm}{\centerline{
  \begin{fmfgraph*}(12,13)
  \setval
  \fmfforce{0w,0h}{i1}
  \fmfforce{0w,1h}{i2}
  \fmfforce{1w,0h}{o1}
  \fmfforce{1w,1h}{o2}
  \fmfforce{3.5/12w,3.5/13h}{v1}
  \fmfforce{3.5/12w,9.5/13h}{v2}
  \fmfforce{8.5/12w,9.5/13h}{v3}
  \fmfforce{8.5/12w,3.5/13h}{v4}
  \fmf{plain}{i1,v1}
  \fmf{plain}{i2,v2}
  \fmf{plain}{o1,v4}
  \fmf{plain}{o2,v3}
  \fmf{plain,left=1}{v1,v3,v1}
  \fmf{plain,right=0.3}{v1,v2}
  \fmf{plain,right=0.3}{v3,v4}
  \fmfdot{v1,v2,v3,v4}
  \fmfv{decor.size=0, label=${\scs 1}$, l.dist=1mm, l.angle=-135}{i1}
  \fmfv{decor.size=0, label=${\scs 3}$, l.dist=1mm, l.angle=135}{i2}
  \fmfv{decor.size=0, label=${\scs 4}$, l.dist=1mm, l.angle=45}{o2}
  \fmfv{decor.size=0, label=${\scs 2}$, l.dist=1mm, l.angle=-45}{o1}
  \end{fmfgraph*} } }
\hspace*{0.5mm} + \hspace*{1.5mm} 4 \hs
  \parbox{14mm}{\centerline{
  \begin{fmfgraph*}(12,13)
  \setval
  \fmfforce{0w,0h}{i1}
  \fmfforce{0w,1h}{i2}
  \fmfforce{1w,0h}{o1}
  \fmfforce{1w,1h}{o2}
  \fmfforce{3.5/12w,3.5/13h}{v1}
  \fmfforce{3.5/12w,9.5/13h}{v2}
  \fmfforce{8.5/12w,9.5/13h}{v3}
  \fmfforce{8.5/12w,3.5/13h}{v4}
  \fmf{plain}{i1,v1}
  \fmf{plain}{i2,v2}
  \fmf{plain}{o1,v4}
  \fmf{plain}{o2,v3}
  \fmf{plain,left=1}{v1,v3,v1}
  \fmf{plain,right=0.3}{v1,v2}
  \fmf{plain,right=0.3}{v3,v4}
  \fmfdot{v1,v2,v3,v4}
  \fmfv{decor.size=0, label=${\scs 1}$, l.dist=1mm, l.angle=-135}{i1}
  \fmfv{decor.size=0, label=${\scs 4}$, l.dist=1mm, l.angle=135}{i2}
  \fmfv{decor.size=0, label=${\scs 3}$, l.dist=1mm, l.angle=45}{o2}
  \fmfv{decor.size=0, label=${\scs 2}$, l.dist=1mm, l.angle=-45}{o1}
  \end{fmfgraph*} } }
\hspace*{1.5mm} . \la{SYMMM} \\ \no
\eeq 
Note that the weights of the diagrams of the connected
two- and four-point function in Table \ref{tab1} and \ref{tab2} have been determined by solving the graphical recursion relations.
However, as a cross-check, they also follow from the formula \ci{SYM,Neu,Verena}
\beq
\la{WEIG}
w^{(n)} = \frac{n!}{2!^{S+D}3!^T P N} \, ,
\eeq
where $n$ stands for the number of external legs and
$S,D,T$ denote the number of self-, double, triple connections between vertices. Furthermore, $P$ stands for the number
of vertex permutations leaving the diagrams unchanged.
\subsection{Connected Vacuum Diagrams} \la{VAA}
The connected vacuum diagrams of $\phi^4$-theory can be generated order by order together with their weights in two
ways. In this section we show that they follow from short-circuiting the external legs of
the diagrams of the connected two- and four-point function,
respectively. Thus our present approach is complementary to Ref.~\cite{SYM}, where a nonlinear functional integrodifferential
equation for the vacuum energy was recursively solved in a graphical way in order to directly generate the connected vacuum
diagrams.
\subsubsection{Relation to the Diagrams of the Connected Two-Point Function}
Our first approach is based on the connected two-point function $\fullg_{12}$. We start with concluding 
\beq
\la{GIT}
\fullg_{12}  = -2 \, \frac{\delta W}{\delta G^{-1}_{12}} \, ,
\eeq
which follows from (\ref{PF}), (\ref{VAS}), and (\ref{2P}) by taking into account (\ref{SR}). We can read off from (\ref{GIT})
that cutting a line of the connected diagrams of the vacuum energy in all possible ways leads to all diagrams of the
connected two-point function. Here, however, we want to 
regard (\ref{GIT}) as a functional differential equation for the vacuum energy $W$.
If the interaction $V$ vanishes, Eq.~(\ref{GIT}) is solved by the free contribution of the vacuum energy
\beq
W^{(0)} \hs = \hs - \hs \frac{1}{2} \hs {\rm Tr} \hs {\rm ln} \hs G^{-1}  \, .
\label{28}
\eeq
Here the trace of the logarithm of the kernel $G^{-1}$ is defined by the series \ci[p.~16]{Kleinert4}
\beq
{\rm Tr} \hs {\rm ln} \hs G^{-1} \hs \equiv \hs - \sum_{n=1}^{\infty} 
\hs \frac{1}{n} \int_{1...n} (\delta_{12} - G^{-1}_{12})...
(\delta_{n-1,n} - G^{-1}_{n-1,n}) (\delta_{n1} - G^{-1}_{n1})  \, ,
\label{29}
\eeq
so that we get
\beq
\label{0ID}
- 2 \, \frac{\delta W^{(0)}}{\delta G^{-1}_{12}} = G_{12} \, .
\eeq
For a non-vanishing interaction $V$, the functional differential equation (\ref{GIT}) produces
corrections to (\ref{28}), which we shall denote with $W^{({\rm int})}$. Thus the vacuum energy decomposes according to
\beq
W \hs = \hs W^{(0)} \hs + W^{({\rm int})} \hspace*{3mm} ,
\label{34}
\eeq
and we obtain together with (\ref{GIT}) and (\ref{0ID})
\beq
\label{GITb}
- \, \int_{12} G^{-1}_{12}\, \frac{\delta W^{({\rm int})}}{\delta G^{-1}_{12}} = 
\frac{1}{2} \int_{12} G^{-1}_{12} \left( \fullg_{12} - G_{12} \right) \, .
\eeq
In the following, we aim at recursively determining 
$W^{({\rm int})}$ in a graphical way. To this end we perform a perturbative expansion
of the interaction part of the vacuum energy
\beq
W^{({\rm int})} \hs = \hs \sum_{p=1}^{\infty} \hs W^{(p)} \hspace*{3mm} ,
\label{39}
\eeq
and a corresponding one of the connected two-point function
\beq
\fullg_{12} = G_{12} + \sum_{p=1}^{\infty}\fullg_{12}^{(p)} \, .
\label{34b}
\eeq
Inserting (\ref{39}) and (\ref{34b}) into (\ref{GITb}) and using the functional chain rule (\ref{13}), we obtain for $p \ge 1$
\beq
\label{34c}
\int_{12} G_{12} \hs \frac{ \delta W^{(p)} }{ \delta G_{12} } \hspace*{2mm}
= \hs \frac{1}{2} \int_{12} G^{-1}_{12} \hs \Gz^{(p)}_{12}
\hspace*{3mm} .
\eeq
The contributions $W^{(p)}$ of the vacuum energy obey the following eigenvalue problem for $p \ge 1$
\beq
\label{EIG}
\int_{12} G_{12} \hs \frac{ \delta W^{(p)} }{ \delta G_{12} } \hspace*{2mm}
= 2 p \,W^{(p)} \, .
\eeq
Indeed, in the $p$th perturbative order
the functional derivative $\delta / \delta G_{12}$ generates diagrams in each of which one of the $2p$ lines of the original
vacuum diagram is removed and, subsequently, the removed line is again reinserted. Thus the left-hand
side of (\ref{EIG}) results in counting
the lines of the vacuum diagrams in $W^{(p)}$, so that we obtain the eigenvalue $2p$. 
With (\ref{EIG}) we can explicitly solve (\ref{34c}) for
the respective perturbative contributions of the vacuum energy and obtain for $p \ge 1$
\beq
W^{(p)} \hs = \hs \frac{1}{4 \hs p} \hs \int_{12} G^{-1}_{12} \hs 
\Gz^{(p)}_{12}  \hspace*{3mm} .
\label{83}
\eeq
Now we supplement the above-mentioned Feynman rules with a graphical representation for the contributions of the vacuum energy 
\beq
W^{(p)} \,\, \equiv \,\,
\parbox{10mm}{\centerline{
  \begin{fmfgraph*}(8,6)
  \setval
  \fmfforce{0w,1/2h}{v1}
  \fmfforce{1w,1/2h}{v2}
  \fmfforce{1/2w,1h}{v3}
  \fmfforce{1/2w,0h}{v4}
  \fmfforce{1/2w,1/2h}{v5}
  \fmf{plain,left=0.4}{v1,v3,v2,v4,v1}
  \fmfv{decor.size=0, label=${\scriptstyle p}$, l.dist=0mm, l.angle=0}{v5}
  \end{fmfgraph*} } }  
\eeq
and for the kernel 
\beq
G^{-1}_{12}  \hspace*{2mm} \equiv \hspace*{2mm}
 \parbox{9mm}{\centerline{
  \begin{fmfgraph*}(5,3)
  \setval
  \fmfforce{0w,1/2h}{v1}
  \fmfforce{1w,1/2h}{v2}
  \fmfforce{1/2w,1/2h}{v3}
   \fmf{plain}{v1,v2}
  \fmfblob{1.5mm}{v3}
  \fmfv{decor.size=0, label=${\scs 1}$, l.dist=1mm, l.angle=180}{v1}
  \fmfv{decor.size=0, label=${\scs 2}$, l.dist=1mm, l.angle=0}{v2}
  \end{fmfgraph*} } }  
\hspace*{5mm} .    
\label{2Vert} \\ \no
\eeq 
The latter graphical element serves for gluing two lines together according to 
\beq
\int_{12} G_{31} G^{-1}_{12} G_{24} \hspace*{2mm} \equiv \hspace*{3mm}
  \parbox{10mm}{\centerline{
  \begin{fmfgraph*}(7,3)
  \setval
  \fmfforce{0w,1/2h}{v1}
  \fmfforce{1w,1/2h}{v2}
  \fmf{plain}{v1,v2}
  \fmfv{decor.size=0, label=${\scs 3}$, l.dist=1mm, l.angle=180}{v1}
  \fmfv{decor.size=0, label=${\scs 1}$, l.dist=1mm, l.angle=0}{v2}
  \end{fmfgraph*} } }  
\hspace*{3mm}
  \parbox{9mm}{\centerline{
  \begin{fmfgraph*}(5,3)
  \setval
  \fmfforce{0w,1/2h}{v1}
  \fmfforce{1w,1/2h}{v2}
  \fmfforce{1/2w,1/2h}{v3}
  \fmf{plain}{v1,v2}
  \fmfblob{1.5mm}{v3}
  \fmfv{decor.size=0, label=${\scs 1}$, l.dist=0.5mm, l.angle=180}{v1}
  \fmfv{decor.size=0, label=${\scs 2}$, l.dist=0.5mm, l.angle=0}{v2}
  \end{fmfgraph*} } }  
\hspace*{3mm}
  \parbox{10mm}{\centerline{
  \begin{fmfgraph*}(7,3)
  \setval
  \fmfforce{0w,1/2h}{v1}
  \fmfforce{1w,1/2h}{v2}
  \fmf{plain}{v1,v2}
  \fmfv{decor.size=0, label=${\scs 2}$, l.dist=1mm, l.angle=180}{v1}
  \fmfv{decor.size=0, label=${\scs 4}$, l.dist=1mm, l.angle=0}{v2}
  \end{fmfgraph*} } }  
\hspace*{4mm} = \hspace*{4mm}
  \parbox{10mm}{\centerline{
  \begin{fmfgraph*}(7,3)
  \setval
  \fmfforce{0w,1/2h}{v1}
  \fmfforce{1w,1/2h}{v2}
  \fmf{plain}{v1,v2}
  \fmfv{decor.size=0, label=${\scs 3}$, l.dist=1mm, l.angle=180}{v1}
  \fmfv{decor.size=0, label=${\scs 4}$, l.dist=1mm, l.angle=0}{v2}
  \end{fmfgraph*} } }  
\hspace*{6mm} ,
\label{84}
\eeq
which follows from
\beq
\int_{12} G_{31} \, G^{-1}_{12}\, G_{24} = G_{34} \, .
\eeq
Thus our result (\ref{83}) can be depicted graphically as follows:
\beq
\parbox{10mm}{\centerline{
  \begin{fmfgraph*}(8,6)
  \setval
  \fmfforce{0w,1/2h}{v1}
  \fmfforce{1w,1/2h}{v2}
  \fmfforce{1/2w,1h}{v3}
  \fmfforce{1/2w,0h}{v4}
  \fmfforce{1/2w,1/2h}{v5}
  \fmf{plain,left=0.4}{v1,v3,v2,v4,v1}
  \fmfv{decor.size=0, label=${\scriptstyle p}$, l.dist=0mm, l.angle=0}{v5}
  \end{fmfgraph*} } }  
 \hspace*{3mm} = \hspace*{3mm} \frac{1}{4 p} \hspace*{3mm}
  \parbox{10mm}{\centerline{
  \begin{fmfgraph*}(7,7)
  \setval
  \fmfforce{1/2w,0h}{v1}
  \fmfforce{1/2w,1h}{v2}
  \fmf{double,width=0.2mm,left=1}{v1,v2,v1}
  \fmfblob{1.5mm}{v2}
  \fmfv{decor.size=0, label=${\scs (p)}$, l.dist=1mm, l.angle=-90}{v1}
  \end{fmfgraph*} } }  
\hspace*{2mm} . \label{gr83} \\ \no
\eeq
It states that closing the perturbative contributions of the connected two-point function yields the corresponding
contributions of the vacuum energy. From (\ref{69}) and (\ref{gr83}) we yield for $p=1$
\beq
\parbox{10mm}{\centerline{
  \begin{fmfgraph*}(8,6)
  \setval
  \fmfforce{0w,1/2h}{v1}
  \fmfforce{1w,1/2h}{v2}
  \fmfforce{1/2w,1h}{v3}
  \fmfforce{1/2w,0h}{v4}
  \fmfforce{1/2w,1/2h}{v5}
  \fmf{plain,left=0.4}{v1,v3,v2,v4,v1}
  \fmfv{decor.size=0, label=${\scriptstyle 1}$, l.dist=0mm, l.angle=0}{v5}
  \end{fmfgraph*} } }  
\hs = \hspace*{2mm} \frac{1}{8} \hspace*{3mm}
  \parbox{10mm}{\centerline{
  \begin{fmfgraph}(10,5)
  \setval
  \fmfforce{0w,1/2h}{v1}
  \fmfforce{1/2w,1/2h}{v2}
  \fmfforce{1w,1/2h}{v3}
  \fmf{plain,left=1}{v1,v2,v1}
  \fmf{plain,left=1}{v3,v2,v3}
  \fmfdot{v2}
  \end{fmfgraph} } }  
\hspace*{3mm} . 
\label{85}
\eeq
Correspondingly, the second-order result follows from (\ref{73}) and (\ref{gr83}):
\beq
\parbox{10mm}{\centerline{
  \begin{fmfgraph*}(8,6)
  \setval
  \fmfforce{0w,1/2h}{v1}
  \fmfforce{1w,1/2h}{v2}
  \fmfforce{1/2w,1h}{v3}
  \fmfforce{1/2w,0h}{v4}
  \fmfforce{1/2w,1/2h}{v5}
  \fmf{plain,left=0.4}{v1,v3,v2,v4,v1}
  \fmfv{decor.size=0, label=${\scriptstyle 2}$, l.dist=0mm, l.angle=0}{v5}
  \end{fmfgraph*} } }  \hspace*{2mm} =  \hspace*{2mm} \frac{1}{16} \hspace*{2mm}
  \parbox{18mm}{\centerline{
  \begin{fmfgraph}(15,5)
  \setval
  \fmfforce{0w,1/2h}{v1}
  \fmfforce{1/3w,1/2h}{v2}
  \fmfforce{2/3w,1/2h}{v3}
  \fmfforce{1w,1/2h}{v4}
  \fmf{plain,left=1}{v1,v2,v1}
  \fmf{plain,left=1}{v2,v3,v2}
  \fmf{plain,left=1}{v3,v4,v3}
  \fmfdot{v2}
  \fmfdot{v3}
  \end{fmfgraph} } }  
\hspace{1mm} + \hspace*{2mm} \frac{1}{48} \hspace*{3mm}
  \parbox{10mm}{\centerline{
  \begin{fmfgraph}(8,8)
  \setval
  \fmfforce{0w,1/2h}{v1}
  \fmfforce{1w,1/2h}{v2}
  \fmf{plain,left=1}{v1,v2}
  \fmf{plain,left=0.4}{v1,v2}
  \fmf{plain,right=0.4}{v1,v2}
  \fmf{plain,right=1}{v1,v2}
  \fmfdot{v1}
  \fmfdot{v2}
  \end{fmfgraph} } }  
\hspace*{2mm} .
\label{86}
\eeq
\subsubsection{Relation to the Diagrams of the Connected Four-Point Function}
Now we elaborate a second approach which is based on closing the diagrams of the connected 
four-point function $\Gvc_{1234}$. 
To this end we insert (\ref{GIT}) into the left-hand side of the Schwinger-Dyson equation (\ref{Idy}) and use 
(\ref{13}), (\ref{0ID}) as well as the decomposition (\ref{34}) to obtain
a functional differential equation for the interaction part of the vacuum energy:
\beq
\int_{12} G_{12} \frac{\delta W^{({\rm int})}}{\delta G_{12}}  
= - \hs \frac{1}{12} \int_{1234} V_{1234} \hs \Gvc_{1234} \hs - 
\hs \frac{1}{4} \int_{1234} V_{1234} \hs \Gz_{12} \Gz_{34}   \hspace*{3mm} . 
\label{81}
\eeq
Combining the perturbation expansions (\ref{39}), (\ref{34b}) for $W^{({\rm int})}$,  $\fullg_{12}$ 
with the corresponding one for $\Gvc_{1234}$, i.e.
\beq
\Gvc_{1234} = \sum_{p=1}^{\infty} \Gz^{c,(p)}_{1234}  \, ,
\eeq
we obtain together with the eigenvalue problem (\ref{EIG}) from (\ref{81})
\beq
W^{(p+1)}   = - \hs \frac{1}{24(p+1)} \int_{1234} V_{1234} \hs \Gz^{c,(p)}_{1234} \hs - 
\hs \frac{1}{8(p+1)} \sum_{q=0}^{p} \int_{1234} V_{1234} \hs 
\Gz_{12}^{(q)} \Gz_{34}^{(p-q)}   \hspace*{3mm} . 
\label{81b}
\eeq
The graphical representation of this result reads
\beq
\hspace{2mm}
  \parbox{10mm}{\centerline{
  \begin{fmfgraph*}(9,6.8)
  \setval
  \fmfforce{0w,1/2h}{v1}
  \fmfforce{1w,1/2h}{v2}
  \fmfforce{1/2w,1h}{v3}
  \fmfforce{1/2w,0h}{v4}
  \fmfforce{1/2w,1/2h}{v5}
  \fmf{plain,left=0.4}{v1,v3,v2,v4,v1}
  \fmfv{decor.size=0, label=${\scs {p+1}}$, l.dist=0mm, l.angle=90}{v5}
  \end{fmfgraph*} } }  
\hspace*{3mm} = \hspace*{3mm} \frac{1}{24(p+1)} \hspace*{3mm}
  \parbox{14mm}{\centerline{
  \begin{fmfgraph*}(13,10)
  \setval
  \fmfforce{0w,0.5h}{v1}
  \fmfforce{10.5/13w,7.5/10h}{v3}
  \fmfforce{10.5/13w,2.5/10h}{v4}
  \fmfforce{8.5/13w,6.3/10h}{v5}
  \fmfforce{8.5/13w,3.7/10h}{v6}
  \fmfforce{10.5/13w,0.5h}{v8}
  \fmf{plain,left=0.75}{v1,v3}
  \fmf{plain,left=0.3}{v1,v5}
  \fmf{plain,right=0.75}{v1,v4}
  \fmf{plain,right=0.3}{v1,v6}
  \fmf{plain,width=0.2mm,left=1}{v3,v4,v3}
  \fmfv{decor.size=0, label=${\scs p}$, l.dist=0mm, l.angle=90}{v8}
  \fmfdot{v1}
  \end{fmfgraph*} } }
\hspace*{3mm} + \hspace*{3mm} \frac{1}{8(p+1)} \hspace*{2mm} 
\sum_{q=0}^{p} \hspace*{3mm}
  \parbox{12mm}{\centerline{
  \begin{fmfgraph*}(10,5)
  \setval
  \fmfforce{0w,1/2h}{v1}
  \fmfforce{1/2w,1/2h}{v2}
  \fmfforce{1w,1/2h}{v3}
  \fmf{double,width=0.2mm,left=1}{v1,v2,v1}
  \fmf{double,width=0.2mm,left=1}{v3,v2,v3}
  \fmfv{decor.size=0, label=${\scs (q)}$, l.dist=1mm, l.angle=180}{v1}
  \fmfv{decor.size=0, label=${\scs {(p-q)}}$, l.dist=1mm, l.angle=0}{v3}
  \fmfdot{v2}
  \end{fmfgraph*} } }  
\hspace*{8mm} .
\la{VRE} \\ \no
\eeq
Inserting the diagrams of the connected two- and four-point function from Tab. \ref{tab1} and \ref{tab2}, we obtain
the corresponding vacuum diagrams. For instance, Eq. (\ref{VRE}) reduces for $p=2$ 
\beq
  \parbox{10mm}{\centerline{
  \begin{fmfgraph*}(9,6.8)
  \setval
  \fmfforce{0w,1/2h}{v1}
  \fmfforce{1w,1/2h}{v2}
  \fmfforce{1/2w,1h}{v3}
  \fmfforce{1/2w,0h}{v4}
  \fmfforce{1/2w,1/2h}{v5}
  \fmf{plain,left=0.4}{v1,v3,v2,v4,v1}
  \fmfv{decor.size=0, label=${\scs 3}$, l.dist=0mm, l.angle=90}{v5}
  \end{fmfgraph*} } }  
\hspace*{2.5mm} = \hspace*{2.5mm} \frac{1}{72} \hspace*{3mm}
  \parbox{14mm}{\centerline{
  \begin{fmfgraph*}(13,10)
  \setval
  \fmfforce{0w,0.5h}{v1}
  \fmfforce{10.5/13w,7.5/10h}{v3}
  \fmfforce{10.5/13w,2.5/10h}{v4}
  \fmfforce{8.5/13w,6.3/10h}{v5}
  \fmfforce{8.5/13w,3.7/10h}{v6}
  \fmfforce{10.5/13w,0.5h}{v8}
  \fmf{plain,left=0.75}{v1,v3}
  \fmf{plain,left=0.3}{v1,v5}
  \fmf{plain,right=0.75}{v1,v4}
  \fmf{plain,right=0.3}{v1,v6}
  \fmf{plain,width=0.2mm,left=1}{v3,v4,v3}
  \fmfv{decor.size=0, label=${\scs 2}$, l.dist=0mm, l.angle=90}{v8}
  \fmfdot{v1}
  \end{fmfgraph*} } }
\hspace*{2.5mm} + \hspace*{2mm} \frac{1}{12} \hspace*{4mm}
  \parbox{12mm}{\centerline{
  \begin{fmfgraph*}(10,5)
  \setval
  \fmfforce{0w,1/2h}{v1}
  \fmfforce{1/2w,1/2h}{v2}
  \fmfforce{1w,1/2h}{v3}
  \fmf{double,width=0.2mm,left=1}{v1,v2,v1}
  \fmf{double,width=0.2mm,left=1}{v3,v2,v3}
  \fmfv{decor.size=0, label=${\scs (0)}$, l.dist=1mm, l.angle=180}{v1}
  \fmfv{decor.size=0, label=${\scs (2)}$, l.dist=1mm, l.angle=0}{v3}
  \fmfdot{v2}
  \end{fmfgraph*} } }  
\hspace*{4mm} + \hspace*{2mm} \frac{1}{24} \hspace*{4mm}
  \parbox{12mm}{\centerline{
  \begin{fmfgraph*}(10,5)
  \setval
  \fmfforce{0w,1/2h}{v1}
  \fmfforce{1/2w,1/2h}{v2}
  \fmfforce{1w,1/2h}{v3}
  \fmf{double,width=0.2mm,left=1}{v1,v2,v1}
  \fmf{double,width=0.2mm,left=1}{v3,v2,v3}
  \fmfv{decor.size=0, label=${\scs (1)}$, l.dist=1mm, l.angle=180}{v1}
  \fmfv{decor.size=0, label=${\scs (1)}$, l.dist=1mm, l.angle=0}{v3}
  \fmfdot{v2}
  \end{fmfgraph*} } }  
\label{82.1} 
\eeq
with the respective terms
\beq
  \parbox{14mm}{\centerline{
  \begin{fmfgraph*}(13,10)
  \setval
  \fmfforce{0w,0.5h}{v1}
  \fmfforce{10.5/13w,7.5/10h}{v3}
  \fmfforce{10.5/13w,2.5/10h}{v4}
  \fmfforce{8.5/13w,6.3/10h}{v5}
  \fmfforce{8.5/13w,3.7/10h}{v6}
  \fmfforce{10.5/13w,0.5h}{v8}
  \fmf{plain,left=0.75}{v1,v3}
  \fmf{plain,left=0.3}{v1,v5}
  \fmf{plain,right=0.75}{v1,v4}
  \fmf{plain,right=0.3}{v1,v6}
  \fmf{plain,width=0.2mm,left=1}{v3,v4,v3}
  \fmfv{decor.size=0, label=${\scs 2}$, l.dist=0mm, l.angle=90}{v8}
  \fmfdot{v1}
  \end{fmfgraph*} } }
& = &
\frac{3}{2} 
\hspace*{2mm}
  \parbox{11mm}{\centerline{
  \begin{fmfgraph}(8,8)
  \setval
  \fmfforce{0.5w,0h}{v1}
  \fmfforce{0.5w,1h}{v2}
  \fmfforce{0.066987w,0.25h}{v3}
  \fmfforce{0.93301w,0.25h}{v4}
  \fmf{plain,left=1}{v1,v2,v1}
  \fmf{plain}{v2,v3}
  \fmf{plain}{v3,v4}
  \fmf{plain}{v2,v4}
  \fmfdot{v2,v3,v4}
  \end{fmfgraph}}} 
\hspace*{1.5mm} + \hspace*{2mm} 2 \hspace*{2mm}
  \parbox{10.5mm}{\centerline{
  \begin{fmfgraph}(7.5,12.5)
  \setval
  \fmfforce{0w,0.3h}{v1}
  \fmfforce{1w,0.3h}{v2}
  \fmfforce{0.5w,0.6h}{v3}
  \fmfforce{0.5w,1h}{v4}
  \fmf{plain,left=1}{v1,v2,v1}
  \fmf{plain,left=0.4}{v1,v2,v1}
  \fmf{plain,left=1}{v3,v4,v3}
  \fmfdot{v1,v2,v3}
  \end{fmfgraph}}} 
\hspace*{2mm} , \\
  \parbox{12mm}{\centerline{
  \begin{fmfgraph*}(10,5)
  \setval
  \fmfforce{0w,1/2h}{v1}
  \fmfforce{1/2w,1/2h}{v2}
  \fmfforce{1w,1/2h}{v3}
  \fmf{double,width=0.2mm,left=1}{v1,v2,v1}
  \fmf{double,width=0.2mm,left=1}{v3,v2,v3}
  \fmfv{decor.size=0, label=${\scs (0)}$, l.dist=1mm, l.angle=180}{v1}
  \fmfv{decor.size=0, label=${\scs (2)}$, l.dist=1mm, l.angle=0}{v3}
  \fmfdot{v2}
  \end{fmfgraph*} } }  
\hspace*{4mm} &=&  
\frac{1}{6} \hspace*{3mm}
  \parbox{10.5mm}{\centerline{
  \begin{fmfgraph}(7.5,12.5)
  \setval
  \fmfforce{0w,0.3h}{v1}
  \fmfforce{1w,0.3h}{v2}
  \fmfforce{0.5w,0.6h}{v3}
  \fmfforce{0.5w,1h}{v4}
  \fmf{plain,left=1}{v1,v2,v1}
  \fmf{plain,left=0.4}{v1,v2,v1}
  \fmf{plain,left=1}{v3,v4,v3}
  \fmfdot{v1,v2,v3}
  \end{fmfgraph}}} 
\hspace*{2mm} + \hspace*{2mm} \frac{1}{4} \hspace*{2.5mm}
  \parbox{20mm}{\centerline{
  \begin{fmfgraph}(20,5)
  \setval
  \fmfleft{i1}
  \fmfright{o1}
  \fmf{plain,left=1}{i1,v1,i1}
  \fmf{plain,left=1}{v1,v2,v1}
  \fmf{plain,left=1}{v2,v3,v2}
  \fmf{plain,left=1}{o1,v3,o1}
  \fmfdot{v1,v2,v3}
  \end{fmfgraph}}}
\hspace*{3mm} + \hspace*{2mm} \frac{1}{4} \hspace*{2mm}
\parbox{17mm}{\centerline{
\begin{fmfgraph}(13,11)
\setval
\fmfforce{1/2w,1/11h}{v1}
\fmfforce{1/2w,6/11h}{v2}
\fmfforce{1/2w,1h}{v3}
\fmfforce{4.3/13w,2.4/11h}{v4}
\fmfforce{8.7/13w,2.4/11h}{v5}
\fmfforce{0w,0h}{v6}
\fmfforce{1w,0h}{v7}
\fmf{plain,left=1}{v1,v2,v1}
\fmf{plain,left=1}{v2,v3,v2}
\fmf{plain,left=1}{v4,v6,v4}
\fmf{plain,left=1}{v5,v7,v5}
\fmfdot{v2,v4,v5}
\end{fmfgraph}}} 
\hspace*{3mm} , \\[2mm]
\parbox{12mm}{\centerline{
  \begin{fmfgraph*}(10,5)
  \setval
  \fmfforce{0w,1/2h}{v1}
  \fmfforce{1/2w,1/2h}{v2}
  \fmfforce{1w,1/2h}{v3}
  \fmf{double,width=0.2mm,left=1}{v1,v2,v1}
  \fmf{double,width=0.2mm,left=1}{v3,v2,v3}
  \fmfv{decor.size=0, label=${\scs (1)}$, l.dist=1mm, l.angle=180}{v1}
  \fmfv{decor.size=0, label=${\scs (1)}$, l.dist=1mm, l.angle=0}{v3}
  \fmfdot{v2}
  \end{fmfgraph*} } }  
\hspace*{4mm} &=&\frac{1}{4} \hspace*{3mm} 
  \parbox{20mm}{\centerline{
  \begin{fmfgraph}(20,5)
  \setval
  \fmfleft{i1}
  \fmfright{o1}
  \fmf{plain,left=1}{i1,v1,i1}
  \fmf{plain,left=1}{v1,v2,v1}
  \fmf{plain,left=1}{v2,v3,v2}
  \fmf{plain,left=1}{o1,v3,o1}
  \fmfdot{v1,v2,v3}
  \end{fmfgraph}}}
\hspace*{3mm} .
\eeq
Thus we obtain the connected vacuum diagrams shown in Tab. \ref{tab3} for $p=3$:
\beq
  \parbox{10mm}{\centerline{
  \begin{fmfgraph*}(8,6)
  \setval
  \fmfforce{0w,1/2h}{v1}
  \fmfforce{1w,1/2h}{v2}
  \fmfforce{1/2w,1h}{v3}
  \fmfforce{1/2w,0h}{v4}
  \fmfforce{1/2w,1/2h}{v5}
  \fmf{plain,left=0.4}{v1,v3,v2,v4,v1}
  \fmfv{decor.size=0, label=${\scs 3}$, l.dist=0mm, l.angle=90}{v5}
  \end{fmfgraph*} } }  
\hspace*{2mm} = \hspace*{2.5mm} \frac{1}{48} \hspace*{3mm} 
  \parbox{11mm}{\centerline{
  \begin{fmfgraph}(8,8)
  \setval
  \fmfforce{0.5w,0h}{v1}
  \fmfforce{0.5w,1h}{v2}
  \fmfforce{0.066987w,0.25h}{v3}
  \fmfforce{0.93301w,0.25h}{v4}
  \fmf{plain,left=1}{v1,v2,v1}
  \fmf{plain}{v2,v3}
  \fmf{plain}{v3,v4}
  \fmf{plain}{v2,v4}
  \fmfdot{v2,v3,v4}
  \end{fmfgraph}}} 
\hspace*{2mm} + \hspace*{2mm} \frac{1}{24} \hspace*{3mm}
  \parbox{10.5mm}{\centerline{
  \begin{fmfgraph}(7.5,12.5)
  \setval
  \fmfforce{0w,0.3h}{v1}
  \fmfforce{1w,0.3h}{v2}
  \fmfforce{0.5w,0.6h}{v3}
  \fmfforce{0.5w,1h}{v4}
  \fmf{plain,left=1}{v1,v2,v1}
  \fmf{plain,left=0.4}{v1,v2,v1}
  \fmf{plain,left=1}{v3,v4,v3}
  \fmfdot{v1,v2,v3}
  \end{fmfgraph}}} 
\hspace*{2mm} + \hspace*{2mm} \frac{1}{32} \hspace*{3mm}
  \parbox{20mm}{\centerline{
  \begin{fmfgraph}(20,5)
  \setval
  \fmfleft{i1}
  \fmfright{o1}
  \fmf{plain,left=1}{i1,v1,i1}
  \fmf{plain,left=1}{v1,v2,v1}
  \fmf{plain,left=1}{v2,v3,v2}
  \fmf{plain,left=1}{o1,v3,o1}
  \fmfdot{v1,v2,v3}
  \end{fmfgraph}}}
\hspace*{3mm} + \hspace*{2mm} \frac{1}{48} \hspace*{2mm}
  \parbox{17mm}{\centerline{
  \begin{fmfgraph}(13,11)
  \setval
  \fmfforce{1/2w,1/11h}{v1}
  \fmfforce{1/2w,6/11h}{v2}
  \fmfforce{1/2w,1h}{v3}
  \fmfforce{4.3/13w,2.4/11h}{v4}
  \fmfforce{8.7/13w,2.4/11h}{v5}
  \fmfforce{0w,0h}{v6}
  \fmfforce{1w,0h}{v7}
  \fmf{plain,left=1}{v1,v2,v1}
  \fmf{plain,left=1}{v2,v3,v2}
  \fmf{plain,left=1}{v4,v6,v4}
  \fmf{plain,left=1}{v5,v7,v5}
  \fmfdot{v2,v4,v5}
  \end{fmfgraph}}} 
\hspace*{3mm} .
\label{82.2}
\eeq
The results (\ref{85}), (\ref{86}), and (\ref{82.2}) 
are listed in Table \ref{tab3} which shows also one further perturbative order.
Note that the weights of the connected vacuum diagrams obey a formula similar to (\ref{WEIG})
\beq
\la{WEIGT}
w = \frac{1}{2!^{S+D}3!^T 4!^F P} \, ,
\eeq
where $S,D,T,F$ denote the number of self-, double, triple, fourfold connections and $P$ stands for the number of
vertex permutations leaving the vacuum diagram unchanged \ci{SYM,Neu,Verena}.
\end{fmffile}
\begin{fmffile}{graph2}
\section{One-Particle Irreducible Feynman Diagrams}\la{IRRED}
So far, we have explained how to generate 
connected Feynman diagrams of the $\phi^4$-theory. We shall now eliminate from them
the reducible contributions. To this end we derive in Subsection \ref{NANU} a closed set of Schwinger-Dyson equations
for the one-particle irreducible two- and four-point function. In Subsection \ref{NANA} they are converted into graphical recursion
relations for the corresponding one-particle irreducible Feynman diagrams needed for renormalizing the $\phi^4$-theory.
Subsection \ref{IVVA} discusses how these Feynman diagrams of the one-particle irreducible two- and four-point functions
are related to the connected vacuum diagrams which are also one-particle-irreducible.
\subsection{Closed Set of Schwinger-Dyson Equations for One-Particle Irreducible Two- and Four-Point Functions}\la{NANU}
In this subsection we revisit the functional identity (\ref{Idy}) which immediately followed from the definition of the functional
integral. We show that it can be also used to derive a closed set of Schwinger-Dyson equations for the one-particle
irreducible two- and four-point functions.
\subsubsection{Field-Theoretic Definitions}
From Table \ref{tab1} we can distinguish two classes of diagrams contributing to the connected two-point function $\fullg_{12}$.
The first one contains one-particle irreducible diagrams which remain connected after amputating one arbitrary line.
The second one consists of the remaining diagrams which are called reducible. The diagrams \# 2.1 and \# 2.2, for instance,
are one-particle irreducible, whereas \# 2.3 is reducible. 
When considering the self-energy
\beq
\Sig_{12} \hspace*{2mm} = \hspace*{2mm} G^{-1}_{12}   - \Gz^{-1}_{12} \, ,
\label{sigma}
\eeq
where $\Gz^{-1}_{12}$ is the functional inverse of $\Gz_{12}$:
\beq
\int_3 \Gz_{13} \hs \Gz^{-1}_{32} \hspace*{2mm} \equiv \hspace*{2mm} \delta_{12}  \, ,
\label{101}
\eeq
we will see later on that this quantity is graphically represented by all one-particle irreducible
diagrams of the connected two-point function $\fullg_{12}$ where the external legs are omitted.
Reversely, all connected diagrams of the connected two-point function $\fullg_{12}$ are reconstructed from the one-particle
irreducible ones of the self-energy $\Sig_{12}$ according to the Dyson equation  
\beq
\Gz_{12} \hs = \hs G_{12} \hs + \hs \int_{34} G_{13} \Sig_{34}
\Gz_{42} \, ,
\label{Dyson}
\eeq
which immediately follows from (\ref{sigma}) by taking into account (\ref{FP}) and (\ref{101}). 
When the self-energy is graphically represented in Feynman diagrams by a big open dot with two legs
\beq
\Sig_{12} \hspace*{2mm} \equiv \hspace*{3mm}
 \parbox{11mm}{\centerline{
  \begin{fmfgraph*}(8,3)
  \setval
  \fmfforce{0w,1/2h}{v1}
  \fmfforce{2.5/8w,1/2h}{v2}
  \fmfforce{5.5/8w,1/2h}{v3}
  \fmfforce{1w,1/2h}{v4}
  \fmfforce{1/2w,1h}{v5}
  \fmfforce{1/2w,0h}{v6}
  \fmf{plain,width=0.2mm}{v1,v2}
  \fmf{plain,width=0.2mm}{v3,v4}
  \fmf{double,width=0.2mm,left=1}{v5,v6,v5}
  \fmfv{decor.size=0, label=${\scs 1}$, l.dist=1mm, l.angle=-180}{v1}
  \fmfv{decor.size=0, label=${\scs 2}$, l.dist=1mm, l.angle=0}{v4}
  \end{fmfgraph*} } } 
\hspace*{5mm} ,
\eeq
the Dyson equation (\ref{Dyson}) reads graphically
\beq
  \parbox{10mm}{\centerline{
  \begin{fmfgraph*}(7,3)
  \setval
  \fmfforce{0w,1/2h}{v1}
  \fmfforce{1w,1/2h}{v2}
  \fmf{double,width=0.2mm}{v2,v1}
  \fmfv{decor,size=0, label=${\scs 1}$, l.dist=1mm, l.angle=-180}{v1}
  \fmfv{decor,size=0, label=${\scs 2}$, l.dist=1mm, l.angle=0}{v2}
  \end{fmfgraph*} } } 
\hspace*{4mm} = \hspace*{4mm}
  \parbox{10mm}{\centerline{
  \begin{fmfgraph*}(7,3)
  \setval
  \fmfforce{0w,1/2h}{v1}
  \fmfforce{1w,1/2h}{v2}
  \fmf{plain,width=0.2mm}{v2,v1}
  \fmfv{decor,size=0, label=${\scs 1}$, l.dist=1mm, l.angle=-180}{v1}
  \fmfv{decor,size=0, label=${\scs 2}$, l.dist=1mm, l.angle=0}{v2}
  \end{fmfgraph*} } } 
\hspace*{4mm} + \hspace*{4mm}
  \parbox{15mm}{\centerline{
  \begin{fmfgraph*}(13,3)
  \setval
  \fmfforce{0w,1/2h}{v1}
  \fmfforce{5/13w,1/2h}{v2}
  \fmfforce{8/13w,1/2h}{v3}
  \fmfforce{1w,1/2h}{v4}
  \fmfforce{1/2w,1h}{v5}
  \fmfforce{1/2w,0h}{v6}
  \fmf{plain,width=0.2mm}{v1,v2}
  \fmf{double,width=0.2mm}{v3,v4}
  \fmf{double,width=0.2mm,left=1}{v5,v6,v5}
  \fmfv{decor.size=0, label=${\scs 1}$, l.dist=1mm, l.angle=-180}{v1}
  \fmfv{decor.size=0, label=${\scs 2}$, l.dist=1mm, l.angle=0}{v4}
  \end{fmfgraph*} } } 
\hspace*{5mm} .
\label{VOLL}
\eeq
In a similar way, Table \ref{tab2} shows that the diagrams of the connected four-point function $\Gz_{1234}$
are either one-particle irreducible, as for instance diagram \# 2.1, or reducible such as \# 2.2. Defining
\beq
  \Gam_{1234} \hs = - \hs \int_{5678} \Gz_{5678}^{\rm c} \hs \Gz^{-1}_{51}
  \Gz^{-1}_{62} \Gz^{-1}_{73} \Gz^{-1}_{84}  \hspace*{3mm} ,
  \label{gamma}
\eeq
we will see later on that this quantity consists of all one-particle irreducible diagrams of the connected 
four-point function $\Gz_{1234}$ where the external legs are omitted. Therefore this quantity  $\Gam_{1234}$
is called the one-particle irreducible four-point function. Inverting relation (\ref{gamma}) yields
\beq
  \Gz_{1234}^{\rm c} \hs = \hs - \int_{5678} \Gam_{5678} \hs \Gz_{51} \Gz_{62}
  \Gz_{73} \Gz_{84}  \hspace*{3mm} .
  \label{100}
\eeq 
When the one-particle four-point function is depicted by using a big open dot with four legs \\
\beq
-\Gam_{1234} \hspace*{2mm} \equiv \hspace*{2mm}  
  \parbox{10mm}{\centerline{
  \begin{fmfgraph*}(6,6)
  \setval
  \fmfforce{0w,0h}{v1}
  \fmfforce{0w,1h}{v2}
  \fmfforce{1w,1h}{v3}
  \fmfforce{1w,0h}{v4}
  \fmfforce{1/3w,1/3h}{v5}
  \fmfforce{1/3w,2/3h}{v6}
  \fmfforce{2/3w,2/3h}{v7}
  \fmfforce{2/3w,1/3h}{v8}
  \fmf{plain}{v1,v5}
  \fmf{plain}{v2,v6}
  \fmf{plain}{v3,v7}
  \fmf{plain}{v4,v8}
  \fmf{double,width=0.2mm,left=1}{v5,v7,v5}
  \fmfv{decor.size=0, label=${\scs 1}$, l.dist=1mm, l.angle=-135}{v1}
  \fmfv{decor.size=0, label=${\scs 2}$, l.dist=1mm, l.angle=135}{v2}
  \fmfv{decor.size=0, label=${\scs 3}$, l.dist=1mm, l.angle=45}{v3}
  \fmfv{decor.size=0, label=${\scs 4}$, l.dist=1mm, l.angle=-45}{v4}
  \end{fmfgraph*} } } 
\hspace*{5mm} \, ,
\eeq
the identity (\ref{100}) reads graphically
\beq
  \parbox{13mm}{\centerline{
  \begin{fmfgraph*}(9,9)
  \setval
  \fmfforce{0w,0h}{v1}
  \fmfforce{0w,1h}{v2}
  \fmfforce{1w,1h}{v3}
  \fmfforce{1w,0h}{v4}
  \fmfforce{3.5/9w,3.5/9h}{v5}
  \fmfforce{3.5/9w,5.5/9h}{v6}
  \fmfforce{5.5/9w,5.5/9h}{v7}
  \fmfforce{5.5/9w,3.5/9h}{v8}
  \fmf{plain}{v1,v5}
  \fmf{plain}{v2,v6}
  \fmf{plain}{v3,v7}
  \fmf{plain}{v4,v8}
  \fmf{plain,left=1}{v5,v7,v5}
  \fmfv{decor.size=0, label=${\scs 1}$, l.dist=1mm, l.angle=-135}{v1}
  \fmfv{decor.size=0, label=${\scs 2}$, l.dist=1mm, l.angle=135}{v2}
  \fmfv{decor.size=0, label=${\scs 3}$, l.dist=1mm, l.angle=45}{v3}
  \fmfv{decor.size=0, label=${\scs 4}$, l.dist=1mm, l.angle=-45}{v4}
  \end{fmfgraph*} } } 
\hspace*{3mm} = \hspace*{3mm} 
  \parbox{13mm}{\centerline{
  \begin{fmfgraph*}(9,9)
  \setval
  \fmfforce{0w,0h}{v1}
  \fmfforce{0w,1h}{v2}
  \fmfforce{1w,1h}{v3}
  \fmfforce{1w,0h}{v4}
  \fmfforce{3.5/9w,3.5/9h}{v5}
  \fmfforce{3.5/9w,5.5/9h}{v6}
  \fmfforce{5.5/9w,5.5/9h}{v7}
  \fmfforce{5.5/9w,3.5/9h}{v8}
  \fmf{double,width=0.2mm}{v1,v5}
  \fmf{double,width=0.2mm}{v2,v6}
  \fmf{double,width=0.2mm}{v3,v7}
  \fmf{double,width=0.2mm}{v4,v8}
  \fmf{double,width=0.2mm,left=1}{v5,v7,v5}
  \fmfv{decor.size=0, label=${\scs 1}$, l.dist=1mm, l.angle=-135}{v1}
  \fmfv{decor.size=0, label=${\scs 2}$, l.dist=1mm, l.angle=135}{v2}
  \fmfv{decor.size=0, label=${\scs 3}$, l.dist=1mm, l.angle=45}{v3}
  \fmfv{decor.size=0, label=${\scs 4}$, l.dist=1mm, l.angle=-45}{v4}
  \end{fmfgraph*} } } 
\hspace*{4mm} . \label{150} \\ \no
\eeq  
\\
In the following we determine a closed set of Schwinger-Dyson equations for the self-energy and the one-particle irreducible
four-point function.
\subsubsection{Self-Energy}
Multiplying the identity (\ref{Idy}) with $\fullg_{27}^{-1}$ and integrating with respect to the index $2$, we
take into account the self-energy (\ref{sigma})
as well as the connected four-point function (\ref{100}). Thus we yield the Schwinger-Dyson equation for the self-energy:
\beq
\Sig_{12} \hs = \hs - \hs \frac{1}{2} \int_{34} V_{1234} \hs \Gz_{34} \hs
+ \hs \frac{1}{6} \int_{345678} V_{1345} \hs \Gz_{36} \Gz_{47} \Gz_{58}
\hs \Gam_{6782}   \hspace*{3mm} .
\label{103}
\eeq
Its graphical representation reads
\beq
  \parbox{11mm}{\centerline{
  \begin{fmfgraph*}(9,3)
  \setval
  \fmfforce{0w,1/2h}{v1}
  \fmfforce{3/9w,1/2h}{v2}
  \fmfforce{6/9w,1/2h}{v3}
  \fmfforce{1w,1/2h}{v4}
  \fmfforce{1/2w,1h}{v5}
  \fmfforce{1/2w,0h}{v6}
  \fmf{plain,width=0.2mm}{v1,v2}
  \fmf{plain,width=0.2mm}{v3,v4}
  \fmf{double,width=0.2mm,left=1}{v5,v6,v5}
  \fmfv{decor.size=0, label=${\scs 1}$, l.dist=1mm, l.angle=-180}{v1}
  \fmfv{decor.size=0, label=${\scs 2}$, l.dist=1mm, l.angle=0}{v4}
  \end{fmfgraph*} } } 
\hspace*{4mm} = \hspace*{2mm} \frac{1}{2} \hspace*{3mm}
  \parbox{9mm}{\centerline{
  \begin{fmfgraph*}(6,5)
  \setval
  \fmfforce{0w,0h}{v1}
  \fmfforce{1/2w,0h}{v2}
  \fmfforce{1w,0h}{v3}
  \fmfforce{1/12w,1/2h}{v4}
  \fmfforce{11/12w,1/2h}{v5}
  \fmf{plain}{v1,v3}
  \fmf{double,width=0.2mm,left=1}{v4,v5,v4}
  \fmfv{decor.size=0, label=${\scs 1}$, l.dist=1mm, l.angle=-180}{v1}
  \fmfv{decor.size=0, label=${\scs 2}$, l.dist=1mm, l.angle=0}{v3}
  \fmfdot{v2}
  \end{fmfgraph*} } }
\hspace*{3mm} + \hspace*{2mm} \frac{1}{6} \hspace*{3mm}
  \parbox{17mm}{\centerline{
  \begin{fmfgraph*}(14,6)
  \setval
  \fmfforce{0w,1/2h}{v1}
  \fmfforce{3/14w,1/2h}{v2}
  \fmfforce{8/14w,1/2h}{v3}
  \fmfforce{11/14w,1/2h}{v4}
  \fmfforce{1w,1/2h}{v5}
  \fmfforce{9.2/14w,2.2/3h}{v6}
  \fmfforce{9.2/14w,0.8/3h}{v7}
  \fmf{plain}{v1,v2}
  \fmf{plain}{v4,v5}
  \fmf{double,width=0.2mm,left=0.8}{v2,v6}
  \fmf{double,width=0.2mm}{v2,v3}
  \fmf{double,width=0.2mm,right=0.8}{v2,v7}
  \fmf{double,width=0.2mm,right=1}{v3,v4,v3}
  \fmfv{decor.size=0, label=${\scs 1}$, l.dist=1mm, l.angle=-180}{v1}
  \fmfv{decor.size=0, label=${\scs 2}$, l.dist=1mm, l.angle=0}{v5}
  \fmfdot{v2}
  \end{fmfgraph*} } }
\hspace*{5mm} . \label{106} \\[-3mm] \nonumber
\eeq
It contains on the right-hand side the connected two-point function which is determined by the Dyson equation (\ref{VOLL}).
Iteratively solving the integral equations (\ref{VOLL}) and (\ref{106}) necessitates, however, the knowledge of the one-particle
irreducible four-point function $\Gam_{1234}$. 
\subsubsection{One-Particle Irreducible Four-Point Function}
In principle, one could determine $\Gam_{1234}$ from the self-energy $\Sig_{12}$.
To this end we insert relation (\ref{GDE}) into the definition (\ref{gamma}) of the one-particle irreducible four-point function: 
\beq
  \Gam_{1234} \hs = \hs 2 \hs \int_{5678} \frac{\delta \Gz_{56}}
  {\delta G^{-1}_{78}} \hs \Gz^{-1}_{51} \Gz^{-1}_{62} \Gz^{-1}_{73} 
  \Gz^{-1}_{84} \hs + \hs \Gz^{-1}_{13} \Gz^{-1}_{24} \hs + \hs
  \Gz^{-1}_{14} \Gz^{-1}_{23}   \hspace*{3mm} .
  \label{107}
\eeq
The functional derivative on the right-hand side follows from applying a functional derivative with respect to the kernel $G^{-1}$
to the identity (\ref{101}):
\beq
  \frac{\delta \Gz_{56}}{\delta G^{-1}_{78}} \hspace*{2mm} = \hs - 
  \hs \int_{90} \Gz_{59} \hs \frac{\delta \Gz^{-1}_{90}}{\delta G^{-1}_{78}} 
  \hs \Gz_{06} \hspace*{3mm} .
\eeq
Using the definition of the self-energy (\ref{sigma}) and Eq. (\ref{DR1}), the one-particle irreducible four-point function
(\ref{107}) reduces to
\beq
  \Gam_{1234} \hspace*{2mm} = \hspace*{2mm} 2 \int_{56} \Gz^{-1}_{35} \hs 
  \frac{\delta \Sig_{12}}{\delta G^{-1}_{56}} \hs \Gz^{-1}_{64} \, .
  \label{109}
\eeq
Applying again (\ref{sigma}) and the functional chain rule (\ref{13}), this yields
\beq
  -\Gam_{1234}  = 2 \hspace*{2mm} \frac{\delta \Sig_{12}}{\delta G_{34}} 
  \hspace*{2mm} - \hs 2 \int_{56} 
  \Sig_{35} \hs G_{56} \hs \frac{\delta \Sig_{12}}{\delta G_{64}} 
  - \hs 2 \int_{56} \frac{\delta \Sig_{12}}{\delta G_{35}} 
  \hs G_{56} \hs \Sig_{64} \hspace*{1mm} + \hs 2 \int_{5678} \Sig_{35} \hs G_{57} \hs \frac{\delta \Sig_{12}}
  {\delta G_{78}} \hs G_{86} \hs \Sig_{64} 
  \hspace*{3mm} ,
  \label{109*}
\eeq
which is represented graphically as\\
\beq
  \parbox{10mm}{\centerline{
  \begin{fmfgraph*}(6,6)
  \setval
  \fmfforce{0w,0h}{v1}
  \fmfforce{0w,1h}{v2}
  \fmfforce{1w,1h}{v3}
  \fmfforce{1w,0h}{v4}
  \fmfforce{1/3w,1/3h}{v5}
  \fmfforce{1/3w,2/3h}{v6}
  \fmfforce{2/3w,2/3h}{v7}
  \fmfforce{2/3w,1/3h}{v8}
  \fmf{plain}{v1,v5}
  \fmf{plain}{v2,v6}
  \fmf{plain}{v3,v7}
  \fmf{plain}{v4,v8}
  \fmf{double,width=0.2mm,left=1}{v5,v7,v5}
  \fmfv{decor.size=0, label=${\scs 1}$, l.dist=1mm, l.angle=-135}{v1}
  \fmfv{decor.size=0, label=${\scs 2}$, l.dist=1mm, l.angle=135}{v2}
  \fmfv{decor.size=0, label=${\scs 3}$, l.dist=1mm, l.angle=45}{v3}
  \fmfv{decor.size=0, label=${\scs 4}$, l.dist=1mm, l.angle=-45}{v4}
  \end{fmfgraph*} } } 
\hs & = & \hspace*{0.5mm} 2 \hspace*{3mm}
%
\dphi{
  \parbox{14mm}{\centerline{ \hs
  \begin{fmfgraph*}(8,3)
  \setval
  \fmfforce{0w,3/4h}{v1}
  \fmfforce{2.5/8w,3/4h}{v2}
  \fmfforce{5.5/8w,3/4h}{v3}
  \fmfforce{1w,3/4h}{v4}
  \fmfforce{1/2w,5/4h}{v5}
  \fmfforce{1/2w,1/4h}{v6}
  \fmf{plain}{v1,v2}
  \fmf{plain}{v3,v4}
  \fmf{double,width=0.2mm,left=1}{v5,v6,v5}
  \fmfv{decor.size=0, label=${\scs 1}$, l.dist=0.5mm, l.angle=-180}{v1}
  \fmfv{decor.size=0, label=${\scs 2}$, l.dist=0.5mm, l.angle=0}{v4}
  \end{fmfgraph*} } }
}{3}{4} 
\hspace*{3mm} + \hspace*{2mm} 2 \hspace*{3mm}
  \parbox{14mm}{\centerline{
  \begin{fmfgraph*}(11,3)
  \setval
  \fmfforce{0w,1/2h}{v1}
  \fmfforce{1w,1/2h}{v2}
  \fmfforce{3/11w,1/2h}{v3}
  \fmfforce{6/11w,1/2h}{v4}
  \fmf{plain}{v1,v3}
  \fmf{plain}{v2,v4}  
  \fmf{double,width=0.2mm,left=1}{v3,v4,v3}
  \fmfv{decor.size=0, label=${\scs 3}$, l.dist=1mm, l.angle=180}{v1}
  \fmfv{decor.size=0, label=${\scs 5}$, l.dist=1mm, l.angle=0}{v2}
  \end{fmfgraph*} } } 
\hspace*{3mm}  
\dphi{
  \parbox{14mm}{\centerline{
  \begin{fmfgraph*}(8,3)
  \setval
  \fmfforce{0w,3/4h}{v1}
  \fmfforce{2.5/8w,3/4h}{v2}
  \fmfforce{5.5/8w,3/4h}{v3}
  \fmfforce{1w,3/4h}{v4}
  \fmfforce{1/2w,5/4h}{v5}
  \fmfforce{1/2w,1/4h}{v6}
  \fmf{plain}{v1,v2}
  \fmf{plain}{v3,v4}
  \fmf{double,width=0.2mm,left=1}{v5,v6,v5}
  \fmfv{decor.size=0, label=${\scs 1}$, l.dist=0.5mm, l.angle=-180}{v1}
  \fmfv{decor.size=0, label=${\scs 2}$, l.dist=0.5mm, l.angle=0}{v4}
  \end{fmfgraph*} } }
}{5}{6}
\hspace*{3mm} 
  \parbox{14mm}{\centerline{
  \begin{fmfgraph*}(11,3)
  \setval
  \fmfforce{0w,1/2h}{v1}
  \fmfforce{1w,1/2h}{v2}
  \fmfforce{5/11w,1/2h}{v3}
  \fmfforce{8/11w,1/2h}{v4}
  \fmf{plain}{v1,v3}
  \fmf{plain}{v2,v4}  
  \fmf{double,width=0.2mm,left=1}{v3,v4,v3}
  \fmfv{decor.size=0, label=${\scs 6}$, l.dist=1mm, l.angle=180}{v1}
  \fmfv{decor.size=0, label=${\scs 4}$, l.dist=1mm, l.angle=0}{v2}
  \end{fmfgraph*} } }  
\no \\*[4mm]
& &- \hspace*{2mm} 2 \hspace*{3mm} 
  \parbox{14mm}{\centerline{
  \begin{fmfgraph*}(11,3)
  \setval
  \fmfforce{0w,1/2h}{v1}
  \fmfforce{1w,1/2h}{v2}
  \fmfforce{3/11w,1/2h}{v3}
  \fmfforce{6/11w,1/2h}{v4}
  \fmf{plain}{v1,v3}
  \fmf{plain}{v2,v4}  
  \fmf{double,width=0.2mm,left=1}{v3,v4,v3}
  \fmfv{decor.size=0, label=${\scs 3}$, l.dist=1mm, l.angle=180}{v1}
  \fmfv{decor.size=0, label=${\scs 5}$, l.dist=1mm, l.angle=0}{v2}
  \end{fmfgraph*} } } 
\hspace*{3mm}  
\dphi{
  \parbox{14mm}{\centerline{
  \begin{fmfgraph*}(8,3)
  \setval
  \fmfforce{0w,3/4h}{v1}
  \fmfforce{2.5/8w,3/4h}{v2}
  \fmfforce{5.5/8w,3/4h}{v3}
  \fmfforce{1w,3/4h}{v4}
  \fmfforce{1/2w,5/4h}{v5}
  \fmfforce{1/2w,1/4h}{v6}
  \fmf{plain}{v1,v2}
  \fmf{plain}{v3,v4}
  \fmf{double,width=0.2mm,left=1}{v5,v6,v5}
  \fmfv{decor.size=0, label=${\scs 1}$, l.dist=0.5mm, l.angle=-180}{v1}
  \fmfv{decor.size=0, label=${\scs 2}$, l.dist=0.5mm, l.angle=0}{v4}
  \end{fmfgraph*} } }
}{5}{4} 
\hspace*{2mm} - \hspace*{2mm} 2 \hspace*{3mm}  
\dphi{
  \parbox{14mm}{\centerline{
  \begin{fmfgraph*}(8,3)
  \setval
  \fmfforce{0w,3/4h}{v1}
  \fmfforce{2.5/8w,3/4h}{v2}
  \fmfforce{5.5/8w,3/4h}{v3}
  \fmfforce{1w,3/4h}{v4}
  \fmfforce{1/2w,5/4h}{v5}
  \fmfforce{1/2w,1/4h}{v6}
  \fmf{plain}{v1,v2}
  \fmf{plain}{v3,v4}
  \fmf{double,width=0.2mm,left=1}{v5,v6,v5}
  \fmfv{decor.size=0, label=${\scs 1}$, l.dist=0.5mm, l.angle=-180}{v1}
  \fmfv{decor.size=0, label=${\scs 2}$, l.dist=0.5mm, l.angle=0}{v4}
  \end{fmfgraph*} } }
}{3}{5} 
\hspace*{3mm} 
  \parbox{14mm}{\centerline{
  \begin{fmfgraph*}(11,3)
  \setval
  \fmfforce{0w,1/2h}{v1}
  \fmfforce{1w,1/2h}{v2}
  \fmfforce{5/11w,1/2h}{v3}
  \fmfforce{8/11w,1/2h}{v4}
  \fmf{plain}{v1,v3}
  \fmf{plain}{v2,v4}  
  \fmf{double,width=0.2mm,left=1}{v3,v4,v3}
  \fmfv{decor.size=0, label=${\scs 5}$, l.dist=1mm, l.angle=180}{v1}
  \fmfv{decor.size=0, label=${\scs 4}$, l.dist=1mm, l.angle=0}{v2}
  \end{fmfgraph*} } } 
\hspace*{4mm} .
\label{111}
\eeq
Thus amputating a line from the self energy $\Sig_{12}$ leads to the one-particle irreducible diagrams of $\Gam_{1234}$. However, this
procedure has the disadvantage that the first two terms
yield reducible diagrams which are later on removed in the last two terms in (\ref{111}).
As the number of undesired one-particle reducible diagrams occuring at an intermediate step of the calculation increases with
the loop order, the procedure of determining $\Gam_{1234}$ via relation (\ref{111}) is quite inefficient. Therefore we aim at deriving
another equation for $\Gam_{1234}$ whose iterative solution only involves one-particle irreducible diagrams.\\

Going back to Eq. (\ref{109}), we insert the Schwinger-Dyson equation (\ref{103}) for the self-energy and obtain
\beq
  \Gam_{1234} &=&  \hs V_{1234} \hs - \hs \frac{g}{2} \int_{5678} 
  V_{1256} \hs \Gz_{57} \Gz_{68} \Gam_{7834} \hs  - \hs \frac{1}{2} 
  \int_{5678} V_{1356} \hs \Gz_{57} \Gz_{68} \Gam_{7824}
\no  \\*[3mm]
  & & - \hs \frac{1}{2} \int_{5678} V_{1456} \hs \Gz_{57} \Gz_{68}  
  \Gam_{7823}  \hs + \hs \frac{1}{2} \int_{567890\bar{1}\bar{2}} 
  V_{5167} \hs \Gz_{69} \Gz_{70} \Gam_{902\bar{1}} \Gz_{\bar{1}\bar{2}}
  \Gam_{\bar{2}348} \Gz_{85} 
\no  \\*[3mm]
  & & + \hs \frac{1}{3} \int_{567890\bar{1}\bar{2}} V_{1567} \hs \Gz_{58} 
  \Gz_{69} \Gz_{70} \hs 
  \frac{\delta \Gam_{8902} }{\delta G^{-1}_{\bar{1}\bar{2}}} \hspace*{2mm}
  \Gz^{-1}_{\bar{1}3} \Gz^{-1}_{\bar{2}4} 
  \hspace*{3mm} .
  \label{113}
\eeq
The last term is still problematic, as inserting the definition (\ref{sigma}) of the self-energy would yield again an inefficient
equation. This time, however, we can circumvent the inefficiency problem by deriving from (\ref{109}) the commutator relation
\beq
 \int_{78} \frac{\delta \Gam_{1234}}{\delta G^{-1}_{78}} 
  \hs \Gz^{-1}_{75} \Gz^{-1}_{86} \hs - \hs \int_{78} 
  \frac{\delta \Gam_{1256}}{\delta G^{-1}_{78}} \hs \Gz^{-1}_{73} 
  \Gz^{-1}_{84} \hspace*{2mm}
&=& 2 \int_{7890\bar{1}\bar{2}} \Gz^{-1}_{57} \Gz^{-1}_{68} \hs
  \frac{\delta}{\delta G^{-1}_{78}}  \left(  \Gz^{-1}_{39} \Gz^{-1}_{40}
  \right)  \Gz_{9\bar{1}} \Gz_{0\bar{2}} \Gam_{\bar{1}\bar{2}12}
\no  \\*[3mm]  & &
- \hspace*{2mm} 2 \int_{7890\bar{1}\bar{2}} \Gz^{-1}_{37} \Gz^{-1}_{48} 
  \hs \frac{\delta}{\delta G^{-1}_{78}}  \left(  \Gz^{-1}_{59} \Gz^{-1}_{60}
  \right)  \Gz_{9\bar{1}} \Gz_{0\bar{2}} \Gam_{\bar{1}\bar{2}12}
  \hspace*{3mm} .
  \label{115}
\eeq
Applying then
\beq
\frac{\delta \Gz^{-1}_{12}}{\delta G^{-1}_{34}} \hs = \hs \frac{1}{2} 
\left(  \delta_{13} \delta_{24} + \delta_{14} \delta_{23}  \right)
- \hs \frac{1}{2} \int_{56} \Gz_{35} \Gam_{5126} \Gz_{64}
\hspace*{3mm} ,
\label{116}
\eeq
which follows from inserting (\ref{109}) in (\ref{sigma}), the 
intermediate commutator relation (\ref{115}) is converted to the final one
\beq
  \int_{78} \frac{\delta \Gam_{1234}}{\delta G^{-1}_{78}} \hs \Gz^{-1}_{75}
  \Gz^{-1}_{86} \hs - \hs  \int_{78} \frac{\delta \Gam_{1256}}
  {\delta G^{-1}_{78}} \hs \Gz^{-1}_{73} \Gz^{-1}_{84}  \hspace*{1mm} &=& \hspace*{1mm} 
  \frac{1}{2} \int_{78} \Gam_{3457} \Gz_{78} \Gam_{8126} \hs + \hs 
  \frac{1}{2} \int_{78} \Gam_{3467} \Gz_{78} \Gam_{8125} 
\no  \\*[3mm]
&& \hspace*{1mm} -  \frac{1}{2} \int_{78} \Gam_{5637} \Gz_{78} \Gam_{8124}  
  -\frac{1}{2} \int_{78} \Gam_{5647} \Gz_{78} \Gam_{8123} 
  \hspace*{3mm} .
  \label{117}
\eeq
Note that the present commutator relation (\ref{117}) is equivalent to the former one (\ref{KO1}) in Section \ref{PHI}. Indeed,
Eq. (\ref{117}) follows also from inserting (\ref{100}) into (\ref{KO1}) after a lengthy but straight-forward calculation. Now we can
treat the last problematic term in (\ref{113}) with the commutator relation (\ref{117}) and thus yield a 
nonlinear functional integrodifferential equation for the one-particle irreducible four-point function:
\beq 
-  \Gam_{1234} & = & - V_{1234}   +  \frac{1}{3}  \int_{567890} 
V_{1567} \Gz_{58} G_{69} G_{70}   \frac{\delta \Gam_{8234} }
{\delta G_{90}} 
+  \frac{1}{2}   \int_{5678} V_{1256}  \Gz_{57} \Gz_{68} \Gam_{7834}  
\no  \\*[3mm] 
& & +  \,  \frac{1}{2} \int_{5678} V_{1356}  \Gz_{57} \Gz_{68} \Gam_{7824}
+  \frac{1}{2} \int_{5678} V_{1456}  \Gz_{57} \Gz_{68} \Gam_{7823} 
-  \frac{1}{6} \int_{567890\bar{1}\bar{2}}   V_{5167} \Gz_{69} \Gz_{70} \Gam_{902\bar{1}} \Gz_{\bar{1}\bar{2}}
\Gam_{\bar{2}348} \Gz_{85}  
\no  \\*[3mm]    & &
- \,\frac{1}{6}   \int_{567890\bar{1}\bar{2}} V_{5167}  \Gz_{69} \Gz_{70} 
\Gam_{903\bar{1}} \Gz_{\bar{1}\bar{2}} \Gam_{\bar{2}248} \Gz_{85}    
-  \frac{1}{6} \int_{567890\bar{1}\bar{2}} 
V_{5167}  \Gz_{69} \Gz_{70} \Gam_{904\bar{1}} \Gz_{\bar{1}\bar{2}}
\Gam_{\bar{2}238} \Gz_{85}  \, .
\label{119}
\eeq
Its graphical representation reads
\beq
  \parbox{10mm}{\centerline{
  \begin{fmfgraph*}(6,6)
  \setval
  \fmfforce{0w,0h}{v1}
  \fmfforce{0w,1h}{v2}
  \fmfforce{1w,1h}{v3}
  \fmfforce{1w,0h}{v4}
  \fmfforce{1/3w,1/3h}{v5}
  \fmfforce{1/3w,2/3h}{v6}
  \fmfforce{2/3w,2/3h}{v7}
  \fmfforce{2/3w,1/3h}{v8}
  \fmf{plain}{v1,v5}
  \fmf{plain}{v2,v6}
  \fmf{plain}{v3,v7}
  \fmf{plain}{v4,v8}
  \fmf{double,width=0.2mm,left=1}{v5,v7,v5}
  \fmfv{decor.size=0, label=${\scs 1}$, l.dist=1mm, l.angle=-135}{v1}
  \fmfv{decor.size=0, label=${\scs 2}$, l.dist=1mm, l.angle=135}{v2}
  \fmfv{decor.size=0, label=${\scs 3}$, l.dist=1mm, l.angle=45}{v3}
  \fmfv{decor.size=0, label=${\scs 4}$, l.dist=1mm, l.angle=-45}{v4}
  \end{fmfgraph*} } } 
\hspace*{3mm} = \hspace*{3mm}
  \parbox{7mm}{\centerline{
  \begin{fmfgraph*}(4.5,4.5)
  \setval
  \fmfforce{0w,0h}{v1}
  \fmfforce{0w,1h}{v2}
  \fmfforce{1w,1h}{v3}
  \fmfforce{1w,0h}{v4}
  \fmfforce{1/2w,1/2h}{v5}
  \fmf{plain}{v1,v3}
  \fmf{plain}{v2,v4}
  \fmfv{decor.size=0, label=${\scs 1}$, l.dist=1mm, l.angle=-135}{v1}
  \fmfv{decor.size=0, label=${\scs 2}$, l.dist=1mm, l.angle=135}{v2}
  \fmfv{decor.size=0, label=${\scs 3}$, l.dist=1mm, l.angle=45}{v3}
  \fmfv{decor.size=0, label=${\scs 4}$, l.dist=1mm, l.angle=-45}{v4}
  \fmfdot{v5}
  \end{fmfgraph*} } } 
\hspace*{3mm} + \hspace*{2mm} \frac{1}{3} \hspace*{3mm}
  \parbox{10mm}{\centerline{
  \begin{fmfgraph*}(8,8)
  \setval
  \fmfforce{0w,4/8h}{v1}
  \fmfforce{3/8w,4/8h}{v2}
  \fmfforce{1w,4/8h}{v3}
  \fmfforce{1w,0h}{v4}
  \fmfforce{1w,1h}{v5}
  \fmf{plain}{v1,v2}
  \fmf{plain}{v2,v3}
  \fmf{plain,right=0.3}{v2,v4}
  \fmf{double,width=0.2mm,left=0.3}{v2,v5}
  \fmfv{decor.size=0, label=${\scs 1}$, l.dist=1mm, l.angle=-180}{v1}
  \fmfv{decor.size=0, label=${\scs 6}$, l.dist=1mm, l.angle=0}{v3}
  \fmfv{decor.size=0, label=${\scs 7}$, l.dist=1mm, l.angle=0}{v4}
  \fmfv{decor.size=0, label=${\scs 5}$, l.dist=1mm, l.angle=0}{v5}
  \fmfdot{v2}
  \end{fmfgraph*} } } 
\hspace*{4mm} 
\raisebox{2mm}{\begin{minipage}{1.4cm}
\beq
\dphi{
  \parbox{11mm}{\centerline{
  \begin{fmfgraph*}(6,7)
  \setval
  \fmfforce{0w,2/7h}{v1}
  \fmfforce{0w,8/7h}{v2}
  \fmfforce{1w,8/7h}{v3}
  \fmfforce{1w,2/7h}{v4}
  \fmfforce{1/3w,4/7h}{v5}
  \fmfforce{1/3w,6/7h}{v6}
  \fmfforce{2/3w,6/7h}{v7}
  \fmfforce{2/3w,4/7h}{v8}
  \fmf{plain}{v1,v5}
  \fmf{plain}{v2,v6}
  \fmf{plain}{v3,v7}
  \fmf{plain}{v4,v8}
  \fmf{double,width=0.2mm,left=1}{v5,v7,v5}
  \fmfv{decor.size=0, label=${\scs 5}$, l.dist=0.5mm, l.angle=-180}{v1}
  \fmfv{decor.size=0, label=${\scs 2}$, l.dist=0.5mm, l.angle=180}{v2}
  \fmfv{decor.size=0, label=${\scs 3}$, l.dist=0.5mm, l.angle=0}{v3}
  \fmfv{decor.size=0, label=${\scs 4}$, l.dist=0.5mm, l.angle=0}{v4}
  \end{fmfgraph*} } } 
}{6}{7}
\no  \eeq 
\end{minipage}}
\hspace*{3mm} + \hspace*{2mm} \frac{1}{2} \hspace*{3mm}
  \parbox{14mm}{\centerline{
  \begin{fmfgraph*}(11,6)
  \setval
  \fmfforce{0w,1/6h}{v1}
  \fmfforce{0w,5/6h}{v2}
  \fmfforce{1w,1h}{v3}
  \fmfforce{1w,0h}{v4}
  \fmfforce{7/11w,1/3h}{v5}
  \fmfforce{7/11w,2/3h}{v6}
  \fmfforce{9/11w,2/3h}{v7}
  \fmfforce{9/11w,1/3h}{v8}
  \fmfforce{2/11w,1/2h}{v9}
  \fmf{plain}{v1,v9}
  \fmf{plain}{v2,v9}
  \fmf{plain}{v3,v7}
  \fmf{plain}{v4,v8}
  \fmf{double,width=0.2mm,left=1}{v5,v7,v5}
  \fmf{double,width=0.2mm,left=0.7}{v9,v6}
  \fmf{double,width=0.2mm,right=0.7}{v9,v5}
  \fmfv{decor.size=0, label=${\scs 1}$, l.dist=1mm, l.angle=-135}{v1}
  \fmfv{decor.size=0, label=${\scs 2}$, l.dist=1mm, l.angle=135}{v2}
  \fmfv{decor.size=0, label=${\scs 3}$, l.dist=1mm, l.angle=25}{v3}
  \fmfv{decor.size=0, label=${\scs 4}$, l.dist=1mm, l.angle=-25}{v4}
  \fmfdot{v9}
  \end{fmfgraph*} } } 
\hspace*{3mm} + \hspace*{2mm} \frac{1}{2} \hspace*{3mm}
  \parbox{14mm}{\centerline{
  \begin{fmfgraph*}(11,6)
  \setval
  \fmfforce{0w,1/6h}{v1}
  \fmfforce{0w,5/6h}{v2}
  \fmfforce{1w,1h}{v3}
  \fmfforce{1w,0h}{v4}
  \fmfforce{7/11w,1/3h}{v5}
  \fmfforce{7/11w,2/3h}{v6}
  \fmfforce{9/11w,2/3h}{v7}
  \fmfforce{9/11w,1/3h}{v8}
  \fmfforce{2/11w,1/2h}{v9}
  \fmf{plain}{v1,v9}
  \fmf{plain}{v2,v9}
  \fmf{plain}{v3,v7}
  \fmf{plain}{v4,v8}
  \fmf{double,width=0.2mm,left=1}{v5,v7,v5}
  \fmf{double,width=0.2mm,left=0.7}{v9,v6}
  \fmf{double,width=0.2mm,right=0.7}{v9,v5}
  \fmfv{decor.size=0, label=${\scs 1}$, l.dist=1mm, l.angle=-135}{v1}
  \fmfv{decor.size=0, label=${\scs 3}$, l.dist=1mm, l.angle=135}{v2}
  \fmfv{decor.size=0, label=${\scs 4}$, l.dist=1mm, l.angle=25}{v3}
  \fmfv{decor.size=0, label=${\scs 2}$, l.dist=1mm, l.angle=-25}{v4}
  \fmfdot{v9}
  \end{fmfgraph*} } } 
\no 
\eeq
\beq
\hspace*{2.6cm} + \hspace*{2mm} \frac{1}{2} \hspace*{3mm}
  \parbox{14mm}{\centerline{
  \begin{fmfgraph*}(11,6)
  \setval
  \fmfforce{0w,1/6h}{v1}
  \fmfforce{0w,5/6h}{v2}
  \fmfforce{1w,1h}{v3}
  \fmfforce{1w,0h}{v4}
  \fmfforce{7/11w,1/3h}{v5}
  \fmfforce{7/11w,2/3h}{v6}
  \fmfforce{9/11w,2/3h}{v7}
  \fmfforce{9/11w,1/3h}{v8}
  \fmfforce{2/11w,1/2h}{v9}
  \fmf{plain}{v1,v9}
  \fmf{plain}{v2,v9}
  \fmf{plain}{v3,v7}
  \fmf{plain}{v4,v8}
  \fmf{double,width=0.2mm,left=1}{v5,v7,v5}
  \fmf{double,width=0.2mm,left=0.7}{v9,v6}
  \fmf{double,width=0.2mm,right=0.7}{v9,v5}
  \fmfv{decor.size=0, label=${\scs 1}$, l.dist=1mm, l.angle=-135}{v1}
  \fmfv{decor.size=0, label=${\scs 4}$, l.dist=1mm, l.angle=135}{v2}
  \fmfv{decor.size=0, label=${\scs 2}$, l.dist=1mm, l.angle=25}{v3}
  \fmfv{decor.size=0, label=${\scs 3}$, l.dist=1mm, l.angle=-25}{v4}
  \fmfdot{v9}
  \end{fmfgraph*} } } 
\hspace*{3mm} + \hspace*{2mm} \frac{1}{6} \hspace*{2mm}
  \parbox{14mm}{\centerline{
  \begin{fmfgraph*}(11,11)
  \setval
  \fmfforce{1/11w,0h}{v1}
  \fmfforce{0w,1h}{v2}
  \fmfforce{3/11w,2/11h}{v3}
  \fmfforce{2/11w,9/11h}{v4}
  \fmfforce{3/11w,6.5/11h}{v5}
  \fmfforce{1.5/11w,7.5/11h}{v6}
  \fmfforce{4.5/11w,8.5/11h}{v7}
  \fmfforce{7/11w,4.5/11h}{v8}
  \fmfforce{7/11w,6.5/11h}{v9}
  \fmfforce{9.2/11w,6.7/11h}{v10}
  \fmfforce{9/11w,4.5/11h}{v11}
  \fmfforce{1w,8.5/11h}{v12}
  \fmfforce{1w,2.5/11h}{v13}
  \fmf{plain}{v1,v3}
  \fmf{plain}{v2,v4}
  \fmf{plain}{v10,v12}
  \fmf{plain}{v11,v13}
  \fmf{double,width=0.2mm,left=1}{v6,v7,v6}
  \fmf{double,width=0.2mm,left=1}{v9,v11,v9}
  \fmf{double,width=0.2mm}{v3,v5}
  \fmf{double,width=0.2mm,left=0.5}{v3,v6}
  \fmf{double,width=0.2mm,left=0.3}{v7,v9}
  \fmf{double,width=0.2mm,right=0.4}{v3,v8}
  \fmfv{decor.size=0, label=${\scs 1}$, l.dist=1mm, l.angle=-180}{v1}
  \fmfv{decor.size=0, label=${\scs 2}$, l.dist=1mm, l.angle=180}{v2}
  \fmfv{decor.size=0, label=${\scs 3}$, l.dist=1mm, l.angle=45}{v12}
  \fmfv{decor.size=0, label=${\scs 4}$, l.dist=1mm, l.angle=-45}{v13}
  \fmfdot{v3}
  \end{fmfgraph*} } } 
\hspace*{3mm} + \hspace*{2mm} \frac{1}{6} \hspace*{2mm}
  \parbox{14mm}{\centerline{
  \begin{fmfgraph*}(11,11)
  \setval
  \fmfforce{1/11w,0h}{v1}
  \fmfforce{0w,1h}{v2}
  \fmfforce{3/11w,2/11h}{v3}
  \fmfforce{2/11w,9/11h}{v4}
  \fmfforce{3/11w,6.5/11h}{v5}
  \fmfforce{1.5/11w,7.5/11h}{v6}
  \fmfforce{4.5/11w,8.5/11h}{v7}
  \fmfforce{7/11w,4.5/11h}{v8}
  \fmfforce{7/11w,6.5/11h}{v9}
  \fmfforce{9.2/11w,6.7/11h}{v10}
  \fmfforce{9/11w,4.5/11h}{v11}
  \fmfforce{1w,8.5/11h}{v12}
  \fmfforce{1w,2.5/11h}{v13}
  \fmf{plain}{v1,v3}
  \fmf{plain}{v2,v4}
  \fmf{plain}{v10,v12}
  \fmf{plain}{v11,v13}
  \fmf{double,width=0.2mm,left=1}{v6,v7,v6}
  \fmf{double,width=0.2mm,left=1}{v9,v11,v9}
  \fmf{double,width=0.2mm}{v3,v5}
  \fmf{double,width=0.2mm,left=0.5}{v3,v6}
  \fmf{double,width=0.2mm,left=0.3}{v7,v9}
  \fmf{double,width=0.2mm,right=0.4}{v3,v8}
  \fmfv{decor.size=0, label=${\scs 1}$, l.dist=1mm, l.angle=-180}{v1}
  \fmfv{decor.size=0, label=${\scs 3}$, l.dist=1mm, l.angle=180}{v2}
  \fmfv{decor.size=0, label=${\scs 4}$, l.dist=1mm, l.angle=45}{v12}
  \fmfv{decor.size=0, label=${\scs 2}$, l.dist=1mm, l.angle=-45}{v13}
  \fmfdot{v3}
  \end{fmfgraph*} } } 
\hspace*{3mm} + \hspace*{2mm} \frac{1}{6} \hspace*{2mm}
  \parbox{14mm}{\centerline{
  \begin{fmfgraph*}(11,11)
  \setval
  \fmfforce{1/11w,0h}{v1}
  \fmfforce{0w,1h}{v2}
  \fmfforce{3/11w,2/11h}{v3}
  \fmfforce{2/11w,9/11h}{v4}
  \fmfforce{3/11w,6.5/11h}{v5}
  \fmfforce{1.5/11w,7.5/11h}{v6}
  \fmfforce{4.5/11w,8.5/11h}{v7}
  \fmfforce{7/11w,4.5/11h}{v8}
  \fmfforce{7/11w,6.5/11h}{v9}
  \fmfforce{9.2/11w,6.7/11h}{v10}
  \fmfforce{9/11w,4.5/11h}{v11}
  \fmfforce{1w,8.5/11h}{v12}
  \fmfforce{1w,2.5/11h}{v13}
  \fmf{plain}{v1,v3}
  \fmf{plain}{v2,v4}
  \fmf{plain}{v10,v12}
  \fmf{plain}{v11,v13}
  \fmf{double,width=0.2mm,left=1}{v6,v7,v6}
  \fmf{double,width=0.2mm,left=1}{v9,v11,v9}
  \fmf{double,width=0.2mm}{v3,v5}
  \fmf{double,width=0.2mm,left=0.5}{v3,v6}
  \fmf{double,width=0.2mm,left=0.3}{v7,v9}
  \fmf{double,width=0.2mm,right=0.4}{v3,v8}
  \fmfv{decor.size=0, label=${\scs 1}$, l.dist=1mm, l.angle=-180}{v1}
  \fmfv{decor.size=0, label=${\scs 4}$, l.dist=1mm, l.angle=180}{v2}
  \fmfv{decor.size=0, label=${\scs 2}$, l.dist=1mm, l.angle=45}{v12}
  \fmfv{decor.size=0, label=${\scs 3}$, l.dist=1mm, l.angle=-45}{v13}
  \fmfdot{v3}
  \end{fmfgraph*} } } 
\hspace*{5mm} .
\label{120}
\eeq
Thus the closed set of Schwinger-Dyson equations for the connected two-point function, the self-energy and the one-particle
irreducible four-point function is given by (\ref{VOLL}), (\ref{106}), and (\ref{120}). In the subsequent section we show
that its recursive graphical solution leads to all one-particle irreducible Feynman diagrams needed for renormalizing the 
$\phi^4$-theory.
\subsection{Graphical Recursion Relations}\la{NANA}
To this end we supplement the perturbative expansion (\ref{G2P}) of the connected two-point function with corresponding
ones for the self-energy
\beq
  \parbox{12mm}{\centerline{
  \begin{fmfgraph*}(9,3)
  \setval
  \fmfforce{0w,1/2h}{v1}
  \fmfforce{3/9w,1/2h}{v2}
  \fmfforce{6/9w,1/2h}{v3}
  \fmfforce{1w,1/2h}{v4}
  \fmfforce{1/2w,1h}{v5}
  \fmfforce{1/2w,0h}{v6}
  \fmf{plain}{v1,v2}
  \fmf{plain}{v3,v4}
  \fmf{double,width=0.2mm,left=1}{v5,v6,v5}
  \fmfv{decor.size=0, label=${\scs 1}$, l.dist=1mm, l.angle=-180}{v1}
  \fmfv{decor.size=0, label=${\scs 2}$, l.dist=1mm, l.angle=0}{v4}
  \end{fmfgraph*} } }
\hspace*{3mm} = \hspace*{3mm} \sum _{p=1} ^\infty \hspace*{3mm} 
  \parbox{14mm}{\centerline{
  \begin{fmfgraph*}(11,5)
  \setval
  \fmfforce{0w,1/2h}{v1}
  \fmfforce{3/11w,1/2h}{v2}
  \fmfforce{8/11w,1/2h}{v3}
  \fmfforce{1w,1/2h}{v4}
  \fmfforce{1/2w,1h}{v5}
  \fmfforce{1/2w,0h}{v6}
  \fmfforce{1/2w,1/2h}{v7}
  \fmf{plain}{v1,v2}
  \fmf{plain}{v3,v4}
  \fmf{double,width=0.2mm,left=1}{v5,v6,v5}
  \fmfv{decor.size=0, label=${\scs 1}$, l.dist=1mm, l.angle=-180}{v1}
  \fmfv{decor.size=0, label=${\scs 2}$, l.dist=1mm, l.angle=0}{v4}
  \fmfv{decor.size=0, label=${\scs p}$, l.dist=0mm, l.angle=90}{v7}
  \end{fmfgraph*} } }
\eeq
and for the one-particle irreducible four-point function
\beq
\parbox{10mm}{\centerline{
  \begin{fmfgraph*}(6,6)
  \setval
  \fmfforce{0w,0h}{v1}
  \fmfforce{0w,1h}{v2}
  \fmfforce{1w,1h}{v3}
  \fmfforce{1w,0h}{v4}
  \fmfforce{1/3w,1/3h}{v5}
  \fmfforce{1/3w,2/3h}{v6}
  \fmfforce{2/3w,2/3h}{v7}
  \fmfforce{2/3w,1/3h}{v8}
  \fmf{plain}{v1,v5}
  \fmf{plain}{v2,v6}
  \fmf{plain}{v3,v7}
  \fmf{plain}{v4,v8}
  \fmf{double,width=0.2mm,left=1}{v5,v7,v5}
  \fmfv{decor.size=0, label=${\scs 1}$, l.dist=1mm, l.angle=-135}{v1}
  \fmfv{decor.size=0, label=${\scs 2}$, l.dist=1mm, l.angle=135}{v2}
  \fmfv{decor.size=0, label=${\scs 3}$, l.dist=1mm, l.angle=45}{v3}
  \fmfv{decor.size=0, label=${\scs 4}$, l.dist=1mm, l.angle=-45}{v4}
  \end{fmfgraph*} } } 
\hspace*{2mm} = \hspace*{3mm} \sum _{p=1} ^\infty \hspace*{3mm} 
  \parbox{10mm}{\centerline{
  \begin{fmfgraph*}(7.5,7.5)
  \setval
  \fmfforce{0w,0h}{v1}
  \fmfforce{0w,1h}{v2}
  \fmfforce{1w,1h}{v3}
  \fmfforce{1w,0h}{v4}
  \fmfforce{1/4w,1/4h}{v5}
  \fmfforce{1/4w,3/4h}{v6}
  \fmfforce{3/4w,3/4h}{v7}
  \fmfforce{3/4w,1/4h}{v8}
  \fmfforce{1/2w,1/2h}{v9}
  \fmf{plain}{v1,v5}
  \fmf{plain}{v2,v6}
  \fmf{plain}{v3,v7}
  \fmf{plain}{v4,v8}
  \fmf{double,width=0.2mm,left=1}{v5,v7,v5}
  \fmfv{decor.size=0, label=${\scs 1}$, l.dist=1mm, l.angle=-135}{v1}
  \fmfv{decor.size=0, label=${\scs 2}$, l.dist=1mm, l.angle=135}{v2}
  \fmfv{decor.size=0, label=${\scs 3}$, l.dist=1mm, l.angle=45}{v3}
  \fmfv{decor.size=0, label=${\scs 4}$, l.dist=1mm, l.angle=-45}{v4}
  \fmfv{decor.size=0, label=${\scs p}$, l.dist=0mm, l.angle=90}{v9}
  \end{fmfgraph*} } } 
\hspace*{5mm} .
\label{122}
\eeq
As a result we obtain the following closed set of graphical recursion relations: \\
\beq
\la{REK4}
\hspace*{-6cm}
  \parbox{14mm}{\centerline{
  \begin{fmfgraph*}(11,5)
  \setval
  \fmfforce{0w,1/2h}{v1}
  \fmfforce{3/11w,1/2h}{v2}
  \fmfforce{8/11w,1/2h}{v3}
  \fmfforce{1w,1/2h}{v4}
  \fmfforce{1/2w,1h}{v5}
  \fmfforce{1/2w,0h}{v6}
  \fmfforce{1/2w,1/2h}{v7}
  \fmf{plain}{v1,v2}
  \fmf{plain}{v3,v4}
  \fmf{double,width=0.2mm,left=1}{v5,v6,v5}
  \fmfv{decor.size=0, label=${\scs 1}$, l.dist=1mm, l.angle=-180}{v1}
  \fmfv{decor.size=0, label=${\scs 2}$, l.dist=1mm, l.angle=0}{v4}
  \fmfv{decor.size=0, label=${\scs p}$, l.dist=0mm, l.angle=90}{v7}
  \end{fmfgraph*} } }
\hspace*{2mm} &= &\hspace*{2mm} \frac{1}{2} \hspace*{3mm}
  \parbox{9mm}{\centerline{
  \begin{fmfgraph*}(6,5)
  \setval
  \fmfforce{0w,0h}{v1}
  \fmfforce{1/2w,0h}{v2}
  \fmfforce{1w,0h}{v3}
  \fmfforce{1/2w,1h}{v4}
  \fmf{plain}{v1,v2}
  \fmf{plain}{v2,v3}
  \fmf{double,width=0.2mm,left=1}{v4,v2,v4}
  \fmfv{decor.size=0, label=${\scs 1}$, l.dist=1mm, l.angle=-180}{v1}
  \fmfv{decor.size=0, label=${\scs 2}$, l.dist=1mm, l.angle=0}{v3}
  \fmfv{decor.size=0, label=${\scs {(p-1)}}$, l.dist=1mm, l.angle=90}{v4}
  \fmfdot{v2}
  \end{fmfgraph*} } }
\hspace*{3mm} + \hspace*{2mm} \frac{1}{6} \hspace*{1.5mm}
\sum_{q=1} ^{p-1} \sum_{r=1} ^q \sum_{s=1} ^r \hspace*{4mm} 
  \parbox{22mm}{\centerline{
  \begin{fmfgraph*}(24,8)
  \setval
  \fmfforce{0w,1/2h}{v1}
  \fmfforce{3/24w,1/2h}{v2}
  \fmfforce{15/24w,1/2h}{v3}
  \fmfforce{21/24w,1/2h}{v4}
  \fmfforce{24/24w,1/2h}{v5}
  \fmfforce{17.7/24w,6.9/8h}{v6}
  \fmfforce{17.7/24w,1.1/8h}{v7}
  \fmfforce{10/24w,1/2h}{v8}
  \fmfforce{10/24w,1.1h}{v9} 
  \fmfforce{11/24w,-0.1h}{v10}
  \fmfforce{18/24w,1/2h}{v11}
  \fmf{plain}{v1,v2}
  \fmf{plain}{v4,v5}
  \fmf{double,width=0.2mm,left=0.6}{v2,v6}
  \fmf{double,width=0.2mm}{v2,v3}
  \fmf{double,width=0.2mm,right=0.6}{v2,v7}
  \fmf{double,width=0.2mm,right=1}{v3,v4,v3}
  \fmfv{decor.size=0, label=${\scs 1}$, l.dist=0.5mm, l.angle=-180}{v1}
  \fmfv{decor.size=0, label=${\scs 2}$, l.dist=0.5mm, l.angle=0}{v5}
  \fmfv{decor.size=0, label=${\scs {(q-r)}}$, l.dist=1mm, l.angle=90}{v8}
  \fmfv{decor.size=0, label=${\scs {(p-q-1)}}$, l.dist=2.2mm, l.angle=90}{v9}
  \fmfv{decor.size=0, label=${\scs {(r-s)}}$, l.dist=0.5mm, l.angle=90}{v10}
  \fmfv{decor.size=0, label=${\scs s}$, l.dist=0mm, l.angle=90}{v11}
  \fmfdot{v2}
  \end{fmfgraph*} } }
\hspace*{5mm} , \\
\la{REK5}
  \hspace*{-6cm}
\parbox{10mm}{\centerline{
  \begin{fmfgraph*}(7,3)
  \setval
  \fmfleft{v1}
  \fmfright{v2}
  \fmfforce{1/2w,1/2h}{v3}
  \fmf{double,width=0.2mm}{v2,v1}
  \fmfv{decor.size=0, label=${\scs 1}$, l.dist=1mm, l.angle=-180}{v1}
  \fmfv{decor.size=0, label=${\scs 2}$, l.dist=1mm, l.angle=0}{v2}
  \fmfv{decor.size=0, label=${\scs (p)}$, l.dist=1.5mm, l.angle=90}{v3}
  \end{fmfgraph*}}}
\hspace*{4mm}& =& \hspace*{3mm} \sum _{q=1} ^p \hspace*{2mm}
  \parbox{17mm}{\centerline{
  \begin{fmfgraph*}(15,3)
  \setval
  \fmfforce{0w,1/2h}{v1}
  \fmfforce{5/15w,1/2h}{v2}
  \fmfforce{10/15w,1/2h}{v3}
  \fmfforce{1w,1/2h}{v4}
  \fmfforce{1/2w,1/2h}{v5}
  \fmfforce{13.5/15w,1/2h}{v6}
  \fmf{plain}{v1,v2}
  \fmf{double,width=0.2mm}{v3,v4}
  \fmf{double,width=0.2mm,left=1}{v2,v3,v2}
  \fmfv{decor.size=0, label=${\scs 1}$, l.dist=1mm, l.angle=-180}{v1}
  \fmfv{decor.size=0, label=${\scs 2}$, l.dist=1.5mm, l.angle=0}{v4}
  \fmfv{decor.size=0, label=${\scs q}$, l.dist=0mm, l.angle=90}{v5}
  \fmfv{decor.size=0, label=${\scs {(p-q)}}$, l.dist=1.5mm, l.angle=90}{v6}
  \end{fmfgraph*} } } 
\hspace*{6mm} , 
\eeq
\beq
\la{REK6}
\hspace*{0cm}
  \parbox{10mm}{\centerline{
  \begin{fmfgraph*}(8,8)
  \setval
  \fmfforce{0w,0h}{v1}
  \fmfforce{0w,1h}{v2}
  \fmfforce{1w,1h}{v3}
  \fmfforce{1w,0h}{v4}
  \fmfforce{2/8w,2/8h}{v5}
  \fmfforce{2/8w,6/8h}{v6}
  \fmfforce{6/8w,6/8h}{v7}
  \fmfforce{6/8w,2/8h}{v8}
  \fmfforce{1/2w,1/2h}{v9}
  \fmf{plain}{v1,v5}
  \fmf{plain}{v2,v6}
  \fmf{plain}{v3,v7}
  \fmf{plain}{v4,v8}
  \fmf{double,width=0.2mm,left=1}{v5,v7,v5}
  \fmfv{decor.size=0, label=${\scs 1}$, l.dist=1mm, l.angle=-135}{v1}
  \fmfv{decor.size=0, label=${\scs 2}$, l.dist=1mm, l.angle=135}{v2}
  \fmfv{decor.size=0, label=${\scs 3}$, l.dist=1mm, l.angle=45}{v3}
  \fmfv{decor.size=0, label=${\scs 4}$, l.dist=1mm, l.angle=-45}{v4}
  \fmfv{decor.size=0, label=${\scs {p+1}}$, l.dist=0mm, l.angle=90}{v9}
  \end{fmfgraph*} } } 
\hspace*{4mm} = \hspace*{2mm} \frac{1}{3} \hspace*{2mm}
\sum _{q=1} ^p \hspace*{3mm}
  \parbox{11mm}{\centerline{
  \begin{fmfgraph*}(8,8)
  \setval
  \fmfforce{0w,4/8h}{v1}
  \fmfforce{3/8w,4/8h}{v2}
  \fmfforce{1w,4/8h}{v3}
  \fmfforce{1w,0h}{v4}
  \fmfforce{1w,1h}{v5}
  \fmfforce{4.5/8w,8.5/8h}{v6}
  \fmf{plain}{v1,v2}
  \fmf{plain}{v2,v3}
  \fmf{plain,right=0.3}{v2,v4}
  \fmf{double,width=0.2mm,left=0.3}{v2,v5}
  \fmfv{decor.size=0, label=${\scs 1}$, l.dist=1mm, l.angle=-180}{v1}
  \fmfv{decor.size=0, label=${\scs 6}$, l.dist=1mm, l.angle=0}{v3}
  \fmfv{decor.size=0, label=${\scs 7}$, l.dist=1mm, l.angle=0}{v4}
  \fmfv{decor.size=0, label=${\scs 5}$, l.dist=1mm, l.angle=0}{v5}
  \fmfv{decor.size=0, label=${\scs {(p-q)}}$, l.dist=0.5mm, l.angle=90}{v6}
  \fmfdot{v2}
  \end{fmfgraph*} } } 
\hspace*{3mm} 
\raisebox{2mm}{\begin{minipage}{1.8cm}
\beq
\dphi{
  \parbox{14mm}{\centerline{
  \begin{fmfgraph*}(7.5,7.5)
  \setval
  \fmfforce{0w,2.5/7.5h}{v1}
  \fmfforce{0w,10/7.5h}{v2}
  \fmfforce{1w,10/7.5h}{v3}
  \fmfforce{1w,2.5/7.5h}{v4}
  \fmfforce{2/7.5w,4.5/7.5h}{v5}
  \fmfforce{2/7.5w,8/7.5h}{v6}
  \fmfforce{5.5/7.5w,8/7.5h}{v7}
  \fmfforce{5.5/7.5w,4.5/7.5h}{v8}
  \fmfforce{1/2w,6.25/7.5h}{v9}
  \fmf{plain}{v1,v5}
  \fmf{plain}{v2,v6}
  \fmf{plain}{v3,v7}
  \fmf{plain}{v4,v8}
  \fmf{double,width=0.2mm,left=1}{v5,v7,v5}
  \fmfv{decor.size=0, label=${\scs 5}$, l.dist=0.5mm, l.angle=-135}{v1}
  \fmfv{decor.size=0, label=${\scs 2}$, l.dist=0.5mm, l.angle=135}{v2}
  \fmfv{decor.size=0, label=${\scs 3}$, l.dist=0.5mm, l.angle=45}{v3}
  \fmfv{decor.size=0, label=${\scs 4}$, l.dist=0.5mm, l.angle=-45}{v4}
  \fmfv{decor.size=0, label=${\scs q}$, l.dist=0mm, l.angle=90}{v9}
  \end{fmfgraph*} } } 
}{6}{7}
\no  \eeq 
\end{minipage}}
\hspace*{2mm} + \hspace*{2mm} \frac{1}{2} \hspace*{2mm}
\sum _{q=1} ^p \sum _{r=1} ^q \hspace*{4mm}
  \parbox{14mm}{\centerline{
  \begin{fmfgraph*}(12.5,7.6)
  \setval
  \fmfforce{0w,1.8/7.6h}{v1}
  \fmfforce{0w,5.8/7.6h}{v2}
  \fmfforce{1w,1h}{v3}
  \fmfforce{1w,0h}{v4}
  \fmfforce{7/12.5w,2/7.6h}{v5}
  \fmfforce{7/12.5w,5.6/7.6h}{v6}
  \fmfforce{10.6/12.5w,5.6/7.6h}{v7}
  \fmfforce{10.6/12.5w,2/7.6h}{v8}
  \fmfforce{2/12.5w,1/2h}{v9}
  \fmfforce{8.8/12.5w,1/2h}{v10}
  \fmfforce{4.5/12.5w,1h}{v11}
  \fmfforce{4.5/12.5w,-0.5/7.6h}{v12}
  \fmf{plain}{v1,v9}
  \fmf{plain}{v2,v9}
  \fmf{plain}{v3,v7}
  \fmf{plain}{v4,v8}
  \fmf{double,width=0.2mm,left=1}{v5,v7,v5}
  \fmf{double,width=0.2mm,left=0.6}{v9,v6}
  \fmf{double,width=0.2mm,right=0.6}{v9,v5}
  \fmfv{decor.size=0, label=${\scs 1}$, l.dist=1mm, l.angle=-135}{v1}
  \fmfv{decor.size=0, label=${\scs 2}$, l.dist=1mm, l.angle=135}{v2}
  \fmfv{decor.size=0, label=${\scs 3}$, l.dist=1mm, l.angle=25}{v3}
  \fmfv{decor.size=0, label=${\scs 4}$, l.dist=1mm, l.angle=-25}{v4}
  \fmfv{decor.size=0, label=${\scs r}$, l.dist=0mm, l.angle=0}{v10}
  \fmfv{decor.size=0, label=${\scs {(p-q)}}$, l.dist=0.05mm, l.angle=90}{v11}
  \fmfv{decor.size=0, label=${\scs {(q-r)}}$, l.dist=0mm, l.angle=-90}{v12}
  \fmfdot{v9}
  \end{fmfgraph*} } } 
%
\hspace*{2mm} + \hspace*{2mm} \frac{1}{2} \hspace*{2mm}
\sum _{q=1} ^p \sum _{r=1} ^q \hspace*{4mm}
  \parbox{14mm}{\centerline{
  \begin{fmfgraph*}(12.5,7.6)
  \setval
  \fmfforce{0w,1.8/7.6h}{v1}
  \fmfforce{0w,5.8/7.6h}{v2}
  \fmfforce{1w,1h}{v3}
  \fmfforce{1w,0h}{v4}
  \fmfforce{7/12.5w,2/7.6h}{v5}
  \fmfforce{7/12.5w,5.6/7.6h}{v6}
  \fmfforce{10.6/12.5w,5.6/7.6h}{v7}
  \fmfforce{10.6/12.5w,2/7.6h}{v8}
  \fmfforce{2/12.5w,1/2h}{v9}
  \fmfforce{8.8/12.5w,1/2h}{v10}
  \fmfforce{4.5/12.5w,1h}{v11}
  \fmfforce{4.5/12.5w,0h}{v12}
  \fmf{plain}{v1,v9}
  \fmf{plain}{v2,v9}
  \fmf{plain}{v3,v7}
  \fmf{plain}{v4,v8}
  \fmf{double,width=0.2mm,left=1}{v5,v7,v5}
  \fmf{double,width=0.2mm,left=0.6}{v9,v6}
  \fmf{double,width=0.2mm,right=0.6}{v9,v5}
  \fmfv{decor.size=0, label=${\scs 1}$, l.dist=1mm, l.angle=-135}{v1}
  \fmfv{decor.size=0, label=${\scs 4}$, l.dist=1mm, l.angle=135}{v2}
  \fmfv{decor.size=0, label=${\scs 2}$, l.dist=1mm, l.angle=25}{v3}
  \fmfv{decor.size=0, label=${\scs 3}$, l.dist=1mm, l.angle=-25}{v4}
  \fmfv{decor.size=0, label=${\scs r}$, l.dist=0mm, l.angle=0}{v10}
  \fmfv{decor.size=0, label=${\scs {(p-q)}}$, l.dist=0.05mm, l.angle=90}{v11}
  \fmfv{decor.size=0, label=${\scs {(q-r)}}$, l.dist=0.01mm, l.angle=-90}{v12}
  \fmfdot{v9}
  \end{fmfgraph*} } } 
\no  
\eeq
\beq
\hspace*{1.6cm} &&+ \hspace*{2mm} \frac{1}{2} \hspace*{2mm}
\sum _{q=1} ^p \sum _{r=1} ^q \hspace*{4mm}
  \parbox{14mm}{\centerline{
  \begin{fmfgraph*}(12.5,7.6)
  \setval
  \fmfforce{0w,1.8/7.6h}{v1}
  \fmfforce{0w,5.8/7.6h}{v2}
  \fmfforce{1w,1h}{v3}
  \fmfforce{1w,0h}{v4}
  \fmfforce{7/12.5w,2/7.6h}{v5}
  \fmfforce{7/12.5w,5.6/7.6h}{v6}
  \fmfforce{10.6/12.5w,5.6/7.6h}{v7}
  \fmfforce{10.6/12.5w,2/7.6h}{v8}
  \fmfforce{2/12.5w,1/2h}{v9}
  \fmfforce{8.8/12.5w,1/2h}{v10}
  \fmfforce{4.5/12.5w,1h}{v11}
  \fmfforce{4.5/12.5w,0h}{v12}
  \fmf{plain}{v1,v9}
  \fmf{plain}{v2,v9}
  \fmf{plain}{v3,v7}
  \fmf{plain}{v4,v8}
  \fmf{double,width=0.2mm,left=1}{v5,v7,v5}
  \fmf{double,width=0.2mm,left=0.6}{v9,v6}
  \fmf{double,width=0.2mm,right=0.6}{v9,v5}
  \fmfv{decor.size=0, label=${\scs 1}$, l.dist=1mm, l.angle=-135}{v1}
  \fmfv{decor.size=0, label=${\scs 3}$, l.dist=1mm, l.angle=135}{v2}
  \fmfv{decor.size=0, label=${\scs 4}$, l.dist=1mm, l.angle=25}{v3}
  \fmfv{decor.size=0, label=${\scs 2}$, l.dist=1mm, l.angle=-25}{v4}
  \fmfv{decor.size=0, label=${\scs r}$, l.dist=0mm, l.angle=0}{v10}
  \fmfv{decor.size=0, label=${\scs {(p-q)}}$, l.dist=0.05mm, l.angle=90}{v11}
  \fmfv{decor.size=0, label=${\scs {(q-r)}}$, l.dist=0.01mm, l.angle=-90}{v12}
  \fmfdot{v9}
  \end{fmfgraph*} } } 
\hspace*{2mm}+ \hspace*{2mm} \frac{1}{6} \hspace*{1mm}
\sum _{q=2} ^p \sum _{r=2} ^q \sum _{s=2} ^r 
\sum _{t=1} ^{s-1} \sum _{k=0} ^{t-1} \hspace*{10mm}
  \parbox{14mm}{\centerline{
  \begin{fmfgraph*}(16,13.5)
  \setval
  \fmfforce{2/16w,0/13.5h}{v1}
  \fmfforce{0w,14.5/13.5h}{v2}
  \fmfforce{4/16w,2/13.5h}{v3}
  \fmfforce{2/16w,12.5/13.5h}{v4}
  \fmfforce{4/16w,7.5/13.5h}{v5}
  \fmfforce{1.3/16w,9.5/13.5h}{v6}
  \fmfforce{6.7/16w,11/13.5h}{v7}
  \fmfforce{10/16w,4.5/13.5h}{v8}
  \fmfforce{10/16w,8.5/13.5h}{v9}
  \fmfforce{14.2/16w,8.7/13.5h}{v10}
  \fmfforce{14/16w,4.5/13.5h}{v11}
  \fmfforce{1w,10.5/13.5h}{v12}
  \fmfforce{1w,2.5/13.5h}{v13}
  \fmfforce{4/16w,13.5/13.5h}{v14}
  \fmfforce{4/16w,10.5/13.5h}{v20}
  \fmfforce{12/16w,6.5/13.5h}{v21}
  \fmfforce{4/16w,5.5/13.5h}{v22}
  \fmfforce{-1/16w,5.5/13.5h}{v23}
  \fmfforce{12/16w,11/13.5h}{v24}
  \fmfforce{7.5/16w,0h}{v25}
  \fmf{plain}{v1,v3}
  \fmf{plain}{v2,v4}
  \fmf{plain}{v10,v12}
  \fmf{plain}{v11,v13}
  \fmf{double,width=0.2mm,left=1}{v5,v14,v5}
  \fmf{double,width=0.2mm,left=1}{v9,v11,v9}
  \fmf{double,width=0.2mm}{v3,v5}
  \fmf{double,width=0.2mm,left=0.6}{v3,v6}
  \fmf{double,width=0.2mm,left=0.15}{v7,v9}
  \fmf{double,width=0.2mm,right=0.25}{v3,v8}
  \fmfv{decor.size=0, label=${\scs 1}$, l.dist=1.5mm, l.angle=-180}{v1}
  \fmfv{decor.size=0, label=${\scs 2}$, l.dist=0.5mm, l.angle=180}{v2}
  \fmfv{decor.size=0, label=${\scs 3}$, l.dist=1mm, l.angle=45}{v12}
  \fmfv{decor.size=0, label=${\scs 4}$, l.dist=1mm, l.angle=-45}{v13}
  \fmfv{decor.size=0, label=${\scs {t-k}}$, l.dist=0mm, l.angle=90}{v20}
  \fmfv{decor.size=0, label=${\scs {s-t}}$, l.dist=0mm, l.angle=90}{v21}
  \fmfv{decor.size=0, label=${\scs (k)}$, l.dist=0.5mm, l.angle=0}{v22}
  \fmfv{decor.size=0, label=${\scs {(q-r)}}$, l.dist=0.1mm, l.angle=180}{v23}
  \fmfv{decor.size=0, label=${\scs {(r-s)}}$, l.dist=0.5mm, l.angle=90}{v24}
  \fmfv{decor.size=0, label=${\scs {(p-q)}}$, l.dist=0mm, l.angle=-90}{v25}
  \fmfdot{v3}
  \end{fmfgraph*} } } 
\hspace*{2cm}
\no \\*[0.5cm]
\hspace*{1.6cm} && +  \hspace*{2mm} \frac{1}{6} \hspace*{1mm}
\sum _{q=2} ^p \sum _{r=2} ^q \sum _{s=2} ^r 
\sum _{t=1} ^{s-1} \sum _{k=0} ^{t-1} \hspace*{10mm}
  \parbox{14mm}{\centerline{
  \begin{fmfgraph*}(16,13.5)
  \setval
  \fmfforce{2/16w,0/13.5h}{v1}
  \fmfforce{0w,14.5/13.5h}{v2}
  \fmfforce{4/16w,2/13.5h}{v3}
  \fmfforce{2/16w,12.5/13.5h}{v4}
  \fmfforce{4/16w,7.5/13.5h}{v5}
  \fmfforce{1.3/16w,9.5/13.5h}{v6}
  \fmfforce{6.7/16w,11/13.5h}{v7}
  \fmfforce{10/16w,4.5/13.5h}{v8}
  \fmfforce{10/16w,8.5/13.5h}{v9}
  \fmfforce{14.2/16w,8.7/13.5h}{v10}
  \fmfforce{14/16w,4.5/13.5h}{v11}
  \fmfforce{1w,10.5/13.5h}{v12}
  \fmfforce{1w,2.5/13.5h}{v13}
  \fmfforce{4/16w,13.5/13.5h}{v14}
  \fmfforce{4/16w,10.5/13.5h}{v20}
  \fmfforce{12/16w,6.5/13.5h}{v21}
  \fmfforce{4/16w,5.5/13.5h}{v22}
  \fmfforce{-1/16w,5.5/13.5h}{v23}
  \fmfforce{12/16w,11/13.5h}{v24}
  \fmfforce{7.5/16w,0h}{v25}
  \fmf{plain}{v1,v3}
  \fmf{plain}{v2,v4}
  \fmf{plain}{v10,v12}
  \fmf{plain}{v11,v13}
  \fmf{double,width=0.2mm,left=1}{v5,v14,v5}
  \fmf{double,width=0.2mm,left=1}{v9,v11,v9}
  \fmf{double,width=0.2mm}{v3,v5}
  \fmf{double,width=0.2mm,left=0.6}{v3,v6}
  \fmf{double,width=0.2mm,left=0.15}{v7,v9}
  \fmf{double,width=0.2mm,right=0.25}{v3,v8}
  \fmfv{decor.size=0, label=${\scs 1}$, l.dist=1.5mm, l.angle=-180}{v1}
  \fmfv{decor.size=0, label=${\scs 4}$, l.dist=0.5mm, l.angle=180}{v2}
  \fmfv{decor.size=0, label=${\scs 2}$, l.dist=1mm, l.angle=45}{v12}
  \fmfv{decor.size=0, label=${\scs 3}$, l.dist=1mm, l.angle=-45}{v13}
  \fmfv{decor.size=0, label=${\scs {t-k}}$, l.dist=0mm, l.angle=90}{v20}
  \fmfv{decor.size=0, label=${\scs {s-t}}$, l.dist=0mm, l.angle=90}{v21}
  \fmfv{decor.size=0, label=${\scs (k)}$, l.dist=0.5mm, l.angle=0}{v22}
  \fmfv{decor.size=0, label=${\scs {(q-r)}}$, l.dist=0.1mm, l.angle=180}{v23}
  \fmfv{decor.size=0, label=${\scs {(r-s)}}$, l.dist=0.5mm, l.angle=90}{v24}
  \fmfv{decor.size=0, label=${\scs {(p-q)}}$, l.dist=0mm, l.angle=-90}{v25}
  \fmfdot{v3}
  \end{fmfgraph*} } } 
\hspace*{4mm}+ \hspace*{2mm} \frac{1}{6} \hspace*{1mm}
\sum _{q=2} ^p \sum _{r=2} ^q \sum _{s=2} ^r 
\sum _{t=1} ^{s-1} \sum _{k=0} ^{t-1} \hspace*{10mm}
  \parbox{14mm}{\centerline{
  \begin{fmfgraph*}(16,13.5)
  \setval
  \fmfforce{2/16w,0/13.5h}{v1}
  \fmfforce{0w,14.5/13.5h}{v2}
  \fmfforce{4/16w,2/13.5h}{v3}
  \fmfforce{2/16w,12.5/13.5h}{v4}
  \fmfforce{4/16w,7.5/13.5h}{v5}
  \fmfforce{1.3/16w,9.5/13.5h}{v6}
  \fmfforce{6.7/16w,11/13.5h}{v7}
  \fmfforce{10/16w,4.5/13.5h}{v8}
  \fmfforce{10/16w,8.5/13.5h}{v9}
  \fmfforce{14.2/16w,8.7/13.5h}{v10}
  \fmfforce{14/16w,4.5/13.5h}{v11}
  \fmfforce{1w,10.5/13.5h}{v12}
  \fmfforce{1w,2.5/13.5h}{v13}
  \fmfforce{4/16w,13.5/13.5h}{v14}
  \fmfforce{4/16w,10.5/13.5h}{v20}
  \fmfforce{12/16w,6.5/13.5h}{v21}
  \fmfforce{4/16w,5.5/13.5h}{v22}
  \fmfforce{-1/16w,5.5/13.5h}{v23}
  \fmfforce{12/16w,11/13.5h}{v24}
  \fmfforce{7.5/16w,0h}{v25}
  \fmf{plain}{v1,v3}
  \fmf{plain}{v2,v4}
  \fmf{plain}{v10,v12}
  \fmf{plain}{v11,v13}
  \fmf{double,width=0.2mm,left=1}{v5,v14,v5}
  \fmf{double,width=0.2mm,left=1}{v9,v11,v9}
  \fmf{double,width=0.2mm}{v3,v5}
  \fmf{double,width=0.2mm,left=0.6}{v3,v6}
  \fmf{double,width=0.2mm,left=0.15}{v7,v9}
  \fmf{double,width=0.2mm,right=0.25}{v3,v8}
  \fmfv{decor.size=0, label=${\scs 1}$, l.dist=1.5mm, l.angle=-180}{v1}
  \fmfv{decor.size=0, label=${\scs 3}$, l.dist=0.5mm, l.angle=180}{v2}
  \fmfv{decor.size=0, label=${\scs 4}$, l.dist=1mm, l.angle=45}{v12}
  \fmfv{decor.size=0, label=${\scs 2}$, l.dist=1mm, l.angle=-45}{v13}
  \fmfv{decor.size=0, label=${\scs {t-k}}$, l.dist=0mm, l.angle=90}{v20}
  \fmfv{decor.size=0, label=${\scs {s-t}}$, l.dist=0mm, l.angle=90}{v21}
  \fmfv{decor.size=0, label=${\scs (k)}$, l.dist=0.5mm, l.angle=0}{v22}
  \fmfv{decor.size=0, label=${\scs {(q-r)}}$, l.dist=0.1mm, l.angle=180}{v23}
  \fmfv{decor.size=0, label=${\scs {(r-s)}}$, l.dist=0.5mm, l.angle=90}{v24}
  \fmfv{decor.size=0, label=${\scs {(p-q)}}$, l.dist=0mm, l.angle=-90}{v25}
  \fmfdot{v3}
  \end{fmfgraph*} } } 
\hspace*{5mm} . \\*[4mm] && \no 
\eeq
This is to be solved starting from 
\beq
\la{G0}
  \parbox{10mm}{\centerline{
  \begin{fmfgraph*}(7,3)
  \setval
  \fmfforce{0w,1/2h}{v1}
  \fmfforce{1w,1/2h}{v2}
  \fmfforce{1/2w,1/2h}{v3}
  \fmf{double,width=0.2mm}{v2,v1}
  \fmfv{decor.size=0, label=${\scs 1}$, l.dist=1mm, l.angle=-180}{v1}
  \fmfv{decor.size=0, label=${\scs 2}$, l.dist=1mm, l.angle=0}{v2}
  \fmfv{decor.size=0, label=${\scs (0)}$, l.dist=1.5mm, l.angle=90}{v3}
  \end{fmfgraph*}}}
\hspace*{3mm} &=& \hspace*{3mm}
  \parbox{10mm}{\centerline{
  \begin{fmfgraph*}(7,3)
  \setval
  \fmfleft{v1}
  \fmfright{v2}
  \fmf{plain}{v2,v1}
  \fmfv{decor.size=0, label=${\scs 1}$, l.dist=1mm, l.angle=-180}{v1}
  \fmfv{decor.size=0, label=${\scs 2}$, l.dist=1mm, l.angle=0}{v2}
  \end{fmfgraph*}}}
\hspace*{5mm} ,\\ && \no \\
 \parbox{10mm}{\centerline{
  \begin{fmfgraph*}(7.5,7.5)
  \setval
  \fmfforce{0w,0h}{v1}
  \fmfforce{0w,1h}{v2}
  \fmfforce{1w,1h}{v3}
  \fmfforce{1w,0h}{v4}
  \fmfforce{2/7.5w,2/7.5h}{v5}
  \fmfforce{2/7.5w,5.5/7.5h}{v6}
  \fmfforce{5.5/7.5w,5.5/7.5h}{v7}
  \fmfforce{5.5/7.5w,2/7.5h}{v8}
  \fmfforce{1/2w,1/2h}{v9}
  \fmf{plain}{v1,v5}
  \fmf{plain}{v2,v6}
  \fmf{plain}{v3,v7}
  \fmf{plain}{v4,v8}
  \fmf{double,width=0.2mm,left=1}{v5,v7,v5}
  \fmfv{decor.size=0, label=${\scs 1}$, l.dist=1mm, l.angle=-135}{v1}
  \fmfv{decor.size=0, label=${\scs 2}$, l.dist=1mm, l.angle=135}{v2}
  \fmfv{decor.size=0, label=${\scs 3}$, l.dist=1mm, l.angle=45}{v3}
  \fmfv{decor.size=0, label=${\scs 4}$, l.dist=1mm, l.angle=-45}{v4}
  \fmfv{decor.size=0, label=${\scs 1}$, l.dist=0mm, l.angle=90}{v9}
  \end{fmfgraph*} } } 
\hspace*{3mm} &=& \hspace*{3mm}
  \parbox{7mm}{\centerline{
  \begin{fmfgraph*}(4.5,4.5)
  \setval
  \fmfforce{0w,0h}{v1}
  \fmfforce{0w,1h}{v2}
  \fmfforce{1w,1h}{v3}
  \fmfforce{1w,0h}{v4}
  \fmfforce{1/2w,1/2h}{v5}
  \fmf{plain}{v1,v3}
  \fmf{plain}{v2,v4}
  \fmfv{decor.size=0, label=${\scs 1}$, l.dist=1mm, l.angle=-135}{v1}
  \fmfv{decor.size=0, label=${\scs 2}$, l.dist=1mm, l.angle=135}{v2}
  \fmfv{decor.size=0, label=${\scs 3}$, l.dist=1mm, l.angle=45}{v3}
  \fmfv{decor.size=0, label=${\scs 4}$, l.dist=1mm, l.angle=-45}{v4}
  \fmfdot{v5}
  \end{fmfgraph*} } } 
\hspace*{5mm} .\la{GAM1} \\ && \nonumber
\eeq
Note that these graphical recursion relations (\ref{REK4})--(\ref{GAM1})
allow to prove via complete induction that all diagrams which contribute to the
self-energy and the one-particle irreducible four-point function are, indeed, one-particle irreducible \ci{Glaum}.\\

The first few perturbative contributions to $\Sig_{12}$ and $\Gam_{1234}$ are determined as follows. Inserting 
(\ref{G0}) in (\ref{REK4}) and (\ref{REK5}) yield for $p=1$ the self-energy
\beq
  \parbox{14mm}{\centerline{
  \begin{fmfgraph*}(11,5)
  \setval
  \fmfforce{0w,1/2h}{v1}
  \fmfforce{3/11w,1/2h}{v2}
  \fmfforce{8/11w,1/2h}{v3}
  \fmfforce{1w,1/2h}{v4}
  \fmfforce{1/2w,1h}{v5}
  \fmfforce{1/2w,0h}{v6}
  \fmfforce{1/2w,1/2h}{v7}
  \fmf{plain}{v1,v2}
  \fmf{plain}{v3,v4}
  \fmf{double,width=0.2mm,left=1}{v5,v6,v5}
  \fmfv{decor.size=0, label=${\scs 1}$, l.dist=1mm, l.angle=-180}{v1}
  \fmfv{decor.size=0, label=${\scs 2}$, l.dist=1mm, l.angle=0}{v4}
  \fmfv{decor.size=0, label=${\scs 1}$, l.dist=0mm, l.angle=90}{v7}
  \end{fmfgraph*} } }
\hspace*{4mm} = \hspace*{3mm} \frac{1}{2} \hspace*{3mm}
  \parbox{9mm}{\centerline{
  \begin{fmfgraph*}(6,5)
  \setval
  \fmfforce{0w,0h}{v1}
  \fmfforce{1/2w,0h}{v2}
  \fmfforce{1w,0h}{v3}
  \fmfforce{1/2w,1h}{v4}
  \fmf{plain}{v1,v2}
  \fmf{plain}{v2,v3}
  \fmf{plain,left=1}{v4,v2,v4}
  \fmfv{decor.size=0, label=${\scs 1}$, l.dist=1mm, l.angle=-180}{v1}
  \fmfv{decor.size=0, label=${\scs 2}$, l.dist=1mm, l.angle=0}{v3}
  \fmfdot{v2}
  \end{fmfgraph*} } }
\hspace*{6mm} ,
\label{125}
\eeq
and the connected two-point function
\beq
  \parbox{10mm}{\centerline{
  \begin{fmfgraph*}(7,3)
  \setval
  \fmfleft{v1}
  \fmfright{v2}
  \fmfforce{1/2w,1/2h}{v3}
  \fmf{double,width=0.2mm}{v2,v1}
  \fmfv{decor.size=0, label=${\scs 1}$, l.dist=1mm, l.angle=-180}{v1}
  \fmfv{decor.size=0, label=${\scs 2}$, l.dist=1mm, l.angle=0}{v2}
  \fmfv{decor.size=0, label=${\scs (1)}$, l.dist=1.5mm, l.angle=90}{v3}
  \end{fmfgraph*}}}
\hspace*{5mm} = \hspace*{3mm}
\parbox{17mm}{\centerline{
  \begin{fmfgraph*}(15,3)
  \setval
  \fmfforce{0w,1/2h}{v1}
  \fmfforce{5/15w,1/2h}{v2}
  \fmfforce{10/15w,1/2h}{v3}
  \fmfforce{1w,1/2h}{v4}
  \fmfforce{1/2w,1/2h}{v5}
  \fmfforce{13.5/15w,1/2h}{v6}
  \fmf{plain}{v1,v2}
  \fmf{double,width=0.2mm}{v3,v4}
  \fmf{double,width=0.2mm,left=1}{v2,v3,v2}
  \fmfv{decor.size=0, label=${\scs 1}$, l.dist=1mm, l.angle=-180}{v1}
  \fmfv{decor.size=0, label=${\scs 2}$, l.dist=1.5mm, l.angle=0}{v4}
  \fmfv{decor.size=0, label=${\scs 1}$, l.dist=0mm, l.angle=90}{v5}
  \fmfv{decor.size=0, label=${\scs {(0)}}$, l.dist=1.5mm, l.angle=90}{v6}
  \end{fmfgraph*} } } 
\hspace*{5mm} = \hspace*{3mm} \frac{1}{2} \hspace*{3mm}
  \parbox{14mm}{\centerline{
  \begin{fmfgraph*}(10,5)
  \setval
  \fmfforce{0w,0h}{v1}
  \fmfforce{1/2w,0h}{v2}
  \fmfforce{1w,0h}{v3}
  \fmfforce{1/2w,1h}{v4}
  \fmf{plain}{v1,v2}
  \fmf{plain}{v2,v3}
  \fmf{plain,left=1}{v4,v2,v4}
  \fmfv{decor.size=0, label=${\scs 1}$, l.dist=1mm, l.angle=-180}{v1}
  \fmfv{decor.size=0, label=${\scs 2}$, l.dist=1mm, l.angle=0}{v3}
  \fmfdot{v2}
  \end{fmfgraph*} } }
\hspace*{5mm} .
\label{126}
\eeq
As (\ref{GAM1}) represent a bare vertex with no line, we read off
\beq
  \dphi{ }{6}{7} 
\hspace*{3mm}
  \parbox{11mm}{\centerline{
  \begin{fmfgraph*}(7.5,9.5)
  \setval
  \fmfforce{0w,2/9.5h}{v1}
  \fmfforce{0w,9.5/9.5h}{v2}
  \fmfforce{1w,9.5/9.5h}{v3}
  \fmfforce{1w,2/9.5h}{v4}
  \fmfforce{2/7.5w,4/9.5h}{v5}
  \fmfforce{2/7.5w,7.5/9.5h}{v6}
  \fmfforce{5.5/7.5w,7.5/9.5h}{v7}
  \fmfforce{5.5/7.5w,4/9.5h}{v8}
  \fmfforce{1/2w,5.75/9.5h}{v9}
  \fmf{plain}{v1,v5}
  \fmf{plain}{v2,v6}
  \fmf{plain}{v3,v7}
  \fmf{plain}{v4,v8}
  \fmf{double,width=0.2mm,left=1}{v5,v7,v5}
  \fmfv{decor.size=0, label=${\scs 5}$, l.dist=1mm, l.angle=-135}{v1}
  \fmfv{decor.size=0, label=${\scs 2}$, l.dist=1mm, l.angle=135}{v2}
  \fmfv{decor.size=0, label=${\scs 3}$, l.dist=1mm, l.angle=45}{v3}
  \fmfv{decor.size=0, label=${\scs 4}$, l.dist=1mm, l.angle=-45}{v4}
  \fmfv{decor.size=0, label=${\scs 1}$, l.dist=0mm, l.angle=90}{v9}
  \end{fmfgraph*} } } 
\hspace*{3mm} = \hs 0 \hspace*{4mm} .
\label{127}
\eeq
the second-order contribution to the one-particle irreducible four-point function follows from (\ref{REK6}) to be 

\beq
  \parbox{10mm}{\centerline{
  \begin{fmfgraph*}(7.5,7.5)
  \setval
  \fmfforce{0w,0h}{v1}
  \fmfforce{0w,1h}{v2}
  \fmfforce{1w,1h}{v3}
  \fmfforce{1w,0h}{v4}
  \fmfforce{2/7.5w,2/7.5h}{v5}
  \fmfforce{2/7.5w,5.5/7.5h}{v6}
  \fmfforce{5.5/7.5w,5.5/7.5h}{v7}
  \fmfforce{5.5/7.5w,2/7.5h}{v8}
  \fmfforce{1/2w,1/2h}{v9}
  \fmf{plain}{v1,v5}
  \fmf{plain}{v2,v6}
  \fmf{plain}{v3,v7}
  \fmf{plain}{v4,v8}
  \fmf{double,width=0.2mm,left=1}{v5,v7,v5}
  \fmfv{decor.size=0, label=${\scs 1}$, l.dist=1mm, l.angle=-135}{v1}
  \fmfv{decor.size=0, label=${\scs 2}$, l.dist=1mm, l.angle=135}{v2}
  \fmfv{decor.size=0, label=${\scs 3}$, l.dist=1mm, l.angle=45}{v3}
  \fmfv{decor.size=0, label=${\scs 4}$, l.dist=1mm, l.angle=-45}{v4}
  \fmfv{decor.size=0, label=${\scs 2}$, l.dist=0mm, l.angle=90}{v9}
  \end{fmfgraph*} } } 
\hspace*{3mm} = \hspace*{2mm} \frac{1}{2} \hspace*{4mm}
  \parbox{12mm}{\centerline{
  \begin{fmfgraph*}(9,5)
  \setval
  \fmfforce{0w,0h}{v1}
  \fmfforce{0w,1h}{v2}
  \fmfforce{1w,1h}{v3}
  \fmfforce{1w,0h}{v4}
  \fmfforce{2/9w,1/2h}{v5}
  \fmfforce{7/9w,1/2h}{v6}
  \fmf{plain}{v1,v5}
  \fmf{plain}{v2,v5}
  \fmf{plain}{v3,v6}
  \fmf{plain}{v4,v6}
  \fmf{plain,left=1}{v5,v6,v5}
  \fmfv{decor.size=0, label=${\scs 1}$, l.dist=1mm, l.angle=-135}{v1}
  \fmfv{decor.size=0, label=${\scs 2}$, l.dist=1mm, l.angle=135}{v2}
  \fmfv{decor.size=0, label=${\scs 3}$, l.dist=1mm, l.angle=45}{v3}
  \fmfv{decor.size=0, label=${\scs 4}$, l.dist=1mm, l.angle=-45}{v4}
  \fmfdot{v5,v6}
  \end{fmfgraph*} } } 
\hspace*{2mm} + \hspace*{2mm} \frac{1}{2} \hspace*{4mm}
  \parbox{12mm}{\centerline{
  \begin{fmfgraph*}(9,5)
  \setval
  \fmfforce{0w,0h}{v1}
  \fmfforce{0w,1h}{v2}
  \fmfforce{1w,1h}{v3}
  \fmfforce{1w,0h}{v4}
  \fmfforce{2/9w,1/2h}{v5}
  \fmfforce{7/9w,1/2h}{v6}
  \fmf{plain}{v1,v5}
  \fmf{plain}{v2,v5}
  \fmf{plain}{v3,v6}
  \fmf{plain}{v4,v6}
  \fmf{plain,left=1}{v5,v6,v5}
  \fmfv{decor.size=0, label=${\scs 1}$, l.dist=1mm, l.angle=-135}{v1}
  \fmfv{decor.size=0, label=${\scs 4}$, l.dist=1mm, l.angle=135}{v2}
  \fmfv{decor.size=0, label=${\scs 2}$, l.dist=1mm, l.angle=45}{v3}
  \fmfv{decor.size=0, label=${\scs 3}$, l.dist=1mm, l.angle=-45}{v4}
  \fmfdot{v5,v6}
  \end{fmfgraph*} } } 
\hspace*{2mm} + \hspace*{2mm}  \frac{1}{2} \hspace*{4mm}
  \parbox{12mm}{\centerline{
  \begin{fmfgraph*}(9,5)
  \setval
  \fmfforce{0w,0h}{v1}
  \fmfforce{0w,1h}{v2}
  \fmfforce{1w,1h}{v3}
  \fmfforce{1w,0h}{v4}
  \fmfforce{2/9w,1/2h}{v5}
  \fmfforce{7/9w,1/2h}{v6}
  \fmf{plain}{v1,v5}
  \fmf{plain}{v2,v5}
  \fmf{plain}{v3,v6}
  \fmf{plain}{v4,v6}
  \fmf{plain,left=1}{v5,v6,v5}
  \fmfv{decor.size=0, label=${\scs 1}$, l.dist=1mm, l.angle=-135}{v1}
  \fmfv{decor.size=0, label=${\scs 3}$, l.dist=1mm, l.angle=135}{v2}
  \fmfv{decor.size=0, label=${\scs 4}$, l.dist=1mm, l.angle=45}{v3}
  \fmfv{decor.size=0, label=${\scs 2}$, l.dist=1mm, l.angle=-45}{v4}
  \fmfdot{v5,v6}
  \end{fmfgraph*} } } 
\hspace*{5mm} .\label{128} \\ \nonumber
\eeq
The results in (\ref{125}), (\ref{126}), and (\ref{128}) are then used to determine for $p=2$ the self-energy (\ref{REK4}) 
\beq
  \parbox{14mm}{\centerline{
  \begin{fmfgraph*}(11,5)
  \setval
  \fmfforce{0w,1/2h}{v1}
  \fmfforce{3/11w,1/2h}{v2}
  \fmfforce{8/11w,1/2h}{v3}
  \fmfforce{1w,1/2h}{v4}
  \fmfforce{1/2w,1h}{v5}
  \fmfforce{1/2w,0h}{v6}
  \fmfforce{1/2w,1/2h}{v7}
  \fmf{plain}{v1,v2}
  \fmf{plain}{v3,v4}
  \fmf{double,width=0.2mm,left=1}{v5,v6,v5}
  \fmfv{decor.size=0, label=${\scs 1}$, l.dist=1mm, l.angle=-180}{v1}
  \fmfv{decor.size=0, label=${\scs 2}$, l.dist=1mm, l.angle=0}{v4}
  \fmfv{decor.size=0, label=${\scs 2}$, l.dist=0mm, l.angle=90}{v7}
  \end{fmfgraph*} } }
 \hspace*{3mm} = \hspace*{3mm} \frac{1}{4}  \hspace*{2mm} 
  \parbox{9mm}{\centerline{
  \begin{fmfgraph*}(5,10)
  \setval
  \fmfforce{0w,2/10h}{v1}
  \fmfforce{1/2w,2/10h}{v2}
  \fmfforce{1w,2/10h}{v3}
  \fmfforce{1/2w,7/10h}{v4}
  \fmfforce{1/2w,12/10h}{v5}
  \fmf{plain}{v1,v3}
  \fmf{plain,left=1}{v2,v4,v2}
  \fmf{plain,left=1}{v4,v5,v4}
  \fmfv{decor.size=0, label=${\scs 1}$, l.dist=1mm, l.angle=-180}{v1}
  \fmfv{decor.size=0, label=${\scs 2}$, l.dist=1mm, l.angle=0}{v3}
  \fmfdot{v2}
  \fmfdot{v4}
  \end{fmfgraph*}}}
\hspace*{3mm} + \hspace*{1.5mm} \frac{1}{6} \hspace*{2.5mm}
  \parbox{15mm}{\centerline{
  \begin{fmfgraph*}(11,6)
  \setval
  \fmfforce{0w,1/2h}{v1}
  \fmfforce{2.5/11w,1/2h}{v2}
  \fmfforce{8.5/11w,1/2h}{v3}
  \fmfforce{1w,1/2h}{v4}
  \fmf{plain}{v1,v4}
  \fmf{plain,left=1}{v2,v3,v2}
  \fmfv{decor.size=0, label=${\scs 1}$, l.dist=1mm, l.angle=-180}{v1}
  \fmfv{decor.size=0, label=${\scs 2}$, l.dist=1mm, l.angle=0}{v4}
  \fmfdot{v2}
  \fmfdot{v3}
  \end{fmfgraph*}}}
\label{129}
\eeq
and the connected two-point function (\ref{REK5})
\beq
\hspace*{5mm}
  \parbox{10mm}{\centerline{
  \begin{fmfgraph*}(7,3)
  \setval
  \fmfleft{v1}
  \fmfright{v2}
  \fmfforce{1/2w,1/2h}{v3}
  \fmf{double,width=0.2mm}{v2,v1}
  \fmfv{decor.size=0, label=${\scs 1}$, l.dist=1mm, l.angle=-180}{v1}
  \fmfv{decor.size=0, label=${\scs 2}$, l.dist=1mm, l.angle=0}{v2}
  \fmfv{decor.size=0, label=${\scs (2)}$, l.dist=1.5mm, l.angle=90}{v3}
  \end{fmfgraph*}}}
\hspace*{2mm}= \hspace*{3mm}
\parbox{17mm}{\centerline{
  \begin{fmfgraph*}(15,3)
  \setval
  \fmfforce{0w,1/2h}{v1}
  \fmfforce{5/15w,1/2h}{v2}
  \fmfforce{10/15w,1/2h}{v3}
  \fmfforce{1w,1/2h}{v4}
  \fmfforce{1/2w,1/2h}{v5}
  \fmfforce{13.5/15w,1/2h}{v6}
  \fmf{plain}{v1,v2}
  \fmf{double,width=0.2mm}{v3,v4}
  \fmf{double,width=0.2mm,left=1}{v2,v3,v2}
  \fmfv{decor.size=0, label=${\scs 1}$, l.dist=1mm, l.angle=-180}{v1}
  \fmfv{decor.size=0, label=${\scs 2}$, l.dist=1.5mm, l.angle=0}{v4}
  \fmfv{decor.size=0, label=${\scs 2}$, l.dist=0mm, l.angle=90}{v5}
  \fmfv{decor.size=0, label=${\scs {(0)}}$, l.dist=1.5mm, l.angle=90}{v6}
  \end{fmfgraph*} } } 
\hspace*{3mm}+ \hspace*{3mm}
\parbox{17mm}{\centerline{
  \begin{fmfgraph*}(15,3)
  \setval
  \fmfforce{0w,1/2h}{v1}
  \fmfforce{5/15w,1/2h}{v2}
  \fmfforce{10/15w,1/2h}{v3}
  \fmfforce{1w,1/2h}{v4}
  \fmfforce{1/2w,1/2h}{v5}
  \fmfforce{13.5/15w,1/2h}{v6}
  \fmf{plain}{v1,v2}
  \fmf{double,width=0.2mm}{v3,v4}
  \fmf{double,width=0.2mm,left=1}{v2,v3,v2}
  \fmfv{decor.size=0, label=${\scs 1}$, l.dist=1mm, l.angle=-180}{v1}
  \fmfv{decor.size=0, label=${\scs 2}$, l.dist=1.5mm, l.angle=0}{v4}
  \fmfv{decor.size=0, label=${\scs 1}$, l.dist=0mm, l.angle=90}{v5}
  \fmfv{decor.size=0, label=${\scs {(1)}}$, l.dist=1.5mm, l.angle=90}{v6}
  \end{fmfgraph*} } } 
\hspace*{3mm}= \hspace*{2mm} \frac{1}{4}  \hspace*{2mm} 
  \parbox{20mm}{\centerline{
  \begin{fmfgraph*}(17,5)
  \setval
  \fmfforce{0w,0h}{v1}
  \fmfforce{5/17w,0h}{v2}
  \fmfforce{12/17w,0h}{v3}
  \fmfforce{1w,0h}{v4}
  \fmfforce{5/17w,1h}{v5}
  \fmfforce{12/17w,1h}{v6}
  \fmf{plain}{v1,v4}
  \fmf{plain,left=1}{v2,v5,v2}
  \fmf{plain,left=1}{v3,v6,v3}
  \fmfv{decor.size=0, label=${\scs 1}$, l.dist=1mm, l.angle=-180}{v1}
  \fmfv{decor.size=0, label=${\scs 2}$, l.dist=1mm, l.angle=0}{v4}
  \fmfdot{v2}
  \fmfdot{v3}
  \end{fmfgraph*}}}
\hspace*{3mm} + \hspace*{1.5mm} \frac{1}{4} \hspace*{2mm}
  \parbox{13mm}{\centerline{
  \begin{fmfgraph*}(10,10)
  \setval
  \fmfforce{0w,2/10h}{v1}
  \fmfforce{1/2w,2/10h}{v2}
  \fmfforce{1w,2/10h}{v3}
  \fmfforce{1/2w,7/10h}{v4}
  \fmfforce{1/2w,12/10h}{v5}
  \fmf{plain}{v1,v3}
  \fmf{plain,left=1}{v2,v4,v2}
  \fmf{plain,left=1}{v4,v5,v4}
  \fmfv{decor.size=0, label=${\scs 1}$, l.dist=1mm, l.angle=-180}{v1}
  \fmfv{decor.size=0, label=${\scs 2}$, l.dist=1mm, l.angle=0}{v3}
  \fmfdot{v2}
  \fmfdot{v4}
  \end{fmfgraph*}}}
\hspace*{3mm} + \hspace*{1.5mm} \frac{1}{6} \hspace*{2.5mm}
  \parbox{20mm}{\centerline{
  \begin{fmfgraph*}(16,9)
  \setval
  \fmfforce{0w,1/2h}{v1}
  \fmfforce{5/16w,1/2h}{v2}
  \fmfforce{11/16w,1/2h}{v3}
  \fmfforce{1w,1/2h}{v4}
  \fmf{plain}{v1,v4}
  \fmf{plain,left=1}{v2,v3,v2}
  \fmfv{decor.size=0, label=${\scs 1}$, l.dist=1mm, l.angle=-180}{v1}
  \fmfv{decor.size=0, label=${\scs 2}$, l.dist=1mm, l.angle=0}{v4}
  \fmfdot{v2}
  \fmfdot{v3}
  \end{fmfgraph*}}}
\hspace*{3mm} . \hspace*{4mm}
\label{130}
\eeq
Amputating one line from (\ref{128}),
\beq
\raisebox{2mm}{\begin{minipage}{1.8cm}
\beq
\dphi{
  \parbox{14mm}{\centerline{
  \begin{fmfgraph*}(7.5,7.5)
  \setval
  \fmfforce{0w,2.5/7.5h}{v1}
  \fmfforce{0w,10/7.5h}{v2}
  \fmfforce{1w,10/7.5h}{v3}
  \fmfforce{1w,2.5/7.5h}{v4}
  \fmfforce{2/7.5w,4.5/7.5h}{v5}
  \fmfforce{2/7.5w,8/7.5h}{v6}
  \fmfforce{5.5/7.5w,8/7.5h}{v7}
  \fmfforce{5.5/7.5w,4.5/7.5h}{v8}
  \fmfforce{1/2w,6.25/7.5h}{v9}
  \fmf{plain}{v1,v5}
  \fmf{plain}{v2,v6}
  \fmf{plain}{v3,v7}
  \fmf{plain}{v4,v8}
  \fmf{double,width=0.2mm,left=1}{v5,v7,v5}
  \fmfv{decor.size=0, label=${\scs 5}$, l.dist=0.5mm, l.angle=-135}{v1}
  \fmfv{decor.size=0, label=${\scs 2}$, l.dist=0.5mm, l.angle=135}{v2}
  \fmfv{decor.size=0, label=${\scs 3}$, l.dist=0.5mm, l.angle=45}{v3}
  \fmfv{decor.size=0, label=${\scs 4}$, l.dist=0.5mm, l.angle=-45}{v4}
  \fmfv{decor.size=0, label=${\scs 2}$, l.dist=0mm, l.angle=90}{v9}
  \end{fmfgraph*} } } 
}{6}{7}
\no  \eeq 
\end{minipage}}
\hspace*{1.5mm} = \hspace*{3mm} \frac{1}{2} \hspace*{2mm}
  \dphi{ }{6}{7}
\hspace*{1.5mm}  \left(  \hspace*{3mm}
  \parbox{12mm}{\centerline{
  \begin{fmfgraph*}(9,5)
  \setval
  \fmfforce{0w,0h}{v1}
  \fmfforce{0w,1h}{v2}
  \fmfforce{1w,1h}{v3}
  \fmfforce{1w,0h}{v4}
  \fmfforce{2/9w,1/2h}{v5}
  \fmfforce{7/9w,1/2h}{v6}
  \fmf{plain}{v1,v5}
  \fmf{plain}{v2,v5}
  \fmf{plain}{v3,v6}
  \fmf{plain}{v4,v6}
  \fmf{plain,left=1}{v5,v6,v5}
  \fmfv{decor.size=0, label=${\scs 5}$, l.dist=1mm, l.angle=-135}{v1}
  \fmfv{decor.size=0, label=${\scs 2}$, l.dist=1mm, l.angle=135}{v2}
  \fmfv{decor.size=0, label=${\scs 3}$, l.dist=1mm, l.angle=45}{v3}
  \fmfv{decor.size=0, label=${\scs 4}$, l.dist=1mm, l.angle=-45}{v4}
  \fmfdot{v5,v6}
  \end{fmfgraph*} } } 
\hspace*{3mm} + \hspace*{3mm}
  \parbox{12mm}{\centerline{
  \begin{fmfgraph*}(9,5)
  \setval
  \fmfforce{0w,0h}{v1}
  \fmfforce{0w,1h}{v2}
  \fmfforce{1w,1h}{v3}
  \fmfforce{1w,0h}{v4}
  \fmfforce{2/9w,1/2h}{v5}
  \fmfforce{7/9w,1/2h}{v6}
  \fmf{plain}{v1,v5}
  \fmf{plain}{v2,v5}
  \fmf{plain}{v3,v6}
  \fmf{plain}{v4,v6}
  \fmf{plain,left=1}{v5,v6,v5}
  \fmfv{decor.size=0, label=${\scs 5}$, l.dist=1mm, l.angle=-135}{v1}
  \fmfv{decor.size=0, label=${\scs 4}$, l.dist=1mm, l.angle=135}{v2}
  \fmfv{decor.size=0, label=${\scs 2}$, l.dist=1mm, l.angle=45}{v3}
  \fmfv{decor.size=0, label=${\scs 3}$, l.dist=1mm, l.angle=-45}{v4}
  \fmfdot{v5,v6}
  \end{fmfgraph*} } } 
\hspace*{3mm} + \hspace*{3mm}
  \parbox{12mm}{\centerline{
  \begin{fmfgraph*}(9,5)
  \setval
  \fmfforce{0w,0h}{v1}
  \fmfforce{0w,1h}{v2}
  \fmfforce{1w,1h}{v3}
  \fmfforce{1w,0h}{v4}
  \fmfforce{2/9w,1/2h}{v5}
  \fmfforce{7/9w,1/2h}{v6}
  \fmf{plain}{v1,v5}
  \fmf{plain}{v2,v5}
  \fmf{plain}{v3,v6}
  \fmf{plain}{v4,v6}
  \fmf{plain,left=1}{v5,v6,v5}
  \fmfv{decor.size=0, label=${\scs 5}$, l.dist=1mm, l.angle=-135}{v1}
  \fmfv{decor.size=0, label=${\scs 3}$, l.dist=1mm, l.angle=135}{v2}
  \fmfv{decor.size=0, label=${\scs 4}$, l.dist=1mm, l.angle=45}{v3}
  \fmfv{decor.size=0, label=${\scs 2}$, l.dist=1mm, l.angle=-45}{v4}
  \fmfdot{v5,v6}
  \end{fmfgraph*} } } 
\hspace*{3mm}  \right) 
\no  
\eeq
\beq
= \hspace*{1.5mm} \frac{1}{2} \hspace*{1.5mm}  \left(   \hspace*{1.5mm}
  \parbox{15mm}{\centerline{
  \begin{fmfgraph*}(11,5)
  \setval
  \fmfforce{0w,1/2h}{v1}
  \fmfforce{1w,1/2h}{v2}
  \fmfforce{2.5/11w,1h}{v3}
  \fmfforce{2.5/11w,0h}{v4}
  \fmfforce{8.5/11w,1h}{v5}
  \fmfforce{8.5/11w,0h}{v6}
  \fmfforce{2.5/11w,1/2h}{v7}
  \fmfforce{8.5/11w,1/2h}{v8}
  \fmf{plain}{v1,v2}
  \fmf{plain}{v3,v4}
  \fmf{plain}{v5,v6}
  \fmfv{decor.size=0, label=${\scs 5}$, l.dist=1mm, l.angle=180}{v1}
  \fmfv{decor.size=0, label=${\scs 4}$, l.dist=1mm, l.angle=0}{v2}
  \fmfv{decor.size=0, label=${\scs 2}$, l.dist=1mm, l.angle=90}{v3}
  \fmfv{decor.size=0, label=${\scs 6}$, l.dist=1mm, l.angle=-90}{v4}
  \fmfv{decor.size=0, label=${\scs 3}$, l.dist=1mm, l.angle=90}{v5}
  \fmfv{decor.size=0, label=${\scs 7}$, l.dist=1mm, l.angle=-90}{v6}
  \fmfdot{v7,v8}
  \end{fmfgraph*} } } 
\hspace*{2mm} + \hspace*{2mm} 
  \parbox{15mm}{\centerline{
  \begin{fmfgraph*}(11,5)
  \setval
  \fmfforce{0w,1/2h}{v1}
  \fmfforce{1w,1/2h}{v2}
  \fmfforce{2.5/11w,1h}{v3}
  \fmfforce{2.5/11w,0h}{v4}
  \fmfforce{8.5/11w,1h}{v5}
  \fmfforce{8.5/11w,0h}{v6}
  \fmfforce{2.5/11w,1/2h}{v7}
  \fmfforce{8.5/11w,1/2h}{v8}
  \fmf{plain}{v1,v2}
  \fmf{plain}{v3,v4}
  \fmf{plain}{v5,v6}
  \fmfv{decor.size=0, label=${\scs 5}$, l.dist=1mm, l.angle=180}{v1}
  \fmfv{decor.size=0, label=${\scs 4}$, l.dist=1mm, l.angle=0}{v2}
  \fmfv{decor.size=0, label=${\scs 2}$, l.dist=1mm, l.angle=90}{v3}
  \fmfv{decor.size=0, label=${\scs 7}$, l.dist=1mm, l.angle=-90}{v4}
  \fmfv{decor.size=0, label=${\scs 3}$, l.dist=1mm, l.angle=90}{v5}
  \fmfv{decor.size=0, label=${\scs 6}$, l.dist=1mm, l.angle=-90}{v6}
  \fmfdot{v7,v8}
  \end{fmfgraph*} } } 
\hspace*{2mm} + \hspace*{2mm} 
  \parbox{15mm}{\centerline{
  \begin{fmfgraph*}(11,5)
  \setval
  \fmfforce{0w,1/2h}{v1}
  \fmfforce{1w,1/2h}{v2}
  \fmfforce{2.5/11w,1h}{v3}
  \fmfforce{2.5/11w,0h}{v4}
  \fmfforce{8.5/11w,1h}{v5}
  \fmfforce{8.5/11w,0h}{v6}
  \fmfforce{2.5/11w,1/2h}{v7}
  \fmfforce{8.5/11w,1/2h}{v8}
  \fmf{plain}{v1,v2}
  \fmf{plain}{v3,v4}
  \fmf{plain}{v5,v6}
  \fmfv{decor.size=0, label=${\scs 5}$, l.dist=1mm, l.angle=180}{v1}
  \fmfv{decor.size=0, label=${\scs 3}$, l.dist=1mm, l.angle=0}{v2}
  \fmfv{decor.size=0, label=${\scs 4}$, l.dist=1mm, l.angle=90}{v3}
  \fmfv{decor.size=0, label=${\scs 6}$, l.dist=1mm, l.angle=-90}{v4}
  \fmfv{decor.size=0, label=${\scs 2}$, l.dist=1mm, l.angle=90}{v5}
  \fmfv{decor.size=0, label=${\scs 7}$, l.dist=1mm, l.angle=-90}{v6}
  \fmfdot{v7,v8}
  \end{fmfgraph*} } } 
\hspace*{2mm} + \hspace*{2mm} 
  \parbox{15mm}{\centerline{
  \begin{fmfgraph*}(11,5)
  \setval
  \fmfforce{0w,1/2h}{v1}
  \fmfforce{1w,1/2h}{v2}
  \fmfforce{2.5/11w,1h}{v3}
  \fmfforce{2.5/11w,0h}{v4}
  \fmfforce{8.5/11w,1h}{v5}
  \fmfforce{8.5/11w,0h}{v6}
  \fmfforce{2.5/11w,1/2h}{v7}
  \fmfforce{8.5/11w,1/2h}{v8}
  \fmf{plain}{v1,v2}
  \fmf{plain}{v3,v4}
  \fmf{plain}{v5,v6}
  \fmfv{decor.size=0, label=${\scs 5}$, l.dist=1mm, l.angle=180}{v1}
  \fmfv{decor.size=0, label=${\scs 3}$, l.dist=1mm, l.angle=0}{v2}
  \fmfv{decor.size=0, label=${\scs 4}$, l.dist=1mm, l.angle=90}{v3}
  \fmfv{decor.size=0, label=${\scs 7}$, l.dist=1mm, l.angle=-90}{v4}
  \fmfv{decor.size=0, label=${\scs 2}$, l.dist=1mm, l.angle=90}{v5}
  \fmfv{decor.size=0, label=${\scs 6}$, l.dist=1mm, l.angle=-90}{v6}
  \fmfdot{v7,v8}
  \end{fmfgraph*} } } 
\hspace*{2mm} + \hspace*{2mm} 
  \parbox{15mm}{\centerline{
  \begin{fmfgraph*}(11,5)
  \setval
  \fmfforce{0w,1/2h}{v1}
  \fmfforce{1w,1/2h}{v2}
  \fmfforce{2.5/11w,1h}{v3}
  \fmfforce{2.5/11w,0h}{v4}
  \fmfforce{8.5/11w,1h}{v5}
  \fmfforce{8.5/11w,0h}{v6}
  \fmfforce{2.5/11w,1/2h}{v7}
  \fmfforce{8.5/11w,1/2h}{v8}
  \fmf{plain}{v1,v2}
  \fmf{plain}{v3,v4}
  \fmf{plain}{v5,v6}
  \fmfv{decor.size=0, label=${\scs 5}$, l.dist=1mm, l.angle=180}{v1}
  \fmfv{decor.size=0, label=${\scs 2}$, l.dist=1mm, l.angle=0}{v2}
  \fmfv{decor.size=0, label=${\scs 3}$, l.dist=1mm, l.angle=90}{v3}
  \fmfv{decor.size=0, label=${\scs 6}$, l.dist=1mm, l.angle=-90}{v4}
  \fmfv{decor.size=0, label=${\scs 4}$, l.dist=1mm, l.angle=90}{v5}
  \fmfv{decor.size=0, label=${\scs 7}$, l.dist=1mm, l.angle=-90}{v6}
  \fmfdot{v7,v8}
  \end{fmfgraph*} } } 
\hspace*{2mm} + \hspace*{2mm} 
  \parbox{15mm}{\centerline{
  \begin{fmfgraph*}(11,5)
  \setval
  \fmfforce{0w,1/2h}{v1}
  \fmfforce{1w,1/2h}{v2}
  \fmfforce{2.5/11w,1h}{v3}
  \fmfforce{2.5/11w,0h}{v4}
  \fmfforce{8.5/11w,1h}{v5}
  \fmfforce{8.5/11w,0h}{v6}
  \fmfforce{2.5/11w,1/2h}{v7}
  \fmfforce{8.5/11w,1/2h}{v8}
  \fmf{plain}{v1,v2}
  \fmf{plain}{v3,v4}
  \fmf{plain}{v5,v6}
  \fmfv{decor.size=0, label=${\scs 5}$, l.dist=1mm, l.angle=180}{v1}
  \fmfv{decor.size=0, label=${\scs 2}$, l.dist=1mm, l.angle=0}{v2}
  \fmfv{decor.size=0, label=${\scs 3}$, l.dist=1mm, l.angle=90}{v3}
  \fmfv{decor.size=0, label=${\scs 6}$, l.dist=1mm, l.angle=-90}{v4}
  \fmfv{decor.size=0, label=${\scs 4}$, l.dist=1mm, l.angle=90}{v5}
  \fmfv{decor.size=0, label=${\scs 7}$, l.dist=1mm, l.angle=-90}{v6}
  \fmfdot{v7,v8}
  \end{fmfgraph*} } } 
\hspace*{1.5mm}  \right)  \hspace*{2mm} ,   \label{132} \\ && \no
\eeq
we find the third-order contribution to the one-particle irreducible four-point function from (\ref{REK6}):\\
\beq
& & \hspace*{1.7cm}
  \parbox{10mm}{\centerline{
  \begin{fmfgraph*}(7.5,7.5)
  \setval
  \fmfforce{0w,0h}{v1}
  \fmfforce{0w,1h}{v2}
  \fmfforce{1w,1h}{v3}
  \fmfforce{1w,0h}{v4}
  \fmfforce{2/7.5w,2/7.5h}{v5}
  \fmfforce{2/7.5w,5.5/7.5h}{v6}
  \fmfforce{5.5/7.5w,5.5/7.5h}{v7}
  \fmfforce{5.5/7.5w,2/7.5h}{v8}
  \fmfforce{1/2w,1/2h}{v9}
  \fmf{plain}{v1,v5}
  \fmf{plain}{v2,v6}
  \fmf{plain}{v3,v7}
  \fmf{plain}{v4,v8}
  \fmf{double,width=0.2mm,left=1}{v5,v7,v5}
  \fmfv{decor.size=0, label=${\scs 1}$, l.dist=1mm, l.angle=-135}{v1}
  \fmfv{decor.size=0, label=${\scs 2}$, l.dist=1mm, l.angle=135}{v2}
  \fmfv{decor.size=0, label=${\scs 3}$, l.dist=1mm, l.angle=45}{v3}
  \fmfv{decor.size=0, label=${\scs 4}$, l.dist=1mm, l.angle=-45}{v4}
  \fmfv{decor.size=0, label=${\scs 3}$, l.dist=0mm, l.angle=90}{v9}
  \end{fmfgraph*} } } 
\hspace*{2.5mm} = \hspace*{2mm} \frac{1}{4} \hspace*{2mm}  
\left(   \hspace*{3mm}
  \parbox{17mm}{\centerline{
  \begin{fmfgraph*}(14,5)
  \setval
  \fmfforce{0w,0h}{v1}
  \fmfforce{0w,1h}{v2}
  \fmfforce{1w,1h}{v3}
  \fmfforce{1w,0h}{v4}
  \fmfforce{2/14w,1/2h}{v5}
  \fmfforce{7/14w,1/2h}{v6}
  \fmfforce{12/14w,1/2h}{v7}
  \fmf{plain}{v1,v5}
  \fmf{plain}{v2,v5}
  \fmf{plain}{v3,v7}
  \fmf{plain}{v4,v7}
  \fmf{plain,left=1}{v5,v6,v5}
  \fmf{plain,left=1}{v7,v6,v7}
  \fmfv{decor.size=0, label=${\scs 1}$, l.dist=1mm, l.angle=-135}{v1}
  \fmfv{decor.size=0, label=${\scs 2}$, l.dist=1mm, l.angle=135}{v2}
  \fmfv{decor.size=0, label=${\scs 3}$, l.dist=1mm, l.angle=45}{v3}
  \fmfv{decor.size=0, label=${\scs 4}$, l.dist=1mm, l.angle=-45}{v4}
  \fmfdot{v5,v6,v7}
  \end{fmfgraph*} } } 
\hspace*{2.5mm} + \hspace*{2.5mm}
  \parbox{17mm}{\centerline{
  \begin{fmfgraph*}(14,5)
  \setval
  \fmfforce{0w,0h}{v1}
  \fmfforce{0w,1h}{v2}
  \fmfforce{1w,1h}{v3}
  \fmfforce{1w,0h}{v4}
  \fmfforce{2/14w,1/2h}{v5}
  \fmfforce{7/14w,1/2h}{v6}
  \fmfforce{12/14w,1/2h}{v7}
  \fmf{plain}{v1,v5}
  \fmf{plain}{v2,v5}
  \fmf{plain}{v3,v7}
  \fmf{plain}{v4,v7}
  \fmf{plain,left=1}{v5,v6,v5}
  \fmf{plain,left=1}{v7,v6,v7}
  \fmfv{decor.size=0, label=${\scs 1}$, l.dist=1mm, l.angle=-135}{v1}
  \fmfv{decor.size=0, label=${\scs 4}$, l.dist=1mm, l.angle=135}{v2}
  \fmfv{decor.size=0, label=${\scs 2}$, l.dist=1mm, l.angle=45}{v3}
  \fmfv{decor.size=0, label=${\scs 3}$, l.dist=1mm, l.angle=-45}{v4}
  \fmfdot{v5,v6,v7}
  \end{fmfgraph*} } } 
\hspace*{2.5mm} + \hspace*{2.5mm}
  \parbox{17mm}{\centerline{
  \begin{fmfgraph*}(14,5)
  \setval
  \fmfforce{0w,0h}{v1}
  \fmfforce{0w,1h}{v2}
  \fmfforce{1w,1h}{v3}
  \fmfforce{1w,0h}{v4}
  \fmfforce{2/14w,1/2h}{v5}
  \fmfforce{7/14w,1/2h}{v6}
  \fmfforce{12/14w,1/2h}{v7}
  \fmf{plain}{v1,v5}
  \fmf{plain}{v2,v5}
  \fmf{plain}{v3,v7}
  \fmf{plain}{v4,v7}
  \fmf{plain,left=1}{v5,v6,v5}
  \fmf{plain,left=1}{v7,v6,v7}
  \fmfv{decor.size=0, label=${\scs 1}$, l.dist=1mm, l.angle=-135}{v1}
  \fmfv{decor.size=0, label=${\scs 3}$, l.dist=1mm, l.angle=135}{v2}
  \fmfv{decor.size=0, label=${\scs 4}$, l.dist=1mm, l.angle=45}{v3}
  \fmfv{decor.size=0, label=${\scs 2}$, l.dist=1mm, l.angle=-45}{v4}
  \fmfdot{v5,v6,v7}
  \end{fmfgraph*} } } 
\hspace*{3mm}  \right)
\no  \\*[5mm]
& &\hspace*{-8mm} + \hspace*{2mm} \frac{1}{2} \hspace*{2mm} 
\left(  \hspace*{2mm}
  \parbox{14mm}{\centerline{
  \begin{fmfgraph*}(10,7)
  \setval
  \fmfforce{0w,2.5/7h}{v1}
  \fmfforce{1/4w,2.5/7h}{v2}
  \fmfforce{3/4w,2.5/7h}{v3}
  \fmfforce{1w,2.5/7h}{v4}
  \fmfforce{1/2w,5/7h}{v5}
  \fmfforce{3/10w,1h}{v6}
  \fmfforce{7/10w,1h}{v7}
  \fmf{plain}{v1,v4}
  \fmf{plain}{v5,v6}
  \fmf{plain}{v5,v7}
  \fmf{plain,left=1}{v2,v3,v2}
  \fmfv{decor.size=0, label=${\scs 1}$, l.dist=1mm, l.angle=-180}{v1}
  \fmfv{decor.size=0, label=${\scs 2}$, l.dist=1mm, l.angle=0}{v4}
  \fmfv{decor.size=0, label=${\scs 3}$, l.dist=1mm, l.angle=135}{v6}
  \fmfv{decor.size=0, label=${\scs 4}$, l.dist=1mm, l.angle=45}{v7}
  \fmfdot{v2,v3,v5}
  \end{fmfgraph*}}}
\hspace*{3mm} + \hspace*{2mm}
  \parbox{14mm}{\centerline{
  \begin{fmfgraph*}(10,7)
  \setval
  \fmfforce{0w,2.5/7h}{v1}
  \fmfforce{1/4w,2.5/7h}{v2}
  \fmfforce{3/4w,2.5/7h}{v3}
  \fmfforce{1w,2.5/7h}{v4}
  \fmfforce{1/2w,5/7h}{v5}
  \fmfforce{3/10w,1h}{v6}
  \fmfforce{7/10w,1h}{v7}
  \fmf{plain}{v1,v4}
  \fmf{plain}{v5,v6}
  \fmf{plain}{v5,v7}
  \fmf{plain,left=1}{v2,v3,v2}
  \fmfv{decor.size=0, label=${\scs 1}$, l.dist=1mm, l.angle=-180}{v1}
  \fmfv{decor.size=0, label=${\scs 4}$, l.dist=1mm, l.angle=0}{v4}
  \fmfv{decor.size=0, label=${\scs 2}$, l.dist=1mm, l.angle=135}{v6}
  \fmfv{decor.size=0, label=${\scs 3}$, l.dist=1mm, l.angle=45}{v7}
  \fmfdot{v2,v3,v5}
  \end{fmfgraph*}}}
\hspace*{3mm} + \hspace*{2mm}
  \parbox{14mm}{\centerline{
  \begin{fmfgraph*}(10,7)
  \setval
  \fmfforce{0w,2.5/7h}{v1}
  \fmfforce{1/4w,2.5/7h}{v2}
  \fmfforce{3/4w,2.5/7h}{v3}
  \fmfforce{1w,2.5/7h}{v4}
  \fmfforce{1/2w,5/7h}{v5}
  \fmfforce{3/10w,1h}{v6}
  \fmfforce{7/10w,1h}{v7}
  \fmf{plain}{v1,v4}
  \fmf{plain}{v5,v6}
  \fmf{plain}{v5,v7}
  \fmf{plain,left=1}{v2,v3,v2}
  \fmfv{decor.size=0, label=${\scs 1}$, l.dist=1mm, l.angle=-180}{v1}
  \fmfv{decor.size=0, label=${\scs 3}$, l.dist=1mm, l.angle=0}{v4}
  \fmfv{decor.size=0, label=${\scs 4}$, l.dist=1mm, l.angle=135}{v6}
  \fmfv{decor.size=0, label=${\scs 2}$, l.dist=1mm, l.angle=45}{v7}
  \fmfdot{v2,v3,v5}
  \end{fmfgraph*}}}
\hspace*{3mm} + \hspace*{2mm}
  \parbox{14mm}{\centerline{
  \begin{fmfgraph*}(10,7)
  \setval
  \fmfforce{0w,2.5/7h}{v1}
  \fmfforce{1/4w,2.5/7h}{v2}
  \fmfforce{3/4w,2.5/7h}{v3}
  \fmfforce{1w,2.5/7h}{v4}
  \fmfforce{1/2w,5/7h}{v5}
  \fmfforce{3/10w,1h}{v6}
  \fmfforce{7/10w,1h}{v7}
  \fmf{plain}{v1,v4}
  \fmf{plain}{v5,v6}
  \fmf{plain}{v5,v7}
  \fmf{plain,left=1}{v2,v3,v2}
  \fmfv{decor.size=0, label=${\scs 4}$, l.dist=1mm, l.angle=-180}{v1}
  \fmfv{decor.size=0, label=${\scs 3}$, l.dist=1mm, l.angle=0}{v4}
  \fmfv{decor.size=0, label=${\scs 1}$, l.dist=1mm, l.angle=135}{v6}
  \fmfv{decor.size=0, label=${\scs 2}$, l.dist=1mm, l.angle=45}{v7}
  \fmfdot{v2,v3,v5}
  \end{fmfgraph*}}}
\hspace*{3mm} + \hspace*{2mm}
  \parbox{14mm}{\centerline{
  \begin{fmfgraph*}(10,7)
  \setval
  \fmfforce{0w,2.5/7h}{v1}
  \fmfforce{1/4w,2.5/7h}{v2}
  \fmfforce{3/4w,2.5/7h}{v3}
  \fmfforce{1w,2.5/7h}{v4}
  \fmfforce{1/2w,5/7h}{v5}
  \fmfforce{3/10w,1h}{v6}
  \fmfforce{7/10w,1h}{v7}
  \fmf{plain}{v1,v4}
  \fmf{plain}{v5,v6}
  \fmf{plain}{v5,v7}
  \fmf{plain,left=1}{v2,v3,v2}
  \fmfv{decor.size=0, label=${\scs 3}$, l.dist=1mm, l.angle=-180}{v1}
  \fmfv{decor.size=0, label=${\scs 2}$, l.dist=1mm, l.angle=0}{v4}
  \fmfv{decor.size=0, label=${\scs 1}$, l.dist=1mm, l.angle=135}{v6}
  \fmfv{decor.size=0, label=${\scs 4}$, l.dist=1mm, l.angle=45}{v7}
  \fmfdot{v2,v3,v5}
  \end{fmfgraph*}}}
\hspace*{3mm} + \hspace*{2mm}
  \parbox{14mm}{\centerline{
  \begin{fmfgraph*}(10,7)
  \setval
  \fmfforce{0w,2.5/7h}{v1}
  \fmfforce{1/4w,2.5/7h}{v2}
  \fmfforce{3/4w,2.5/7h}{v3}
  \fmfforce{1w,2.5/7h}{v4}
  \fmfforce{1/2w,5/7h}{v5}
  \fmfforce{3/10w,1h}{v6}
  \fmfforce{7/10w,1h}{v7}
  \fmf{plain}{v1,v4}
  \fmf{plain}{v5,v6}
  \fmf{plain}{v5,v7}
  \fmf{plain,left=1}{v2,v3,v2}
  \fmfv{decor.size=0, label=${\scs 2}$, l.dist=1mm, l.angle=-180}{v1}
  \fmfv{decor.size=0, label=${\scs 4}$, l.dist=1mm, l.angle=0}{v4}
  \fmfv{decor.size=0, label=${\scs 1}$, l.dist=1mm, l.angle=135}{v6}
  \fmfv{decor.size=0, label=${\scs 3}$, l.dist=1mm, l.angle=45}{v7}
  \fmfdot{v2,v3,v5}
  \end{fmfgraph*}}}
\hspace*{3mm} \right)
\no  \\*[4.5mm]
& & \hspace*{3cm} + \hspace*{2mm} \frac{1}{2} \hspace*{2mm}  
\left(  \hspace*{3mm}
  \parbox{12mm}{\centerline{
  \begin{fmfgraph*}(9,10)
  \setval
  \fmfforce{0w,0.3h}{v1}
  \fmfforce{0w,0.7h}{v2}
  \fmfforce{1w,0.7h}{v3}
  \fmfforce{1w,0.3h}{v4}
  \fmfforce{2/9w,0.5h}{v5}
  \fmfforce{7/9w,0.5h}{v6}
  \fmfforce{1/2w,0.75h}{v7}
  \fmfforce{1/2w,1.25h}{v8}
  \fmf{plain}{v1,v5}
  \fmf{plain}{v2,v5}
  \fmf{plain}{v3,v6}
  \fmf{plain}{v4,v6}
  \fmf{plain,left=1}{v5,v6,v5}
  \fmf{plain,left=1}{v7,v8,v7}
  \fmfv{decor.size=0, label=${\scs 1}$, l.dist=1mm, l.angle=-135}{v1}
  \fmfv{decor.size=0, label=${\scs 2}$, l.dist=1mm, l.angle=135}{v2}
  \fmfv{decor.size=0, label=${\scs 3}$, l.dist=1mm, l.angle=45}{v3}
  \fmfv{decor.size=0, label=${\scs 4}$, l.dist=1mm, l.angle=-45}{v4}
  \fmfdot{v5,v6,v7}
  \end{fmfgraph*} } } 
\hspace*{3mm} + \hspace*{3mm}
  \parbox{12mm}{\centerline{
  \begin{fmfgraph*}(9,10)
  \setval
  \fmfforce{0w,0.3h}{v1}
  \fmfforce{0w,0.7h}{v2}
  \fmfforce{1w,0.7h}{v3}
  \fmfforce{1w,0.3h}{v4}
  \fmfforce{2/9w,0.5h}{v5}
  \fmfforce{7/9w,0.5h}{v6}
  \fmfforce{1/2w,0.75h}{v7}
  \fmfforce{1/2w,1.25h}{v8}
  \fmf{plain}{v1,v5}
  \fmf{plain}{v2,v5}
  \fmf{plain}{v3,v6}
  \fmf{plain}{v4,v6}
  \fmf{plain,left=1}{v5,v6,v5}
  \fmf{plain,left=1}{v7,v8,v7}
  \fmfv{decor.size=0, label=${\scs 1}$, l.dist=1mm, l.angle=-135}{v1}
  \fmfv{decor.size=0, label=${\scs 4}$, l.dist=1mm, l.angle=135}{v2}
  \fmfv{decor.size=0, label=${\scs 2}$, l.dist=1mm, l.angle=45}{v3}
  \fmfv{decor.size=0, label=${\scs 3}$, l.dist=1mm, l.angle=-45}{v4}
  \fmfdot{v5,v6,v7}
  \end{fmfgraph*} } } 
\hspace*{3mm} + \hspace*{3mm}
  \parbox{12mm}{\centerline{
  \begin{fmfgraph*}(9,10)
  \setval
  \fmfforce{0w,0.3h}{v1}
  \fmfforce{0w,0.7h}{v2}
  \fmfforce{1w,0.7h}{v3}
  \fmfforce{1w,0.3h}{v4}
  \fmfforce{2/9w,0.5h}{v5}
  \fmfforce{7/9w,0.5h}{v6}
  \fmfforce{1/2w,0.75h}{v7}
  \fmfforce{1/2w,1.25h}{v8}
  \fmf{plain}{v1,v5}
  \fmf{plain}{v2,v5}
  \fmf{plain}{v3,v6}
  \fmf{plain}{v4,v6}
  \fmf{plain,left=1}{v5,v6,v5}
  \fmf{plain,left=1}{v7,v8,v7}
  \fmfv{decor.size=0, label=${\scs 1}$, l.dist=1mm, l.angle=-135}{v1}
  \fmfv{decor.size=0, label=${\scs 3}$, l.dist=1mm, l.angle=135}{v2}
  \fmfv{decor.size=0, label=${\scs 4}$, l.dist=1mm, l.angle=45}{v3}
  \fmfv{decor.size=0, label=${\scs 2}$, l.dist=1mm, l.angle=-45}{v4}
  \fmfdot{v5,v6,v7}
  \end{fmfgraph*} } } 
\hspace*{3mm}  \right)  \hspace*{2mm} .
\label{136}
\eeq
As expected, the diagrams (\ref{126}) and (\ref{130}) for the connected two-point function 
coincide with the ones from Section \ref{GRC}
which are shown in Table \ref{tab1}. Furthermore, the diagrams (\ref{125}), (\ref{129}) and (\ref{GAM1}), (\ref{128}), (\ref{136}) 
for the self-energy and
the one-particle irreducible four-point function are listed in Table \ref{tab4} and \ref{tab5} which show all one-particle
irreducible diagrams up to the order $p=4$ irrespective of their spatial indices (compare the discussion 
before Eq. (\ref{SYMMM})). Note that
these one-particle irreducible 
Feynman diagrams have to be evaluated in order to determine the critical exponents of the $\phi^4$-theory
from renormalizing the field $\phi$, the coupling constant $g$ and the mass $m^2$. 
So far, the results are available up to six and partly to seven loops 
in $d=3$~\cite{Nickel1,Sokolov,Nickel2} and up
to five loops in $d=4-\epsilon$ dimensions  whithin the minimal subtraction scheme \cite{Verena,FIVE}.
\subsection{One-Particle Irreducible Vacuum Diagrams}\la{IVVA}
The vacuum diagrams of $\phi^4$-theory are not only connected but also one-particle irreducible. In this section we elaborate
how they can be generated from short-circuiting the external legs of 
the diagrams of the self-energy and the one-particle irreducible four-point function, respectively.
\subsubsection{Relation to the Diagrams of the Self-Energy}
At first, we consider the approach which is based on the self-energy. Combining the Dyson equation (\ref{Dyson})
with the perturbative expansion of the connected two-point function
(\ref{34b}) and the corresponding one for the self-energy
\beq
\Sig_{12} = \sum_{p=1}^{\infty} \Sig_{12}^{(p)} \, ,
\eeq
we obtain
\beq
\int_{12} \Gz^{(p)}_{12} \hs G^{-1}_{12} \hspace*{2mm} = \hspace*{2mm} 
\sum_{q=1}^{p} \hs \int_{12} \Sig^{(q)}_{12} \hs \Gz^{(p-q)}_{12}   \hspace*{3mm} .
\label{161}
\eeq
Thus our previous equation (\ref{83}) for determining the contributions of the vacuum energy $W^{(p)}$ from the connected
two-point function is converted to 
\beq
W^{(p)} \hspace*{2mm} = \hspace*{2mm} \frac{1}{4p} \hs \sum_{q=1}^{p}\hs \int_{12} 
\Sig^{(q)}_{12} \hs  \Gz^{(p-q)}_{12}   \hspace*{3mm} .
\label{162}
\eeq
This result reads graphically
\beq
\hspace*{5mm}
  \parbox{10mm}{\centerline{
  \begin{fmfgraph*}(8,6)
  \setval
  \fmfforce{0w,1/2h}{v1}
  \fmfforce{1w,1/2h}{v2}
  \fmfforce{1/2w,1h}{v3}
  \fmfforce{1/2w,0h}{v4}
  \fmfforce{1/2w,1/2h}{v5}
  \fmf{plain,left=0.4}{v1,v3,v2,v4,v1}
  \fmfv{decor.size=0, label=${\scs {p}}$, l.dist=0mm, l.angle=90}{v5}
  \end{fmfgraph*} } }  
\hspace*{2mm} = \hspace*{3mm} \frac{1}{4 p} \hspace*{2mm}
\sum_{q=1}^{p} \hspace*{2mm}
  \parbox{12mm}{\centerline{
  \begin{fmfgraph*}(9,8)
  \setval
  \fmfforce{2/9w,7/8h}{v1}
  \fmfforce{7/9w,7/8h}{v2}
  \fmfforce{2/9w,1/8h}{v3}
  \fmfforce{7/9w,1/8h}{v4}
  \fmfforce{1/2w,7/8h}{v5}
  \fmfforce{1/2w,1/8h}{v6}
  \fmf{double,width=0.2mm,left=0.7}{v2,v4}
  \fmf{double,width=0.2mm,right=0.7}{v1,v3}
  \fmf{double,width=0.2mm,right=0.2}{v3,v4}
  \fmf{double,width=0.2mm,left=1}{v1,v2,v1}
  \fmfv{decor.size=0, label=${\scs q}$, l.dist=0mm, l.angle=90}{v5}
  \fmfv{decor.size=0, label=${\scs {(p-q)}}$, l.dist=1.5mm, l.angle=-90}{v6}
  \end{fmfgraph*} } }  
\hspace*{2mm} .
\no \la{SEV} \\ \no
\eeq
We consider one example how the diagrams of the self-energy lead to the corresponding diagrams
of the vacuum energy. For $p=3$ Eq. (\ref{SEV}) reduces to
\beq
  \parbox{10mm}{\centerline{
  \begin{fmfgraph*}(8,6)
  \setval
  \fmfforce{0w,1/2h}{v1}
  \fmfforce{1w,1/2h}{v2}
  \fmfforce{1/2w,1h}{v3}
  \fmfforce{1/2w,0h}{v4}
  \fmfforce{1/2w,1/2h}{v5}
  \fmf{plain,left=0.4}{v1,v3,v2,v4,v1}
  \fmfv{decor.size=0, label=${\scs 3}$, l.dist=0mm, l.angle=90}{v5}
  \end{fmfgraph*} } }  
\hspace*{3mm} = \hspace*{3mm} \frac{1}{12} \hspace*{2mm} 
  \parbox{12mm}{\centerline{
  \begin{fmfgraph*}(9,8)
  \setval
  \fmfforce{2/9w,7/8h}{v1}
  \fmfforce{7/9w,7/8h}{v2}
  \fmfforce{2/9w,1/8h}{v3}
  \fmfforce{7/9w,1/8h}{v4}
  \fmfforce{1/2w,7/8h}{v5}
  \fmfforce{1/2w,1/8h}{v6}
  \fmf{double,width=0.2mm,left=0.7}{v2,v4}
  \fmf{double,width=0.2mm,right=0.7}{v1,v3}
  \fmf{double,width=0.2mm,right=0.2}{v3,v4}
  \fmf{double,width=0.2mm,left=1}{v1,v2,v1}
  \fmfv{decor.size=0, label=${\scs 1}$, l.dist=0mm, l.angle=90}{v5}
  \fmfv{decor.size=0, label=${\scs (2)}$, l.dist=1.5mm, l.angle=-90}{v6}
  \end{fmfgraph*} } }  
\hspace*{2mm} + \hspace*{2mm} \frac{1}{12} \hspace*{2mm}
  \parbox{12mm}{\centerline{
  \begin{fmfgraph*}(9,8)
  \setval
  \fmfforce{2/9w,7/8h}{v1}
  \fmfforce{7/9w,7/8h}{v2}
  \fmfforce{2/9w,1/8h}{v3}
  \fmfforce{7/9w,1/8h}{v4}
  \fmfforce{1/2w,7/8h}{v5}
  \fmfforce{1/2w,1/8h}{v6}
  \fmf{double,width=0.2mm,left=0.7}{v2,v4}
  \fmf{double,width=0.2mm,right=0.7}{v1,v3}
  \fmf{double,width=0.2mm,right=0.2}{v3,v4}
  \fmf{double,width=0.2mm,left=1}{v1,v2,v1}
  \fmfv{decor.size=0, label=${\scs 2}$, l.dist=0mm, l.angle=90}{v5}
  \fmfv{decor.size=0, label=${\scs (1)}$, l.dist=1.5mm, l.angle=-90}{v6}
  \end{fmfgraph*} } }  
\hspace*{2mm} + \hspace*{2mm} \frac{1}{12} \hspace*{2mm} 
  \parbox{12mm}{\centerline{
  \begin{fmfgraph*}(9,8)
  \setval
  \fmfforce{2/9w,7/8h}{v1}
  \fmfforce{7/9w,7/8h}{v2}
  \fmfforce{2/9w,1/8h}{v3}
  \fmfforce{7/9w,1/8h}{v4}
  \fmfforce{1/2w,7/8h}{v5}
  \fmfforce{1/2w,1/8h}{v6}
  \fmf{double,width=0.2mm,left=0.7}{v2,v4}
  \fmf{double,width=0.2mm,right=0.7}{v1,v3}
  \fmf{double,width=0.2mm,right=0.2}{v3,v4}
  \fmf{double,width=0.2mm,left=1}{v1,v2,v1}
  \fmfv{decor.size=0, label=${\scs 3}$, l.dist=0mm, l.angle=90}{v5}
  \fmfv{decor.size=0, label=${\scs (0)}$, l.dist=1.5mm, l.angle=-90}{v6}
  \end{fmfgraph*} } }  
\hspace*{3mm} 
\label{163}
\eeq
with the respective terms
\beq
  \parbox{12mm}{\centerline{
  \begin{fmfgraph*}(9,8)
  \setval
  \fmfforce{2/9w,7/8h}{v1}
  \fmfforce{7/9w,7/8h}{v2}
  \fmfforce{2/9w,1/8h}{v3}
  \fmfforce{7/9w,1/8h}{v4}
  \fmfforce{1/2w,7/8h}{v5}
  \fmfforce{1/2w,1/8h}{v6}
  \fmf{double,width=0.2mm,left=0.7}{v2,v4}
  \fmf{double,width=0.2mm,right=0.7}{v1,v3}
  \fmf{double,width=0.2mm,right=0.2}{v3,v4}
  \fmf{double,width=0.2mm,left=1}{v1,v2,v1}
  \fmfv{decor.size=0, label=${\scs 1}$, l.dist=0mm, l.angle=90}{v5}
  \fmfv{decor.size=0, label=${\scs (2)}$, l.dist=1.5mm, l.angle=-90}{v6}
  \end{fmfgraph*} } }  
&=&
\frac{1}{12} \hspace*{3mm}
  \parbox{10.5mm}{\centerline{
  \begin{fmfgraph}(8,11)
  \setval
  \fmfforce{0w,3/11h}{v1}
  \fmfforce{1w,3/11h}{v2}
  \fmfforce{0.5w,7/11h}{v3}
  \fmfforce{0.5w,12/11h}{v4}
  \fmf{plain,left=1}{v1,v2,v1}
  \fmf{plain,left=0.4}{v1,v2,v1}
  \fmf{plain,left=1}{v3,v4,v3}
  \fmfdot{v1,v2,v3}
  \end{fmfgraph}}} 
\hspace*{3mm} + \hspace*{2mm} \frac{1}{8} \hspace*{2.5mm}
  \parbox{20mm}{\centerline{
  \begin{fmfgraph}(20,3)
  \setval
  \fmfleft{i1}
  \fmfright{o1}
  \fmf{plain,left=1}{i1,v1,i1}
  \fmf{plain,left=1}{v1,v2,v1}
  \fmf{plain,left=1}{v2,v3,v2}
  \fmf{plain,left=1}{o1,v3,o1}
  \fmfdot{v1,v2,v3}
  \end{fmfgraph}}}
\hspace*{3mm} + \hspace*{2mm} \frac{1}{8} \hspace*{2mm}
\parbox{17mm}{\centerline{
\begin{fmfgraph}(13,9)
\setval
\fmfforce{1/2w,0h}{v1}
\fmfforce{1/2w,5/9h}{v2}
\fmfforce{1/2w,10/9h}{v3}
\fmfforce{4.3/13w,1.4/9h}{v4}
\fmfforce{8.7/13w,1.4/9h}{v5}
\fmfforce{0w,-1/9h}{v6}
\fmfforce{1w,-1/9h}{v7}
\fmf{plain,left=1}{v1,v2,v1}
\fmf{plain,left=1}{v2,v3,v2}
\fmf{plain,left=1}{v4,v6,v4}
\fmf{plain,left=1}{v5,v7,v5}
\fmfdot{v2,v4,v5}
\end{fmfgraph}}} 
\hspace*{4mm} ,
\label{164} \\[4mm]
\parbox{12mm}{\centerline{
  \begin{fmfgraph*}(9,8)
  \setval
  \fmfforce{2/9w,7/8h}{v1}
  \fmfforce{7/9w,7/8h}{v2}
  \fmfforce{2/9w,1/8h}{v3}
  \fmfforce{7/9w,1/8h}{v4}
  \fmfforce{1/2w,7/8h}{v5}
  \fmfforce{1/2w,1/8h}{v6}
  \fmf{double,width=0.2mm,left=0.7}{v2,v4}
  \fmf{double,width=0.2mm,right=0.7}{v1,v3}
  \fmf{double,width=0.2mm,right=0.2}{v3,v4}
  \fmf{double,width=0.2mm,left=1}{v1,v2,v1}
  \fmfv{decor.size=0, label=${\scs 2}$, l.dist=0mm, l.angle=90}{v5}
  \fmfv{decor.size=0, label=${\scs (1)}$, l.dist=1.5mm, l.angle=-90}{v6}
  \end{fmfgraph*} } }  
&=&
\frac{1}{12} \hspace*{3mm}
  \parbox{10.5mm}{\centerline{
  \begin{fmfgraph}(8,11)
  \setval
  \fmfforce{0w,3/11h}{v1}
  \fmfforce{1w,3/11h}{v2}
  \fmfforce{0.5w,7/11h}{v3}
  \fmfforce{0.5w,12/11h}{v4}
  \fmf{plain,left=1}{v1,v2,v1}
  \fmf{plain,left=0.4}{v1,v2,v1}
  \fmf{plain,left=1}{v3,v4,v3}
  \fmfdot{v1,v2,v3}
  \end{fmfgraph}}} 
\hspace*{2mm} + \hspace*{2mm} \frac{1}{8} \hspace*{3.5mm}
  \parbox{20mm}{\centerline{
  \begin{fmfgraph}(20,3)
  \setval
  \fmfleft{i1}
  \fmfright{o1}
  \fmf{plain,left=1}{i1,v1,i1}
  \fmf{plain,left=1}{v1,v2,v1}
  \fmf{plain,left=1}{v2,v3,v2}
  \fmf{plain,left=1}{o1,v3,o1}
  \fmfdot{v1,v2,v3}
  \end{fmfgraph}}}
\hspace*{5mm} , 
\label{165}\\[4mm]
\parbox{12mm}{\centerline{
  \begin{fmfgraph*}(9,8)
  \setval
  \fmfforce{2/9w,7/8h}{v1}
  \fmfforce{7/9w,7/8h}{v2}
  \fmfforce{2/9w,1/8h}{v3}
  \fmfforce{7/9w,1/8h}{v4}
  \fmfforce{1/2w,7/8h}{v5}
  \fmfforce{1/2w,1/8h}{v6}
  \fmf{double,width=0.2mm,left=0.7}{v2,v4}
  \fmf{double,width=0.2mm,right=0.7}{v1,v3}
  \fmf{double,width=0.2mm,right=0.2}{v3,v4}
  \fmf{double,width=0.2mm,left=1}{v1,v2,v1}
  \fmfv{decor.size=0, label=${\scs 3}$, l.dist=0mm, l.angle=90}{v5}
  \fmfv{decor.size=0, label=${\scs (0)}$, l.dist=1.5mm, l.angle=-90}{v6}
  \end{fmfgraph*} } }  
&=&
\frac{1}{4} \hspace*{3mm}
  \parbox{11mm}{\centerline{
  \begin{fmfgraph}(8,6)
  \setval
  \fmfforce{0.5w,-1/6h}{v1}
  \fmfforce{0.5w,7/6h}{v2}
  \fmfforce{0.066987w,1/6h}{v3}
  \fmfforce{0.93301w,1/6h}{v4}
  \fmf{plain,left=1}{v1,v2,v1}
  \fmf{plain}{v2,v3}
  \fmf{plain}{v3,v4}
  \fmf{plain}{v2,v4}
  \fmfdot{v2,v3,v4}
  \end{fmfgraph}}}
\hspace*{3mm} + \hspace*{2mm} \frac{1}{3} \hspace*{3mm}
  \parbox{10.5mm}{\centerline{
  \begin{fmfgraph}(8,11)
  \setval
  \fmfforce{0w,3/11h}{v1}
  \fmfforce{1w,3/11h}{v2}
  \fmfforce{0.5w,7/11h}{v3}
  \fmfforce{0.5w,12/11h}{v4}
  \fmf{plain,left=1}{v1,v2,v1}
  \fmf{plain,left=0.4}{v1,v2,v1}
  \fmf{plain,left=1}{v3,v4,v3}
  \fmfdot{v1,v2,v3}
  \end{fmfgraph}}} 
\hspace*{3mm} + \hspace*{2mm} \frac{1}{8} \hspace*{3mm}
  \parbox{20mm}{\centerline{
  \begin{fmfgraph}(20,3)
  \setval
  \fmfleft{i1}
  \fmfright{o1}
  \fmf{plain,left=1}{i1,v1,i1}
  \fmf{plain,left=1}{v1,v2,v1}
  \fmf{plain,left=1}{v2,v3,v2}
  \fmf{plain,left=1}{o1,v3,o1}
  \fmfdot{v1,v2,v3}
  \end{fmfgraph}}}
\hspace*{3mm} + \hspace*{2mm} \frac{1}{8} \hspace*{2mm}
  \parbox{17mm}{\centerline{
  \begin{fmfgraph}(13,9)
  \setval
  \fmfforce{1/2w,0h}{v1}
  \fmfforce{1/2w,5/9h}{v2}
  \fmfforce{1/2w,10/9h}{v3}
  \fmfforce{4.3/13w,1.4/9h}{v4}
  \fmfforce{8.7/13w,1.4/9h}{v5}
  \fmfforce{0w,-1/9h}{v6}
  \fmfforce{1w,-1/9h}{v7}
  \fmf{plain,left=1}{v1,v2,v1}
  \fmf{plain,left=1}{v2,v3,v2}
  \fmf{plain,left=1}{v4,v6,v4}
  \fmf{plain,left=1}{v5,v7,v5}
  \fmfdot{v2,v4,v5}
  \end{fmfgraph}}} 
\hspace*{3mm} .
\label{166} \\[-1mm] \no
\eeq
Thus we reobtain the connected vacuum diagrams (\ref{82.2}) shown in Tab. \ref{tab3} for $p=3$.
\subsubsection{Relation to the Diagrams of the One-Particle Irreducible Four-Point Function}
For the sake of completeness we also mention that the vacuum diagrams can be generated from closing the diagrams of the
one-particle irreducible four-point function. To this end we insert in the equation (\ref{SEV})
determining the contributions of the vacuum energy the recursion relation (\ref{REK4}) for the self-energy and yield\\[2mm]
\beq
\hspace*{4mm}
  \parbox{10mm}{\centerline{
  \begin{fmfgraph*}(9,6.8)
  \setval
  \fmfforce{0w,1/2h}{v1}
  \fmfforce{1w,1/2h}{v2}
  \fmfforce{1/2w,1h}{v3}
  \fmfforce{1/2w,0h}{v4}
  \fmfforce{1/2w,1/2h}{v5}
  \fmf{plain,left=0.4}{v1,v3,v2,v4,v1}
  \fmfv{decor.size=0, label=${\scs {p+1}}$, l.dist=0mm, l.angle=90}{v5}
  \end{fmfgraph*} } }  
\hspace*{2mm} = \hspace*{2mm} \frac{1}{8(p+1)} \hspace*{3mm}
\sum_{q=0}^{p} \hspace*{5mm}
  \parbox{12mm}{\centerline{
  \begin{fmfgraph*}(10,5)
  \setval
  \fmfforce{0w,1/2h}{v1}
  \fmfforce{1/2w,1/2h}{v2}
  \fmfforce{1w,1/2h}{v3}
  \fmf{double,width=0.2mm,left=1}{v1,v2,v1}
  \fmf{double,width=0.2mm,left=1}{v3,v2,v3}
  \fmfv{decor.size=0, label=${\scs (q)}$, l.dist=1mm, l.angle=180}{v1}
  \fmfv{decor.size=0, label=${\scs {(p-q)}}$, l.dist=1mm, l.angle=0}{v3}
  \fmfdot{v2}
  \end{fmfgraph*} } }  
\hspace*{8mm} + \hspace*{1.5mm} \frac{1}{24(p+1)} \hspace*{2mm}
\sum_{q=1}^p  \sum_{r=1}^q  \sum_{s=1}^r  \sum_{t=1}^s  \hspace*{4mm}
  \parbox{14mm}{\centerline{
  \begin{fmfgraph*}(16,10)
  \setval
  \fmfforce{-1/16w,5/10h}{v1}
  \fmfforce{13.5/16w,7.5/10h}{v3}
  \fmfforce{13.5/16w,2.5/10h}{v4}
  \fmfforce{11.5/16w,6.3/10h}{v5}
  \fmfforce{11.5/16w,3.7/10h}{v6}
  \fmfforce{13.5/16w,0.5h}{v8}
  \fmfforce{7/16w,12/10h}{v10}
  \fmfforce{7/16w,8/10h}{v11}
  \fmfforce{7/16w,2/10h}{v12}
  \fmfforce{7/16w,-2/10h}{v13}
  \fmf{double,width=0.2mm,left=0.8}{v1,v3}
  \fmf{double,width=0.2mm,left=0.28}{v1,v5}
  \fmf{double,width=0.2mm,right=0.8}{v1,v4}
  \fmf{double,width=0.2mm,right=0.28}{v1,v6}
  \fmf{double,width=0.2mm,width=0.2mm,left=1}{v3,v4,v3}
  \fmfv{decor.size=0, label=${\scs t}$, l.dist=0mm, l.angle=90}{v8}
  \fmfv{decor.size=0, label=${\scs {(s-t)}}$, l.dist=1.3mm, l.angle=90}{v10}
  \fmfv{decor.size=0, label=${\scs {(r-s)}}$, l.dist=0.5mm, l.angle=90}{v11}
  \fmfv{decor.size=0, label=${\scs {(q-r)}}$, l.dist=0.3mm, l.angle=-90}{v12}
  \fmfv{decor.size=0, label=${\scs {(p-q)}}$, l.dist=1mm, l.angle=-90}{v13}
  \fmfdot{v1}
  \end{fmfgraph*} } }
\hspace*{3mm} .
\label{NEWV} \\[2mm]  \no
\eeq
Note that this equation follows also directly from the previous recursion relation (\ref{VRE}) by taking into account
the identity (\ref{150}) and the perturbative expansions (\ref{G2P}), (\ref{122}).
\end{fmffile}
\section{Summary and Outlook}
In this paper we have derived a closed set of Schwinger-Dyson equations in the 
disordered, symmetric phase of the $\phi^4$-theory. In particular, we supplemented the well-known integral equations
(\ref{SDG1}) and (\ref{106}) for the connected two-point function and the self-energy by the new 
functional integrodifferential equations (\ref{SDG2}) and (\ref{120})
for the connected and one-particle irreducible four-point function. 
Their conversion to graphical recursion relations has allowed us to
systematically generate the corresponding connected and one-particle irreducible Feynman diagrams. Furthermore,
we have discussed how the short-circuiting of their external legs 
leads to the associated connected vacuum diagrams. In the subsequent paper \ci{new}
we shall elaborate how tadpoles and, more generally, corrections to the connected two-point function as well as the vertex
can be successively eliminated by introducing higher Legendre transformations \ci{Kleinert1,Kleinert2,Vasiliev}. 
This will lead to graphical recursion
relations for the skeleton Feynman diagrams in $\phi^4$-theory.\\

The recursive graphical solution of our closed set of Schwinger-Dyson equations in $\phi^4$-theory is straightforward and
has been carried out in this paper up to the forth perturbative order by hand. Our iterative
procedure can easily be automatized by computer algebra. In Ref. \ci{SYM} it was demonstrated that the basic gra\-phi\-cal
operations as amputating a line or gluing two lines together can be formulated with the help of a unique matrix notation for
Feynman diagrams. It would be interesting to compare the efficiency of our Schwinger-Dyson approach of generating Feynman
diagrams of $n$-point functions together with their weights with already existing computer programs such as 
FEYNARTS \ci{FeynArts1,FeynArts2,FeynArts3} and QGRAF \ci{QGRAF1,QGRAF2}. 
Some of them are based on a combinatorial enumeration of all possible ways of
connecting vertices by lines according to Feynman's rules. Others use a systematic generation of homeomorphically
irreducible star graphs \cite{Heap,Nagle}. The latter approach is quite efficient and popular 
at higher orders. It has, however,  the conceptual disadvantage that it renders at an intermediate stage 
numerous superfluous diagrams with different vertex degrees which have to be discarded at the end.
Further promising methods have been proposed in Refs. \ci{VERSCHELDE,Schroeder}. Whereas the first one is based on a
bootstrap equation that uses only the free field value of the energy as an input, the second one
combines Schwinger-Dyson equations with the two-particle irreducible (``skeleton'') expansion.\\

We believe that our closed set of Schwinger-Dyson equations for the connected and one-particle irreducible two- and four-point
function will turn out to be useful for developing nonperturbative approximations. 
This may proceed, for instance, as in Ref. \ci{Bray} which
proposes a self-consistent solution of the Dyson equation in $\phi^4$-theory
by using a scaling ansatz for the connected two-point function near the phase transition. A similar consideration
in Ginzburg-Landau theory has allowed Ref. \ci{Radzikovsky} to analyze the influence of the thermal fluctuations of the 
order parameter and the vector potential on the superconducting phase transition. A self-consistent solution of our
closed Schwinger-Dyson equation may be found via a phenomenological ansatz for the connected and one-particle
irreducible two- and four-point function, respectively. Within such an ansatz
one has to find how the functional derivative with respect to the free correlation function is approximated,
as this operation is crucial in the new Schwinger-Dyson equations (\ref{SDG2}) and (\ref{120}).
\section*{Acknowledgement}
We thank Hagen Kleinert for stimulating discussions and for reading the manuscript.
\newpage
\begin{fmffile}{tabelle1}
\setlength{\unitlength}{1mm}
\begin{table}[t]
\begin{center}
\begin{tabular}{|c|c|}
\hline\hline  
\,\,\,$p$\,\,\,
&
\begin{tabular}{@{}c} 
$\mbox{}$ \\*[2mm]
$\mbox{}$
\end{tabular}
  \parbox{10mm}{\centerline{
  \begin{fmfgraph*}(7,3)
  \setval
  \fmfforce{0w,1/2h}{v1}
  \fmfforce{1w,1/2h}{v2}
  \fmfforce{1/2w,1/2h}{v3}
  \fmf{double,width=0.2mm}{v1,v2}
  \fmfv{decor.size=0, label=${\scs (p)}$, l.dist=1.5mm, l.angle=90}{v3}
  \end{fmfgraph*} } }  
\\
\hline
$0$ &
  \begin{tabular}{@{}c}
  $\mbox{}$\\
  ${\scs \mbox{\#0.1}}$ \\
  $1$\\ 
  ${\scs ( 0, 0, 0 , 1 ; 2 )}$\\
  $\mbox{}$
  \end{tabular}
\parbox{8mm}{\begin{center}
\begin{fmfgraph}(7,3)
\setval
\fmfforce{0w,1/2h}{v1}
\fmfforce{1w,1/2h}{v2}
\fmf{plain}{v1,v2}
\end{fmfgraph}  \end{center}}
\\
\hline
$1$ &
\hspace{-10pt}
\rule[-10pt]{0pt}{26pt}
  \begin{tabular}{@{}c}
  $\mbox{}$\\
  ${\scs \mbox{\#1.1}}$ \\
  $1/2$\\ 
  ${\scs ( 1, 0, 0 , 1 ; 2 )}$\\
  $\mbox{}$
  \end{tabular}
\parbox{8mm}{\begin{center}
\begin{fmfgraph}(10,5)
\setval
\fmfstraight
\fmfforce{0w,0h}{v1}
\fmfforce{0.5w,0h}{v3}
\fmfforce{1w,0h}{v2}
\fmfforce{0.5w,1h}{v4}
\fmf{plain}{v1,v2}
\fmf{plain,left=1}{v3,v4,v3}
\fmfdot{v3}
\end{fmfgraph}  \end{center}}
\\
\hline
$2$ &
\hspace{-10pt}
\rule[-10pt]{0pt}{26pt}
  \begin{tabular}{@{}c}
  $\mbox{}$\\
  ${\scs \mbox{\#2.1}}$ \\
  $1/6$\\ 
  ${\scs ( 0, 0, 1 , 1 ; 2 )}$\\
  $\mbox{}$
  \end{tabular}
\parbox{18mm}{\begin{center}
\begin{fmfgraph}(15,5)
\setval
\fmfforce{0w,0.5h}{v1}
\fmfforce{1/3w,0.5h}{v2}
\fmfforce{2/3w,0.5h}{v3}
\fmfforce{1w,0.5h}{v4}
\fmf{plain}{v1,v4}
\fmf{plain,left=1}{v3,v2,v3}
\fmfdot{v2,v3}
\end{fmfgraph}  \end{center}}
\quad \,\,
  \begin{tabular}{@{}c}
  ${\scs \mbox{\#2.2}}$ \\
  $1/4$\\ 
  ${\scs ( 1, 1 , 0 , 1 ; 2 )}$
  \end{tabular}
\parbox{13mm}{\begin{center}
\begin{fmfgraph}(10,10)
\setval
\fmfforce{0w,0h}{v1}
\fmfforce{0.5w,0h}{v2}
\fmfforce{1w,0h}{v3}
\fmfforce{0.5w,0.5h}{v4}
\fmfforce{0.5w,1h}{v5}
\fmf{plain}{v1,v3}
\fmf{plain,left=1}{v2,v4,v2}
\fmf{plain,left=1}{v4,v5,v4}
\fmfdot{v2,v4}
\end{fmfgraph} \end{center}}
\quad \,\, 
  \begin{tabular}{@{}c}
  ${\scs \mbox{\#2.3}}$ \\
  $1/4$\\ 
  ${\scs ( 2, 0, 0 , 1 ; 2 )}$
  \end{tabular}
\parbox{20mm}{\begin{center}
\begin{fmfgraph}(17,5)
\setval
\fmfforce{0w,0h}{i1}
\fmfforce{5/17w,0h}{v1}
\fmfforce{5/17w,1h}{v2}
\fmfforce{12/17w,0h}{v3}
\fmfforce{12/17w,1h}{v4}
\fmfforce{1w,0h}{o1}
\fmf{plain}{i1,o1}
\fmf{plain,left=1}{v1,v2,v1}
\fmf{plain,left=1}{v3,v4,v3}
\fmfdot{v1,v3}
\end{fmfgraph} \end{center}}
\\
\hline
$ $ &  \\
$3$ &
\hspace{3mm}
\rule[-10pt]{0pt}{26pt}
  \begin{tabular}{@{}c}
  ${\scs \mbox{\#3.1}}$ \\
  $1/4$\\ 
  ${\scs ( 0, 2, 0 , 1 ; 2 )}$
  \end{tabular}
\parbox{18mm}{\begin{center}
\begin{fmfgraph}(17,8)
\setval
\fmfforce{0w,0.25h}{i1}
\fmfforce{1w,0.25h}{o1}
\fmfforce{0.5w,0h}{v1}
\fmfforce{0.5w,1h}{v2}
\fmfforce{5/17w,0.25h}{v3}
\fmfforce{12/17w,0.25h}{v4}
\fmf{plain,left=1}{v1,v2,v1}
\fmf{plain}{i1,v3}
\fmf{plain}{o1,v4}
\fmf{plain}{v2,v3}
\fmf{plain}{v2,v4}
\fmfdot{v2,v3,v4}
\end{fmfgraph}  \end{center}}
\quad \,\,
  \begin{tabular}{@{}c}
  ${\scs \mbox{\#3.2}}$ \\
  $1/12$\\ 
  ${\scs ( 0, 0, 1 , 2 ; 2 )}$
  \end{tabular}
\parbox{10mm}{\begin{center}
\begin{fmfgraph}(8,8)
\setval
\fmfforce{0w,0.5h}{v1}
\fmfforce{1w,0.5h}{v2}
\fmfforce{0.5w,0h}{v4}
\fmfforce{-1/8w,0h}{i1}
\fmfforce{9/8w,0h}{o1}
\fmf{plain,left=1}{v1,v2,v1}
\fmf{plain,left=0.4}{v1,v2,v1}
\fmf{plain,left=1}{v3,v4,v3}
\fmf{plain}{i1,o1}
\fmfdot{v1,v2,v3}
\end{fmfgraph} \end{center}} 
\quad \,
  \begin{tabular}{@{}c}
  ${\scs \mbox{\#3.3}}$ \\
  $1/4$\\ 
  ${\scs ( 1, 1, 0 , 1 ; 2 )}$
  \end{tabular}
\parbox{16mm}{\begin{center}
\begin{fmfgraph}(15,10)
\setval
\fmfforce{0w,0.25h}{i1}
\fmfforce{1w,0.25h}{o1}
\fmfforce{1/3w,0.25h}{v1}
\fmfforce{2/3w,0.25h}{v2}
\fmfforce{0.5w,0.5h}{v3}
\fmfforce{0.5w,1h}{v4}
\fmf{plain}{i1,o1}
\fmf{plain,left=1}{v1,v2,v1}
\fmf{plain,left=1}{v3,v4,v3}
\fmfdot{v1,v2,v3}
\end{fmfgraph}  \end{center}} 
\quad \,
  \begin{tabular}{@{}c}
  ${\scs \mbox{\#3.4}}$ \\
  $1/8$\\ 
  ${\scs ( 1, 2, 0 , 1 ; 2 )}$
  \end{tabular}
\parbox{8mm}{\begin{center}
\begin{fmfgraph}(10,15)
\setval
\fmfforce{0w,0h}{i1}
\fmfforce{0.5w,0h}{v1}
\fmfforce{1w,0h}{o1}
\fmfforce{0.5w,1/3h}{v2}
\fmfforce{0.5w,2/3h}{v3}
\fmfforce{0.5w,1h}{v4}
\fmf{plain}{i1,v1}
\fmf{plain}{v1,o1}
\fmf{plain,left=1}{v1,v2,v1}
\fmf{plain,left=1}{v2,v3,v2}
\fmf{plain,left=1}{v3,v4,v3}
\fmfdot{v1,v2,v3}
\end{fmfgraph}  \end{center}}
\\ & 
\hspace*{1.5mm}
  \begin{tabular}{@{}c}
  $\mbox{}$\\
  ${\scs \mbox{\#3.5}}$ \\
  $1/8$\\ 
  ${\scs ( 2, 0, 0 , 2 ; 2 )}$\\
  $\mbox{}$
  \end{tabular}
\parbox{12mm}{\begin{center}
\begin{fmfgraph}(14,8)
\setval
\fmfforce{1/7w,0h}{i1}
\fmfforce{6/7w,0h}{o1}
\fmfforce{1/2w,0h}{v1}
\fmfforce{1/2w,5/8h}{v2}
\fmfforce{5/14w,4.25/8h}{v4}
\fmfforce{9/14w,4.25/8h}{v5}
\fmfforce{1.5/14w,7.75/8h}{v6}
\fmfforce{12.5/14w,7.75/8h}{v7}
\fmf{plain,left=1}{v1,v2,v1}
\fmf{plain}{i1,o1}
\fmf{plain,left=1}{v4,v6,v4}
\fmf{plain,left=1}{v5,v7,v5}
\fmfdot{v1,v4,v5}
\end{fmfgraph}\end{center}} 
\quad 
  \begin{tabular}{@{}c}
  ${\scs \mbox{\#3.6}}$ \\
  $1/8$\\ 
  ${\scs ( 3, 0, 0 , 1 ; 2 )}$
  \end{tabular}
\parbox{22mm}{\begin{center}
\begin{fmfgraph}(24,5)
\setval
\fmfforce{0w,0h}{i1}
\fmfforce{5/24w,0h}{v1}
\fmfforce{5/24w,1h}{v2}
\fmfforce{1/2w,0h}{v3}
\fmfforce{1/2w,1h}{v4}
\fmfforce{19/24w,0h}{v5}
\fmfforce{19/24w,1h}{v6}
\fmfforce{1w,0h}{o1}
\fmf{plain}{i1,o1}
\fmf{plain,left=1}{v1,v2,v1}
\fmf{plain,left=1}{v3,v4,v3}
\fmf{plain,left=1}{v5,v6,v5}
\fmfdot{v1,v3,v5}
\end{fmfgraph}  \end{center}}
\quad \,
  \begin{tabular}{@{}c}
  ${\scs \mbox{\#3.7}}$ \\
  $1/6$\\ 
  ${\scs ( 1, 0, 1 , 1 ; 1 )}$
  \end{tabular}
\parbox{17mm}{\begin{center}
\begin{fmfgraph}(20,7.5)
\setval
\fmfforce{0w,1/3h}{i1}
\fmfforce{5/20w,1/3h}{v1}
\fmfforce{10/20w,1/3h}{v2}
\fmfforce{15/20w,1/3h}{v3}
\fmfforce{15/20w,1h}{v4}
\fmfforce{1w,1/3h}{o1}
\fmf{plain}{i1,o1}
\fmf{plain,left=1}{v3,v4,v3}
\fmf{plain,left=1}{v1,v2,v1}
\fmfdot{v1,v2,v3}
\end{fmfgraph}  \end{center}}
\quad \,
  \begin{tabular}{@{}c}
  ${\scs \mbox{\#3.8}}$ \\
  $1/4$\\ 
  ${\scs ( 2, 1, 0 , 1 ; 1 )}$
  \end{tabular}
\parbox{14mm}{\begin{center}
\begin{fmfgraph}(17,10)
\setval
\fmfforce{0w,0h}{i1}
\fmfforce{5/17w,0h}{v1}
\fmfforce{5/17w,0.5h}{v2}
\fmfforce{5/17w,1h}{v5}
\fmfforce{12/17w,0h}{v3}
\fmfforce{12/17w,0.5h}{v4}
\fmfforce{1w,0h}{o1}
\fmf{plain}{i1,o1}
\fmf{plain,left=1}{v1,v2,v1}
\fmf{plain,left=1}{v3,v4,v3}
\fmf{plain,left=1}{v2,v5,v2}
\fmfdot{v1,v2,v3}
\end{fmfgraph}  \end{center}}
\hspace*{6mm}
\\
\hline
$ $ &
\hspace*{4mm}
  \begin{tabular}{@{}c}
  $\mbox{}$ \\
  ${\scs \mbox{\#4.1}}$ \\
  $1/8$\\ 
  ${\scs ( 0, 3, 0 , 1 ; 2 )}$\\
  $\mbox{}$ 
\end{tabular}
\parbox{14mm}{\begin{center}
\begin{fmfgraph}(16,9)
\setval
\fmfforce{0w,1/6h}{i1}
\fmfforce{5/16w,1/6h}{v1}
\fmfforce{5/16w,5/6h}{v2}
\fmfforce{11/16w,5/6h}{v3}
\fmfforce{11/16w,1/6h}{v4}
\fmfforce{1w,1/6h}{o1}
\fmf{plain,left=1}{v1,v3,v1}
\fmf{plain}{v1,v2}
\fmf{plain}{v2,v3}
\fmf{plain}{v3,v4}
\fmf{plain}{i1,v1}
\fmf{plain}{v4,o1}
\fmfdot{v1,v2,v3,v4}
\end{fmfgraph}  \end{center}}
\hspace*{0.3cm} 
  \begin{tabular}{@{}c}
  ${\scs \mbox{\#4.2}}$ \\
  $1/4$\\ 
  ${\scs ( 0, 1, 0 , 2 ; 2 )}$
  \end{tabular}
\parbox{18mm}{\begin{center}
\begin{fmfgraph}(19,9)
\setval
\fmfforce{0w,5/9h}{i1}
\fmfforce{1w,5/9h}{o1}
\fmfforce{5.1/19w,5/9h}{v1}
\fmfforce{13.9/19w,5/9h}{v2}
\fmfforce{1/2w,0h}{v3}
\fmfforce{1/2w,5/9h}{v4}
\fmfforce{1/2w,1h}{v5}
\fmf{plain}{i1,o1}
\fmf{plain,left=1}{v5,v3,v5}
\fmf{plain,left=1}{v3,v4,v3}
\fmfdot{v1,v2,v3,v4}
\end{fmfgraph}
\end{center}}
\hspace*{0.3cm}
  \begin{tabular}{@{}c}
  ${\scs \mbox{\#4.3}}$ \\
  $1/4$\\ 
  ${\scs ( 0, 2, 0 , 1 ; 2 )}$
  \end{tabular}
\parbox{15mm}{\begin{center}
\begin{fmfgraph}(16,9)
\setval
\fmfforce{0w,4/5h}{i1}
\fmfforce{5/18w,4/5h}{v1}
\fmfforce{13/18w,4/5h}{v2}
\fmfforce{5/18w,1/5h}{v3}
\fmfforce{13/18w,1/5h}{v4}
\fmfforce{1w,1/5h}{o1}
\fmf{plain}{i1,v2}
\fmf{plain}{v3,o1}
\fmf{plain,left=1}{v4,v1,v4}
\fmf{plain}{v2,v3}
\fmfdot{v1,v2,v3,v4}
\end{fmfgraph}
\end{center}}
\hspace*{0.3cm}
  \begin{tabular}{@{}c}
  ${\scs \mbox{\#4.4}}$ \\
  $1/12$\\ 
  ${\scs ( 0, 1, 1 , 1 ; 2 )}$
  \end{tabular}
\parbox{15mm}{\begin{center}
\begin{fmfgraph}(15,12.5)
\setval
\fmfforce{1/3w,1/5h}{v1}
\fmfforce{2/3w,1/5h}{v2}
\fmfforce{1/3w,4/5h}{v3}
\fmfforce{2/3w,4/5h}{v4}
\fmfforce{0w,1/5h}{i1}
\fmfforce{1w,1/5h}{o1}
\fmf{plain}{i1,v1}
\fmf{plain}{v2,o1}
\fmf{plain,left=1}{v1,v2,v1}
\fmf{plain}{v3,v4}
\fmf{plain,left=1}{v3,v4,v3}
\fmf{plain,left=0.5}{v1,v3}
\fmf{plain,right=0.5}{v2,v4}
\fmfdot{v1,v2,v3,v4}
\end{fmfgraph} \end{center}}
\hspace*{1.5mm}
\\ 
$4$ &
  \begin{tabular}{@{}c}
  ${\scs \mbox{\#4.5}}$ \\
  $1/8$\\ 
  ${\scs ( 0, 2, 0 , 2 ; 2 )}$
  \end{tabular}
\parbox{11mm}{\begin{center}
\begin{fmfgraph}(8,8)
\setval
\fmfforce{-1/8w,0h}{i1}
\fmfforce{9/8w,0h}{o1}
\fmfforce{0.5w,0h}{v1}
\fmfforce{0.5w,1h}{v2}
\fmfforce{1/16w,2.7/8h}{v3}
\fmfforce{15/16w,2.7/8h}{v4}
\fmf{plain,left=1}{v1,v2,v1}
\fmf{plain}{i1,v1}
\fmf{plain}{o1,v1}
\fmf{plain}{v2,v3}
\fmf{plain}{v2,v4}
\fmf{plain}{v3,v4}
\fmfdot{v1,v2,v3,v4}
\end{fmfgraph}  \end{center}}
\hspace*{4.5mm}
  \begin{tabular}{@{}c}
  ${\scs \mbox{\#4.6}}$ \\
  $1/24$\\ 
  ${\scs ( 0, 1, 1 , 2 ; 2 )}$
  \end{tabular}
\parbox{11mm}{\begin{center}
\begin{fmfgraph}(10,13)
\setval
\fmfforce{0.5w,0h}{v5}
\fmfforce{0w,0h}{i1}
\fmfforce{1w,0h}{o1}
\fmfforce{0.1w,9/13h}{v1}
\fmfforce{0.9w,9/13h}{v2}
\fmfforce{0.5w,5/13h}{v4}
\fmf{plain}{i1,o1}
\fmf{plain,left=1}{v1,v2,v1}
\fmf{plain,left=0.4}{v1,v2,v1}
\fmf{plain,left=1}{v3,v4,v3}
\fmf{plain,left=1}{v5,v4,v5}
\fmf{plain}{i1,o1}
\fmfdot{v1,v2,v3,v5}
\end{fmfgraph} \end{center}} 
\hspace*{0.4cm}
  \begin{tabular}{@{}c}
  ${\scs \mbox{\#4.7}}$ \\
  $1/8$\\ 
  ${\scs ( 2, 1, 0 , 1 ; 2 )}$
  \end{tabular}
\parbox{19mm}{\begin{center}
\begin{fmfgraph}(18,13)
\setval
\fmfforce{0w,4/13h}{i1}
\fmfforce{1w,4/13h}{o1}
\fmfforce{5/18w,4/13h}{v1}
\fmfforce{13/18w,4/13h}{v2}
\fmfforce{6.5/18w,7.5/13h}{v3}
\fmfforce{11.5/18w,7.5/13h}{v4}
\fmfforce{3/18w,11/13h}{v5}
\fmfforce{15/18w,11/13h}{v6}
\fmf{plain}{i1,o1}
\fmf{plain,left=1}{v1,v2,v1}
\fmf{plain,left=1}{v3,v5,v3}
\fmf{plain,left=1}{v4,v6,v4}
\fmfdot{v1,v2,v3,v4}
\end{fmfgraph}  \end{center}} 
\hspace*{0.3cm}
  \begin{tabular}{@{}c}
  ${\scs \mbox{\#4.8}}$ \\
  $1/8$\\ 
  ${\scs ( 2, 0, 0 , 2 ; 2 )}$
  \end{tabular}
\parbox{16mm}{\begin{center}
\begin{fmfgraph}(15,15)
\setval
\fmfforce{0w,0.5h}{i1}
\fmfforce{1w,0.5h}{o1}
\fmfforce{1/3w,0.5h}{v1}
\fmfforce{2/3w,0.5h}{v2}
\fmfforce{0.5w,1h}{v4}
\fmfforce{0.5w,0h}{v5}
\fmfforce{0.5w,1/3h}{v6}
\fmfforce{0.5w,2/3h}{v7}
\fmf{plain,left=1}{v5,v6,v5}
\fmf{plain,left=1}{v7,v4,v7}
\fmf{plain,left=1}{v1,v2,v1}
\fmf{plain}{i1,o1}
\fmfdot{v1,v2,v6,v7}
\end{fmfgraph}  \end{center}} 
\hspace*{0.2cm}
\\  &
  \begin{tabular}{@{}c}
  $\mbox{}$ \\
  ${\scs \mbox{\#4.9}}$ \\
  $1/8$\\ 
  ${\scs ( 1, 2, 0 , 1 ; 2 )}$ \\
  $\mbox{}$ 
  \end{tabular}
\parbox{18mm}{\begin{center}
\begin{fmfgraph}(17,13)
\setval
\fmfforce{0.5w,0h}{v1}
\fmfforce{0.5w,8/13h}{v2}
\fmfforce{5/17w,6/13h}{v3}
\fmfforce{12/17w,6/13h}{v4}
\fmfforce{0.5w,1h}{v5}
\fmfforce{0w,6/13h}{i1}
\fmfforce{1w,6/13h}{o1}
\fmf{plain,left=1}{v1,v2,v1}
\fmf{plain,left=1}{v2,v5,v2}
\fmf{plain}{v1,v3}
\fmf{plain}{i1,v3}
\fmf{plain}{o1,v4}
\fmf{plain}{v1,v4}
\fmfdot{v1,v2,v3,v4}
\end{fmfgraph}\end{center}}
\quad 
  \begin{tabular}{@{}c}
  ${\scs \mbox{\#4.10}}$ \\
  $1/2$\\ 
  ${\scs ( 1, 1, 0 , 1 ; 1 )}$ 
  \end{tabular}
\parbox{18mm}{\begin{center}
\begin{fmfgraph}(17,8)
\setval
\fmfforce{0w,1/8h}{i1}
\fmfforce{5/17w,1/8h}{v1}
\fmfforce{12/17w,1/8h}{v2}
\fmfforce{1w,1/8h}{o1}
\fmfforce{1/2w,-1/8h}{v3}
\fmfforce{1/2w,7/8h}{v5}
\fmfforce{5.2/17w,5/8h}{v4}
\fmfforce{1.7/17w,8.5/8h}{v6}
\fmf{plain,left=1}{v3,v5,v3}
\fmf{plain}{i1,v1}
\fmf{plain}{v2,o1}
\fmf{plain}{v1,v5}
\fmf{plain}{v5,v2}
\fmf{plain,left=1}{v4,v6,v4}
\fmfdot{v1,v2,v5,v4}
\end{fmfgraph}\end{center}}
\quad 
  \begin{tabular}{@{}c}
  ${\scs \mbox{\#4.11}}$ \\
  $1/8$\\ 
  ${\scs ( 1, 1, 0 , 2 ; 2 )}$
  \end{tabular}
\parbox{11mm}{\begin{center}
\begin{fmfgraph}(10,13)
\setval
\fmfforce{0w,0h}{i1}
\fmfforce{1w,0h}{o1}
\fmfforce{0.1w,4/13h}{v1}
\fmfforce{0.9w,4/13h}{v2}
\fmfforce{0.5w,8/13h}{v3}
\fmfforce{0.5w,1h}{v4}
\fmfforce{0.5w,0h}{v5}
\fmfforce{0.5w,1h}{v6}
\fmf{plain}{i1,o1}
\fmf{plain,left=1}{v3,v6,v3}
\fmf{plain,left=1}{v1,v2,v1}
\fmf{plain,left=0.4}{v1,v2,v1}
\fmfdot{v1,v2,v3,v5}
\end{fmfgraph} \end{center}}  
\hspace*{0.4cm}
  \begin{tabular}{@{}c}
  ${\scs \mbox{\#4.12}}$ \\
  $1/12$\\ 
  ${\scs ( 1, 0, 1 , 1 ; 2 )}$
  \end{tabular}
\parbox{18mm}{\begin{center}
\begin{fmfgraph}(18,13)
\setval
\fmfforce{5/18w,4/13h}{v1}
\fmfforce{13/18w,4/13h}{v2}
\fmfforce{6.5/18w,7.5/13h}{v3}
\fmfforce{11.5/18w,7.5/13h}{v4}
\fmfforce{3/18w,11/13h}{v5}
\fmfforce{16.5/18w,7.5/13h}{i1}
\fmfforce{11.5/18w,12.5/13h}{o1}
\fmf{plain,left=0.4}{v1,v2,v1}
\fmf{plain,left=1}{v1,v2,v1}
\fmf{plain,left=1}{v3,v5,v3}
\fmf{plain}{v4,i1}
\fmf{plain}{v4,o1}
\fmfdot{v1,v2,v3,v4}
\end{fmfgraph}  \end{center}} 
\hspace*{0.2cm}\\
$ $ & 
\hspace{-5pt}
\rule[-10pt]{0pt}{26pt}
  \begin{tabular}{@{}c}
  ${\scs \mbox{\#4.13}}$ \\
  $1/8$\\ 
  ${\scs ( 1, 2, 0 , 1 ; 2 )}$ 
  \end{tabular}
\parbox{16mm}{\begin{center}
\begin{fmfgraph}(15,15)
\setval
\fmfforce{0w,1/6h}{i1}
\fmfforce{1w,1/6h}{o1}
\fmfforce{1/3w,1/6h}{v1}
\fmfforce{2/3w,1/6h}{v2}
\fmfforce{1/2w,1/3h}{v3}
\fmfforce{1/2w,2/3h}{v4}
\fmfforce{1/2w,1h}{v5}
\fmf{plain}{i1,o1}
\fmf{plain,left=1}{v1,v2,v1}
\fmf{plain,left=1}{v3,v4,v3}
\fmf{plain,left=1}{v4,v5,v4}
\fmfdot{v1,v2,v3,v4}
\end{fmfgraph} \end{center}} 
\hspace*{0.3cm}
  \begin{tabular}{@{}c}
  ${\scs \mbox{\#4.14}}$ \\
  $1/16$\\ 
  ${\scs ( 2, 1, 0 , 2 ; 2 )}$
  \end{tabular}
\parbox{15mm}{\begin{center}
\begin{fmfgraph}(15,20)
\setval
\fmfforce{1/6w,1/4h}{i1}
\fmfforce{5/6w,1/4h}{o1}
\fmfforce{1/2w,1/4h}{v1}
\fmfforce{1/2w,1/2h}{v8}
\fmfforce{1/2w,3/4h}{v2}
\fmfforce{1/2w,1h}{v3}
\fmfforce{0.355662432w,0.6975h}{v4}
\fmfforce{0.64433568w,0.6975h}{v5}
\fmf{plain,left=1}{v8,v2,v8}
\fmf{plain}{i1,v1}
\fmf{plain}{v1,o1}
\fmfi{plain}{reverse fullcircle scaled 0.33w shifted (0.225w,0.765h)}
\fmfi{plain}{reverse fullcircle scaled 0.33w shifted (0.775w,0.765h)}
\fmf{plain,left=1}{v1,v8,v1}
\fmfdot{v1,v4,v5,v8}
\end{fmfgraph} \end{center}} 
\hspace*{0.4cm}
  \begin{tabular}{@{}c}
  ${\scs \mbox{\#4.15}}$ \\
  $1/8$\\ 
  ${\scs ( 2, 1, 0 , 1 ; 2 )}$
  \end{tabular}
\hs
\parbox{15mm}{\begin{center}
\begin{fmfgraph}(14,11.5)
\setval
\fmfforce{3.75/14w,0h}{i1}
\fmfforce{1w,0h}{o1}
\fmfforce{8.75/14w,0h}{v1}
\fmfforce{8.75/14w,5/11.5h}{v2}
\fmfforce{3/4w,4.25/11.5h}{v3}
\fmfforce{1w,7.75/11.5h}{v4}
\fmfforce{1/2w,4.25/11.5h}{v5}
\fmfforce{1/4w,7.75/11.5h}{v6}
\fmfforce{0w,11.25/11.5h}{v7}
\fmf{plain,left=1}{v1,v2,v1}
\fmf{plain}{i1,o1}
\fmf{plain,left=1}{v3,v4,v3}
\fmf{plain,left=1}{v5,v6,v5}
\fmf{plain,left=1}{v6,v7,v6}
\fmfdot{v1,v3,v5,v6}
\end{fmfgraph} \end{center}} 
\hspace*{5mm}
  \begin{tabular}{@{}c}
  ${\scs\mbox{\#4.16}}$ \\
  $1/16$ \\ 
  ${\scs ( 3, 0, 0 , 2 ; 2 )}$
  \end{tabular}
\parbox{15mm}{\begin{center}
\begin{fmfgraph}(15,10)
\setval
\fmfforce{3.5/15w,-0.2h}{i1}
\fmfforce{11.5/15w,-0.2h}{o1}
\fmfforce{1/3w,1/4h}{v1}
\fmfforce{2/3w,1/4h}{v2}
\fmfforce{1/2w,0h}{v3}
\fmfforce{1/2w,1/2h}{v4}
\fmfforce{1/2w,1h}{v5}
\fmfforce{0w,1/4h}{v7}
\fmfforce{1w,1/4h}{v8}
\fmf{plain}{i1,v3}
\fmf{plain}{v3,o1}
\fmf{plain,left=1}{v3,v4,v3}
\fmf{plain,left=1}{v4,v5,v4}
\fmf{plain,left=1}{v1,v7,v1}
\fmf{plain,left=1}{v2,v8,v2}
\fmfdot{v1,v2,v3,v4}
\end{fmfgraph} \end{center}} 
\hspace*{0.1cm}
\end{tabular}
\end{center}
\end{table}
\newpage
\begin{table}[t]
\begin{center}
\begin{tabular}{|c|c|}
$ $ &
\hs\hs
  \begin{tabular}{@{}c}
  $\mbox{}$ \\
  ${\scs \mbox{\#4.17}}$ \\
  $1/16$\\ 
  ${\scs ( 1, 3, 0 , 1 ; 2 )}$ \\
  $\mbox{}$ 
  \end{tabular}
\parbox{8mm}{\begin{center}
\begin{fmfgraph}(10,20)
\setval
\fmfforce{0w,0h}{i1}
\fmfforce{0.5w,0h}{v1}
\fmfforce{1w,0h}{o1}
\fmfforce{0.5w,1/4h}{v2}
\fmfforce{0.5w,1/2h}{v3}
\fmfforce{0.5w,3/4h}{v4}
\fmfforce{0.5w,1h}{v5}
\fmf{plain}{i1,o1}
\fmf{plain,left=1}{v1,v2,v1}
\fmf{plain,left=1}{v2,v3,v2}
\fmf{plain,left=1}{v3,v4,v3}
\fmf{plain,left=1}{v5,v4,v5}
\fmfdot{v1,v2,v3,v4}
\end{fmfgraph} \end{center}}
\hspace*{0.6cm}
  \begin{tabular}{@{}c}
  ${\scs \mbox{\#4.18}}$ \\
  $1/36$\\ 
  ${\scs ( 0, 0, 2 , 1 ; 2 )}$
  \end{tabular}
\parbox{18mm}{\begin{center}
\begin{fmfgraph}(23,5)
\setval
\fmfforce{0w,0.5h}{i1}
\fmfforce{5/23w,0.5h}{v1}
\fmfforce{10/23w,0.5h}{v2}
\fmfforce{13/23w,0.5h}{v3}
\fmfforce{18/23w,0.5h}{v4}
\fmfforce{1w,0.5h}{o1}
\fmf{plain}{i1,o1}
\fmf{plain,left=1}{v1,v2,v1}
\fmf{plain,left=1}{v3,v4,v3}
\fmfdot{v1,v2,v3,v4}
\end{fmfgraph} \end{center}}
\hspace*{0.6cm}
  \begin{tabular}{@{}c}
  ${\scs \mbox{\#4.19}}$ \\
  $1/4$\\ 
  ${\scs ( 2, 1, 0 , 1 ; 1 )}$
  \end{tabular}
\parbox{17mm}{\begin{center}
\begin{fmfgraph}(20,10)
\setval
\fmfforce{0w,0.25h}{i1}
\fmfforce{1w,0.25h}{o1}
\fmfforce{1/4w,0.25h}{v1}
\fmfforce{1/2w,0.25h}{v2}
\fmfforce{3/8w,0.5h}{v3}
\fmfforce{3/8w,1h}{v4}
\fmfforce{3/4w,0.25h}{v5}
\fmfforce{3/4w,0.75h}{v6}
\fmf{plain}{i1,o1}
\fmf{plain,left=1}{v1,v2,v1}
\fmf{plain,left=1}{v3,v4,v3}
\fmf{plain,left=1}{v5,v6,v5}
\fmfdot{v1,v2,v3,v5}
\end{fmfgraph} \end{center}} 
\hspace*{0.5cm}
  \begin{tabular}{@{}c}
  ${\scs \mbox{\#4.20}}$ \\
  $1/24$\\ 
  ${\scs ( 2, 0, 1 , 1 ; 2 )}$
  \end{tabular}
\parbox{20mm}{\begin{center}
\begin{fmfgraph}(23,5)
\setval
\fmfforce{0w,0.5h}{i1}
\fmfforce{5/23w,0.5h}{v1}
\fmfforce{9/23w,0.5h}{v2}
\fmfforce{14/23w,0.5h}{v3}
\fmfforce{18/23w,0.5h}{v4}
\fmfforce{5/23w,1.5h}{v5}
\fmfforce{18/23w,1.5h}{v6}
\fmfforce{1w,0.5h}{o1}
\fmf{plain}{i1,o1}
\fmf{plain,left=1}{v1,v5,v1}
\fmf{plain,left=1}{v2,v3,v2}
\fmf{plain,left=1}{v4,v6,v4}
\fmfdot{v1,v2,v3,v4}
\end{fmfgraph} \end{center}}
\hspace*{5mm}
\\  
$\,\,\,4\,\,\,$ &
\hs\hs
  \begin{tabular}{@{}c}
  $\mbox{}$ \\
  ${\scs \mbox{\#4.21}}$ \\
  $1/12$\\ 
  ${\scs ( 1, 0, 1 , 2 ; 1 )}$ \\
  $\mbox{}$ 
  \end{tabular}
\parbox{17mm}{\begin{center}
\begin{fmfgraph}(19,8)
\setval
\fmfforce{5/19w,0h}{v5}
\fmfforce{5/19w,5/8h}{v6}
\fmfforce{10/19w,0.5h}{v1}
\fmfforce{18/19w,0.5h}{v2}
\fmfforce{14/19w,0h}{v4}
\fmfforce{0w,0h}{i1}
\fmfforce{1w,0h}{o1}
\fmf{plain,left=1}{v1,v2,v1}
\fmf{plain,left=0.4}{v1,v2,v1}
\fmf{plain,left=1}{v3,v4,v3}
\fmf{plain,left=1}{v5,v6,v5}
\fmf{plain}{i1,o1}
\fmfdot{v1,v2,v3,v5}
\end{fmfgraph} \end{center}} 
\hspace*{0.6cm}
  \begin{tabular}{@{}c}
  ${\scs \mbox{\#4.22}}$ \\
  $1/12$\\ 
  ${\scs ( 1, 1, 1 , 1 ; 1 )}$
  \end{tabular}
\parbox{16mm}{\begin{center}
\begin{fmfgraph}(19,12.5)
\setval
\fmfforce{0w,1/5h}{i1}
\fmfforce{5/19w,1/5h}{v1}
\fmfforce{10/19w,1/5h}{v2}
\fmfforce{14/19w,1/5h}{v3}
\fmfforce{14/19w,3/5h}{v4}
\fmfforce{14/19w,1h}{v5}
\fmfforce{1w,1/5h}{o1}
\fmf{plain}{i1,o1}
\fmf{plain,left=1}{v3,v4,v3}
\fmf{plain,left=1}{v1,v2,v1}
\fmf{plain,left=1}{v4,v5,v4}
\fmfdot{v1,v2,v3,v4}
\end{fmfgraph} \end{center}}
\hspace*{0.6cm}
  \begin{tabular}{@{}c}
  ${\scs \mbox{\#4.23}}$ \\
  $1/16$\\ 
  ${\scs ( 2, 2, 0 , 1 ; 2 )}$
  \end{tabular}
\parbox{17mm}{\begin{center}
\begin{fmfgraph}(18,10)
\setval
\fmfforce{0w,0h}{i1}
\fmfforce{5/18w,0h}{v1}
\fmfforce{5/18w,0.5h}{v2}
\fmfforce{5/18w,1h}{v5}
\fmfforce{13/18w,0h}{v3}
\fmfforce{13/18w,0.5h}{v4}
\fmfforce{13/18w,1h}{v6}
\fmfforce{1w,0h}{o1}
\fmf{plain}{i1,o1}
\fmf{plain,left=1}{v1,v2,v1}
\fmf{plain,left=1}{v3,v4,v3}
\fmf{plain,left=1}{v2,v5,v2}
\fmf{plain,left=1}{v4,v6,v4}
\fmfdot{v1,v2,v3,v4}
\end{fmfgraph} \end{center}}
\hspace*{3mm}
  \begin{tabular}{@{}c}
  ${\scs \mbox{\#4.24}}$ \\
  $1/8$\\ 
  ${\scs ( 2, 2, 0 , 1 ; 1 )}$
  \end{tabular}
\parbox{17mm}{\begin{center}
\begin{fmfgraph}(18,15)
\setval
\fmfforce{0w,0h}{i1}
\fmfforce{5/18w,0h}{v1}
\fmfforce{1w,0h}{o1}
\fmfforce{5/18w,1/3h}{v2}
\fmfforce{5/18w,2/3h}{v3}
\fmfforce{5/18w,1h}{v4}
\fmfforce{13/18w,0h}{v5}
\fmfforce{13/18w,1/3h}{v6}
\fmf{plain}{i1,o1}
\fmf{plain,left=1}{v1,v2,v1}
\fmf{plain,left=1}{v2,v3,v2}
\fmf{plain,left=1}{v3,v4,v3}
\fmf{plain,left=1}{v5,v6,v5}
\fmfdot{v1,v2,v3,v5}
\end{fmfgraph} \end{center}}
\hspace*{4mm}
\\  &
  \begin{tabular}{@{}c}
  ${\scs \mbox{\#4.25}}$ \\
  $1/8$\\ 
  ${\scs ( 3, 0, 0 , 2 ; 1 )}$ 
  \end{tabular}
\parbox{20mm}{\begin{center}
\begin{fmfgraph}(20,8)
\setval
\fmfforce{0w,0h}{i1}
\fmfforce{1w,0h}{o1}
\fmfforce{3/4w,0h}{v8}
\fmfforce{3/4w,5/8h}{v9}
\fmfforce{1/4w,0h}{v1}
\fmfforce{1/4w,5/8h}{v2}
\fmfforce{3/20w,1/2h}{v3}
\fmfforce{-0.5/20w,7.5/8h}{v4}
\fmfforce{7/20w,1/2h}{v5}
\fmfforce{10.5/20w,7.5/8h}{v6}
\fmf{plain,left=1}{v1,v2,v1}
\fmf{plain,left=1}{v8,v9,v8}
\fmf{plain}{i1,o1}
\fmf{plain,left=1}{v3,v4,v3}
\fmf{plain,left=1}{v5,v6,v5}
\fmfdot{v1,v3,v5,v8}
\end{fmfgraph} \end{center}} 
\hspace*{3mm}
  \begin{tabular}{@{}c}
  ${\scs \mbox{\#4.26}}$ \\
  $1/12$\\ 
  ${\scs ( 2, 0, 1 , 1 ; 1 )}$ 
  \end{tabular}
\parbox{26mm}{\begin{center}
\begin{fmfgraph}(26,5)
\setval
\fmfforce{0w,0.5h}{i1}
\fmfforce{5/26w,0.5h}{v1}
\fmfforce{10/26w,0.5h}{v2}
\fmfforce{14/26w,0.5h}{v3}
\fmfforce{14/26w,1.5h}{v4}
\fmfforce{21/26w,0.5h}{v5}
\fmfforce{21/26w,1.5h}{v6}
\fmfforce{1w,0.5h}{o1}
\fmf{plain}{i1,o1}
\fmf{plain,left=1}{v1,v2,v1}
\fmf{plain,left=1}{v3,v4,v3}
\fmf{plain,left=1}{v5,v6,v5}
\fmfdot{v1,v2,v3,v5}
\end{fmfgraph}  \end{center}}
\hspace*{0.3cm}
  \begin{tabular}{@{}c}
  ${\scs \mbox{\#4.27}}$ \\
  $1/4$\\ 
  ${\scs ( 1, 2, 0 , 1 ; 1 )}$
  \end{tabular}
\parbox{22mm}{\begin{center}
\begin{fmfgraph}(22,8)
\setval
\fmfforce{0w,0.25h}{i1}
\fmfforce{5/22w,0.25h}{v1}
\fmfforce{5/22w,7/8h}{v5}
\fmfforce{1w,0.25h}{o1}
\fmfforce{10/22w,0.25h}{v2}
\fmfforce{13.5/22w,0h}{v3}
\fmfforce{13.5/22w,1h}{v4}
\fmfforce{17/22w,0.25h}{v6}
\fmf{plain,left=1}{v3,v4,v3}
\fmf{plain,left=1}{v1,v5,v1}
\fmf{plain}{i1,v2}
\fmf{plain}{v2,v4}
\fmf{plain}{v4,v6}
\fmf{plain}{v6,o1}
\fmfdot{v1,v2,v4,v6}
\end{fmfgraph}  \end{center}}
\hspace*{0.3cm}
\\  &
  \begin{tabular}{@{}c}
  $\mbox{}$ \\
  ${\scs \mbox{\#4.28}}$ \\
  $1/16$\\ 
  ${\scs ( 3, 1, 0 , 1 ; 2 )}$ \\
  $\mbox{}$ 
  \end{tabular}
\parbox{24mm}{\begin{center}
\begin{fmfgraph}(24,10)
\setval
\fmfforce{0w,0h}{i1}
\fmfforce{5/24w,0h}{v1}
\fmfforce{5/24w,1/2h}{v2}
\fmfforce{1/2w,0h}{v3}
\fmfforce{1/2w,1/2h}{v4}
\fmfforce{1/2w,1h}{v7}
\fmfforce{19/24w,0h}{v5}
\fmfforce{19/24w,1/2h}{v6}
\fmfforce{1w,0h}{o1}
\fmf{plain}{i1,o1}
\fmf{plain,left=1}{v1,v2,v1}
\fmf{plain,left=1}{v3,v4,v3}
\fmf{plain,left=1}{v5,v6,v5}
\fmf{plain,left=1}{v4,v7,v4}
\fmfdot{v1,v3,v5,v4}
\end{fmfgraph}  \end{center}}
\hspace*{0.4cm}
  \begin{tabular}{@{}c}
  ${\scs \mbox{\#4.29}}$ \\
  $1/8$\\ 
  ${\scs ( 3, 1, 0 , 1 ; 1 )}$
  \end{tabular}
\parbox{25mm}{\begin{center}
\begin{fmfgraph}(24,10)
\setval
\fmfforce{0w,0h}{i1}
\fmfforce{5/24w,0h}{v1}
\fmfforce{5/24w,0.5h}{v2}
\fmfforce{5/24w,1h}{v5}
\fmfforce{1/2w,0h}{v3}
\fmfforce{1/2w,0.5h}{v4}
\fmfforce{19/24w,0h}{v7}
\fmfforce{19/24w,0.5h}{v8}
\fmfforce{1w,0h}{o1}
\fmf{plain}{i1,o1}
\fmf{plain,left=1}{v1,v2,v1}
\fmf{plain,left=1}{v3,v4,v3}
\fmf{plain,left=1}{v2,v5,v2}
\fmf{plain,left=1}{v7,v8,v7}
\fmfdot{v1,v2,v3,v7}
\end{fmfgraph}  \end{center}}
\hspace*{3mm}
  \begin{tabular}{@{}c}
  ${\scs \mbox{\#4.30}}$ \\
  $1/16$\\ 
  ${\scs ( 4, 0, 0 , 1 ; 2 )}$
  \end{tabular}
\parbox{31mm}{\begin{center}
\begin{fmfgraph}(31,5)
\setval
\fmfforce{0w,0h}{i1}
\fmfforce{5/31w,0h}{v1}
\fmfforce{5/31w,1h}{v2}
\fmfforce{12/31w,0h}{v3}
\fmfforce{12/31w,1h}{v4}
\fmfforce{19/31w,0h}{v5}
\fmfforce{19/31w,1h}{v6}
\fmfforce{26/31w,0h}{v7}
\fmfforce{26/31w,1h}{v8}
\fmfforce{1w,0h}{o1}
\fmf{plain}{i1,o1}
\fmf{plain,left=1}{v1,v2,v1}
\fmf{plain,left=1}{v3,v4,v3}
\fmf{plain,left=1}{v5,v6,v5}
\fmf{plain,left=1}{v7,v8,v7}
\fmfdot{v1,v3,v5,v7}
\end{fmfgraph}  \end{center}}
\hs
\\
\hline\hline
\end{tabular}
\end{center}
\caption{\la{tab1} Diagrams of the connected two-point function and their 
weights of the $\phi^4$-theory up to four loops characterized by the vector $(S,D,T,P;N$).
Its components $S,D,T$ specify the number of self-, double,
triple connections, $P$ stands for the number of vertex permutations leaving the diagram unchanged, and $N$ denotes the symmetry
degree.}
\end{table}
%
\newpage
\begin{table}[t]
\begin{center}
\begin{tabular}{|c|c|}
\hline\hline
\,\,\,$p$\,\,\, &
  \begin{tabular}{@{}c}
  $\mbox{}$ \\*[5mm]
  $\mbox{}$
  \end{tabular}
  \parbox{11mm}{\centerline{
  \begin{fmfgraph*}(10.5,10.5)
  \setval
  \fmfforce{0w,0h}{v1}
  \fmfforce{0w,1h}{v2}
  \fmfforce{1w,1h}{v3}
  \fmfforce{1w,0h}{v4}
  \fmfforce{3.5/10.5w,3.5/10.5h}{v5}
  \fmfforce{3.5/10.5w,7/10.5h}{v6}
  \fmfforce{7/10.5w,7/10.5h}{v7}
  \fmfforce{7/10.5w,3.5/10.5h}{v8}
  \fmfforce{1/2w,1/2h}{v9}
  \fmf{plain}{v1,v5}
  \fmf{plain}{v2,v6}
  \fmf{plain}{v3,v7}
  \fmf{plain}{v4,v8}
  \fmf{plain,width=0.2mm,left=1}{v5,v7,v5}
  \fmfv{decor.size=0, label=${\scs p}$, l.dist=0mm, l.angle=90}{v9}
  \end{fmfgraph*} } }
\\
\hline
$1$ &
\hspace*{2mm}
  \begin{tabular}{@{}c}
  $\mbox{}$\\
  ${\scs \#1.1}$ \\
  $1$\\ 
  ${\scs ( 0, 0, 0 , 1 ; 24 )}$\\
  $\mbox{}$
  \end{tabular}
\parbox{8mm}{\begin{center}
\begin{fmfgraph}(7,7)
\setval
\fmfstraight
\fmfforce{0w,0h}{i1}
\fmfforce{0w,1h}{i2}
\fmfforce{1w,0h}{o1}
\fmfforce{1w,1h}{o2}
\fmfforce{0.5w,0.5h}{v1}
\fmf{plain}{i1,o2}
\fmf{plain}{i2,o1}
\fmfdot{v1}
\end{fmfgraph}
\end{center}}
\\ \hline
$2$ &
\hspace*{2mm}
  \begin{tabular}{@{}c}
  $\mbox{}$\\
  ${\scs \#2.1}$ \\
  $3/2$\\ 
  ${\scs ( 0, 1, 0 , 1 ; 8 )}$\\
  $\mbox{}$
  \end{tabular}
\parbox{14mm}{\begin{center}
\begin{fmfgraph}(12,7)
\setval
\fmfstraight
\fmfforce{0w,0h}{i1}
\fmfforce{0w,1h}{i2}
\fmfforce{1w,0h}{o1}
\fmfforce{1w,1h}{o2}
\fmfforce{3.5/12w,0.5h}{v1}
\fmfforce{8.5/12w,0.5h}{v2}
\fmf{plain}{i1,v1}
\fmf{plain}{v2,o1}
\fmf{plain}{i2,v1}
\fmf{plain}{v2,o2}
\fmf{plain,left=1}{v1,v2,v1}
\fmfdot{v1,v2}
\end{fmfgraph}
\end{center}}
\quad
  \begin{tabular}{@{}c}
  ${\scs \#2.2}$ \\ 
  $2$\\
  ${\scs ( 1, 0, 0 , 1 ; 6 )}$
  \end{tabular}
\parbox{17mm}{\begin{center}
\begin{fmfgraph}(15,10)
\setval
\fmfstraight
\fmfforce{0w,1/2h}{i1}
\fmfforce{1/3w,1h}{i2}
\fmfforce{1/3w,0h}{i3}
\fmfforce{1w,1/2h}{o1}
\fmfforce{2/3w,1/2h}{v1}
\fmfforce{2/3w,1h}{v2}
\fmfforce{1/3w,1/2h}{v3}
\fmf{plain}{i1,o1}
\fmf{plain}{i2,v3}
\fmf{plain}{i3,v3}
\fmf{plain,left=1}{v1,v2,v1}
\fmfdot{v1,v3}
\end{fmfgraph}
\end{center}}
\\ \hline
$ $ & \\
$3$ &
\hspace{-10pt}
\rule[-10pt]{0pt}{26pt}
  \begin{tabular}{@{}c}
  ${\scs \#3.1}$ \\ 
  $3/4$\\
  ${\scs ( 0, 2, 0 , 1 ; 8 )}$
  \end{tabular}
\parbox{17mm}{\begin{center}
\begin{fmfgraph}(17,7)
\setval
\fmfstraight
\fmfforce{0w,0h}{i1}
\fmfforce{0w,1h}{i2}
\fmfforce{1w,0h}{o1}
\fmfforce{1w,1h}{o2}
\fmfforce{3.5/17w,0.5h}{v1}
\fmfforce{1/2w,0.5h}{v2}
\fmfforce{13.5/17w,0.5h}{v3}
\fmf{plain}{i1,v1}
\fmf{plain}{v3,o1}
\fmf{plain}{i2,v1}
\fmf{plain}{v3,o2}
\fmf{plain,left=1}{v1,v2,v1}
\fmf{plain,left=1}{v2,v3,v2}
\fmfdot{v1,v2,v3}
\end{fmfgraph} \end{center}}
\quad
  \begin{tabular}{@{}c}
  ${\scs \#3.2}$ \\ 
  $3$\\
  ${\scs ( 0, 1, 0 , 1 ; 4 )}$
  \end{tabular}
\parbox{14mm}{\begin{center}
\begin{fmfgraph}(15,8.5)
\setval
\fmfstraight
\fmfforce{0w,2.5/8.5h}{i1}
\fmfforce{1w,2.5/8.5h}{o1}
\fmfforce{1/3w,2.5/8.5h}{v1}
\fmfforce{2/3w,2.5/8.5h}{v2}
\fmfforce{1/2w,5/8.5h}{v3}
\fmfforce{11/15w,1h}{i2}
\fmfforce{4/15w,1h}{o2}
\fmf{plain}{i1,o1}
\fmf{plain}{i2,v3}
\fmf{plain}{o2,v3}
\fmf{plain,left=1}{v1,v2,v1}
\fmfdot{v1,v2,v3}
\end{fmfgraph} \end{center}}
\quad
\hspace*{1.5mm}
  \begin{tabular}{@{}c}
  ${\scs \#3.3}$ \\
  $3/2$\\
  ${\scs ( 1, 0, 0 , 1  ; 8 )}$
  \end{tabular}
\parbox{12mm}{\begin{center}
\begin{fmfgraph}(12,10)
\setval
\fmfstraight
\fmfforce{0w,-0.1h}{i1}
\fmfforce{0w,0.6h}{i2}
\fmfforce{1w,-0.1h}{o1}
\fmfforce{1w,0.6h}{o2}
\fmfforce{3.5/12w,1/4h}{v1}
\fmfforce{8.5/12w,1/4h}{v2}
\fmfforce{1/2w,1/2h}{v3}
\fmfforce{1/2w,1h}{v4}
\fmf{plain}{i1,v1}
\fmf{plain}{v2,o1}
\fmf{plain}{i2,v1}
\fmf{plain}{v2,o2}
\fmf{plain,left=1}{v1,v2,v1}
\fmf{plain,left=1}{v3,v4,v3}
\fmfdot{v1,v2,v3}
\end{fmfgraph} \end{center}}
\hspace*{4mm}
  \begin{tabular}{@{}c}
  ${\scs \#3.4}$ \\ 
  $2/3$\\
  ${\scs ( 0, 0, 1 , 1 ; 6 )}$
  \end{tabular}
\parbox{16mm}{\begin{center}
\begin{fmfgraph}(18,10)
\setval
\fmfstraight
\fmfforce{0w,1/2h}{i1}
\fmfforce{5/18w,0h}{i2}
\fmfforce{5/18w,1h}{i3}
\fmfforce{1w,1/2h}{o1}
\fmfforce{5/18w,1/2h}{v1}
\fmfforce{8/18w,1/2h}{v2}
\fmfforce{13/18w,1/2h}{v3}
\fmf{plain}{i1,o1}
\fmf{plain}{i2,v1}
\fmf{plain}{i3,v1}
\fmf{plain,left=1}{v2,v3,v2}
\fmfdot{v1,v3,v2}
\end{fmfgraph}  \end{center}}
\quad
\\   &
  \begin{tabular}{@{}c}
  $\mbox{}$\\
  ${\scs \#3.5}$ \\ 
  $3$\\
  ${\scs ( 1, 1, 0 , 1 ; 2 )}$\\
  $\mbox{}$
  \end{tabular}
\parbox{15mm}{\begin{center}
\begin{fmfgraph}(18,5)
\setval
\fmfstraight
\fmfforce{0w,0h}{i1}
\fmfforce{1w,0h}{o1}
\fmfforce{1.5/18w,8.5/5h}{i2}
\fmfforce{8.5/18w,8.5/5h}{o2}
\fmfforce{5/18w,0h}{v1}
\fmfforce{5/18w,1h}{v2}
\fmfforce{13/18w,0h}{v3}
\fmfforce{13/18w,1h}{v4}
\fmf{plain,left=1}{v1,v2,v1}
\fmf{plain,left=1}{v3,v4,v3}
\fmf{plain}{i1,o1}
\fmf{plain}{i2,v2}
\fmf{plain}{o2,v2}
\fmfdot{v1,v2,v3}
\end{fmfgraph} \end{center}}
\quad \,
  \begin{tabular}{@{}c}
  ${\scs \#3.6}$ \\
  $3/2$\\
  ${\scs ( 2, 0, 0 , 1 ; 4 )}$
  \end{tabular}
\parbox{18mm}{\begin{center}
\begin{fmfgraph}(20,10)
\setval
\fmfstraight
\fmfforce{0w,1/2h}{i1}
\fmfforce{1/2w,1h}{i2}
\fmfforce{1/2w,0h}{o1}
\fmfforce{1w,1/2h}{o2}
\fmfforce{1/4w,1/2h}{v1}
\fmfforce{1/4w,1h}{v2}
\fmfforce{3/4w,1/2h}{v3}
\fmfforce{3/4w,1h}{v4}
\fmfforce{1/2w,1/2h}{v5}
\fmf{plain}{i1,o2}
\fmf{plain}{i2,v5}
\fmf{plain}{o1,v5}
\fmf{plain,left=1}{v1,v2,v1}
\fmf{plain,left=1}{v3,v4,v3}
\fmfdot{v1,v3,v5}
\end{fmfgraph} \end{center}}
\quad
  \begin{tabular}{@{}c}
  ${\scs \#3.7}$ \\ 
  $1$\\
  ${\scs ( 2, 0, 0 , 1 ; 6 )}$
  \end{tabular}
\parbox{18mm}{\begin{center}
\begin{fmfgraph}(22,10)
\setval
\fmfstraight
\fmfforce{0w,1/2h}{i1}
\fmfforce{5/22w,1h}{i2}
\fmfforce{5/22w,0h}{i3}
\fmfforce{1w,1/2h}{o1}
\fmfforce{10/22w,1/2h}{v1}
\fmfforce{10/22w,1h}{v2}
\fmfforce{5/22w,1/2h}{v3}
\fmfforce{17/22w,1/2h}{v4}
\fmfforce{17/22w,1h}{v5}
\fmf{plain}{i1,o1}
\fmf{plain}{i2,v3}
\fmf{plain}{i3,v3}
\fmf{plain,left=1}{v1,v2,v1}
\fmf{plain,left=1}{v4,v5,v4}
\fmfdot{v1,v3,v4}
\end{fmfgraph} \end{center}}
\quad \,
  \begin{tabular}{@{}c}
  ${\scs \#3.8}$ \\ 
  $1$\\
  ${\scs ( 1, 1, 0 , 1 ; 6 )}$
  \end{tabular}
\parbox{12mm}{\begin{center}
\begin{fmfgraph}(15,15)
\setval
\fmfstraight
\fmfforce{0w,1/3h}{i1}
\fmfforce{1/3w,2/3h}{i2}
\fmfforce{1/3w,0h}{i3}
\fmfforce{1w,1/3h}{o1}
\fmfforce{2/3w,1/3h}{v1}
\fmfforce{2/3w,2/3h}{v2}
\fmfforce{1/3w,1/3h}{v3}
\fmfforce{2/3w,1h}{v4}
\fmf{plain}{i1,o1}
\fmf{plain}{i2,v3}
\fmf{plain}{i3,v3}
\fmf{plain,left=1}{v1,v2,v1}
\fmf{plain,left=1}{v4,v2,v4}
\fmfdot{v1,v2,v3}
\end{fmfgraph} \end{center}}
\hspace*{2mm} \\ 
\hline
$ $ &
\hspace*{2mm}
  \begin{tabular}{@{}c}
  $\mbox{}$\\
  ${\scs \#4.1}$ \\ 
  $3/8$\\
  ${\scs ( 0, 3, 0 , 1 ; 8 )}$\\
  $\mbox{}$
  \end{tabular}
\parbox{22mm}{\begin{center}
\begin{fmfgraph}(22,7)
\setval
\fmfstraight
\fmfforce{0w,0h}{i1}
\fmfforce{0w,1h}{i2}
\fmfforce{1w,0h}{o1}
\fmfforce{1w,1h}{o2}
\fmfforce{3.5/22w,0.5h}{v1}
\fmfforce{8.5/22w,0.5h}{v2}
\fmfforce{13.5/22w,0.5h}{v3}
\fmfforce{18.5/22w,0.5h}{v4}
\fmf{plain}{i1,v1}
\fmf{plain}{v4,o1}
\fmf{plain}{i2,v1}
\fmf{plain}{v4,o2}
\fmf{plain,left=1}{v1,v2,v1}
\fmf{plain,left=1}{v2,v3,v2}
\fmf{plain,left=1}{v4,v3,v4}
\fmfdot{v1,v2,v3,v4}
\end{fmfgraph} \end{center}}
\quad
  \begin{tabular}{@{}c}
  ${\scs \#4.2}$ \\
  $1$\\
  ${\scs ( 0, 0, 0 , 1 ; 24 )}$
  \end{tabular}
\parbox{10mm}{\begin{center}
\begin{fmfgraph}(12.5,12.5)
\setval
\fmfforce{0w,0h}{i1}
\fmfforce{0w,1h}{i2}
\fmfforce{1w,1h}{o1}
\fmfforce{1w,0h}{o2}
\fmfforce{3.5/12.5w,3.5/12.5h}{v1}
\fmfforce{3.5/12.5w,9/12.5h}{v2}
\fmfforce{9/12.5w,9/12.5h}{v3}
\fmfforce{9/12.5w,3.5/12.5h}{v4}
\fmf{plain}{i1,o1}
\fmf{plain}{i2,o2}
\fmf{plain,left=1}{v1,v3,v1}
\fmfdot{v1,v2,v3,v4}
\end{fmfgraph} \end{center}}
\quad \,
  \begin{tabular}{@{}c}
  ${\scs \#4.3}$ \\ 
  $3/2$\\ 
  ${\scs ( 0, 2, 0 , 1 ; 4 )}$
  \end{tabular}
\parbox{18mm}{\begin{center}
\begin{fmfgraph}(17,13)
\setval
\fmfforce{0.5w,0h}{v1}
\fmfforce{0.5w,8/13h}{v2}
\fmfforce{5/17w,6/13h}{v3}
\fmfforce{12/17w,6/13h}{v4}
\fmfforce{0w,6/13h}{i1}
\fmfforce{1w,6/13h}{i2}
\fmfforce{5.5/17w,11.5/13h}{o1}
\fmfforce{11.5/17w,11.5/13h}{o2}
\fmf{plain,left=1}{v1,v2,v1}
\fmf{plain}{v2,o1}
\fmf{plain}{v2,o2}
\fmf{plain}{v1,v3}
\fmf{plain}{i1,v3}
\fmf{plain}{i2,v4}
\fmf{plain}{v1,v4}
\fmfdot{v1,v2,v3,v4}
\end{fmfgraph} \end{center}}
\hspace*{2mm}
  \begin{tabular}{@{}c}
  ${\scs \#4.4}$ \\
  $3/4$\\
  ${\scs ( 0, 1, 0 , 2 ; 8)}$
  \end{tabular}
\parbox{16mm}{\begin{center}
\begin{fmfgraph}(15,8)
\setval
\fmfforce{0w,0.5/8h}{i1}
\fmfforce{0w,7.5/8h}{i2}
\fmfforce{1w,0.5/8h}{o1}
\fmfforce{1w,7.5/8h}{o2}
\fmfforce{3.5/15w,1/2h}{v1}
\fmfforce{1/2w,0h}{v2}
\fmfforce{1/2w,1h}{v3}
\fmfforce{11.5/15w,1/2h}{v4}
\fmf{plain}{i1,v1}
\fmf{plain}{i2,v1}
\fmf{plain}{o1,v4}
\fmf{plain}{o2,v4}
\fmf{plain,left=1}{v1,v4,v1}
\fmf{plain,left=0.4}{v2,v3,v2}
\fmfdot{v1,v2,v3,v4}
\end{fmfgraph} \end{center}} 
\quad
\\  
$4$ &
  \begin{tabular}{@{}c}
  ${\scs \#4.5}$ \\
  $3/2$\\
  ${\scs ( 0, 2, 0 , 1 ; 4 )}$
  \end{tabular}
\parbox{14mm}{\begin{center}
\begin{fmfgraph}(12,13)
\setval
\fmfstraight
\fmfforce{0w,0h}{i1}
\fmfforce{0w,1h}{i2}
\fmfforce{1w,0h}{o1}
\fmfforce{1w,1h}{o2}
\fmfforce{3.5/12w,3.5/13h}{v1}
\fmfforce{3.5/12w,9.5/13h}{v2}
\fmfforce{8.5/12w,9.5/13h}{v3}
\fmfforce{8.5/12w,3.5/13h}{v4}
\fmf{plain}{i1,v1}
\fmf{plain}{i2,v2}
\fmf{plain}{o1,v4}
\fmf{plain}{o2,v3}
\fmf{plain,left=1}{v1,v3,v1}
\fmf{plain,right=0.3}{v1,v2}
\fmf{plain,right=0.3}{v3,v4}
\fmfdot{v1,v2,v3,v4}
\end{fmfgraph} \end{center}}
\quad
  \begin{tabular}{@{}c}
  ${\scs \#4.6}$ \\
  $6$\\
  ${\scs ( 0, 1, 0 , 1 ; 2)}$
  \end{tabular}
\parbox{18mm}{\begin{center}
\begin{fmfgraph}(16.5,11.5)
\setval
\fmfforce{11.5/16.5w,5.5/11.5h}{v1}
\fmfforce{3.5/16.5w,4/11.5h}{v2}
\fmfforce{5/16.5w,7.7/11.5h}{v3}
\fmfforce{6.5/16.5w,1/11.5h}{v4}
\fmfforce{1.5/16.5w,11/11.5h}{i1}
\fmfforce{1w,5.5/11.5h}{i2}
\fmfforce{0w,0.5/11.5h}{o1}
\fmfforce{0w,7.5/11.5h}{o2}
\fmf{plain,left=1}{v1,v2,v1}
\fmf{plain}{v2,o1}
\fmf{plain}{v2,o2}
\fmf{plain}{i1,v3}
\fmf{plain}{i2,v1}
\fmf{plain}{v1,v4}
\fmf{plain}{v3,v4}
\fmfdot{v1,v2,v3,v4}
\end{fmfgraph}\end{center}}
\hspace*{2mm} 
  \begin{tabular}{@{}c}
  ${\scs \#4.7}$ \\ 
  $3/2$\\
  ${\scs ( 0, 2, 0 , 1 ; 4 )}$
  \end{tabular}
\parbox{16mm}{\begin{center}
\begin{fmfgraph}(15,13.5)
\setval
\fmfstraight
\fmfforce{0w,2.5/13.5h}{i1}
\fmfforce{1w,2.5/13.5h}{o1}
\fmfforce{1/3w,2.5/13.5h}{v1}
\fmfforce{2/3w,2.5/13.5h}{v2}
\fmfforce{1/2w,5/13.5h}{v3}
\fmfforce{1/2w,10/13.5h}{v4}
\fmfforce{4/15w,1h}{i2}
\fmfforce{11/15w,1h}{o2}
\fmf{plain}{i1,o1}
\fmf{plain}{i2,v4}
\fmf{plain}{o2,v4}
\fmf{plain,left=1}{v1,v2,v1}
\fmf{plain,left=1}{v3,v4,v3}
\fmfdot{v1,v2,v3,v4}
\end{fmfgraph} \end{center}}
\quad
  \begin{tabular}{@{}c}
  ${\scs \#4.8}$ \\ 
  $1/2$\\
  ${\scs ( 0, 0, 1 , 1 ; 8 )}$
  \end{tabular}
\parbox{15mm}{\begin{center}
\begin{fmfgraph}(14,12)
\setval
\fmfforce{3/14w,4/12h}{v1}
\fmfforce{11/14w,4/12h}{v2}
\fmfforce{4.5/14w,7.5/12h}{v3}
\fmfforce{9.5/14w,7.5/12h}{v4}
\fmfforce{0w,7.5/12h}{i1}
\fmfforce{4.5/14w,1h}{i2}
\fmfforce{1w,7.5/12h}{o1}
\fmfforce{9.5/14w,1h}{o2}
\fmfdot{v1,v2}
\fmf{plain,left=1}{v1,v2,v1}
\fmf{plain,left=0.4}{v1,v2,v1}
\fmf{plain}{i1,v3}
\fmf{plain}{i2,v3}
\fmf{plain}{o1,v4}
\fmf{plain}{o2,v4}
\fmfdot{v1,v2,v3,v4}
\end{fmfgraph} \end{center}}
\quad
\\  &
\hspace*{2mm} 
  \begin{tabular}{@{}c}
  $\mbox{}$\\
  ${\scs \#4.9}$ \\
  $3$\\
  ${\scs ( 1, 0, 0 , 1 ; 4)}$ \\
  $\mbox{}$\\
  \end{tabular}
\parbox{15mm}{\begin{center}
\begin{fmfgraph}(15,13.5)
\setval
\fmfforce{0w,7.5/13.5h}{i1}
\fmfforce{1w,7.5/13.5h}{o1}
\fmfforce{1/3w,7.5/13.5h}{v1}
\fmfforce{2/3w,7.5/13.5h}{v2}
\fmfforce{1/2w,10/13.5h}{v3}
\fmfforce{1/2w,5/13.5h}{v4}
\fmfforce{1/2w,0h}{v5}
\fmfforce{4/15w,1h}{i2}
\fmfforce{11/15w,1h}{o2}
\fmf{plain}{i1,o1}
\fmf{plain}{i2,v3}
\fmf{plain}{o2,v3}
\fmf{plain,left=1}{v1,v2,v1}
\fmf{plain,left=1}{v4,v5,v4}
\fmfdot{v1,v2,v3,v4}
\end{fmfgraph} \end{center}}
\hspace*{5mm}
  \begin{tabular}{@{}c}
  ${\scs \#4.10}$ \\
  $3/2$\\
  ${\scs ( 1, 1, 0 , 1 ; 4)}$
  \end{tabular}
\parbox{18mm}{\begin{center}
\begin{fmfgraph}(17,10)
\setval
\fmfforce{0w,-0.1h}{i1}
\fmfforce{0w,0.6h}{i2}
\fmfforce{1w,-0.1h}{o1}
\fmfforce{1w,0.6h}{o2}
\fmfforce{3.5/17w,1/4h}{v1}
\fmfforce{1/2w,1/4h}{v2}
\fmfforce{13.5/17w,1/4h}{v3}
\fmfforce{11/17w,1/2h}{v4}
\fmfforce{11/17w,1h}{v5}
\fmf{plain}{i1,v1}
\fmf{plain}{v3,o1}
\fmf{plain}{i2,v1}
\fmf{plain}{v3,o2}
\fmf{plain,left=1}{v1,v2,v1}
\fmf{plain,left=1}{v2,v3,v2}
\fmf{plain,left=1}{v4,v5,v4}
\fmfdot{v1,v2,v3,v4}
\end{fmfgraph} \end{center}}
\quad
  \begin{tabular}{@{}c}
  ${\scs \#4.11}$ \\
  $3$\\
  ${\scs ( 1, 1, 0 , 1 ; 2)}$
  \end{tabular}
\parbox{19mm}{\begin{center}
\begin{fmfgraph}(18,12)
\setval
\fmfstraight
\fmfforce{0w,1/3h}{i1}
\fmfforce{1w,1/3h}{o1}
\fmfforce{5/18w,1/3h}{v1}
\fmfforce{13/18w,1/3h}{v2}
\fmfforce{6.5/18w,7/12h}{v3}
\fmfforce{11.5/18w,7/12h}{v4}
\fmfforce{15/18w,10.5/12h}{v5}
\fmfforce{1.5/18w,7/12h}{i2}
\fmfforce{6.5/18w,1h}{o2}
\fmf{plain}{i1,o1}
\fmf{plain}{i2,v3}
\fmf{plain}{o2,v3}
\fmf{plain,left=1}{v4,v5,v4}
\fmf{plain,left=1}{v1,v2,v1}
\fmfdot{v1,v2,v3,v4}
\end{fmfgraph} \end{center}}
\quad
  \begin{tabular}{@{}c}
  ${\scs \#4.12}$ \\
  $3/4$\\
  ${\scs ( 2, 0, 0 , 1 ; 8)}$
  \end{tabular}
\parbox{15mm}{\begin{center}
\begin{fmfgraph}(15,10.5)
\setval
\fmfstraight
\fmfforce{0w,0.5/10.5h}{i1}
\fmfforce{0w,7.5/10.5h}{i2}
\fmfforce{1w,0.5/10.5h}{o1}
\fmfforce{1w,7.5/10.5h}{o2}
\fmfforce{3.5/15w,4/10.5h}{v1}
\fmfforce{11.5/15w,4/10.5h}{v2}
\fmfforce{5.5/15w,7.5/10.5h}{v3}
\fmfforce{9.5/15w,7.5/10.5h}{v4}
\fmfforce{2/15w,11/10.5h}{v5}
\fmfforce{13/15w,11/10.5h}{v6}
\fmf{plain}{i1,v1}
\fmf{plain}{i2,v1}
\fmf{plain}{o1,v2}
\fmf{plain}{o2,v2}
\fmf{plain,left=1}{v4,v6,v4}
\fmf{plain,left=1}{v3,v5,v3}
\fmf{plain,left=1}{v1,v2,v1}
\fmfdot{v1,v2,v3,v4}
\end{fmfgraph} \end{center}}
\hspace*{2mm}
\\ & 
  \begin{tabular}{@{}c}
  ${\scs \#4.13}$ \\
  $3/4$\\
  ${\scs ( 1, 1, 0 , 1 ; 8)}$
  \end{tabular}
\parbox{12mm}{\begin{center}
\begin{fmfgraph}(12,15)
\setval
\fmfstraight
\fmfforce{0w,-1/15h}{i1}
\fmfforce{0w,6/15h}{i2}
\fmfforce{1w,-1/15h}{o1}
\fmfforce{1w,6/15h}{o2}
\fmfforce{3.5/12w,1/6h}{v1}
\fmfforce{8.5/12w,1/6h}{v2}
\fmfforce{1/2w,1/3h}{v3}
\fmfforce{1/2w,2/3h}{v4}
\fmfforce{1/2w,1h}{v5}
\fmf{plain}{i1,v1}
\fmf{plain}{v2,o1}
\fmf{plain}{i2,v1}
\fmf{plain}{v2,o2}
\fmf{plain,left=1}{v1,v2,v1}
\fmf{plain,left=1}{v3,v4,v3}
\fmf{plain,left=1}{v5,v4,v5}
\fmfdot{v1,v2,v3,v4}
\end{fmfgraph}  \end{center}}
\hspace*{4mm}
  \begin{tabular}{@{}c}
  ${\scs \#4.14}$ \\
  $3/8$\\
  ${\scs ( 2, 0, 0 , 2 ; 8)}$
  \end{tabular}
\parbox{12mm}{\begin{center}
\begin{fmfgraph}(12,15)
\setval
\fmfstraight
\fmfforce{0w,4/15h}{i1}
\fmfforce{0w,11/15h}{i2}
\fmfforce{1w,11/15h}{o1}
\fmfforce{1w,4/15h}{o2}
\fmfforce{1/2w,0h}{v1}
\fmfforce{1/2w,1/3h}{v2}
\fmfforce{1/2w,2/3h}{v3}
\fmfforce{1/2w,1h}{v4}
\fmfforce{3.5/12w,1/2h}{v5}
\fmfforce{8.5/12w,1/2h}{v6}
\fmf{plain}{i1,v5}
\fmf{plain}{v6,o1}
\fmf{plain}{i2,v5}
\fmf{plain}{v6,o2}
\fmf{plain,left=1}{v1,v2,v1}
\fmf{plain,left=1}{v2,v3,v2}
\fmf{plain,left=1}{v3,v4,v3}
\fmfdot{v2,v3,v5,v6}
\end{fmfgraph}  \end{center}}
\quad
  \begin{tabular}{@{}c}
  ${\scs \#4.15}$ \\
  $1$\\
  ${\scs ( 0, 2, 0 , 1 ; 6)}$
  \end{tabular}
\parbox{16mm}{\begin{center}
\begin{fmfgraph}(22,8)
\setval
\fmfforce{0w,0.25h}{i1}
\fmfforce{5/22w,0.25h}{v1}
\fmfforce{5/22w,7/8h}{v5}
\fmfforce{5/22w,-3/8h}{v7}
\fmfforce{1w,0.25h}{o1}
\fmfforce{10/22w,0.25h}{v2}
\fmfforce{13.5/22w,0h}{v3}
\fmfforce{13.5/22w,1h}{v4}
\fmfforce{17/22w,0.25h}{v6}
\fmf{plain,left=1}{v3,v4,v3}
\fmf{plain}{v1,v5}
\fmf{plain}{v1,v7}
\fmf{plain}{i1,v2}
\fmf{plain}{v2,v4}
\fmf{plain}{v4,v6}
\fmf{plain}{v6,o1}
\fmfdot{v1,v2,v4,v6}
\end{fmfgraph}  \end{center}}
\hspace*{5mm}
  \begin{tabular}{@{}c}
  ${\scs \#4.16}$ \\
  $1$\\
  ${\scs ( 0, 1,1 , 1 ; 2)}$
  \end{tabular}
\parbox{18mm}{\begin{center}
\begin{fmfgraph}(20,11)
\setval
\fmfforce{0w,2.5/11h}{i1}
\fmfforce{1w,2.5/11h}{o1}
\fmfforce{1/4w,2.5/11h}{v1}
\fmfforce{1/4w,7.5/11h}{v2}
\fmfforce{1.5/20w,1h}{i2}
\fmfforce{8.5/20w,1h}{o2}
\fmfforce{1/2w,2.5/11h}{v3}
\fmfforce{3/4w,2.5/11h}{v4}
\fmf{plain}{i1,o1}
\fmf{plain}{v2,i2}
\fmf{plain}{v2,o2}
\fmf{plain,left=1}{v1,v2,v1}
\fmf{plain,left=1}{v3,v4,v3}
\fmfdot{v1,v2,v3,v4}
\end{fmfgraph} \end{center}}
\quad
\\
$ $ & \hspace*{3mm}
  \begin{tabular}{@{}c}
  ${\scs \#4.17}$ \\
  $3$\\ 
  ${\scs ( 1, 1, 0 , 1 ; 2)}$
  \end{tabular}
\parbox{18mm}{\begin{center}
\begin{fmfgraph}(20,7.5)
\setval
\fmfstraight
\fmfforce{0w,1/3h}{i1}
\fmfforce{1w,1/3h}{o1}
\fmfforce{1/4w,1/3h}{v1}
\fmfforce{1/2w,1/3h}{v2}
\fmfforce{3/8w,2/3h}{v3}
\fmfforce{3/4w,1/3h}{v4}
\fmfforce{3/4w,1h}{v5}
\fmfforce{1/5w,8.5/7.5h}{i2}
\fmfforce{11/20w,8.5/7.5h}{o2}
\fmf{plain}{i1,o1}
\fmf{plain}{i2,v3}
\fmf{plain}{o2,v3}
\fmf{plain,left=1}{v1,v2,v1}
\fmf{plain,left=1}{v4,v5,v4}
\fmfdot{v1,v2,v3,v4}
\end{fmfgraph}  \end{center}}
\hspace*{3mm}
  \begin{tabular}{@{}c}
  ${\scs \#4.18}$ \\
  $3$\\ 
  ${\scs ( 1, 1, 0 , 1 ; 2)}$
  \end{tabular}
\parbox{17mm}{\begin{center}
\begin{fmfgraph}(17.5,10)
\setval
\fmfstraight
\fmfforce{0w,1/4h}{i1}
\fmfforce{15/17.5w,1/4h}{o1}
\fmfforce{5/17.5w,1/4h}{v1}
\fmfforce{10/17.5w,1/4h}{v2}
\fmfforce{7.5/17.5w,1/2h}{v3}
\fmfforce{12.5/17.5w,1/2h}{v5}
\fmfforce{12.5/17.5w,1h}{v6}
\fmfforce{2.5/17.5w,1/2h}{i2}
\fmfforce{1w,1/2h}{o2}
\fmf{plain}{i1,o1}
\fmf{plain}{i2,o2}
\fmf{plain,left=1}{v1,v2,v1}
\fmf{plain,left=1}{v5,v6,v5}
\fmfdot{v1,v2,v3,v5}
\end{fmfgraph}  \end{center}}
\hspace*{3mm}
  \begin{tabular}{@{}c}
  ${\scs \#4.19}$ \\
  $3$\\
  ${\scs ( 2, 0, 0 , 1 ; 2)}$
  \end{tabular}
\parbox{21mm}{\begin{center}
\begin{fmfgraph}(20,8)
\setval
\fmfforce{0w,0h}{i1}
\fmfforce{1w,0h}{o1}
\fmfforce{3/4w,0h}{v8}
\fmfforce{3/4w,5/8h}{v9}
\fmfforce{1/4w,0h}{v1}
\fmfforce{1/4w,5/8h}{v2}
\fmfforce{3/20w,1/2h}{v3}
\fmfforce{-2/20w,1/2h}{v4}
\fmfforce{3/20w,9/8h}{v7}
\fmfforce{7/20w,1/2h}{v5}
\fmfforce{10.5/20w,7.5/8h}{v6}
\fmf{plain,left=1}{v1,v2,v1}
\fmf{plain,left=1}{v8,v9,v8}
\fmf{plain}{i1,o1}
\fmf{plain}{v3,v4}
\fmf{plain}{v3,v7}
\fmf{plain,left=1}{v5,v6,v5}
\fmfdot{v1,v3,v5,v8}
\end{fmfgraph} \end{center}} 
\hspace*{2mm}
  \begin{tabular}{@{}c}
  ${\scs \#4.20}$ \\
  $3/2$\\
  ${\scs ( 1, 2, 0 , 1 ; 2)}$
  \end{tabular}
\parbox{18mm}{\begin{center}
\begin{fmfgraph}(17,13.5)
\setval
\fmfstraight
\fmfforce{0w,0h}{i1}
\fmfforce{1w,0h}{o1}
\fmfforce{1.5/17w,1h}{i2}
\fmfforce{1/2w,1h}{o2}
\fmfforce{5/17w,0h}{v1}
\fmfforce{5/17w,5/13.5h}{v2}
\fmfforce{12/17w,0h}{v3}
\fmfforce{12/17w,5/13.5h}{v4}
\fmfforce{5/17w,10/13.5h}{v5}
\fmf{plain,left=1}{v1,v2,v1}
\fmf{plain,left=1}{v3,v4,v3}
\fmf{plain,left=1}{v2,v5,v2}
\fmf{plain}{i1,o1}
\fmf{plain}{i2,v5}
\fmf{plain}{o2,v5}
\fmfdot{v1,v2,v3,v5}
\end{fmfgraph}  \end{center}}
\hspace*{0.1mm}
\end{tabular}
\end{center}
\end{table}
\newpage
\begin{table}
\begin{center}
\begin{tabular}{|c|c|}  
$ $ & \hspace*{3mm}
  \begin{tabular}{@{}c}
  $\mbox{}$\\
  ${\scs \#4.21}$ \\
  $3/2$\\
  ${\scs( 1, 2, 0 , 1 ; 2)}$ \\
  $\mbox{}$
  \end{tabular}
\parbox{15mm}{\begin{center}
\begin{fmfgraph}(18,10)
\setval
\fmfstraight
\fmfforce{0w,0h}{i1}
\fmfforce{1w,0h}{o1}
\fmfforce{1.5/18w,8.5/10h}{i2}
\fmfforce{8.5/18w,8.5/10h}{o2}
\fmfforce{5/18w,0h}{v1}
\fmfforce{5/18w,1/2h}{v2}
\fmfforce{13/18w,0h}{v3}
\fmfforce{13/18w,1/2h}{v4}
\fmfforce{13/18w,1h}{v5}
\fmf{plain,left=1}{v1,v2,v1}
\fmf{plain,left=1}{v3,v4,v3}
\fmf{plain,left=1}{v4,v5,v4}
\fmf{plain}{i1,o1}
\fmf{plain}{i2,v2}
\fmf{plain}{o2,v2}
\fmfdot{v1,v2,v3,v4}
\end{fmfgraph} \end{center}}
\hspace*{4mm}
  \begin{tabular}{@{}c}
  ${\scs \#4.22}$ \\
  $3/2$\\
  ${\scs ( 2, 1, 0 , 1 ; 2)}$
  \end{tabular}
\parbox{21mm}{\begin{center}
\begin{fmfgraph}(24,8.5)
\setval
\fmfstraight
\fmfforce{0w,0h}{i1}
\fmfforce{1w,0h}{o1}
\fmfforce{1.5/24w,1h}{i2}
\fmfforce{8.5/24w,1h}{o2}
\fmfforce{5/24w,0h}{v1}
\fmfforce{5/24w,5/8.5h}{v2}
\fmfforce{1/2w,0h}{v3}
\fmfforce{1/2w,5/8.5h}{v4}
\fmfforce{19/24w,0h}{v5}
\fmfforce{19/24w,5/8.5h}{v6}
\fmf{plain,left=1}{v1,v2,v1}
\fmf{plain,left=1}{v3,v4,v3}
\fmf{plain,left=1}{v5,v6,v5}
\fmf{plain}{i1,o1}
\fmf{plain}{i2,v2}
\fmf{plain}{o2,v2}
\fmfdot{v1,v2,v3,v5}
\end{fmfgraph} \end{center}}
\quad \,
  \begin{tabular}{@{}c}
  ${\scs \#4.23}$ \\
  $3/4$\\
  ${\scs( 2, 1, 0 , 1 ; 4)}$
  \end{tabular}
\parbox{20mm}{\begin{center}
\begin{fmfgraph}(24,8.5)
\setval
\fmfforce{0w,0h}{i1}
\fmfforce{5/24w,0h}{v1}
\fmfforce{5/24w,5/8.5h}{v2}
\fmfforce{1/2w,0h}{v3}
\fmfforce{1/2w,5/8.5h}{v4}
\fmfforce{19/24w,0h}{v5}
\fmfforce{19/24w,5/8.5h}{v6}
\fmfforce{1w,0h}{o1}
\fmfforce{8.5/24w,1h}{i2}
\fmfforce{15.5/24w,1h}{o2}
\fmf{plain,left=1}{v1,v2,v1}
\fmf{plain,left=1}{v3,v4,v3}
\fmf{plain,left=1}{v5,v6,v5}
\fmf{plain}{i1,o1}
\fmf{plain}{i2,v4}
\fmf{plain}{o2,v4}
\fmfdot{v1,v3,v4,v5}
\end{fmfgraph} \end{center}}
\quad \,
  \begin{tabular}{@{}c}
  ${\scs \#4.24}$ \\
  $1$\\
  ${\scs ( 1, 1, 0 , 1 ; 6)}$
  \end{tabular}
\hspace*{-1mm}
\parbox{13mm}{\begin{center}
\begin{fmfgraph}(18,12.5)
\setval
\fmfforce{0w,5/12.5h}{i1}
\fmfforce{5/18w,10/12.5h}{i2}
\fmfforce{5/18w,0h}{o1}
\fmfforce{1w,5/12.5h}{o2}
\fmfforce{8/18w,5/12.5h}{v1}
\fmfforce{13/18w,5/12.5h}{v2}
\fmfforce{10.5/18w,3/5h}{v4}
\fmfforce{10.5/18w,1h}{v7}
\fmfforce{5/18w,5/12.5h}{v6}
\fmf{plain}{i1,o2}
\fmf{plain}{i2,o1}
\fmf{plain}{v2,o2}
\fmf{plain,left=1}{v7,v4,v7}
\fmf{plain,left=1}{v1,v2,v1}
\fmfdot{v1,v2,v4,v6}
\end{fmfgraph} \end{center}} 
\hspace*{7mm}
\\ 
$\,\,\, 4 \,\,\,$ &
\hspace*{0mm}
  \begin{tabular}{@{}c}
  ${\scs \#4.25}$ \\
  $3/2$\\
  ${\scs ( 2, 1, 0 , 1 ; 2)}$
  \end{tabular}
\parbox{19mm}{\begin{center}
\begin{fmfgraph}(19,10.5)
\setval
\fmfstraight
\fmfforce{0w,1h}{i1}
\fmfforce{1w,1h}{o1}
\fmfforce{3.5/19w,0h}{i2}
\fmfforce{15.5/19w,0h}{o2}
\fmfforce{3.5/19w,2/3h}{v1}
\fmfforce{0w,1/3h}{v5}
\fmfforce{7/19w,1/3h}{v2}
\fmfforce{12/19w,1/3h}{v3}
\fmfforce{15.5/19w,2/3h}{v4}
\fmfforce{1w,1/3h}{v6}
\fmf{plain}{i1,v2}
\fmf{plain}{v3,o1}
\fmf{plain}{i2,v2}
\fmf{plain}{o2,v3}
\fmf{plain,left=1}{v1,v5,v1}
\fmf{plain,left=1}{v2,v3,v2}
\fmf{plain,left=1}{v4,v6,v4}
\fmfdot{v1,v3,v2,v4}
\end{fmfgraph}  \end{center}}
\quad \,
  \begin{tabular}{@{}c}
  ${\scs \#4.26}$ \\
  $1/2$\\
  ${\scs ( 2, 0, 0 , 2 ; 6)}$
  \end{tabular}
\parbox{20mm}{\begin{center}
\begin{fmfgraph}(19,13)
\setval
\fmfforce{0w,5/13h}{i1}
\fmfforce{1w,5/13h}{o1}
\fmfforce{14/19w,5/13h}{v8}
\fmfforce{14/19w,10/13h}{v9}
\fmfforce{14/19w,0h}{v10}
\fmfforce{5/19w,5/13h}{v1}
\fmfforce{5/19w,10/13h}{v2}
\fmfforce{3/19w,9/13h}{v3}
\fmfforce{-0.5/19w,12.5/13h}{v4}
\fmfforce{7/19w,9/13h}{v5}
\fmfforce{10.5/19w,12.5/13h}{v6}
\fmf{plain,left=1}{v1,v2,v1}
\fmf{plain}{i1,o1}
\fmf{plain,left=1}{v3,v4,v3}
\fmf{plain}{v9,v10} 
\fmf{plain,left=1}{v5,v6,v5}
\fmfdot{v1,v3,v5,v8}
\end{fmfgraph} \end{center}} 
\quad
  \begin{tabular}{@{}c}
  ${\scs \#4.27}$ \\
  $1/2$\\
  ${\scs ( 3, 0, 0 , 1 ; 6)}$
  \end{tabular}
\parbox{18mm}{\begin{center}
\begin{fmfgraph}(20,15)
\setval
\fmfforce{0w,2/3h}{i1}
\fmfforce{1/2w,1h}{i2}
\fmfforce{1w,2/3h}{o1}
\fmfforce{1/2w,0h}{o2}
\fmfforce{1/4w,2/3h}{v1}
\fmfforce{1/4w,1h}{v2}
\fmfforce{1/2w,2/3h}{v3}
\fmfforce{1/2w,1/3h}{v4}
\fmfforce{3/4w,1/3h}{v5}
\fmfforce{3/4w,2/3h}{v6}
\fmfforce{3/4w,1h}{v7}
\fmf{plain}{i1,o1}
\fmf{plain}{i2,o2}
\fmf{plain,left=1}{v1,v2,v1}
\fmf{plain,left=1}{v4,v5,v4}
\fmf{plain,left=1}{v6,v7,v6}
\fmfdot{v1,v3,v4,v6}
\end{fmfgraph} \end{center}}
\hspace*{3mm}
  \begin{tabular}{@{}c}
  ${\scs \#4.28}$ \\
  $1/3$\\
  ${\scs ( 0, 0, 1 , 2 ; 6)}$
  \end{tabular}
\hspace*{-1mm}
\parbox{14mm}{\begin{center}
\begin{fmfgraph}(17,13)
\setval
\fmfforce{8/17w,9/13h}{v1}
\fmfforce{16/17w,9/13h}{v2}
\fmfforce{12/17w,5/13h}{v4}
\fmfforce{0w,5/13h}{i1}
\fmfforce{5/17w,10/13h}{i2}
\fmfforce{5/17w,0h}{o2}
\fmfforce{1w,5/13h}{o1}
\fmfforce{5/17w,5/13h}{v5}
\fmf{plain,left=1}{v1,v2,v1}
\fmf{plain,left=0.4}{v1,v2,v1}
\fmf{plain,left=1}{v3,v4,v3}
\fmf{plain}{i1,o1}
\fmf{plain}{i2,o2}
\fmfdot{v1,v2,v3,v5}
\end{fmfgraph} \end{center}} 
\hs \hs 
\\  &
  \begin{tabular}{@{}c}
  $\mbox{}$\\
  ${\scs \#4.29}$ \\
  $1/3$\\
  ${\scs ( 1, 0, 1 , 1 ; 6)}$\\
  $\mbox{}$
  \end{tabular}
\parbox{23mm}{\begin{center}
\begin{fmfgraph}(23,10)
\setval
\fmfforce{0w,1/2h}{i1}
\fmfforce{5/23w,0h}{i2}
\fmfforce{5/23w,1h}{i3}
\fmfforce{1w,1/2h}{o1}
\fmfforce{5/23w,1/2h}{v1}
\fmfforce{8/23w,1/2h}{v2}
\fmfforce{13/23w,1/2h}{v3}
\fmfforce{18/23w,1/2h}{v4}
\fmfforce{18/23w,1h}{v5}
\fmf{plain}{i1,o1}
\fmf{plain}{i2,v1}
\fmf{plain}{i3,v1}
\fmf{plain,left=1}{v2,v3,v2}
\fmf{plain,left=1}{v4,v5,v4}
\fmfdot{v1,v3,v2,v4}
\end{fmfgraph} \end{center}}
\hspace*{3mm}
  \begin{tabular}{@{}c}
  ${\scs \#4.30}$ \\
  $3/2$\\
  ${\scs ( 3, 0, 0 , 1 ; 2)}$
  \end{tabular}
\parbox{27mm}{\begin{center}
\begin{fmfgraph}(27,10)
\setval
\fmfforce{0w,1/2h}{i1}
\fmfforce{10/27w,1h}{i2}
\fmfforce{10/27w,0h}{o1}
\fmfforce{1w,1/2h}{o2}
\fmfforce{5/27w,1/2h}{v1}
\fmfforce{5/27w,1h}{v2}
\fmfforce{15/27w,1/2h}{v3}
\fmfforce{15/27w,1h}{v4}
\fmfforce{10/27w,1/2h}{v5}
\fmfforce{22/27w,1/2h}{v6}
\fmfforce{22/27w,1h}{v7}
\fmf{plain}{i1,o2}
\fmf{plain}{i2,v5}
\fmf{plain}{o1,v5}
\fmf{plain,left=1}{v1,v2,v1}
\fmf{plain,left=1}{v3,v4,v3}
\fmf{plain,left=1}{v6,v7,v6}
\fmfdot{v1,v3,v5,v6}
\end{fmfgraph} \end{center}}
\hspace*{3mm}
  \begin{tabular}{@{}c}
  ${\scs \#4.31}$ \\
  $3/2$\\
  ${\scs ( 2, 1, 0 , 1 ; 2)}$
  \end{tabular}
\parbox{20mm}{\begin{center}
\begin{fmfgraph}(20,15)
\setval
\fmfforce{0w,1/3h}{i1}
\fmfforce{1/2w,2/3h}{i2}
\fmfforce{1/2w,0h}{o1}
\fmfforce{1w,1/3h}{o2}
\fmfforce{1/4w,1/3h}{v1}
\fmfforce{1/4w,2/3h}{v2}
\fmfforce{3/4w,1/3h}{v3}
\fmfforce{3/4w,2/3h}{v4}
\fmfforce{1/2w,1/3h}{v5}
\fmfforce{3/4w,1h}{v6}
\fmf{plain}{i1,o2}
\fmf{plain}{i2,v5}
\fmf{plain}{o1,v5}
\fmf{plain,left=1}{v1,v2,v1}
\fmf{plain,left=1}{v3,v4,v3}
\fmf{plain,left=1}{v6,v4,v6}
\fmfdot{v1,v3,v5,v4}
\end{fmfgraph} \end{center}}
\quad
\\ &
  \begin{tabular}{@{}c}
  ${\scs \#4.32}$ \\
  $1/2$\\
  ${\scs ( 2, 1, 0 , 1 ; 6)}$
  \end{tabular}
\parbox{22mm}{\begin{center}
\begin{fmfgraph}(22,15)
\setval
\fmfforce{0w,1/3h}{i1}
\fmfforce{5/22w,2/3h}{i2}
\fmfforce{5/22w,0h}{i3}
\fmfforce{1w,1/3h}{o1}
\fmfforce{10/22w,1/3h}{v1}
\fmfforce{10/22w,2/3h}{v2}
\fmfforce{5/22w,1/3h}{v3}
\fmfforce{17/22w,1/3h}{v4}
\fmfforce{17/22w,2/3h}{v5}
\fmfforce{17/22w,1h}{v6}
\fmf{plain}{i1,o1}
\fmf{plain}{i2,v3}
\fmf{plain}{i3,v3}
\fmf{plain,left=1}{v1,v2,v1}
\fmf{plain,left=1}{v4,v5,v4}
\fmf{plain,left=1}{v6,v5,v6}
\fmfdot{v1,v3,v4,v5}
\end{fmfgraph} \end{center}}
\hspace*{3mm}
  \begin{tabular}{@{}c}
  ${\scs \#4.33}$ \\
  $1$\\
  ${\scs ( 1, 0, 1 , 1 ; 2)}$
  \end{tabular}
\parbox{23mm}{\begin{center}
\begin{fmfgraph}(23,10)
\setval
\fmfforce{0w,1/2h}{i1}
\fmfforce{1w,1/2h}{o1}
\fmfforce{10/23w,1h}{i2}
\fmfforce{10/23w,0h}{o2}
\fmfforce{5/23w,1/2h}{v1}
\fmfforce{5/23w,1h}{v5}
\fmfforce{10/23w,1/2h}{v2}
\fmfforce{13/23w,1/2h}{v3}
\fmfforce{18/23w,1/2h}{v4}
\fmf{plain}{i1,o1}
\fmf{plain}{i2,o2}
\fmf{plain,left=1}{v4,v3,v4}
\fmf{plain,left=1}{v1,v5,v1}
\fmfdot{v1,v3,v2,v4}
\end{fmfgraph} \end{center}}
\hspace*{3mm}
  \begin{tabular}{@{}c}
  ${\scs \#4.34}$ \\
  $1/2$\\
  ${\scs ( 2, 1, 0 , 1 ; 6)}$
  \end{tabular}
\parbox{22mm}{\begin{center}
\begin{fmfgraph}(22,15)
\setval
\fmfforce{0w,1/3h}{i1}
\fmfforce{5/22w,2/3h}{i2}
\fmfforce{5/22w,0h}{i3}
\fmfforce{1w,1/3h}{o1}
\fmfforce{10/22w,1/3h}{v1}
\fmfforce{10/22w,2/3h}{v2}
\fmfforce{5/22w,1/3h}{v3}
\fmfforce{17/22w,1/3h}{v4}
\fmfforce{17/22w,2/3h}{v5}
\fmfforce{10/22w,1h}{v6}
\fmf{plain}{i1,o1}
\fmf{plain}{i2,v3}
\fmf{plain}{i3,v3}
\fmf{plain,left=1}{v1,v2,v1}
\fmf{plain,left=1}{v4,v5,v4}
\fmf{plain,left=1}{v6,v2,v6}
\fmfdot{v1,v3,v4,v2}
\end{fmfgraph} \end{center}}
\quad
\\ &
  \begin{tabular}{@{}c}
  $\mbox{}$\\
  ${\scs \#4.35}$ \\
  $1/3$\\
  ${\scs ( 1, 0, 1 , 1 ; 6)}$ \\
  $\mbox{}$
  \end{tabular}
\parbox{25mm}{\begin{center}
\begin{fmfgraph}(25,10)
\setval
\fmfforce{0w,1/2h}{i1}
\fmfforce{1/5w,0h}{i2}
\fmfforce{1/5w,1h}{i3}
\fmfforce{1w,1/2h}{o1}
\fmfforce{1/5w,1/2h}{v1}
\fmfforce{2/5w,1/2h}{v2}
\fmfforce{3/5w,1/2h}{v3}
\fmfforce{4/5w,1/2h}{v4}
\fmfforce{2/5w,1h}{v5}
\fmf{plain}{i1,o1}
\fmf{plain}{i2,v1}
\fmf{plain}{i3,v1}
\fmf{plain,left=1}{v3,v4,v3}
\fmf{plain,left=1}{v2,v5,v2}
\fmfdot{v1,v3,v2,v4}
\end{fmfgraph}  \end{center}}
\hspace*{3mm}
  \begin{tabular}{@{}c}
  ${\scs \#4.36}$ \\
  $1/2$\\
  ${\scs ( 3, 0, 0 , 1 ; 6)}$
  \end{tabular}
\parbox{29mm}{\begin{center}
\begin{fmfgraph}(29,10)
\setval
\fmfstraight
\fmfforce{0w,1/2h}{i1}
\fmfforce{5/29w,1h}{i2}
\fmfforce{5/29w,0h}{i3}
\fmfforce{1w,1/2h}{o1}
\fmfforce{10/29w,1/2h}{v1}
\fmfforce{10/29w,1h}{v2}
\fmfforce{5/29w,1/2h}{v3}
\fmfforce{17/29w,1/2h}{v4}
\fmfforce{17/29w,1h}{v5}
\fmfforce{24/29w,1/2h}{v6}
\fmfforce{24/29w,1h}{v7}
\fmf{plain}{i1,o1}
\fmf{plain}{i2,v3}
\fmf{plain}{i3,v3}
\fmf{plain,left=1}{v1,v2,v1}
\fmf{plain,left=1}{v4,v5,v4}
\fmf{plain,left=1}{v6,v7,v6}
\fmfdot{v1,v3,v4,v6}
\end{fmfgraph} \end{center}}
\hspace*{3mm}
  \begin{tabular}{@{}c}
  ${\scs \#4.37}$ \\
  $1/2$\\
  ${\scs ( 1, 2, 0 , 1 ; 6)}$
  \end{tabular}
\parbox{15mm}{\begin{center}
\begin{fmfgraph}(15,20)
\setval
\fmfstraight
\fmfforce{0w,1/4h}{i1}
\fmfforce{1/3w,1/2h}{i2}
\fmfforce{1/3w,0h}{i3}
\fmfforce{1w,1/4h}{o1}
\fmfforce{2/3w,1/4h}{v1}
\fmfforce{2/3w,2/4h}{v2}
\fmfforce{1/3w,1/4h}{v3}
\fmfforce{2/3w,3/4h}{v4}
\fmfforce{2/3w,1h}{v5}
\fmf{plain}{i1,o1}
\fmf{plain}{i2,v3}
\fmf{plain}{i3,v3}
\fmf{plain,left=1}{v1,v2,v1}
\fmf{plain,left=1}{v4,v2,v4}
\fmf{plain,left=1}{v4,v5,v4}
\fmfdot{v1,v2,v3,v4}
\end{fmfgraph} \end{center}}
\\
\hline\hline
\end{tabular}
\end{center}
\caption{\la{tab2} Diagrams of the connected four-point function and their weights of the $\phi^4$-theory
up to three loops characterized by the vector $(S,D,T,P;N$).
Its components $S,D,T$ specify the number of self-, double,
triple connections, $P$ stands for the number of vertex permutations leaving the diagram unchanged, and $N$ denotes the symmetry
degree.}
\end{table}

\newpage
\begin{table}[t]
\begin{center}
\begin{tabular}{|c|c|}
\hline\hline  
\,\,\,$p$\,\,\,
&
\begin{tabular}{@{}c}
$\mbox{}$ \\*[4mm]
$\mbox{}$
\end{tabular}
  \parbox{10mm}{\centerline{
  \begin{fmfgraph*}(8,6)
  \setval
  \fmfforce{0w,1/2h}{v1}
  \fmfforce{1w,1/2h}{v2}
  \fmfforce{1/2w,1h}{v3}
  \fmfforce{1/2w,0h}{v4}
  \fmfforce{1/2w,1/2h}{v5}
  \fmf{plain,left=0.4}{v1,v3,v2,v4,v1}
  \fmfv{decor.size=0, label=${\scs p}$, l.dist=0mm, l.angle=0}{v5}
  \end{fmfgraph*} } }  
\\
\hline
$1$ &
\hspace{-10pt}
\rule[-10pt]{0pt}{26pt}
  \begin{tabular}{@{}c}
  $\mbox{}$\\
  ${\scs \mbox{\#1.1}}$ \\
  $1/8$ \\ ${\scs ( 2, 0, 0 , 0 , 1 )}$\\
  $\mbox{}$
  \end{tabular}
\parbox{10mm}{\begin{center}
\begin{fmfgraph}(10,5)
\setval
\fmfforce{0w,1/2h}{v1}
\fmfforce{1/2w,1/2h}{v2}
\fmfforce{1w,1/2h}{v3}
\fmf{plain,left=1}{v1,v2,v1}
\fmf{plain,left=1}{v3,v2,v3}
\fmfdot{v2}
\end{fmfgraph}\end{center}}
\\
\hline
$2$ &
\hspace{-10pt}
\rule[-10pt]{0pt}{26pt}
  \begin{tabular}{@{}c}
  $\mbox{}$\\
  ${\scs \mbox{\#2.1}}$ \\
  $1/48$ \\ 
  ${\scs ( 0, 0, 0 , 1 , 2 )}$\\
  $\mbox{}$
  \end{tabular}
\parbox{10.5mm}{\begin{center}
\begin{fmfgraph}(7.5,5)
\setval
\fmfforce{0w,0.5h}{v1}
\fmfforce{1w,0.5h}{v2}
\fmf{plain,left=1}{v1,v2,v1}
\fmf{plain,left=0.4}{v1,v2,v1}
\fmfdot{v1,v2}
\end{fmfgraph}\end{center}} 
\hspace*{7mm}
  \begin{tabular}{@{}c}
  ${\scs \mbox{\#2.2}}$ \\
  $1/16$ \\ 
  ${\scs ( 2, 1, 0 , 0 , 2 )}$
  \end{tabular}
\parbox{16mm}{\begin{center}
\begin{fmfgraph}(15,5)
\setval
\fmfleft{i1}
\fmfright{o1}
\fmf{plain,left=1}{i1,v1,i1}
\fmf{plain,left=1}{v1,v2,v1}
\fmf{plain,left=1}{o1,v2,o1}
\fmfdot{v1,v2}
\end{fmfgraph}\end{center}}
\\
\hline
$3$ &
\hspace*{2mm}
  \begin{tabular}{@{}c}
  $\mbox{}$\\
  ${\scs \mbox{\#3.1}}$ \\
  $1/48$ \\ ${\scs ( 0, 3, 0 , 0 , 6 )}$ \\
  $\mbox{}$
  \end{tabular}
\parbox{11mm}{\begin{center}
\begin{fmfgraph}(8,8)
\setval
\fmfforce{0.5w,0h}{v1}
\fmfforce{0.5w,1h}{v2}
\fmfforce{0.066987w,0.25h}{v3}
\fmfforce{0.93301w,0.25h}{v4}
\fmf{plain,left=1}{v1,v2,v1}
\fmf{plain}{v2,v3}
\fmf{plain}{v3,v4}
\fmf{plain}{v2,v4}
\fmfdot{v2,v3,v4}
\end{fmfgraph}
\end{center}} 
\quad
  \begin{tabular}{@{}c}
  ${\scs \mbox{\#3.2}}$ \\
  $1/24$\\ 
  ${\scs ( 1, 0, 1 , 0 , 2 )}$
  \end{tabular}
\parbox{10.5mm}{\begin{center}
\begin{fmfgraph}(7.5,12.5)
\setval
\fmfforce{0w,0.3h}{v1}
\fmfforce{1w,0.3h}{v2}
\fmfforce{0.5w,0.6h}{v3}
\fmfforce{0.5w,1h}{v4}
\fmf{plain,left=1}{v1,v2,v1}
\fmf{plain,left=0.4}{v1,v2,v1}
\fmf{plain,left=1}{v3,v4,v3}
\fmfdot{v1,v2,v3}
\end{fmfgraph}\end{center}} 
\quad
  \begin{tabular}{@{}c}
  ${\scs \mbox{\#3.3}}$ \\
  $1/48$\\ 
  ${\scs ( 3, 0, 0 , 0 , 6 )}$
  \end{tabular}
\parbox{18mm}{\begin{center}
\begin{fmfgraph}(15,15)
\setval
\fmfforce{1/2w,1/3h}{v1}
\fmfforce{1/2w,2/3h}{v2}
\fmfforce{1/2w,1h}{v3}
\fmfforce{0.355662432w,0.416666666h}{v4}
\fmfforce{0.64433568w,0.416666666h}{v5}
\fmfforce{0.067w,1/4h}{v6}
\fmfforce{0.933w,1/4h}{v7}
\fmf{plain,left=1}{v1,v2,v1}
\fmf{plain,left=1}{v2,v3,v2}
\fmf{plain,left=1}{v4,v6,v4}
\fmf{plain,left=1}{v5,v7,v5}
\fmfdot{v2,v4,v5}
\end{fmfgraph}\end{center}} 
\quad
  \begin{tabular}{@{}c}
  ${\scs \mbox{\#3.4}}$ \\
  $1/32$\\ 
  ${\scs ( 2, 2, 0 , 0 , 2 )}$
  \end{tabular}
\parbox{20mm}{\begin{center}
\begin{fmfgraph}(20,5)
\setval
\fmfleft{i1}
\fmfright{o1}
\fmf{plain,left=1}{i1,v1,i1}
\fmf{plain,left=1}{v1,v2,v1}
\fmf{plain,left=1}{v2,v3,v2}
\fmf{plain,left=1}{o1,v3,o1}
\fmfdot{v1,v2,v3}
\end{fmfgraph}\end{center}}
\hspace*{2mm}
\\
\hline
& \\
&
  \begin{tabular}{@{}c}
  ${\scs \mbox{\#4.1}}$ \\
  $1/128$ \\ ${\scs ( 0, 4 , 0 , 0 , 8 )}$ 
  \end{tabular} 
\parbox{13mm}{\begin{center}
\begin{fmfgraph*}(10,10)
\setval
\fmfforce{0.1464466w,0.1464466h}{v1}
\fmfforce{0.1464466w,0.8535534h}{v2}
\fmfforce{0.8535534w,0.8535534h}{v3}
\fmfforce{0.8535534w,0.1464466h}{v4}
\fmfforce{1/2w,0h}{v5}
\fmfforce{1/2w,1h}{v6}
\fmf{plain,left=1}{v5,v6,v5}
\fmf{plain}{v1,v2}
\fmf{plain}{v2,v3}
\fmf{plain}{v3,v4}
\fmf{plain}{v4,v1}
\fmfdot{v1,v2,v3,v4}
\end{fmfgraph*} \end{center}}
\quad
  \begin{tabular}{@{}c}
  ${\scs \mbox{\#4.2}}$ \\
  $1/32$ \\ 
  ${\scs ( 0, 2 , 0 , 0 , 8 )}$
  \end{tabular}
\parbox{13mm}{\begin{center}
\begin{fmfgraph*}(10,10)
\setval
\fmfforce{0w,1/2h}{v1}
\fmfforce{1w,1/2h}{v2}
\fmfforce{1/2w,1/4h}{v3}
\fmfforce{1/2w,3/4h}{v4}
\fmf{plain,left=1}{v1,v2,v1}
\fmf{plain,left=1}{v3,v4,v3}
\fmf{plain,right=0.5}{v1,v2,v1}
\fmfdot{v1,v2,v3,v4}
\end{fmfgraph*} \end{center}}
\quad
  \begin{tabular}{@{}c}
  ${\scs \mbox{\#4.3}}$ \\
  $1/144$ \\ 
  ${\scs ( 0, 0 , 2 , 0 , 4 )}$
  \end{tabular}
\parbox{13mm}{\begin{center}
\begin{fmfgraph*}(10,12.5)
\setval
\fmfforce{1/4w,1/5h}{v1}
\fmfforce{3/4w,1/5h}{v2}
\fmfforce{1/4w,4/5h}{v3}
\fmfforce{3/4w,4/5h}{v4}
\fmf{plain}{v1,v2}
\fmf{plain,left=1}{v1,v2,v1}
\fmf{plain}{v3,v4}
\fmf{plain,left=1}{v3,v4,v3}
\fmf{plain,left=0.5}{v1,v3}
\fmf{plain,right=0.5}{v2,v4}
\fmfdot{v1,v2,v3,v4}
\end{fmfgraph*} \end{center}}
\quad
  \begin{tabular}{@{}c}
  ${\scs \mbox{\#4.4}}$ \\
  $1/16$ \\ 
  ${\scs ( 1, 2  , 0 , 0 , 2 )}$
  \end{tabular}
\parbox{11mm}{\begin{center}
\begin{fmfgraph*}(8,13)
\setval
\fmfforce{0.5w,0h}{v1}
\fmfforce{0.5w,8/13h}{v2}
\fmfforce{0.0669873w,0.46154h}{v3}
\fmfforce{0.933w,0.46154h}{v4}
\fmfforce{0.5w,1h}{v5}
\fmf{plain,left=1}{v1,v2,v1}
\fmf{plain,left=1}{v2,v5,v2}
\fmf{plain}{v1,v3}
\fmf{plain}{v3,v4}
\fmf{plain}{v1,v4}
\fmfdot{v1,v2,v3,v4}
\end{fmfgraph*}\end{center}}
\quad
\\
  $4$ & \begin{tabular}{@{}c}
  ${\scs \mbox{\#4.5}}$ \\
  $1/48$ \\ 
  ${\scs ( 2 , 0 , 1 , 0 , 2 )}$
  \end{tabular}
\parbox{15mm}{\begin{center}
\begin{fmfgraph*}(8,13)
\setval
\fmfforce{0w,0.3h}{v1}
\fmfforce{1w,0.3h}{v2}
\fmfforce{0.2w,0.55h}{v3}
\fmfforce{0.8w,0.55h}{v4}
\fmf{plain,left=1}{v1,v2,v1}
\fmf{plain,left=0.4}{v1,v2,v1}
\fmfi{plain}{reverse fullcircle scaled 0.625w shifted (1w,0.7h)}
\fmfi{plain}{reverse fullcircle scaled 0.625w shifted (0w,0.7h)}
\fmfdot{v1,v2,v3,v4}
\end{fmfgraph*}\end{center}}
\quad
  \begin{tabular}{@{}c}
  ${\scs \mbox{\#4.6}}$ \\
  $1/32$ \\ 
  ${\scs ( 2,1, 0 , 0 , 4 )}$
  \end{tabular}
\parbox{10.5mm}{\begin{center}
\begin{fmfgraph*}(7.5,17.5)
\setval
\fmfforce{0w,0.5h}{v1}
\fmfforce{1w,0.5h}{v2}
\fmfforce{0.5w,h}{v3}
\fmfforce{0.5w,1h}{v4}
\fmfforce{0.5w,0h}{v5}
\fmfforce{0.5w,0.2857h}{v6}
\fmfforce{0.5w,0.71429h}{v7}
\fmf{plain,left=1}{v5,v6,v5}
\fmf{plain,left=1}{v7,v4,v7}
\fmf{plain,left=1}{v1,v2,v1}
\fmf{plain,left=0.4}{v1,v2,v1}
\fmfdot{v1,v2,v6,v7}
\end{fmfgraph*}\end{center}} 
\quad
  \begin{tabular}{@{}c}
  ${\scs \mbox{\#4.7}}$ \\
  $1/48$ \\ 
  ${\scs ( 1,1, 1 , 0 , 2 )}$
  \end{tabular}
\parbox{10.5mm}{\begin{center}
\begin{fmfgraph*}(7.5,17.5)
\setval
\fmfforce{0w,0.21429h}{v1}
\fmfforce{1w,0.21429h}{v2}
\fmfforce{0.5w,0.42857h}{v3}
\fmfforce{0.5w,0.71429h}{v4}
\fmfforce{0.5w,1h}{v5}
\fmf{plain,left=1}{v1,v2,v1}
\fmf{plain,left=1}{v3,v4,v3}
\fmf{plain,left=1}{v4,v5,v4}
\fmf{plain,left=0.4}{v1,v2,v1}
\fmfdot{v1,v2,v3,v4}
\end{fmfgraph*}\end{center}} 
\quad
  \begin{tabular}{@{}c}
  ${\scs \mbox{\#4.8}}$ \\
  $1/32$ \\ 
  ${\scs ( 3,1, 0 , 0 , 2 )}$
  \end{tabular}
\parbox{18mm}{\begin{center}
\begin{fmfgraph*}(15,20)
\setval
\fmfforce{1/2w,1/4h}{v1}
\fmfforce{1/2w,1/2h}{v2}
\fmfforce{1/2w,3/4h}{v3}
\fmfforce{1/2w,1h}{v4}
\fmfforce{0.36w,0.3125h}{v5}
\fmfforce{0.64w,0.3125h}{v6}
\fmf{plain,left=1}{v1,v2,v1}
\fmf{plain,left=1}{v2,v3,v2}
\fmf{plain,left=1}{v3,v4,v3}
\fmfi{plain}{reverse fullcircle scaled 0.333333w shifted (0.21132w,0.25h)}
\fmfi{plain}{reverse fullcircle scaled 0.333333w shifted (0.78868w,0.25h)}
\fmfdot{v2,v3,v5,v6}
\end{fmfgraph*}\end{center}} 
\quad
\\
& 
  \begin{tabular}{@{}c}
  ${\scs\mbox{\#4.9}}$ \\
  $1/128$ \\ 
  ${\scs ( 4,0 , 0 , 0 , 8 )}$ 
  \end{tabular}
\parbox{18mm}{\begin{center}
\begin{fmfgraph*}(15,15)
\setval
\fmfforce{1/3w,1/2h}{v1}
\fmfforce{2/3w,1/2h}{v2}
\fmfforce{1/2w,1/3h}{v3}
\fmfforce{1/2w,2/3h}{v4}
\fmfforce{1/2w,1h}{v5}
\fmfforce{1/2w,0h}{v6}
\fmfforce{0w,0.5h}{v7}
\fmfforce{1w,0.5h}{v8}
\fmf{plain,left=1}{v3,v4,v3}
\fmf{plain,left=1}{v4,v5,v4}
\fmf{plain,left=1}{v3,v6,v3}
\fmf{plain,left=1}{v1,v7,v1}
\fmf{plain,left=1}{v2,v8,v2}
\fmfdot{v1,v2,v3,v4}
\end{fmfgraph*}\end{center}} 
\hspace*{7mm}
  \begin{tabular}{@{}c}
  ${\scs \mbox{\#4.10}}$ \\
  $1/64$ \\ 
  ${\scs ( 2,3, 0 , 0 , 2 )}$ 
  \end{tabular}
\parbox{26mm}{\begin{center}
\begin{fmfgraph*}(25,5)
\setval
\fmfleft{i1}
\fmfright{o1}
\fmf{plain,left=1}{i1,v1,i1}
\fmf{plain,left=1}{v1,v2,v1}
\fmf{plain,left=1}{v2,v3,v2}
\fmf{plain,left=1}{v3,v4,v3}
\fmf{plain,left=1}{o1,v4,o1}
\fmfdot{v1,v2,v3,v4}
\end{fmfgraph*} \end{center}}
\\
& \\  
\hline\hline
\end{tabular}
\end{center}
\caption{\la{tab3} Vacuum diagrams and their 
weights of the $\phi^4$-theory
up to five loops. Each diagram is characterized by the
vector $(S,D,T,F,P$) whose components specify the number of self-, double,
triple and fourfold connections, and of the vertex permutations leaving the
vacuum diagram unchanged, respectively.}
\end{table}

\end{fmffile}  
%
\begin{fmffile}{tabelle2}
\setlength{\unitlength}{1mm}
\begin{table}[t]
\begin{center}
\begin{tabular}{|c|c|}
\hline\hline  
\,\,\,$p$\,\,\,
&
\begin{tabular}{@{}c} 
$\mbox{}$ \\*[3mm]
$\mbox{}$
\end{tabular}
  \parbox{12mm}{\centerline{
  \begin{fmfgraph*}(11,5)
  \setval
  \fmfforce{0w,1/2h}{v1}
  \fmfforce{3/11w,1/2h}{v2}
  \fmfforce{8/11w,1/2h}{v3}
  \fmfforce{1w,1/2h}{v4}
  \fmfforce{1/2w,1h}{v5}
  \fmfforce{1/2w,0h}{v6}
  \fmfforce{1/2w,1/2h}{v7}
  \fmf{plain}{v1,v2}
  \fmf{plain}{v3,v4}
  \fmf{double,width=0.2mm,left=1}{v5,v6,v5}
  \fmfv{decor.size=0, label=${\scs p}$, l.dist=0mm, l.angle=90}{v7}
  \end{fmfgraph*} } }
\\
\hline
$1$ &
\hspace{-10pt}
\rule[-10pt]{0pt}{26pt}
  \begin{tabular}{@{}c}
  $\mbox{}$\\
  ${\scs \mbox{\#1.1}}$ \\
  $1/2$\\ 
  ${\scs ( 1, 0, 0 , 1 ; 2 )}$\\
  $\mbox{}$
  \end{tabular}
\parbox{8mm}{\begin{center}
\begin{fmfgraph}(6,5)
\setval
\fmfstraight
\fmfforce{0w,0h}{v1}
\fmfforce{0.5w,0h}{v3}
\fmfforce{1w,0h}{v2}
\fmfforce{0.5w,1h}{v4}
\fmf{plain}{v1,v2}
\fmf{plain,left=1}{v3,v4,v3}
\fmfdot{v3}
\end{fmfgraph}  \end{center}}
\\
\hline
$2$ &
\hspace{-10pt}
\rule[-10pt]{0pt}{26pt}
  \begin{tabular}{@{}c}
  $\mbox{}$\\
  ${\scs \mbox{\#2.1}}$ \\
  $1/6$\\ 
  ${\scs ( 0, 0, 1 , 1 ; 2 )}$\\
  $\mbox{}$
  \end{tabular}
\parbox{12mm}{\begin{center}
\begin{fmfgraph}(10,5)
\setval
\fmfforce{0w,0.5h}{v1}
\fmfforce{1/4w,0.5h}{v2}
\fmfforce{3/4w,0.5h}{v3}
\fmfforce{1w,0.5h}{v4}
\fmf{plain}{v1,v4}
\fmf{plain,left=1}{v3,v2,v3}
\fmfdot{v2,v3}
\end{fmfgraph}  \end{center}}
\quad \,\,
  \begin{tabular}{@{}c}
  ${\scs \mbox{\#2.2}}$ \\
  $1/4$\\ 
  ${\scs ( 1, 1, 0 , 1 ; 2 )}$
  \end{tabular}
\parbox{13mm}{\begin{center}
\begin{fmfgraph}(5,10)
\setval
\fmfforce{0w,0h}{v1}
\fmfforce{0.5w,0h}{v2}
\fmfforce{1w,0h}{v3}
\fmfforce{0.5w,0.5h}{v4}
\fmfforce{0.5w,1h}{v5}
\fmf{plain}{v1,v3}
\fmf{plain,left=1}{v2,v4,v2}
\fmf{plain,left=1}{v4,v5,v4}
\fmfdot{v2,v4}
\end{fmfgraph} \end{center}}
\\
\hline
$3$ &
\hspace{-5pt}
\rule[-10pt]{0pt}{26pt}
  \begin{tabular}{@{}c}
  $\mbox{}$ \\
  ${\scs \mbox{\#3.1}}$ \\
  $1/4$\\ 
  ${\scs ( 0, 2, 0 , 1 ; 2 )}$ \\
  $\mbox{}$ 
  \end{tabular}
\parbox{14mm}{\begin{center}
\begin{fmfgraph}(13,8)
\setval
\fmfforce{0w,0.25h}{i1}
\fmfforce{1w,0.25h}{o1}
\fmfforce{0.5w,0h}{v1}
\fmfforce{0.5w,1h}{v2}
\fmfforce{3/13w,0.25h}{v3}
\fmfforce{10/13w,0.25h}{v4}
\fmf{plain,left=1}{v1,v2,v1}
\fmf{plain}{i1,v3}
\fmf{plain}{o1,v4}
\fmf{plain}{v2,v3}
\fmf{plain}{v2,v4}
\fmfdot{v2,v3,v4}
\end{fmfgraph}  \end{center}}
\quad 
  \begin{tabular}{@{}c}
  ${\scs \mbox{\#3.2}}$ \\
  $1/12$\\ 
  ${\scs ( 0, 0, 1 , 2 ; 2 )}$
  \end{tabular}
\parbox{9mm}{\begin{center}
\begin{fmfgraph}(8,10)
\setval
\fmfforce{0w,0.6h}{v1}
\fmfforce{1w,0.6h}{v2}
\fmfforce{0.5w,0.2h}{v4}
\fmfforce{2/8w,0h}{i1}
\fmfforce{6/8w,0h}{o1}
\fmf{plain,left=1}{v1,v2,v1}
\fmf{plain,left=0.4}{v1,v2,v1}
\fmf{plain,left=1}{v3,v4,v3}
\fmf{plain}{i1,v4}
\fmf{plain}{o1,v4}
\fmfdot{v1,v2,v3}
\end{fmfgraph} \end{center}} 
\quad \,
  \begin{tabular}{@{}c}
  ${\scs \mbox{\#3.3}}$ \\
  $1/4$\\ 
  ${\scs ( 1, 1, 0 , 1 ; 2 )}$
  \end{tabular}
\parbox{9mm}{\begin{center}
\begin{fmfgraph}(10,10)
\setval
\fmfforce{0w,0.25h}{i1}
\fmfforce{1w,0.25h}{o1}
\fmfforce{1/4w,0.25h}{v1}
\fmfforce{3/4w,0.25h}{v2}
\fmfforce{0.5w,0.5h}{v3}
\fmfforce{0.5w,1h}{v4}
\fmf{plain}{i1,o1}
\fmf{plain,left=1}{v1,v2,v1}
\fmf{plain,left=1}{v3,v4,v3}
\fmfdot{v1,v2,v3}
\end{fmfgraph}  \end{center}} 
\quad \,
  \begin{tabular}{@{}c}
  ${\scs \mbox{\#3.4}}$ \\
  $1/8$\\ 
  ${\scs ( 1, 2, 0 , 1 ; 2 )}$
  \end{tabular}
\parbox{7mm}{\begin{center}
\begin{fmfgraph}(6,15)
\setval
\fmfforce{0w,0h}{i1}
\fmfforce{0.5w,0h}{v1}
\fmfforce{1w,0h}{o1}
\fmfforce{0.5w,1/3h}{v2}
\fmfforce{0.5w,2/3h}{v3}
\fmfforce{0.5w,1h}{v4}
\fmf{plain}{i1,v1}
\fmf{plain}{v1,o1}
\fmf{plain,left=1}{v1,v2,v1}
\fmf{plain,left=1}{v2,v3,v2}
\fmf{plain,left=1}{v3,v4,v3}
\fmfdot{v1,v2,v3}
\end{fmfgraph}  \end{center}}
\hspace*{4mm} 
  \begin{tabular}{@{}c}
  $\mbox{}$\\
  ${\scs \mbox{\#3.5}}$ \\
  $1/8$\\ 
  ${\scs ( 2, 0, 0 , 2 ; 2 )}$\\
  $\mbox{}$
  \end{tabular}
\parbox{10mm}{\begin{center}
\begin{fmfgraph}(12,8)
\setval
\fmfforce{3/12w,0h}{i1}
\fmfforce{9/12w,0h}{o1}
\fmfforce{1/2w,0h}{v1}
\fmfforce{1/2w,5/8h}{v2}
\fmfforce{4/12w,4.25/8h}{v4}
\fmfforce{8/12w,4.25/8h}{v5}
\fmfforce{0.5/12w,7.75/8h}{v6}
\fmfforce{11.5/12w,7.75/8h}{v7}
\fmf{plain,left=1}{v1,v2,v1}
\fmf{plain}{i1,o1}
\fmf{plain,left=1}{v4,v6,v4}
\fmf{plain,left=1}{v5,v7,v5}
\fmfdot{v1,v4,v5}
\end{fmfgraph}\end{center}} 
\hspace*{2mm}
\\ 
\hline
$ $ &
\hspace{-5pt}
\rule[-10pt]{0pt}{26pt}
  \begin{tabular}{@{}c}
  $\mbox{}$ \\
  ${\scs \mbox{\#4.1}}$ \\
  $1/8$\\ 
  ${\scs ( 0, 3, 0 , 1 ; 2 )}$\\
  $\mbox{}$ 
\end{tabular}
\parbox{10mm}{\begin{center}
\begin{fmfgraph}(12,9)
\setval
\fmfforce{0w,1/6h}{i1}
\fmfforce{3/12w,1/6h}{v1}
\fmfforce{3/12w,5/6h}{v2}
\fmfforce{9/12w,5/6h}{v3}
\fmfforce{9/12w,1/6h}{v4}
\fmfforce{1w,1/6h}{o1}
\fmf{plain,left=1}{v1,v3,v1}
\fmf{plain}{v1,v2}
\fmf{plain}{v2,v3}
\fmf{plain}{v3,v4}
\fmf{plain}{i1,v1}
\fmf{plain}{v4,o1}
\fmfdot{v1,v2,v3,v4}
\end{fmfgraph}  \end{center}}
\hspace*{0.3cm} 
  \begin{tabular}{@{}c}
  ${\scs \mbox{\#4.2}}$ \\
  $1/4$\\ 
  ${\scs ( 0, 1, 0 , 2 ; 2 )}$
  \end{tabular}
\parbox{12mm}{\begin{center}
\begin{fmfgraph}(14,9)
\setval
\fmfforce{0w,5/9h}{i1}
\fmfforce{1w,5/9h}{o1}
\fmfforce{2.6/14w,5/9h}{v1}
\fmfforce{11.4/14w,5/9h}{v2}
\fmfforce{1/2w,0h}{v3}
\fmfforce{1/2w,5/9h}{v4}
\fmfforce{1/2w,1h}{v5}
\fmf{plain}{i1,o1}
\fmf{plain,left=1}{v5,v3,v5}
\fmf{plain,left=1}{v3,v4,v3}
\fmfdot{v1,v2,v3,v4}
\end{fmfgraph}
\end{center}}
\hspace*{0.3cm}
  \begin{tabular}{@{}c}
  ${\scs \mbox{\#4.3}}$ \\
  $1/4$\\ 
  ${\scs ( 0, 2, 0 , 1 ; 2 )}$
  \end{tabular}
\parbox{9mm}{\begin{center}
\begin{fmfgraph}(12,9)
\setval
\fmfforce{0w,4/5h}{i1}
\fmfforce{3/12w,4/5h}{v1}
\fmfforce{9/12w,4/5h}{v2}
\fmfforce{3/12w,1/5h}{v3}
\fmfforce{9/12w,1/5h}{v4}
\fmfforce{1w,1/5h}{o1}
\fmf{plain}{i1,v2}
\fmf{plain}{v3,o1}
\fmf{plain,left=1}{v4,v1,v4}
\fmf{plain}{v2,v3}
\fmfdot{v1,v2,v3,v4}
\end{fmfgraph}
\end{center}}
\hspace*{0.3cm}
  \begin{tabular}{@{}c}
  ${\scs \mbox{\#4.4}}$ \\
  $1/12$\\ 
  ${\scs ( 0, 1, 1 , 1 ; 2 )}$
  \end{tabular}
\parbox{10mm}{\begin{center}
\begin{fmfgraph}(10,12.5)
\setval
\fmfforce{1/4w,1/5h}{v1}
\fmfforce{3/4w,1/5h}{v2}
\fmfforce{1/4w,4/5h}{v3}
\fmfforce{3/4w,4/5h}{v4}
\fmfforce{0w,1/5h}{i1}
\fmfforce{1w,1/5h}{o1}
\fmf{plain}{i1,v1}
\fmf{plain}{v2,o1}
\fmf{plain,left=1}{v1,v2,v1}
\fmf{plain}{v3,v4}
\fmf{plain,left=1}{v3,v4,v3}
\fmf{plain,left=0.5}{v1,v3}
\fmf{plain,right=0.5}{v2,v4}
\fmfdot{v1,v2,v3,v4}
\end{fmfgraph} \end{center}}
\hspace*{4mm}
  \begin{tabular}{@{}c}
  ${\scs \mbox{\#4.5}}$ \\
  $1/8$\\ 
  ${\scs ( 0, 2, 0 , 2 ; 2 )}$
  \end{tabular}
\parbox{7mm}{\begin{center}
\begin{fmfgraph}(8,10)
\setval
\fmfforce{2/8w,0h}{i1}
\fmfforce{6/8w,0h}{o1}
\fmfforce{0.5w,0.2h}{v1}
\fmfforce{0.5w,1h}{v2}
\fmfforce{1/16w,4.2/10h}{v3}
\fmfforce{15/16w,4.2/10h}{v4}
\fmf{plain,left=1}{v1,v2,v1}
\fmf{plain}{i1,v1}
\fmf{plain}{o1,v1}
\fmf{plain}{v2,v3}
\fmf{plain}{v2,v4}
\fmf{plain}{v3,v4}
\fmfdot{v1,v2,v3,v4}
\end{fmfgraph}  \end{center}}
\hspace*{2mm}
\\ 
$4$ &
  \begin{tabular}{@{}c}
  ${\scs \mbox{\#4.6}}$ \\
  $1/24$\\ 
  ${\scs ( 0, 1, 1 , 2 ; 2 )}$
  \end{tabular}
\parbox{9mm}{\begin{center}
\begin{fmfgraph}(8,13)
\setval
\fmfforce{0.5w,0h}{v5}
\fmfforce{1/8w,0h}{i1}
\fmfforce{7/8w,0h}{o1}
\fmfforce{0w,9/13h}{v1}
\fmfforce{1w,9/13h}{v2}
\fmfforce{0.5w,5/13h}{v4}
\fmf{plain}{i1,o1}
\fmf{plain,left=1}{v1,v2,v1}
\fmf{plain,left=0.4}{v1,v2,v1}
\fmf{plain,left=1}{v3,v4,v3}
\fmf{plain,left=1}{v5,v4,v5}
\fmf{plain}{i1,o1}
\fmfdot{v1,v2,v3,v5}
\end{fmfgraph} \end{center}} 
\hspace*{0.4cm}
  \begin{tabular}{@{}c}
  ${\scs \mbox{\#4.7}}$ \\
  $1/8$\\ 
  ${\scs ( 2, 1, 0 , 1 ; 2 )}$ 
  \end{tabular}
\parbox{15mm}{\begin{center}
\begin{fmfgraph}(14,13)
\setval
\fmfforce{0w,4/13h}{i1}
\fmfforce{1w,4/13h}{o1}
\fmfforce{3/14w,4/13h}{v1}
\fmfforce{11/14w,4/13h}{v2}
\fmfforce{4.6/14w,7.4/13h}{v3}
\fmfforce{9.4/14w,7.4/13h}{v4}
\fmfforce{1/14w,11/13h}{v5}
\fmfforce{13/14w,11/13h}{v6}
\fmf{plain}{i1,o1}
\fmf{plain,left=1}{v1,v2,v1}
\fmf{plain,left=1}{v3,v5,v3}
\fmf{plain,left=1}{v4,v6,v4}
\fmfdot{v1,v2,v3,v4}
\end{fmfgraph}  \end{center}} 
\quad 
  \begin{tabular}{@{}c}
  ${\scs \mbox{\#4.8}}$ \\
  $1/8$\\ 
  ${\scs ( 2, 0, 0 , 2 ; 2 )}$
  \end{tabular}
\parbox{11mm}{\begin{center}
\begin{fmfgraph}(10,15)
\setval
\fmfforce{0w,0.5h}{i1}
\fmfforce{1w,0.5h}{o1}
\fmfforce{1/4w,0.5h}{v1}
\fmfforce{3/4w,0.5h}{v2}
\fmfforce{0.5w,1h}{v4}
\fmfforce{0.5w,0h}{v5}
\fmfforce{0.5w,1/3h}{v6}
\fmfforce{0.5w,2/3h}{v7}
\fmf{plain,left=1}{v5,v6,v5}
\fmf{plain,left=1}{v7,v4,v7}
\fmf{plain,left=1}{v1,v2,v1}
\fmf{plain}{i1,o1}
\fmfdot{v1,v2,v6,v7}
\end{fmfgraph}  \end{center}} 
\hspace*{0.3cm}
  \begin{tabular}{@{}c}
  ${\scs \mbox{\#4.9}}$ \\
  $1/8$\\ 
  ${\scs ( 1, 2, 0 , 1 ; 2 )}$
  \end{tabular}
\parbox{14mm}{\begin{center}
\begin{fmfgraph}(13,13)
\setval
\fmfforce{0.5w,0h}{v1}
\fmfforce{0.5w,8/13h}{v2}
\fmfforce{3/13w,6/13h}{v3}
\fmfforce{10/13w,6/13h}{v4}
\fmfforce{0.5w,1h}{v5}
\fmfforce{0w,6/13h}{i1}
\fmfforce{1w,6/13h}{o1}
\fmf{plain,left=1}{v1,v2,v1}
\fmf{plain,left=1}{v2,v5,v2}
\fmf{plain}{v1,v3}
\fmf{plain}{i1,v3}
\fmf{plain}{o1,v4}
\fmf{plain}{v1,v4}
\fmfdot{v1,v2,v3,v4}
\end{fmfgraph}\end{center}}
\hspace*{3mm}
\\  & \hspace*{6mm}
  \begin{tabular}{@{}c}
  $\mbox{}$ \\
  ${\scs \mbox{\#4.10}}$ \\
  $1/2$\\ 
  ${\scs ( 1, 1, 0 , 1 ; 1 )}$ \\ 
  $\mbox{}$ 
  \end{tabular}
\parbox{16mm}{\begin{center}
\begin{fmfgraph}(15,10)
\setval
\fmfforce{2/15w,0.2h}{i1}
\fmfforce{5/15w,0.2h}{v1}
\fmfforce{12/15w,0.2h}{v2}
\fmfforce{1w,0.2h}{o1}
\fmfforce{8.5/15w,0h}{v3}
\fmfforce{8.5/15w,0.8h}{v5}
\fmfforce{5.2/17w,0.6h}{v4}
\fmfforce{1.7/17w,9.5/10h}{v6}
\fmf{plain,left=1}{v3,v5,v3}
\fmf{plain}{i1,v1}
\fmf{plain}{v2,o1}
\fmf{plain}{v1,v5}
\fmf{plain}{v5,v2}
\fmf{plain,left=1}{v4,v6,v4}
\fmfdot{v1,v2,v5,v4}
\end{fmfgraph}\end{center}}
\hspace*{4mm}
  \begin{tabular}{@{}c}
  ${\scs \mbox{\#4.11}}$ \\
  $1/8$\\ 
  ${\scs ( 1, 1, 0 , 2 ; 2 )}$
  \end{tabular}
\parbox{8mm}{\begin{center}
\begin{fmfgraph}(8,15)
\setval
\fmfforce{2/8w,0h}{i1}
\fmfforce{6/8w,0h}{o1}
\fmfforce{0w,6/15h}{v1}
\fmfforce{1w,6/15h}{v2}
\fmfforce{0.5w,10/15h}{v3}
\fmfforce{0.5w,1h}{v4}
\fmfforce{0.5w,2/15h}{v5}
\fmfforce{0.5w,1h}{v6}
\fmf{plain}{i1,v5}
\fmf{plain}{o1,v5}
\fmf{plain,left=1}{v3,v6,v3}
\fmf{plain,left=1}{v1,v2,v1}
\fmf{plain,left=0.4}{v1,v2,v1}
\fmfdot{v1,v2,v3,v5}
\end{fmfgraph} \end{center}}  
\hspace*{0.5cm}
  \begin{tabular}{@{}c}
  ${\scs \mbox{\#4.12}}$ \\
  $1/12$\\ 
  ${\scs ( 1, 0, 1 , 1 ; 2 )}$
  \end{tabular}
\parbox{10mm}{\begin{center}
\begin{fmfgraph}(12,11)
\setval
\fmfforce{3/12w,4/11h}{v1}
\fmfforce{11/12w,4/11h}{v2}
\fmfforce{4.5/12w,7.5/11h}{v3}
\fmfforce{9.5/12w,7.5/11h}{v4}
\fmfforce{1/12w,11/11h}{v5}
\fmfforce{12.5/12w,7.5/11h}{i1}
\fmfforce{9.5/12w,10.5/11h}{o1}
\fmf{plain,left=0.4}{v1,v2,v1}
\fmf{plain,left=1}{v1,v2,v1}
\fmf{plain,left=1}{v3,v5,v3}
\fmf{plain}{v4,i1}
\fmf{plain}{v4,o1}
\fmfdot{v1,v2,v3,v4}
\end{fmfgraph}  \end{center}} 
\hspace*{6mm}
  \begin{tabular}{@{}c}
  ${\scs \mbox{\#4.13}}$ \\
  $1/8$\\ 
  ${\scs ( 1, 2, 0 , 1 ; 2 )}$ 
  \end{tabular}
\parbox{11mm}{\begin{center}
\begin{fmfgraph}(10,15)
\setval
\fmfforce{0w,1/6h}{i1}
\fmfforce{1w,1/6h}{o1}
\fmfforce{1/4w,1/6h}{v1}
\fmfforce{3/4w,1/6h}{v2}
\fmfforce{1/2w,1/3h}{v3}
\fmfforce{1/2w,2/3h}{v4}
\fmfforce{1/2w,1h}{v5}
\fmf{plain}{i1,o1}
\fmf{plain,left=1}{v1,v2,v1}
\fmf{plain,left=1}{v3,v4,v3}
\fmf{plain,left=1}{v4,v5,v4}
\fmfdot{v1,v2,v3,v4}
\end{fmfgraph} \end{center}} 
\\  &
  \begin{tabular}{@{}c}
  ${\scs \mbox{\#4.14}}$ \\
  $1/16$\\ 
  ${\scs ( 2, 1, 0 , 2 ; 2 )}$
  \end{tabular}
\parbox{15mm}{\begin{center}
\begin{fmfgraph}(15,20)
\setval
\fmfforce{4.5/15w,1/4h}{i1}
\fmfforce{10.5/15w,1/4h}{o1}
\fmfforce{1/2w,1/4h}{v1}
\fmfforce{1/2w,1/2h}{v8}
\fmfforce{1/2w,3/4h}{v2}
\fmfforce{1/2w,1h}{v3}
\fmfforce{0.355662432w,0.6975h}{v4}
\fmfforce{0.64433568w,0.6975h}{v5}
\fmf{plain,left=1}{v8,v2,v8}
\fmf{plain}{i1,v1}
\fmf{plain}{v1,o1}
\fmfi{plain}{reverse fullcircle scaled 0.33w shifted (0.225w,0.765h)}
\fmfi{plain}{reverse fullcircle scaled 0.33w shifted (0.775w,0.765h)}
\fmf{plain,left=1}{v1,v8,v1}
\fmfdot{v1,v4,v5,v8}
\end{fmfgraph} \end{center}} 
\hspace*{3mm}
  \begin{tabular}{@{}c}
  ${\scs \mbox{\#4.15}}$ \\
  $1/8$\\ 
  ${\scs ( 2, 1, 0 , 1 ; 2 )}$
  \end{tabular}
\hs
\parbox{15mm}{\begin{center}
\begin{fmfgraph}(14,11.5)
\setval
\fmfforce{5.75/14w,0h}{i1}
\fmfforce{11.25/14w,0h}{o1}
\fmfforce{8.75/14w,0h}{v1}
\fmfforce{8.75/14w,5/11.5h}{v2}
\fmfforce{3/4w,4.25/11.5h}{v3}
\fmfforce{1w,7.75/11.5h}{v4}
\fmfforce{1/2w,4.25/11.5h}{v5}
\fmfforce{1/4w,7.75/11.5h}{v6}
\fmfforce{0w,11.25/11.5h}{v7}
\fmf{plain,left=1}{v1,v2,v1}
\fmf{plain}{i1,o1}
\fmf{plain,left=1}{v3,v4,v3}
\fmf{plain,left=1}{v5,v6,v5}
\fmf{plain,left=1}{v6,v7,v6}
\fmfdot{v1,v3,v5,v6}
\end{fmfgraph} \end{center}} 
\hspace*{4mm}
  \begin{tabular}{@{}c}
  ${\scs\mbox{\#4.16}}$ \\
  $1/16$ \\ 
  ${\scs ( 3, 0, 0 , 2 ; 2 )}$
  \end{tabular}
\parbox{16mm}{\begin{center}
\begin{fmfgraph}(15,12)
\setval
\fmfforce{5.5/15w,0h}{i1}
\fmfforce{9.5/15w,0h}{o1}
\fmfforce{1/3w,4.5/12h}{v1}
\fmfforce{2/3w,4.5/12h}{v2}
\fmfforce{1/2w,2/12h}{v3}
\fmfforce{1/2w,7/12h}{v4}
\fmfforce{1/2w,1h}{v5}
\fmfforce{0w,4.5/12h}{v7}
\fmfforce{1w,4.5/12h}{v8}
\fmf{plain}{i1,v3}
\fmf{plain}{v3,o1}
\fmf{plain,left=1}{v3,v4,v3}
\fmf{plain,left=1}{v4,v5,v4}
\fmf{plain,left=1}{v1,v7,v1}
\fmf{plain,left=1}{v2,v8,v2}
\fmfdot{v1,v2,v3,v4}
\end{fmfgraph} \end{center}} 
\hspace*{0.4cm}
  \begin{tabular}{@{}c}
  ${\scs \mbox{\#4.17}}$ \\
  $1/16$\\ 
  ${\scs ( 1, 3, 0 , 1 ; 2 )}$
  \end{tabular}
\parbox{7mm}{\begin{center}
\begin{fmfgraph}(6,20)
\setval
\fmfforce{0w,0h}{i1}
\fmfforce{0.5w,0h}{v1}
\fmfforce{1w,0h}{o1}
\fmfforce{0.5w,1/4h}{v2}
\fmfforce{0.5w,1/2h}{v3}
\fmfforce{0.5w,3/4h}{v4}
\fmfforce{0.5w,1h}{v5}
\fmf{plain}{i1,o1}
\fmf{plain,left=1}{v1,v2,v1}
\fmf{plain,left=1}{v2,v3,v2}
\fmf{plain,left=1}{v3,v4,v3}
\fmf{plain,left=1}{v5,v4,v5}
\fmfdot{v1,v2,v3,v4}
\end{fmfgraph} \end{center}}
\hspace*{0.2cm}
\\ &
\\
\hline\hline
\end{tabular}
\end{center}
\caption{\la{tab4} One-particle irreducible iagrams of the self-energy and their 
weights of the $\phi^4$-theory
up to four loops characterized by the vector $(S,D,T,P;N$).
Its components $S,D,T$ specify the number of self-, double,
triple connections, $P$ stands for the number of vertex permutations leaving the diagram unchanged, and $N$ denotes the symmetry
degree.}
\end{table}
\newpage
\begin{table}[t]
\begin{center}
\begin{tabular}{|c|c|}
\hline\hline
\,\,\,$p$\,\,\, &
  \begin{tabular}{@{}c}
  $\mbox{}$ \\*[5mm]
  $\mbox{}$
  \end{tabular}
  \parbox{10mm}{\centerline{
  \begin{fmfgraph*}(7.5,7.5)
  \setval
  \fmfforce{0w,0h}{v1}
  \fmfforce{0w,1h}{v2}
  \fmfforce{1w,1h}{v3}
  \fmfforce{1w,0h}{v4}
  \fmfforce{1/4w,1/4h}{v5}
  \fmfforce{1/4w,3/4h}{v6}
  \fmfforce{3/4w,3/4h}{v7}
  \fmfforce{3/4w,1/4h}{v8}
  \fmfforce{1/2w,1/2h}{v9}
  \fmf{plain}{v1,v5}
  \fmf{plain}{v2,v6}
  \fmf{plain}{v3,v7}
  \fmf{plain}{v4,v8}
  \fmf{double,width=0.2mm,left=1}{v5,v7,v5}
  \fmfv{decor.size=0, label=${\scs p}$, l.dist=0mm, l.angle=90}{v9}
  \end{fmfgraph*} } } 
\\
\hline
$1$ &
\hspace*{2mm}
  \begin{tabular}{@{}c}
  $\mbox{}$\\
  ${\scs \#1.1}$ \\
  $1$\\ 
  ${\scs ( 0, 0, 0 , 1 ; 24 )}$\\
  $\mbox{}$
  \end{tabular}
\parbox{6mm}{\begin{center}
\begin{fmfgraph}(5,5)
\setval
\fmfstraight
\fmfforce{0w,0h}{i1}
\fmfforce{0w,1h}{i2}
\fmfforce{1w,0h}{o1}
\fmfforce{1w,1h}{o2}
\fmfforce{0.5w,0.5h}{v1}
\fmf{plain}{i1,o2}
\fmf{plain}{i2,o1}
\fmfdot{v1}
\end{fmfgraph}
\end{center}}
\\ \hline
$2$ &
\hspace*{2mm}
  \begin{tabular}{@{}c}
  $\mbox{}$\\
  ${\scs \#2.1}$ \\
  $3/2$\\ 
  ${\scs ( 0, 1, 0 , 1 ; 8 )}$\\
  $\mbox{}$
  \end{tabular}
\parbox{10mm}{\begin{center}
\begin{fmfgraph}(9,4)
\setval
\fmfstraight
\fmfforce{0w,0h}{i1}
\fmfforce{0w,1h}{i2}
\fmfforce{1w,0h}{o1}
\fmfforce{1w,1h}{o2}
\fmfforce{2/9w,0.5h}{v1}
\fmfforce{7/9w,0.5h}{v2}
\fmf{plain}{i1,v1}
\fmf{plain}{v2,o1}
\fmf{plain}{i2,v1}
\fmf{plain}{v2,o2}
\fmf{plain,left=1}{v1,v2,v1}
\fmfdot{v1,v2}
\end{fmfgraph} \end{center}}
\quad
\\ \hline
$3$ &
\hspace{-10pt}
\rule[-10pt]{0pt}{26pt}
  \begin{tabular}{@{}c}
  $\mbox{}$ \\
  ${\scs \#3.1}$ \\ 
  $3/4$\\
  ${\scs ( 0, 2, 0 , 1 ; 8 )}$ \\
  $\mbox{}$
  \end{tabular}
\parbox{14mm}{\begin{center}
\begin{fmfgraph}(14,4)
\setval
\fmfstraight
\fmfforce{0w,0h}{i1}
\fmfforce{0w,1h}{i2}
\fmfforce{1w,0h}{o1}
\fmfforce{1w,1h}{o2}
\fmfforce{2/14w,0.5h}{v1}
\fmfforce{1/2w,0.5h}{v2}
\fmfforce{12/14w,0.5h}{v3}
\fmf{plain}{i1,v1}
\fmf{plain}{v3,o1}
\fmf{plain}{i2,v1}
\fmf{plain}{v3,o2}
\fmf{plain,left=1}{v1,v2,v1}
\fmf{plain,left=1}{v2,v3,v2}
\fmfdot{v1,v2,v3}
\end{fmfgraph} \end{center}}
\hspace*{4mm}
  \begin{tabular}{@{}c}
  ${\scs \#3.2}$ \\ 
  $3$\\
  ${\scs ( 0, 1, 0 , 1 ; 4 )}$
  \end{tabular}
\parbox{9mm}{\begin{center}
\begin{fmfgraph}(10,7)
\setval
\fmfstraight
\fmfforce{0w,2.5/7h}{i1}
\fmfforce{1w,2.5/7h}{o1}
\fmfforce{1/4w,2.5/7h}{v1}
\fmfforce{3/4w,2.5/7h}{v2}
\fmfforce{1/2w,5/7h}{v3}
\fmfforce{0.3w,1h}{i2}
\fmfforce{0.7w,1h}{o2}
\fmf{plain}{i1,o1}
\fmf{plain}{i2,v3}
\fmf{plain}{o2,v3}
\fmf{plain,left=1}{v1,v2,v1}
\fmfdot{v1,v2,v3}
\end{fmfgraph} \end{center}}
\hspace*{5mm}
  \begin{tabular}{@{}c}
  ${\scs \#3.3}$ \\
  $3/2$\\
  ${\scs ( 1, 0, 0 , 1 ; 8 )}$
  \end{tabular}
\parbox{10mm}{\begin{center}
\begin{fmfgraph}(9,10)
\setval
\fmfstraight
\fmfforce{0w,0.05h}{i1}
\fmfforce{0w,0.45h}{i2}
\fmfforce{1w,0.05h}{o1}
\fmfforce{1w,0.45h}{o2}
\fmfforce{2/9w,1/4h}{v1}
\fmfforce{7/9w,1/4h}{v2}
\fmfforce{1/2w,1/2h}{v3}
\fmfforce{1/2w,1h}{v4}
\fmf{plain}{i1,v1}
\fmf{plain}{v2,o1}
\fmf{plain}{i2,v1}
\fmf{plain}{v2,o2}
\fmf{plain,left=1}{v1,v2,v1}
\fmf{plain,left=1}{v3,v4,v3}
\fmfdot{v1,v2,v3}
\end{fmfgraph} \end{center}}
\\
\hline
$ $ &
\hspace*{2mm}
  \begin{tabular}{@{}c}
  $\mbox{}$\\
  ${\scs \#4.1}$ \\ 
  $3/8$\\
  ${\scs ( 0, 3, 0 , 1 ; 8 )}$\\
  $\mbox{}$
  \end{tabular}
\parbox{19mm}{\begin{center}
\begin{fmfgraph}(19,4)
\setval
\fmfstraight
\fmfforce{0w,0h}{i1}
\fmfforce{0w,1h}{i2}
\fmfforce{1w,0h}{o1}
\fmfforce{1w,1h}{o2}
\fmfforce{2/19w,0.5h}{v1}
\fmfforce{7/19w,0.5h}{v2}
\fmfforce{12/19w,0.5h}{v3}
\fmfforce{17/19w,0.5h}{v4}
\fmf{plain}{i1,v1}
\fmf{plain}{v4,o1}
\fmf{plain}{i2,v1}
\fmf{plain}{v4,o2}
\fmf{plain,left=1}{v1,v2,v1}
\fmf{plain,left=1}{v2,v3,v2}
\fmf{plain,left=1}{v4,v3,v4}
\fmfdot{v1,v2,v3,v4}
\end{fmfgraph} \end{center}}
\hspace*{3mm}
  \begin{tabular}{@{}c}
  ${\scs \#4.2}$ \\
  $1$\\
  ${\scs ( 0, 0, 0 , 1 ; 24 )}$
  \end{tabular}
\parbox{7mm}{\begin{center}
\begin{fmfgraph}(9.5,9.5)
\setval
\fmfforce{0w,0h}{i1}
\fmfforce{0w,1h}{i2}
\fmfforce{1w,1h}{o1}
\fmfforce{1w,0h}{o2}
\fmfforce{2/9.5w,2/9.5h}{v1}
\fmfforce{2/9.5w,7.5/9.5h}{v2}
\fmfforce{7.5/9.5w,7.5/9.5h}{v3}
\fmfforce{7.5/9.5w,2/9.5h}{v4}
\fmf{plain}{i1,o1}
\fmf{plain}{i2,o2}
\fmf{plain,left=1}{v1,v3,v1}
\fmfdot{v1,v2,v3,v4}
\end{fmfgraph} \end{center}}
\hspace*{6mm}
  \begin{tabular}{@{}c}
  ${\scs \#4.3}$ \\ 
  $3/2$\\ 
  ${\scs ( 0, 2, 0 , 1 ; 4 )}$
  \end{tabular}
\parbox{13mm}{\begin{center}
\begin{fmfgraph}(13,10)
\setval
\fmfforce{0.5w,0h}{v1}
\fmfforce{0.5w,0.8h}{v2}
\fmfforce{3/13w,0.6h}{v3}
\fmfforce{10/13w,0.6h}{v4}
\fmfforce{0w,0.6h}{i1}
\fmfforce{1w,0.6h}{i2}
\fmfforce{4.5/13w,1h}{o1}
\fmfforce{8.5/13w,1h}{o2}
\fmf{plain,left=1}{v1,v2,v1}
\fmf{plain}{v2,o1}
\fmf{plain}{v2,o2}
\fmf{plain}{v1,v3}
\fmf{plain}{i1,v3}
\fmf{plain}{i2,v4}
\fmf{plain}{v1,v4}
\fmfdot{v1,v2,v3,v4}
\end{fmfgraph} \end{center}}
\hspace*{4mm}
  \begin{tabular}{@{}c}
  ${\scs \#4.4}$ \\
  $3/4$\\
  ${\scs ( 0, 1, 0 , 2 ; 8)}$
  \end{tabular}
\parbox{12mm}{\begin{center}
\begin{fmfgraph}(12,8)
\setval
\fmfforce{0w,2/8h}{i1}
\fmfforce{0w,6/8h}{i2}
\fmfforce{1w,2/8h}{o1}
\fmfforce{1w,6/8h}{o2}
\fmfforce{2/12w,1/2h}{v1}
\fmfforce{1/2w,0h}{v2}
\fmfforce{1/2w,1h}{v3}
\fmfforce{10/12w,1/2h}{v4}
\fmf{plain}{i1,v1}
\fmf{plain}{i2,v1}
\fmf{plain}{o1,v4}
\fmf{plain}{o2,v4}
\fmf{plain,left=1}{v1,v4,v1}
\fmf{plain,left=0.4}{v2,v3,v2}
\fmfdot{v1,v2,v3,v4}
\end{fmfgraph} \end{center}} 
\hspace*{2mm}
\\  
$4$ &
  \begin{tabular}{@{}c}
  ${\scs \#4.5}$ \\
  $3/2$\\
  ${\scs ( 0, 2, 0 , 1 ; 4 )}$
  \end{tabular}
\parbox{7mm}{\begin{center}
\begin{fmfgraph}(9,10)
\setval
\fmfforce{0w,0h}{i1}
\fmfforce{0w,1h}{i2}
\fmfforce{1w,0h}{o1}
\fmfforce{1w,1h}{o2}
\fmfforce{2/9w,0.2h}{v1}
\fmfforce{2/9w,0.8h}{v2}
\fmfforce{7/9w,0.8h}{v3}
\fmfforce{7/9w,0.2h}{v4}
\fmf{plain}{i1,v1}
\fmf{plain}{i2,v2}
\fmf{plain}{o1,v4}
\fmf{plain}{o2,v3}
\fmf{plain,left=1}{v1,v3,v1}
\fmf{plain,right=0.3}{v1,v2}
\fmf{plain,right=0.3}{v3,v4}
\fmfdot{v1,v2,v3,v4}
\end{fmfgraph} \end{center}}
\hspace*{3mm}
  \begin{tabular}{@{}c}
  ${\scs \#4.6}$ \\
  $6$\\
  ${\scs ( 0, 1, 0 , 1 ; 2)}$
  \end{tabular}
\parbox{10mm}{\begin{center}
\begin{fmfgraph}(13,9.5)
\setval
\fmfforce{10/13w,4/9.5h}{v1}
\fmfforce{2/13w,4/9.5h}{v2}
\fmfforce{4/13w,7.5/9.5h}{v3}
\fmfforce{4/13w,0.5/9.5h}{v4}
\fmfforce{2/13w,1h}{i1}
\fmfforce{1w,4/9.5h}{i2}
\fmfforce{0w,2/9.5h}{o1}
\fmfforce{0w,6/9.5h}{o2}
\fmf{plain,left=1}{v1,v2,v1}
\fmf{plain}{v2,o1}
\fmf{plain}{v2,o2}
\fmf{plain}{i1,v3}
\fmf{plain}{i2,v1}
\fmf{plain}{v1,v4}
\fmf{plain}{v3,v4}
\fmfdot{v1,v2,v3,v4}
\end{fmfgraph}\end{center}}
\hspace*{4.5mm} 
  \begin{tabular}{@{}c}
  ${\scs \#4.7}$ \\ 
  $3/2$\\
  ${\scs ( 0, 2, 0 , 1 ; 4 )}$
  \end{tabular}
\parbox{8mm}{\begin{center}
\begin{fmfgraph}(10,12)
\setval
\fmfstraight
\fmfforce{0w,2.5/12h}{i1}
\fmfforce{1w,2.5/12h}{o1}
\fmfforce{1/4w,2.5/12h}{v1}
\fmfforce{3/4w,2.5/12h}{v2}
\fmfforce{1/2w,5/12h}{v3}
\fmfforce{1/2w,10/12h}{v4}
\fmfforce{0.3w,1h}{i2}
\fmfforce{0.7w,1h}{o2}
\fmf{plain}{i1,o1}
\fmf{plain}{i2,v4}
\fmf{plain}{o2,v4}
\fmf{plain,left=1}{v1,v2,v1}
\fmf{plain,left=1}{v3,v4,v3}
\fmfdot{v1,v2,v3,v4}
\end{fmfgraph} \end{center}}
\hspace*{4.5mm}
  \begin{tabular}{@{}c}
  ${\scs \#4.8}$ \\ 
  $1/2$\\
  ${\scs ( 0, 0, 1 , 1 ; 8 )}$
  \end{tabular}
\parbox{9mm}{\begin{center}
\begin{fmfgraph}(11,10)
\setval
\fmfforce{1.5/11w,0.4h}{v1}
\fmfforce{9.5/11w,0.4h}{v2}
\fmfforce{3/11w,7.5/10h}{v3}
\fmfforce{8/11w,7.5/10h}{v4}
\fmfforce{0w,7.5/10h}{i1}
\fmfforce{3/11w,10.5/10h}{i2}
\fmfforce{1w,7.5/10h}{o1}
\fmfforce{8/11w,10.5/10h}{o2}
\fmfdot{v1,v2}
\fmf{plain,left=1}{v1,v2,v1}
\fmf{plain,left=0.4}{v1,v2,v1}
\fmf{plain}{i1,v3}
\fmf{plain}{i2,v3}
\fmf{plain}{o1,v4}
\fmf{plain}{o2,v4}
\fmfdot{v1,v2,v3,v4}
\end{fmfgraph} \end{center}}
\hspace*{4.5mm}
  \begin{tabular}{@{}c}
  ${\scs \#4.9}$ \\
  $3$\\
  ${\scs ( 1, 0, 0 , 1 ; 4)}$ 
  \end{tabular}
\parbox{7mm}{\begin{center}
\begin{fmfgraph}(10,12)
\setval
\fmfforce{0w,7.5/12h}{i1}
\fmfforce{1w,7.5/12h}{o1}
\fmfforce{1/4w,7.5/12h}{v1}
\fmfforce{3/4w,7.5/12h}{v2}
\fmfforce{1/2w,10/12h}{v3}
\fmfforce{1/2w,5/12h}{v4}
\fmfforce{1/2w,0h}{v5}
\fmfforce{0.3w,1h}{i2}
\fmfforce{0.7w,1h}{o2}
\fmf{plain}{i1,o1}
\fmf{plain}{i2,v3}
\fmf{plain}{o2,v3}
\fmf{plain,left=1}{v1,v2,v1}
\fmf{plain,left=1}{v4,v5,v4}
\fmfdot{v1,v2,v3,v4}
\end{fmfgraph} \end{center}}
\hspace*{2.5mm}
\\ &
  \begin{tabular}{@{}c}
  $\mbox{}$ \\
  ${\scs \#4.10}$ \\
  $3/2$\\
  ${\scs ( 1, 1, 0 , 1 ; 4)}$ \\
  $\mbox{}$
  \end{tabular}
\parbox{11mm}{\begin{center}
\begin{fmfgraph}(14,10)
\setval
\fmfforce{0w,0.05h}{i1}
\fmfforce{0w,0.45h}{i2}
\fmfforce{1w,0.05h}{o1}
\fmfforce{1w,0.45h}{o2}
\fmfforce{2/14w,1/4h}{v1}
\fmfforce{1/2w,1/4h}{v2}
\fmfforce{12/14w,1/4h}{v3}
\fmfforce{9.5/14w,1/2h}{v4}
\fmfforce{9.5/14w,1h}{v5}
\fmf{plain}{i1,v1}
\fmf{plain}{v3,o1}
\fmf{plain}{i2,v1}
\fmf{plain}{v3,o2}
\fmf{plain,left=1}{v1,v2,v1}
\fmf{plain,left=1}{v2,v3,v2}
\fmf{plain,left=1}{v4,v5,v4}
\fmfdot{v1,v2,v3,v4}
\end{fmfgraph} \end{center}}
\hspace*{4.5mm}
  \begin{tabular}{@{}c}
  ${\scs \#4.11}$ \\
  $3$\\
  ${\scs ( 1, 1, 0 , 1 ; 2)}$
  \end{tabular}
\parbox{10mm}{\begin{center}
\begin{fmfgraph}(13,10)
\setval
\fmfforce{0w,0.4h}{i1}
\fmfforce{1w,0.4h}{o1}
\fmfforce{2.5/13w,0.4h}{v1}
\fmfforce{10.5/13w,0.4h}{v2}
\fmfforce{4/13w,0.7h}{v3}
\fmfforce{9/13w,0.7h}{v4}
\fmfforce{12.5/13w,10.5/10h}{v5}
\fmfforce{1/13w,0.7h}{i2}
\fmfforce{4/13w,1h}{o2}
\fmf{plain}{i1,o1}
\fmf{plain}{i2,v3}
\fmf{plain}{o2,v3}
\fmf{plain,left=1}{v4,v5,v4}
\fmf{plain,left=1}{v1,v2,v1}
\fmfdot{v1,v2,v3,v4}
\end{fmfgraph} \end{center}}
\hspace*{4.5mm} 
  \begin{tabular}{@{}c}
  ${\scs \#4.12}$ \\
  $3/4$\\
  ${\scs ( 2, 0, 0 , 1 ; 8)}$
  \end{tabular}
\parbox{10mm}{\begin{center}
\begin{fmfgraph}(12,10.5)
\setval
\fmfstraight
\fmfforce{0w,2/10.5h}{i1}
\fmfforce{0w,6/10.5h}{i2}
\fmfforce{1w,2/10.5h}{o1}
\fmfforce{1w,6/10.5h}{o2}
\fmfforce{2/12w,4/10.5h}{v1}
\fmfforce{10/12w,4/10.5h}{v2}
\fmfforce{4/12w,7.5/10.5h}{v3}
\fmfforce{8/12w,7.5/10.5h}{v4}
\fmfforce{0.5/12w,11/10.5h}{v5}
\fmfforce{11.5/12w,11/10.5h}{v6}
\fmf{plain}{i1,v1}
\fmf{plain}{i2,v1}
\fmf{plain}{o1,v2}
\fmf{plain}{o2,v2}
\fmf{plain,left=1}{v4,v6,v4}
\fmf{plain,left=1}{v3,v5,v3}
\fmf{plain,left=1}{v1,v2,v1}
\fmfdot{v1,v2,v3,v4}
\end{fmfgraph} \end{center}}
\hspace*{5mm}
  \begin{tabular}{@{}c}
  ${\scs \#4.13}$ \\
  $3/4$\\
  ${\scs ( 1, 1, 0 , 1 ; 8)}$
  \end{tabular}
\parbox{8mm}{\begin{center}
\begin{fmfgraph}(9,15)
\setval
\fmfstraight
\fmfforce{0w,0.5/15h}{i1}
\fmfforce{0w,4.5/15h}{i2}
\fmfforce{1w,0.5/15h}{o1}
\fmfforce{1w,4.5/15h}{o2}
\fmfforce{2/9w,1/6h}{v1}
\fmfforce{7/9w,1/6h}{v2}
\fmfforce{1/2w,1/3h}{v3}
\fmfforce{1/2w,2/3h}{v4}
\fmfforce{1/2w,1h}{v5}
\fmf{plain}{i1,v1}
\fmf{plain}{v2,o1}
\fmf{plain}{i2,v1}
\fmf{plain}{v2,o2}
\fmf{plain,left=1}{v1,v2,v1}
\fmf{plain,left=1}{v3,v4,v3}
\fmf{plain,left=1}{v5,v4,v5}
\fmfdot{v1,v2,v3,v4}
\end{fmfgraph}  \end{center}}
\hspace*{4.5mm}
  \begin{tabular}{@{}c}
  ${\scs \#4.14}$ \\
  $3/8$\\
  ${\scs ( 2, 0, 0 , 2 ; 8)}$
  \end{tabular}
\parbox{7mm}{\begin{center}
\begin{fmfgraph}(9,15)
\setval
\fmfstraight
\fmfforce{0w,5.5/15h}{i1}
\fmfforce{0w,9.5/15h}{i2}
\fmfforce{1w,9.5/15h}{o1}
\fmfforce{1w,5.5/15h}{o2}
\fmfforce{1/2w,0h}{v1}
\fmfforce{1/2w,1/3h}{v2}
\fmfforce{1/2w,2/3h}{v3}
\fmfforce{1/2w,1h}{v4}
\fmfforce{2/9w,1/2h}{v5}
\fmfforce{7/9w,1/2h}{v6}
\fmf{plain}{i1,v5}
\fmf{plain}{v6,o1}
\fmf{plain}{i2,v5}
\fmf{plain}{v6,o2}
\fmf{plain,left=1}{v1,v2,v1}
\fmf{plain,left=1}{v2,v3,v2}
\fmf{plain,left=1}{v3,v4,v3}
\fmfdot{v2,v3,v5,v6}
\end{fmfgraph}  \end{center}}
\hspace*{2.5mm}
\\ \hline\hline
\end{tabular}
\end{center}
\caption{\la{tab5} Diagrams of the one-particle four-point function
and their weights of the $\phi^4$-theory
up to four loops characterized by the vector $(S,D,T,P;N$).
Its components $S,D,T$ specify the number of self-, double,
triple connections, $P$ stands for the number of vertex permutations leaving the diagram unchanged, and $N$ denotes the symmetry
degree.}
\end{table}
\end{fmffile}


\begin{thebibliography}{199}
%
\bibitem{Drell} 
J.D. Bjorken and S.D. Drell, 
Vol. I {\it Relativistic Quantum Mechanics}, 
Vol. II {\it Relativistic Quantum Fields} 
(McGraw-Hill, New York, 1965).
%
\bibitem{Amit} 
D.J. Amit, 
{\it Field Theory, the Renormalization Group and Critical Phenomena} 
(McGraw-Hill, New York, 1978).
%
\bibitem{Bellac} 
M. Le Bellac, 
{\it Quantum and Statistical Field Theory}
(Oxford Science Publications, Oxford, 1991).
%
\bibitem{Zuber} 
C. Itzykson and J.-B. Zuber, 
{\it Quantum Field Theory}
(McGraw-Hill, New York, 1985).
%
\bibitem{Zinn} 
J. Zinn-Justin, 
{\it Quantum Field Theory and Critical Phenomena}, Third Edition
(Oxford University Press, Oxford, 1996).
%
\bibitem{Peskin} 
M.E. Peskin and D.V. Schroeder, 
{\it Introduction to Quantum Field Theory} 
(Addison-Wesley, Reading, 1995). 
%
\bibitem{Baym}
L.P. Kadanoff and G. Baym, 
{\it Quantum Statistical Mechanics} 
(Benjamin, Menlo Park, 1962).
%
\bibitem{SDQED1}
A. Pelster, H. Kleinert, and M. Bachmann,
Ann. of Phys. (N.Y.)  {\bf 297}, 363 (2002).
%
\bibitem{QED}
M. Bachmann, H. Kleinert, and A. Pelster, 
Phys. Rev. {\bf D 61}, 085017 (2000).
%
\bibitem{SYM}
H. Kleinert, A. Pelster, B. Kastening, and M. Bachmann,
Phys. Rev. {\bf E 62}, 1537 (2000).
%
\bibitem{ASYM1}
B. Kastening, 
Phys. Rev. {\bf E 61}, 3501 (2000).
%
\bibitem{ASYM2}
A. Pelster and H. Kleinert, 
Physica {\bf A} (in press); e-print: {\tt hep-th/0006153}.
%
\bibitem{GINZ}
H. Kleinert, A. Pelster, and B. Van den Bossche, 
Physica {\bf A 312}, 141 (2002).
%
\bibitem{Devreese}
A. Pelster and K. Glaum, 
Phys. Stat. Sol. {\bf B} (in press), eprint: {\tt cond-mat/0211361}.
%
\bibitem{FEST}
A. Pelster and K. Glaum, {\it Recursive Graphical Construction of
Tadpole-Free Feynman Diagrams and Their Weights in $\phi^4$-Theory};
in W. Janke, A. Pelster, H.-J. Schmidt, and M. Bachmann (Editors):
{\it Fluctuating Paths and Fields -- Dedicated to Hagen Kleinert
on the Occasion of His 60th Birthday} (World Scientific, Singapore, 2001), p. 269;
eprint: {\tt hep-th/0105193}.
%
\bibitem{Streater} 
R.F. Streater and A.S. Wightman, 
{\it PCT, Spin and Statistics, and All That} 
(W.A. Benjamin, Reading, Massachusetts, 1964).
%
\bibitem{Schwinger} 
J. Schwinger, {\it Particles, Sources, and Fields},
Vols. I and II (Addison-Wesley, Reading, 1973).
%
\bibitem{Verena}
H. Kleinert and V. Schulte-Frohlinde, 
{\it Critical Properties of $\phi^4$-Theories} 
(World Scientific, Singapore, 2001).
%
\bibitem{Glaum}
K. Glaum, 
MS Thesis (in German), FU-Berlin (2001).
%
\bibitem{Neu}
J. Neu, 
MS Thesis (in German), FU-Berlin (1990).
%
\bibitem{new}
A. Pelster, K. Glaum, and H. Kleinert, 
to be published.
%
\bibitem{Kleinert4} 
H. Kleinert, 
{\it Gauge Fields in Condensed Matter},
Vol. I, {\it Superflow and Vortex Lines} 
(World Scientific, Singapore, 1989).
%
\bibitem{Nickel1}
B.G.~Nickel, D.I.~Meiron, and G.B. Baker Jr., 
University of Guelph, preprint (1977), 
{\tt http://www.physik.fu-ber\-lin.\-de/\~{}klei\-nert/klei\-ner\_reb8/pro\-grams\-/\-pro\-grams.\-html}.
%
\bibitem{Sokolov}
S.A. Antonenko and A.I. Sokolov,
Phys. Rev. {\bf E 51}, 1894 (1995).
%
\bibitem{Nickel2}
D.B. Murray and B.G.~Nickel, 
University of Guelph, preprint (1998).
%
\bibitem{FIVE} 
H. Kleinert, J. Neu, V. Schulte-Frohlinde, K.G. Chetyrkin, and S.A. Larin, 
Phys. Lett. {\bf B 272}, 39 (1991); {\bf 319}, 545 (E) (1993).
%
\bibitem{Kleinert1} 
H. Kleinert, 
Fortschr. Phys. {\bf 30}, 187 (1982).
%
\bibitem{Kleinert2} 
H. Kleinert, 
Fortschr. Phys. {\bf 30}, 351 (1982).
%
\bibitem{Vasiliev}
A.N. Vasiliev, 
{\it Functional Methods in Quantum Field Theory and Statistical Physics}
(Gordon and Breach Science Publishers, New York, 1998); 
translation from the Russian edition
(St. Petersburg University Press, St. Petersburg, 1976).
%
\bibitem{FeynArts1} 
J. K\"ulbeck, M. B\"ohm, and A. Denner, 
Comp. Phys. Comm. {\bf 60}, 165
(1991).
%
\bibitem{FeynArts2} 
T. Hahn, 
Comp. Phys. Comm. {\bf 140}, 418 (2001).
%
\bibitem{FeynArts3} 
{\tt http://www.feynarts.de}. 
%
\bibitem{QGRAF1}
P. Nogueira, 
J. Comput. Phys. {\bf 105}, 279 (1993).
%
\bibitem{QGRAF2}
{\tt ftp://gtae2.ist.utl.pt/pub/qgraf}. 
%
\bibitem{Heap}
B.R. Heap, 
J. Math. Phys. {\bf 7}, 1582 (1966).
%
\bibitem{Nagle}
J.F. Nagle, 
J. Math. Phys. {\bf 7}, 1588 (1966).
%
\bibitem{VERSCHELDE}
S. Schelstraete and H. Verschelde, 
Z. Phys. {\bf C 67}, 343 (1995). 
%
\bibitem{Schroeder}
K. Kajantie, M. Laine, and Y. Schr\"oder, 
Phys. Rev. {\bf D 65}, 045008 (2002).
%
\bibitem{Bray}
A.J. Bray, 
Phys. Rev. Lett. {\bf 32}, 1413 (1974).
%
\bibitem{Radzikovsky}
L. Radzikovsky, 
Europhys. Lett. {\bf 29}, 227 (1995).
%
\end{thebibliography}
\end{document}